\documentclass[11pt]{article}

\usepackage[usenames,dvipsnames,table]{xcolor}
\usepackage[utf8]{inputenc}

\usepackage{jheppub} 
\usepackage{graphicx}
\usepackage{amsmath, amssymb}
\usepackage{youngtab}
\usepackage{changepage}
\usepackage[export]{adjustbox} 
\usepackage{pdflscape} 

\newcommand{\be}{\begin{eqnarray}}
\newcommand{\ee}{\end{eqnarray}}
\newcommand{\bea}{\begin{eqnarray}}
\newcommand{\eea}{\end{eqnarray}}

\newcommand{\nn}{\nonumber}
\newcommand{\bn}{\begin{enumerate}}
\newcommand{\en}{\end{enumerate}}

\def\Tr{\mathop{\text{Tr}}\nolimits}

\def\e{\mathrm{e}}

\def\s{\sigma}

\def\half{\frac{1}{2}}

\newcommand{\ud}{\,\mathrm{d}} 
\newcommand{\udl}[1]{\mathrm{d} #1 \,}

\newcommand{\sbfunc}[1]{s_b\left( #1\right)}

\newcommand{\Gpq}[1]{\Gamma_e\left( #1\right)}

\def\gb{\beta}

\def\Gc{\Gamma}
\def\Gd{\Delta}
\def\gd{\delta}

\def\gs{\sigma}

\def\gr{\rho}

\preprint{\begin{flushright} USTC-ICTS/PCFT-23-08 \\ CTPU-PTC-22-28 \end{flushright}}

\title{\boldmath The $SL(2,\mathbb{Z})$ dualization algorithm at work}

\author[a,b]{Riccardo Comi,}
\author[c,d,e,f]{Chiung Hwang,}
\author[a,b,g]{Fabio Marino,}
\author[a,b]{Sara Pasquetti,}
\author[h]{Matteo Sacchi}

\affiliation[a]{Dipartimento di Fisica, Università di Milano-Bicocca,
Piazza della Scienza 3, I-20126 Milano, Italy}
\affiliation[b]{INFN, sezione di Milano-Bicocca, Piazza della Scienza 3, I-20126 Milano, Italy}
\affiliation[c]{Interdisciplinary Center for Theoretical Study, University of Science and Technology of China, Hefei, Anhui 230026, China}
\affiliation[d]{Peng Huanwu Center for Fundamental Theory, Hefei, Anhui 230026, China}
\affiliation[e]{Center for Theoretical Physics of the Universe, Institute for Basic Science (IBS), Daejeon 34126, Korea}
\affiliation[f]{Department of Applied Mathematics and Theoretical Physics, University of Cambridge, Cambridge CB3 0WA, United Kingdom}
\affiliation[g]{
Department of Mathematics, University of Surrey, Guildford, GU2 7XH, UK}
\affiliation[h]{Mathematical Institute, University of Oxford, Woodstock Road, Oxford, OX2 6GG, United Kingdom}

\emailAdd{r.comi2@campus.unimib.it }
\emailAdd{chiung@ustc.edu.cn}
\emailAdd{f.marino25@campus.unimib.it}
\emailAdd{sara.pasquetti@gmail.com} 
\emailAdd{matteo.sacchi@maths.ox.ac.uk}

\abstract{Recently an algorithm to dualize a theory into its mirror dual has been proposed, both for $3d$ $\mathcal{N}=4$ linear quivers and for their $4d$ $\mathcal{N}=1$ uplift. This mimics the manipulations done at the level of the Type IIB brane setup that engineers the $3d$ theories, where mirror symmetry is realized as $S$-duality, but it is enirely field-theoretic and based on the application of genuine infra-red dualities that implement the local action of $S$-duality on the quiver. In this paper, we generalize the algorithm to the full duality group, which is $SL(2,\mathbb{Z})$ in $3d$ and $PSL(2,\mathbb{Z})$ in $4d$. This also produces dualities for $3d$ $\mathcal{N}=3$ theories with Chern--Simons couplings, some of which have enhanced $\mathcal{N}=4$ supersymmetry, and their new $4d$ $\mathcal{N}=1$ counterpart. In addition, we propose three ways to study the RG flows triggered by possible VEVs appearing at the last step of the algorithm, one of which uses a new duality that implements the Hanany--Witten move in field theory.
\\ 

}

\begin{document} 

\maketitle
\flushbottom

\section{Introduction}

Infra-red (IR) dualities among supersymmetric gauge theories are interesting phenomena that have been extensively studied over the years. These correspond to the situation in which distinct microscopic theories flow to the same fixed point in the IR. The first example due to Seiberg \cite{Seiberg:1994pq} concerns four-dimensional SQCDs with minimal supersymmetry relating those with different gauge groups. After the discovery of the Seiberg duality, many others have been found not only in four dimensions but also in other dimensions. For instance in $3d$ we have analogues of the Seiberg duality (so-called \emph{Seiberg-like dualities}) as well as \emph{mirror symmetry} \cite{Intriligator:1996ex}, which is another class of three-dimensional dualities looking completely different from the former. The $3d$ mirror symmetry can be neatly understood in terms of type IIB brane set-ups \cite{Hanany:1996ie} as the action of $S$-duality, which makes it possible to extend it to a larger class of dualities reflecting more general $SL(2,\mathbb{Z})$ transformations.

Given this proliferation of IR dualities, there are various fundamental questions that one can ask purely from the field theory perspective. For example, among other things, one may ask if we can derive all $3d$ and lower dimensional IR dualities from $4d$ dualities.\footnote{Four dimensions are in some sense the critical number of dimensions for IR dualities because $d > 4$ dimensional gauge theories are free in the IR. Instead, one may consider UV dualities in higher dimensions.} The program of obtaining $3d$ dualities from $4d$ dualities was initiated in \cite{Aharony:2013dha} in the more general context of studying dualities across dimensions (see for example \cite{Aganagic:2001uw,Aharony:2016jki,Aharony:2017adm,Gadde:2015wta,Dedushenko:2017osi,Sacchi:2020pet} for the derivation of $2d$ dualities from higher dimensions) and it was continued in many subsequent works (see \cite{Aharony:2013kma,Csaki:2014cwa,Nii:2014jsa,Amariti:2015vwa,Amariti:2016kat,Benini:2017dud,Hwang:2018uyj,Benvenuti:2018bav,Amariti:2018wht,Nii:2018uck,Pasquetti:2019hxf,Bottini:2021vms,Amariti:2022iaz} for a partial list of references). For a long time, the answer to this question was regarded as negative, especially because of $3d$ mirror symmetry, considered an  inherently three-dimensional phenomenon. However, recently,  $4d$ avatars of  $3d$ mirror dualities
have been discovered \cite{Hwang:2020wpd}. These new $4d$ dualities reduce to $3d$ mirror dualities after circle reduction and suitable deformations,  strengthening the expectation that indeed the physics of $3d$ dualities can be derived
from $4d$.

Another intriguing  question  is whether there is  a fundamental set of dualities in terms of which other dualities can be derived. This second question was addressed in various dimensions by using the notion of \emph{sequential deconfinement} \cite{Berkooz:1995km,Pouliot:1995me,Luty:1996cg,Garcia-Etxebarria:2012ypj,Garcia-Etxebarria:2013tba,Nii:2016jzi,Pasquetti:2019tix,Pasquetti:2019uop,Benvenuti:2020gvy,Etxebarria:2021lmq,Benvenuti:2021nwt,Bottini:2022vpy,Bajeot:2022kwt,Bajeot:2022lah}, which can be used to derive non-trivial dualities for theories with matter fields in tensor representations by only assuming a small set of fundamental dualities. This has a nice counterpart in Mathematics, where various integral identities that can be interpreted as matchings of partition functions of dual theories were derived with a similar strategy \cite{2003math......9252R,spiridonov2004theta,Spiridonov:2009za,Spiridonov:2011hf}.

Furthermore, interestingly, it has recently been shown in \cite{Hwang:2021ulb} (see also \cite{Bottini:2021vms}) that mirror dualities, both in $3d$ and in $4d$, can be derived by an algorithmic procedure based on the iteration of a single fundamental duality, the Aharony duality \cite{Aharony:1997gp} in $3d$ and the Intriligator--Pouliot duality \cite{Intriligator:1995ne} in $4d$, revealing a hidden relation between mirror dualities and Seiberg-like dualities. While the result of \cite{Hwang:2021ulb} only discussed mirror dualities, which as we mentioned are related to the $S$-duality of Type IIB string theory, in this paper we extend the algorithm to the full set of dualities that are inherited from the entire $SL(2,\mathbb{Z})$ duality group of Type IIB string theory.

These dualities typically involve more general $(p,q)$ 5-branes rather than just NS5- and D5-branes. Consequently, the $3d$ theories that are engineered with these brane systems will have Chern--Simons couplings, which a priori preserve only $\mathcal{N}=3$ supersymmetry. However with specific matter contents and superpotentials the amount of supersymmetry can be enhanced to $\mathcal{N}=4$ \cite{Gaiotto:2008sd,Hosomichi:2008jd,Imamura:2008nn,Assel:2022row} and up to $\mathcal{N}=8$ \cite{Aharony:2008ug,Aharony:2008gk}.\footnote{Such enhancements of supersymmetry can be easily detected using the superconformal index, see for example \cite{Bashkirov:2011fr,Cheon:2012be,Evtikhiev:2017heo,Gang:2018huc,Garozzo:2019ejm,Beratto:2020qyk,Beratto:2021xmn,Gang:2021hrd}.} For example, as shown in \cite{Jafferis:2008em} the more general $SL(2,\mathbb{Z})$ dualities can map Yang--Mills fixed points to Chern--Simons theories coupled to matter. In the former $\mathcal{N}=4$ supersymmetry is manifest, meaning that this should be enhanced from the apparent $\mathcal{N}=3$ in the latter.

Our first main result is to uplift these more general dualities for linear quiver theories to $4d$, where we will see that the actual duality group is $PSL(2,\mathbb{Z})$ rather than $SL(2,\mathbb{Z})$ as in the $3d$ case. In this paper, the generators of the $PSL(2,\mathbb Z)$ in $4d$ will be denoted by $\mathsf S$ and $\mathsf T$, while those of the $SL(2,\mathbb Z)$ in $3d$ will be denoted by $\mathcal S$ and $\mathcal T$. These generators satisfy slightly different relations: $\mathsf S^2 = I$ and $(\mathsf{ST})^3 = I$ in $PSL(2,\mathbb{Z})$, while $\mathcal S^2 = -I$ and $(\mathcal{ST})^3 = I$ in $SL(2,\mathbb{Z})$, which basically distinguish these two groups. The second result is to extend the analysis of \cite{Hwang:2021ulb} showing that also the entire $SL(2,\mathbb{Z})$ (in $3d$) and $PSL(2,\mathbb{Z})$ (in $4d$) actions can be defined in field theory as local operations on the quiver theory. The result will lead to an algorithm to derive systematically all of these dualities from a set of fundamental duality moves, which generalizes the idea of \cite{Kapustin:1999ha} for the abelian mirror symmetry to the non-abelian case and beyond mirror dualities.

\begin{figure}[!ht]
\includegraphics[width=1\textwidth,center]{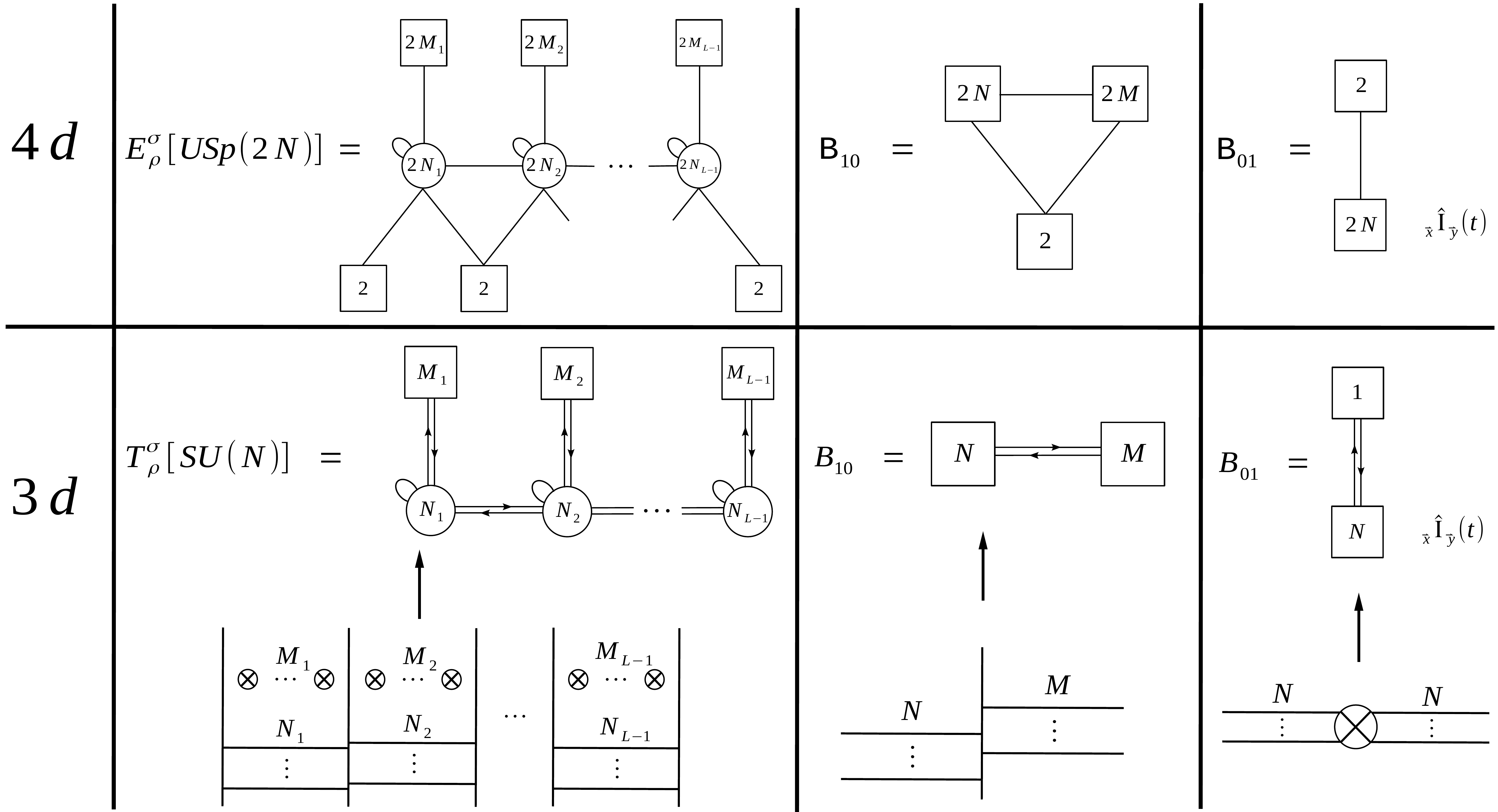}
\caption{The $E_\rho^\sigma[USp(2N)]$ and the $T_\rho^\sigma[U(N)]$ theories with the brane realization of the latter are given in the first column. The partitions $\rho = \left[N^{l_N}, \dots, 1^{l_1}\right]$ and $\sigma = \left[N^{k_N}, \dots, 1^{k_1}\right]$ with $l_n$, $k_m$ satisfying the conditions $\sum_{n = 1}^N n \times l_n = \sum_{m = 1}^N m \times k_m = N$, $ L = l_1+ \dots +l_N $ and $K = k_1+ \dots +k_N$ encode the gauge and flavor ranks as  $ M_{L-i} = k_i$ and $N_{L-i} = \sum_{j = i+1}^L \rho_j-\sum_{j = i+1}^N (j-i) k_j$. In $3d$ all nodes are of unitary type, and each line is an $\mathcal{N}=2$ chiral, with arcs corresponding to adjoint representations and straight lines to bifundamental representations related by the standard $\mathcal{N}=4$ superpotential. In $4d$ all nodes are of sympletic type, and each arc is now an antisymmetric chiral, which couples to the bifundamental chirals on its left and right, while the fields in the ``saw'' are in a cubic superpotential. There are also singlets which we omit in the drawing, see \cite{Hwang:2020wpd} for the details. In the second and third columns the $3d$ and $4d$ QFT blocks associated to NS5 and D5-branes are shown.}
 \label{intro}
\end{figure}

Our results build on some recent ones of \cite{Hwang:2020wpd}, where a family of  $4d$ $\mathcal{N}=1$ theories labelled by partitions $\sigma$ and $\rho$ of  $N$,  the $E_\rho^\sigma[USp(2N)]$ linear quiver theories sketched in Figure \ref{intro}, have been introduced. These theories, upon compactification to $3d$  and suitable real mass and Coulomb branch VEV deformations, reduce to the $3d$ $\mathcal{N}=4$  $T_\rho^\sigma[U(N)]$ linear quiver theories \cite{Gaiotto:2008ak}, which we also depict in Figure \ref{intro}. The  $E_\rho^\sigma[USp(2N)]$ quiver theories were shown to enjoy, like their $3d$ counterpart, a mirror duality relating pairs of theories with partitions $\rho$ and $\sigma$ swapped.

The $3d$ $T_\rho^\sigma[U(N)]$ linear quiver can be realized on an Hanany--Witten  brane set-up \cite{Hanany:1996ie} with $N$ D3-branes suspended between $K$ D5-branes and $L$ NS5-branes, where $K$ and $L$ are the lengths of the partitions $\gs$ and $\gr$ respectively. The entries of the partitions correspond to the net number of D3-branes ending on the D5 and the NS5-branes, respectively, going from the interior to the exterior of the configuration. On the other hand, no brane construction is known for its $4d$ uplift,  the $E_\rho^\sigma[USp(2N)]$ quiver.


The idea of the local dualization approach arises from such brane realization of the $3d$ quiver theories, where the  dualities follow from the $SL(2,\mathbb{Z})$ action on the brane set-up.  As suggested in \cite{Gaiotto:2008ak}, such $SL(2,\mathbb Z)$ actions can be implemented on each 5-brane separately by dualizing that brane only and creating $SL(2,\mathbb{Z})$ duality domain walls on its right and on its left. A QFT analogue of such a local action corresponding to $\mathcal{S} \in SL(2,\mathbb Z)$ was recently formulated in \cite{Hwang:2021ulb}, and we will extend this to the full $SL(2,\mathbb Z)$ duality group in this paper.\footnote{See \cite{Gulotta:2011si,Assel:2014awa} for previous work at the level of the three-sphere partition function.}
 
In order to state the algorithm, first of all, we need to define the notion of QFT blocks to be  {\it associated} to each type of 5-brane.  As sketched in Figure \ref{intro} and discussed in detail in the next section, we can associate QFT blocks
$\mathsf{B}_{10},\mathsf{B}_{01},\mathsf{B}_{11}$  to NS5 (or (1,0)), D5 (or (0,1)) and (1,1)-branes, where the first two were already discussed in \cite{Hwang:2021ulb} while the last one is newly discussed in this paper. We stress the fact that in $4d$ there is no brane realization, so  our $4d$ QFT blocks are defined as the {\it uplift} of the $3d$
QFT blocks $\mathcal{B}_{10},\mathcal{B}_{01},\mathcal{B}_{11}$, in the sense that the former reduces to the latter in the suitable $3d$ limit that we previously mentioned. Notice that the low energy theories of brane set-ups containing $(1,k)$-branes with $k > 1$ are captured by $3d$ theories with Chern--Simons couplings, whose $4d$ uplift will be introduced by generalizing the results of \cite{Hwang:2020wpd}.\footnote{As pointed out already in \cite{Assel:2014awa}, the field theory associated to a brane set-up containing more general $(p,q)$-branes with $p>1$ are non-Lagrangian.}

We then define the QFT realization of the duality walls. In $3d$ each $SL(2,\mathbb{Z})$  element, corresponding to  a duality wall for the $4d$ $\mathcal{N}=4$ SYM, is associated to a certain $3d$ $\mathcal{N}=4$ quiver theory. For example, the $\mathcal{S}$ generator of $SL(2,\mathbb{Z})$ is associated with the $T[SU(N)]$ theory of Gaiotto and Witten \cite{Gaiotto:2008ak}, which is the case with trivial partitions $\rho=\sigma=[1^N]$ of the $T_\rho^\sigma[SU(N)]$ family. In analogy with the three-dimensional  case, we call  ``walls'' also in $4d$  the QFT objects associated to the $PSL(2,\mathbb Z)$  elements, although we do not have their realization as walls in a $5d$ theory. For example, in \cite{Bottini:2021vms}, the $\mathsf{S}$  generator of $PSL(2,\mathbb{Z})$  was associated to  the $FE[USp(2N)]$ quiver theory of \cite{Pasquetti:2019hxf}.\footnote{The $FE[USp(2N)]$ theory was first introduced in \cite{Pasquetti:2019hxf} in the context of the compactifications of the $6d$ $\mathcal{N}=(1,0)$ rank $N$ E-string theory to $4d$ $\mathcal{N}=1$ on Riemann surfaces with flux, and it is also related to various theories enjoying non-trivial symmetry enhancements, see for example \cite{Garozzo:2020pmz,Hwang:2020ddr,Hwang:2021xyw}. This origin of the theory suggests that it should be possible to interpret it as a duality domain wall in a $5d$ $\mathcal{N}=1$ gauge theory similar to those studied in \cite{Gaiotto:2015una,Kim:2017toz,Kim:2018bpg,Kim:2018lfo,Razamat:2022gpm,Sabag:2022hyw}.} This theory is related to the $E[USp(2N)]$ theory by the addition of some singlet flipping fields. Here we will introduce the QFT objects associated to the  $\mathsf{T}$-wall and to the $\mathsf{T}^T$-wall as well.

Finally, we define the action of the duality walls on the QFT blocks. In \cite{Hwang:2021ulb} two basic duality moves were defined. The first one is associated to the action of the $\mathsf{S}$-wall on a (1,0) block, which turns it into a (0,1) block. The second one is associated to  the action of the $\mathsf{S}$-wall on a (0,1) block, which turns it into a (1,0) block.  Crucially  in \cite{Bottini:2021vms,Hwang:2021ulb} it was shown that these basic moves are genuine IR dualities which can be derived by iterative applications of Seiberg-like dualities (Intriligator--Pouliot \cite{Intriligator:1995ne} in $4d$ and Aharony \cite{Aharony:1997gp} in $3d$). We will complete this list by defining the action of the $\mathsf{S}$-wall and of the $\mathsf{T}$-wall on the (1,0), (0,1) and (1,1) blocks. Importantly, these new duality moves will also be consequences of Seiberg-like dualities.

Having defined the QFT blocks, the $(P)SL(2,\mathbb Z)$  generators and the basic duality moves, we can implement the dualization algorithm generating the dual theory of any linear quiver theory that is associated to a given $(P)SL(2,\mathbb{Z})$ generator.\footnote{We remark that starting from these dualities for linear quivers one can algorithmically generate more dualities as done for example in \cite{Dey:2020hfe,Dey:2021rxw}.} The algorithm consists  of the following steps:
\begin{enumerate}
\item Chop a $4d$ or $3d$ quiver into QFT blocks by ungauging every gauge node.
\item Dualize each QFT block  using the basic duality moves corresponding to the chosen $(P)SL(2,\mathbb{Z})$ generator.
\item Glue back the dualized blocks by restoring the original gauge nodes.
\item Follow the RG flow triggered by VEVs that can be generated in the previous step.
\end{enumerate}
We will illustrate these steps in the case of $4d$ SQCD. In particular, we will show that the last step can be implemented in three different ways, one of which consists of another genuine IR duality that is the QFT analogue of the Hanany--Witten move swapping different types of 5-branes \cite{Hanany:1996ie}.

Lastly, we would like to emphasise that, as mentioned above, all the basic moves can be derived by Seiberg-like dualities. Therefore, our algorithm demonstrates that  the full $(P)SL(2,\mathbb Z)$ group of dualities can be derived by Seiberg-like dualities, either Intriligator--Pouliot in $4d$ or Aharony in $3d$.

The rest of the paper is organized as follows. In Section \ref{sec:QFTblocks} we discuss the basic ingredients needed in the dualization algorithm in $4d$, namely the $PSL(2,\mathbb{Z})$ walls, the QFT blocks and the duality moves. In Section \ref{sec:algorithm} we present an example of application of the dualization algorithm in the case of the SQCD. Here we show three equivalent ways to study the RG flow triggered by the VEVs for some operators that can appear at the last step of the algorithm, which consist respectively in studying explicitly the Higgsing with the index, applying sequentially the Intriligator--Pouliot duality or applying iteratively a new duality move that mimics the Hanany--Witten move in the brane setup. In Section \ref{sec:3d} we state the basic ingredient for applying the dualization algorithm in $3d$, which are obtained as a limit of the $4d$ ones. In Section \ref{sec:dualityweb} we discuss a web of dualities for the SQCD that is obtained with $PSL(2,\mathbb{Z})$ dualizations in $4d$ and $SL(2,\mathbb{Z})$ dualizations in $3d$. Finally, in various appendices we summarize some definitions and give derivations for some results used in the main text.

In this paper we will consider only  {\it good} theories in the Gaiotto--Witten sense \cite{Gaiotto:2008ak}, which for the SQCD case implies restricting to $N_f\geq 2N_c$. {\it Bad} theories, containing operators falling below the unitarity bound,  will be studied in \cite{BADpaper} using the dualization algorithm.
Various properties of bad theories have been already studied in \cite{Nanopoulos:2010bv,Kim:2012uz,Yaakov:2013fza,Bashkirov:2013dda,Hwang:2015wna,Hwang:2017kmk,Assel:2017jgo,Dey:2017fqs,Assel:2018exy}. Moreover, it has been observed for example in \cite{Razamat:2014pta,Closset:2020afy,Carta:2022spy,Kang:2022zsl,Akhond:2022jts} that they can arise from the compactification of higher dimensional theories, which makes their study important in the context of understanding the Higgs branch of theories with eight supercharges.

\section{$PSL(2,\mathbb{Z})$ walls, QFT blocks and duality moves}
\label{sec:QFTblocks}

In  this section we introduce several ingredients  involved in the dualization algorithm: the duality-walls, the QFT blocks, and the basic duality moves.

\subsection{$PSL(2,\mathbb{Z})$ operators}

We first introduce the field theory objects, the duality-walls, associated to the generators $\mathsf{S}$ and $\mathsf{T}$ of $PSL(2,\mathbb{Z})$. We will also introduce the Identity-wall and the wall associated to the $\mathsf{T}^T$ generator as it produces interesting dualities although it is not an independent element of $PSL(2,\mathbb Z)$.
As we will see in Section \ref{sec:3d}, the $SL(2,\mathbb{Z})$ group structure in $3d$ can be derived from the $PSL(2,\mathbb Z)$ structure in $4d$ by taking suitable circle reduction and real mass/Coluomb branch (CB) VEV deformations.

\bigskip

\begin{figure}[!ht]
	\centering
	\includegraphics[scale=0.75]{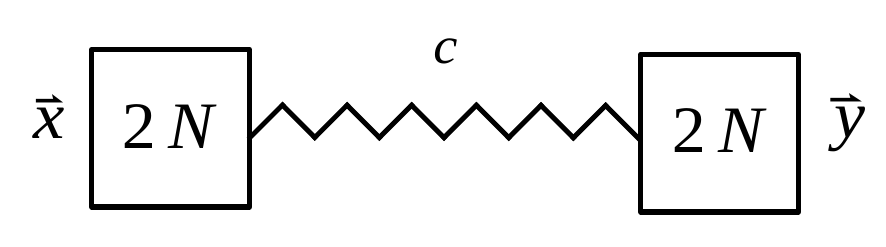}
	\caption{Schematic representation of the $4d$ $\mathsf{S}$-wall.  Here $\vec x$ are the $USp(2N)_x$ fugacities, while $\vec y$ the $USp(2N)_y$ fugacities. We also specify the fugacity $c$ but not $t$.
	}
	\label{Ssch}
\end{figure}

\noindent\textbf{The $\mathsf{S}$-wall.} In $3d$, the  $\mathcal{S}$ generator of $SL(2,\mathbb{Z})$ is associated with the $T[SU(N)]$ theory of Gaiotto and Witten \cite{Gaiotto:2008ak}. Indeed, the way this theory was originally constructed is as a duality domain wall between two copies of the $4d$ $\mathcal{N}=4$ $SU(N)$ SYM theory at values of the coupling $\tau$ and $-\frac{1}{\tau}$, respectively. As it was shown in \cite{Pasquetti:2019hxf}, the $T[SU(N)]$ theory can be obtained from a dimensional reduction of the $4d$ $\mathcal{N}=1$ theory called $E[USp(2N)]$ supplemented by suitable real mass deformations and CB VEVs.
\begin{figure}[t]
	\centering
  	\includegraphics[scale=0.55,center]{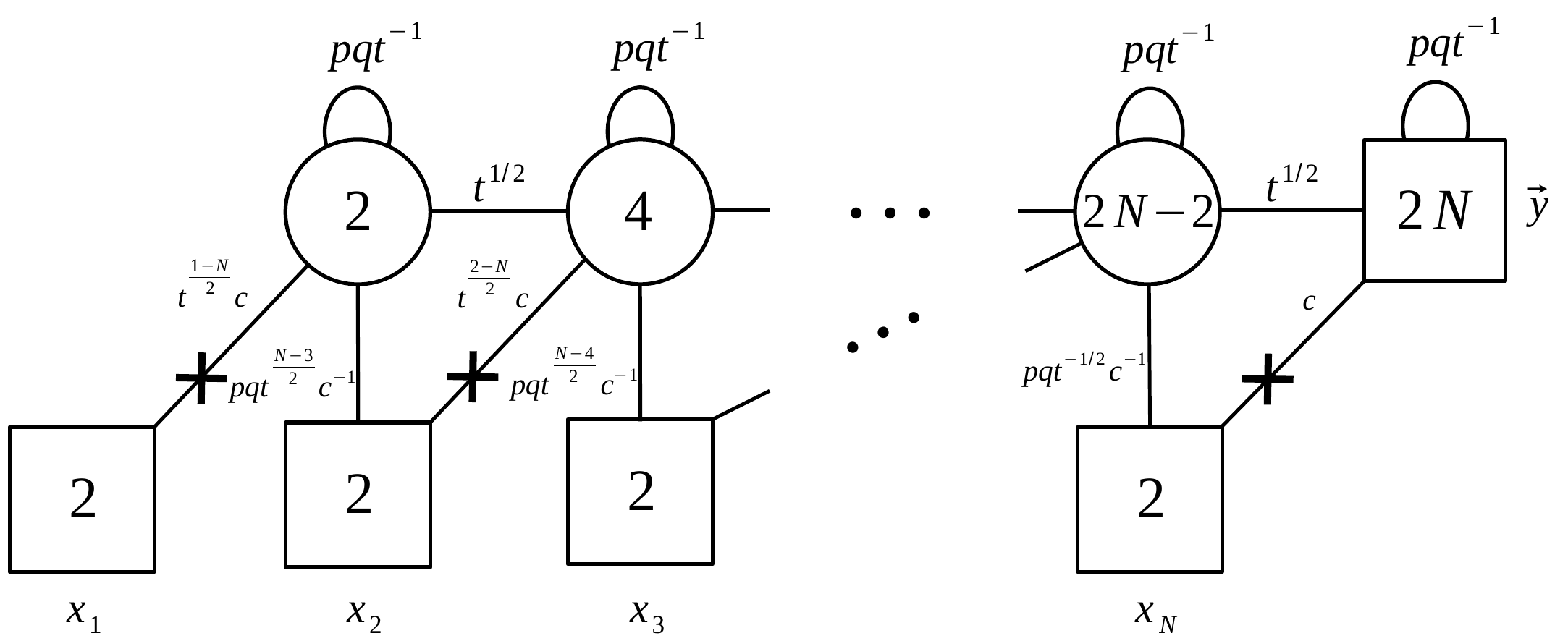} 
   \caption{The $FE[USp(2N)]$  quiver theory. In the drawing we are also specifying our conventions for the fugacities for the global symmetries that we will turn on in the index.  Specifically, $p$ and $q$ are fugacities for combinations of the trial R-charge and angular momenta, and the power of $pq$ represents a half of the R-charge. Similarly, $t$ and  $c$ are the fugacities for the other abelian symmetries, whose powers encode the corresponding charges of the fields. }
\label{FEUSP}
\end{figure}
Following \cite{Bottini:2021vms,Hwang:2021ulb}, in $4d$ we identify $\mathsf{S}$ with a slight variant of this theory which includes additional gauge singlet chiral fields, and that is called the $FE[USp(2N)]$ theory  shown in Figure \ref{FEUSP}. We refer the reader for example to \cite{Bottini:2021vms} for the explicit definition of this theory and to Appendix \ref{sec:FE} for a brief review, while here we only review some of its properties that are relevant for our discussion.
The $FE[USp(2N)]$ theory has a $USp(2N)_x\times USp(2N)_y\times U(1)_t\times U(1)_c$ global symmetry in the IR, 
and we will represent it schematically as in Figure \ref{Ssch}, where we are explicitly displaying its two $USp(2N)$ symmetries.

\noindent
The index of the $\mathsf{S}$-wall theory has the following recursive definition:
\begin{align}\label{indexSwall}
\mathcal{I}_{\mathsf{S}}^{(N)}(\vec{x};\vec{y};t;c) &=\Gpq{pq\,c^{-2}}\Gpq{pq\,t^{-1}}^N\prod_{i<j}^N\Gpq{pq\,t^{-1}x_i^{\pm1}x_j^{\pm1}}\prod_{i=1}^N\Gpq{c\,y_N^{\pm1}x_i^{\pm1}}\nn\\
&\qquad\times\oint\udl{\vec{z}_{N-1}} \Gd_{N-1}(\vec z_{N-1}) \prod_{a=1}^{N-1}\prod_{i=1}^N\Gpq{t^{1/2}z_a^{\pm1}x_i^{\pm1}}\Gpq{pq\,t^{-1/2}c^{-1}y_N^{\pm1}z_a^{\pm1}}\nn \\
&\qquad\times\mathcal{I}_{\mathsf{S}}^{(N-1)}\left(z_1,\cdots,z_{N-1};y_1,\cdots,y_{N-1};t;t^{-1/2}c\right)\, ,
\end{align}
where we defined the contribution of the vector multiplet and the integration measure as
\begin{equation}
\label{eq:vector}
\Gd_{n}(\vec z_{n})=\frac{\left[(p;p)_\infty (q;q)_\infty\right]^n}{\prod_{i=1}^n\Gpq{x_i^{\pm2}}\prod_{i<j}^n\Gpq{x_i^{\pm1} x_j^{\pm1}}},\quad \udl{\vec{z}_{n}}=\frac{1}{n!}\frac{\prod_{i=1}^n\udl{z_i}}{2\pi iz_i}\,.
\end{equation}
This integral function coincides with the \emph{interpolation kernel} of Rains \cite{2014arXiv1408.0305R} and it enjoys various remarkable identities, some of which were interpreted as properties of the $FE[USp(2N)]$ theory in \cite{Pasquetti:2019hxf}. 

\begin{figure}[!ht]
\center
\includegraphics[scale=0.7]{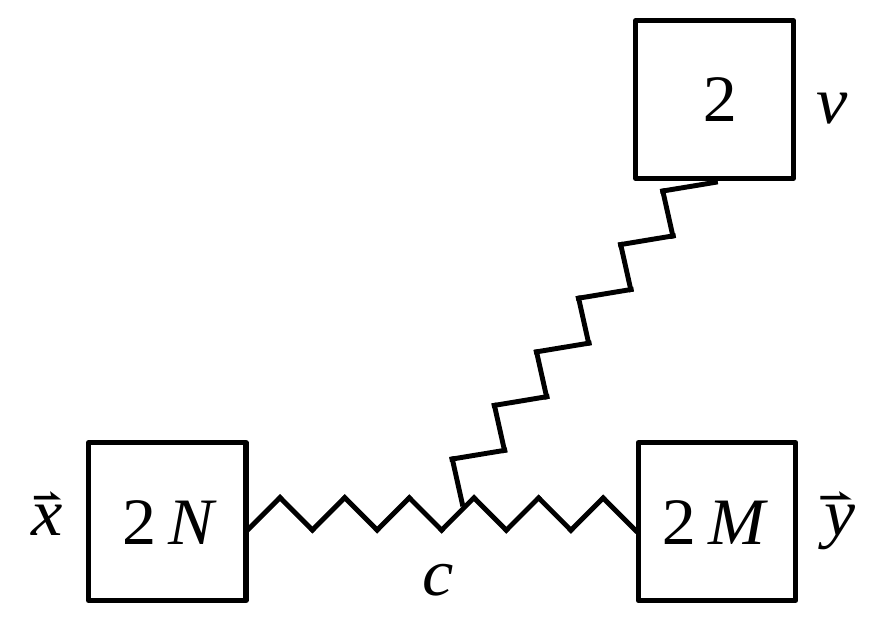}
\caption{The $4d$ asymmetric $\mathsf{S}$-wall.}
\label{fig:asymm_Swall}
\end{figure}

It is also useful to consider, following \cite{Bottini:2021vms,Hwang:2021ulb}, an asymmetric version of the $\mathsf{S}$-wall, in the sense that it displays two non-abelian symmetries of unequal ranks $USp(2N)$ and $USp(2M)$ with $M<N$. We will represent such an asymmetric $\mathsf{S}$-wall schematically as in Figure \ref{fig:asymm_Swall}.

\noindent This can be easily obtained from the $FE[USp(2N)]$ theory by turning on a deformation associated with an operator transforming in the antisymmetric representation of one of the two $USp(2N)$ symmetries, with the effect of breaking it down to $USp(2M)\times SU(2)$ for $M<N$. The details of this deformation can be found in \cite{Hwang:2020wpd,Bottini:2021vms}. At the level of the index, the pattern of breaking of the global symmetry due to the deformation is encoded in a specific specialization of some of the flavor fugacities
\begin{equation}\label{indexASwall}
\mathcal{I}_{\mathsf{S}}^{(N)}(\vec{x};\vec{y},t^{\frac{N-M-1}{2}}v,\cdots,t^{-\frac{N-M-1}{2}}v;t;c)\,.
\end{equation}
More details regarding the asymmetric $\mathsf{S}$-wall can be found  in  Appendix \ref{asyswall}.

\bigskip

\begin{figure}[!ht]
	\includegraphics[scale=0.65,center]{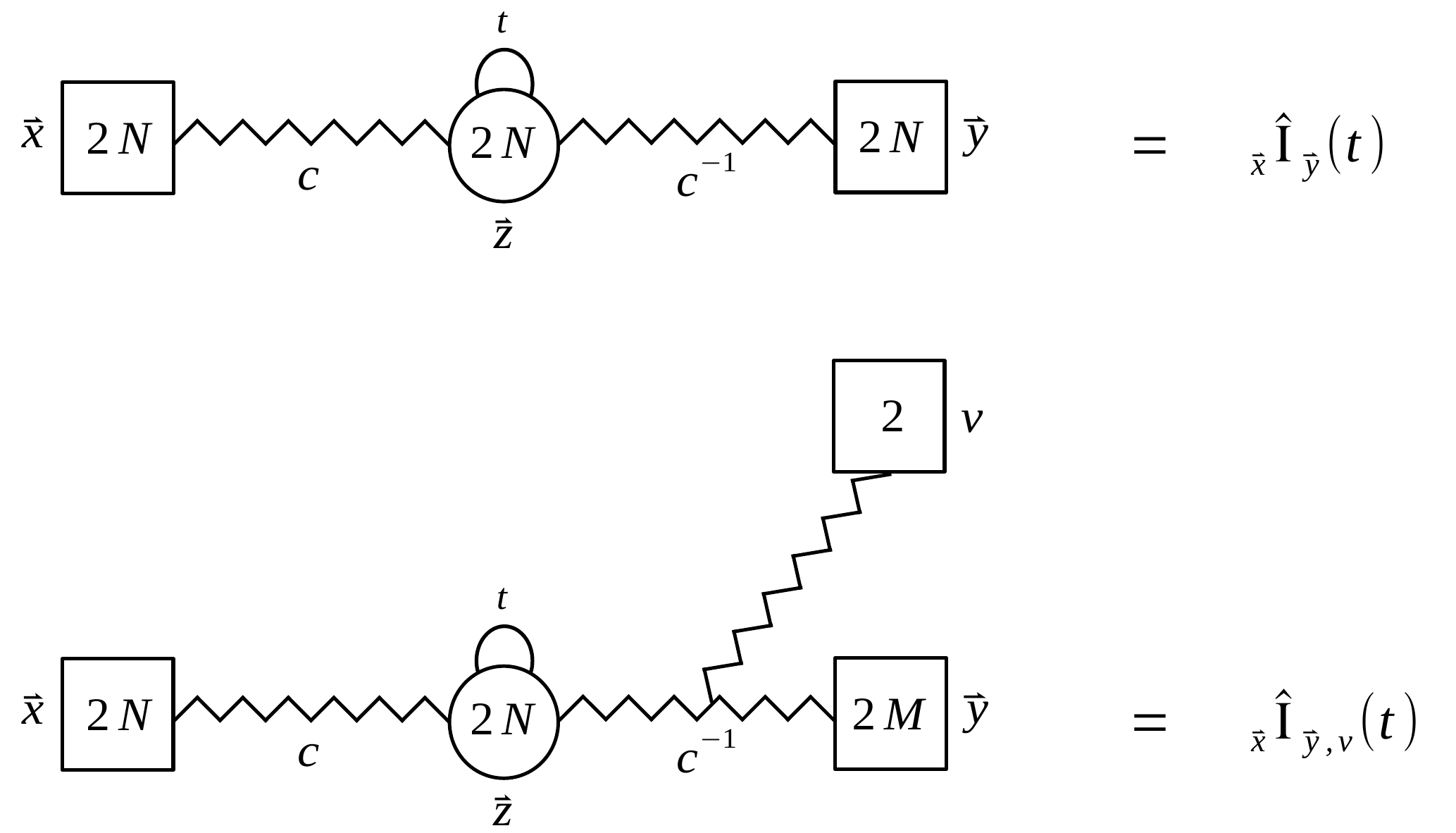}
	\caption{The symmetric and the asymmetric Identity-walls.}
	\label{fSS1}
\end{figure}

\noindent\textbf{The Identity-wall.} One property of the $\mathsf{S}$-wall theory proven in \cite{Bottini:2021vms} is that concatenating two of them, which in field theory amounts to gauging a diagonal combination of a $USp(2N)$ symmetry from each $\mathsf{S}$-wall, these annihilate each other giving an Identity-wall as shown in the first quiver in Figure \ref{fSS1}.

We can make this statement more precise at the level of the index, which, as  shown in \cite{Bottini:2021vms}, is given by a distribution
\begin{eqnarray}
\label{eq:deltaNN}
&&\oint\udl{\vec{z}_N}\Gd_N(\vec{z};t)\mathcal{I}_{\mathsf{S}}^{(N)}(\vec{x};\vec{z};t;c)\mathcal{I}_{\mathsf{S}}^{(N)}(\vec{z};\vec{y};t;c^{-1})={}_{\vec{x}}\hat{\mathbb{I}}_{\vec{y}}(t) \,,
\end{eqnarray}
with
\begin{equation}\label{eq:idopNN}
{}_{\vec x}\hat{\mathbb{I}}_{\vec y}(t)=\frac{\prod_{j=1}^N 2\pi ix_j}{\Gd_N(\vec{x};t)}\sum_{\gs\in S_N}\sum_{\pm}\prod_{j=1}^N\gd\left(x_j- y_{\gs(j)}^\pm\right)\,.
\end{equation}

The sum over elements of the permutation group $S_N$ and over signs $\pm$ corresponds to a sum over the transformation actions of the Weyl group of $USp(2N)$.
In the denominator we have the contribution of the $USp(2N)$ vector multiplet $\Delta_N(\vec x)$, defined in \eqref{eq:vector}, and the antisymmetric chiral multiplet $A_N(\vec{x};t)$:
\begin{align}
\Gd_N(\vec{x};t) =\Gd_N(\vec{x}) A_N(\vec{x};t)\, , \qquad  A_N(\vec{x};t)=\Gpq{t}^N \prod_{i<j}^N\Gpq{t x_i^{\pm1} x_j^{\pm1}} \,. \label{intmes}
\end{align}
Notice that the l.h.s.~of the relation of Figure \ref{fSS1} for $N=1$  coincides, up to flipping fields,  with  the $SU(2)$ theory with four chirals, which is known to have a quantum deformed moduli space of vacua \cite{Seiberg:1994bz} and an index given by a delta distribution \cite{Spiridonov:2014cxa}.

A similar property holds also for the asymmetric $\mathsf{S}$-wall; namely an ordinary $\mathsf{S}$-wall glued to an asymmetric $\mathsf{S}$-wall results in an asymmetric Identity-wall as shown at the bottom of Figure \ref{fSS1}.
In the index we simply implement the specialization of fugacities \eqref{indexASwall} into \eqref{eq:deltaNN}
\begin{eqnarray}
\label{eq:deltaNM}
&&\oint\udl{\vec{z}_N}\Gd_N(\vec{z};t)\mathcal{I}_{\mathsf{S}}^{(N)}(\vec{x};\vec{z};t;c)\mathcal{I}_{\mathsf{S}}^{(N)}(\vec{z};\vec{y},t^{\frac{N-M-1}{2}}v,\cdots,t^{-\frac{N-M-1}{2}}v;t;c^{-1})={}_{\vec{x}}\hat{\mathbb{I}}_{\vec{y},v}(t) \,,\nn\\
\end{eqnarray}
where the index of the asymmetric Identity-wall is given by
\begin{equation}
\label{eq:idop}
{}_{\vec{x}}\hat{\mathbb{I}}_{\vec{y},v}(t)=\frac{{ \prod_{i=1}^{N} 2\pi i x_i }}{\Gd_N (\vec x;t) }\left.\sum_{\sigma \in S_N,\pm}\prod_{i=1}^{N}\delta\left(x_i-y_{\sigma(i)}^{\pm1}\right)\right|_{y_{M+j}=t^{\frac{N-M+1-2j}{2}}v}\,.
\end{equation}

\bigskip

\begin{figure}[!ht]
	\centering
	\includegraphics[scale=0.70]{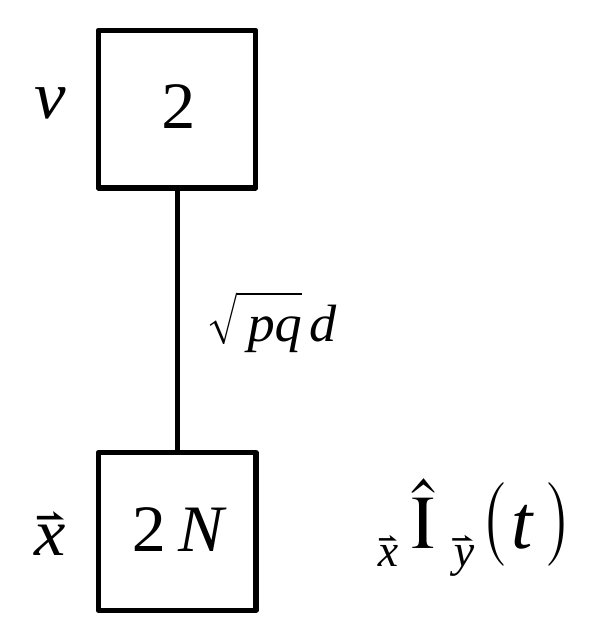}
\caption{The $4d$ $\mathsf{T}$-wall. In the drawing $\vec x$ are the $USp(2N)$ fugacities, $v$ the $SU(2)$ fugacity and $d$ the fugacity for an abelian symmetry. }\label{fig:Twall4d}
\end{figure}

\noindent\textbf{The $\mathsf{T}$-wall.} In $3d$, the $\mathcal{T}$ generator of $SL(2,\mathbb{Z})$ is associated with the addition of a Chern--Simons (CS) coupling at level one. We propose that the $4d$ analogue denoted by $\mathsf{T}$ is a pair of $USp(2N)$ fundamental chiral fields and for convenience we also attach to it an Identity-wall, as shown in Figure \ref{fig:Twall4d}. Indeed, in the $3d$ limit, after giving a CB VEV deformation that breaks the symplectic group to a unitary one, we consider a real mass deformation for the $U(1)_d$ symmetry which integrates out all the fields and produces a CS level $1$ for the $U(N)$ group. We will come back to the connection to $3d$ in Section \ref{sec:3d}.

\noindent
The index expression for the $\mathsf{T}$-wall in $4d$ is simply
\begin{equation}\label{eq:indTwall4d}
\mathcal{I}_{\mathsf{T}}^{(N)}(\vec{x};\vec{y};v;t;d) = \prod_{j=1}^N \Gamma((pq)^\frac12 d x_j^\pm v^\pm) {}_{\vec x}\hat{\mathbb{I}}_{\vec y}(t)\,.
\end{equation}

\bigskip

\noindent\textbf{\boldmath Group multiplication and $PSL(2,\mathbb{Z})$ relations.}
The objects we have introduced can be associated with the corresponding elements of the $PSL(2,\mathbb Z)$ group. To see if those objects respect the group structure, we need to define the group multiplication acting on them. Let us recall eq.~\eqref{eq:deltaNN}. Here, we glue two $\mathsf{S}$-walls by gauging the diagonal $USp(2 N)_z$ of two $USp(2 N)$ symmetries groups, one from each $\mathsf{S}$-wall, to obtain the Identity-wall. Moreover, we add an additional antisymmetric chiral field $\Phi$ with the superpotential
\begin{align}
\label{eq:gluing sup}
\mathrm{Tr}_z \left[\Phi \cdot \left(\mathsf O_\mathsf H^L-\mathsf O_\mathsf H^R\right)\right] \,,
\end{align}
where $\mathsf O_\mathsf H^{L/R}$ are the antisymmetric operators defined in eq. \eqref{aFE} of two glued $USp(2 N)$ symmetries of the $\mathsf{S}$-walls. This superpotential as well as the anomaly-free condition of the gauged $USp(2 N)_z$ group completely determine the abelian charges of the second $\mathsf{S}$-wall in terms of those of the first $\mathsf{S}$-wall, as shown in \eqref{eq:deltaNN}.
Therefore, once we define the group multiplication in this way, the property in Figure \ref{fSS1} can be interpreted as
\begin{align}
\mathsf{S}\mathsf{S} = 1 \,,
\end{align}
indicating that the inverse of the $\mathsf{S}$-wall is the $\mathsf{S} $-wall itself. This is one of the defining relations of $PSL(2,\mathbb Z)$.

\begin{figure}[!ht]
	\includegraphics[width=\textwidth,center]{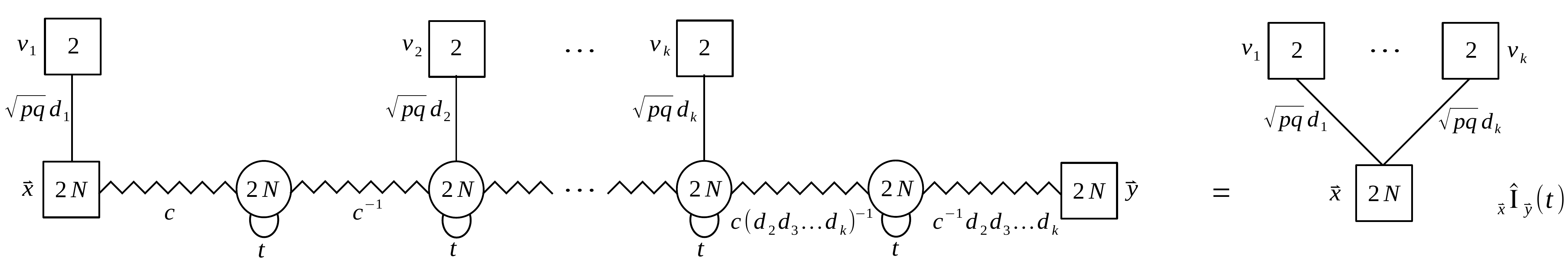}
	\caption{The $\mathsf{T}^{k}$-wall. On the r.h.s.~we used the fact that two glued $\mathsf{S}$-walls give an Identity-wall.}
	\label{tk}
\end{figure}

Next, let us examine the multiplication property of the $\mathsf{T}$-wall as a group element in $PSL(2, \mathbb Z)$. The group multiplication of two $\mathsf{T}$-walls is defined in the same way as that of the $\mathsf{S}$-wall explained above. We introduce an additional chiral $\Phi$ in the antisymmetric representation of the diagonal $USp(2 N)_z$ of two $USp(2 N)$ symmetries, one from each $\mathsf{T}$-wall, with the superpotential \eqref{eq:gluing sup} and then gauge this diagonal $USp(2 N)_z$. The generalization to $\mathsf{T}^k$ is straightforward  and shown in Figure \ref{tk}.

\noindent
For $k = 2$, the corresponding index is given by
\begin{align}
\label{eq:T^2}
&\oint\udl{\vec{z}_N}\Gd_N(\vec{z};t)\mathcal{I}_{\mathsf{T}}^{(N)}(\vec{x};\vec{z};v;t;d_1)\mathcal{I}_{\mathsf{T}}^{(N)}(\vec{z};\vec{y};u;t;d_2)\nn\\
&=\oint\udl{\vec{z}_N}\Gd_N(\vec{z};t) \prod_{i=1}^N \Gamma((pq)^\frac12 d_1 x_i^\pm v^\pm) {}_{\vec x}\hat{\mathbb{I}}_{\vec z}(t) \prod_{j=1}^N \Gamma((pq)^\frac12 d_2 z_j^\pm u^\pm) {}_{\vec z}\hat{\mathbb{I}}_{\vec y}(t) \nonumber \\
&= \prod_{i=1}^N \Gamma((pq)^\frac12 d_1 x_i^\pm v^\pm) \prod_{j=1}^N \Gamma((pq)^\frac12 d_2 x_j^\pm u^\pm) {}_{\vec x}\hat{\mathbb{I}}_{\vec y}(t)=\prod_{i=1}^N\prod_{a=1}^4 \Gamma((pq)^\frac12 d x_i^\pm u_a)  {}_{\vec x}\hat{\mathbb{I}}_{\vec y}(t) \,,
\end{align}
where we defined 
\begin{equation}
d_1=d\,s,\quad d_2=d\,s^{-1},\quad u_a=\begin{cases}s\,v^{\pm1}&a=1,2\\s^{-1}u^{\pm1}&a=3,4\end{cases}
\end{equation}
so that the fugacities $u_a$ satisfying $\prod_{a=1}^4u_a=1$ can be interpreted as $SU(4)$ fugacities. Similarly, for $\mathsf{T}^k$ we get
\begin{align}
\prod_{i=1}^N \prod_{j = 1}^{2 k} \Gamma((pq)^\frac12 d x_i^\pm v_j) {}_{\vec x}\hat{\mathbb{I}}_{\vec y}(t)\,,
\end{align}
where $v_j$ are the $SU(2 k)$ fugacities satisfying $\prod_{j = 1}^{2 k} v_j = 1$, and $d^k = \prod_{i = 1}^k d_i$.\\

\begin{figure}[!ht]
	\centering
	\includegraphics[width=\textwidth]{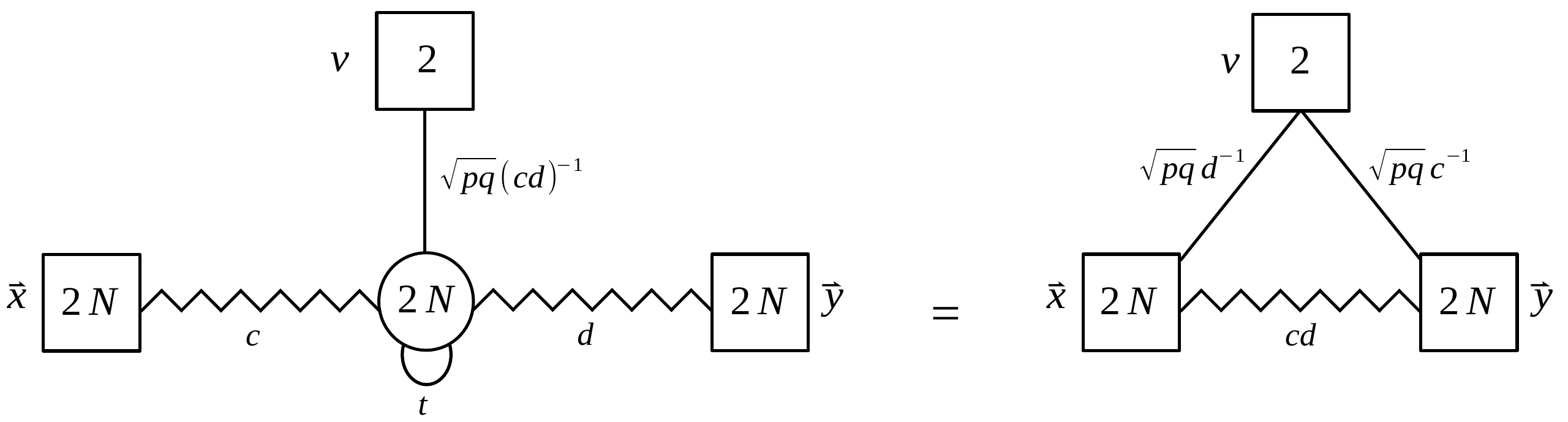}
	\caption{The braid duality move.}
	\label{braid}
\end{figure}

So far, we have defined the group multiplication between the same elements, either $\mathsf{S}$ or $\mathsf{T}$. To complete the definition of the group multiplication, we need to address the multiplication between $\mathsf{S}$ and $\mathsf{T}$. For this purpose, a crucial role will be played by the so-called \emph{braid move}, which is a duality that was proposed in \cite{Pasquetti:2019hxf} and summarize in Figure \ref{braid} (we refer the reader to Section 3.4 of \cite{Pasquetti:2019hxf} for more details). This will also allow us to derive many of the duality moves later needed in the algorithm. \\
\\
\noindent
At the level of the index the braid duality is encoded in the following integral identity:
\begin{align}
\label{eq:braid}
&\oint \udl{\vec{w}_N} \Gd_N(\vec{w};t) \mathcal{I}_{\mathsf{S}}^{(N)}(\vec{x};\vec{w};t;c) \prod_{j=1}^N \Gamma((pq)^\frac12 c^{-1}d^{-1} w_j^\pm v^\pm) \mathcal{I}_{\mathsf{S}}^{(N)}(\vec{w};\vec{y};t;d) \nonumber \\
&\quad=\prod_{j=1}^N \Gamma((pq)^\frac12 d^{-1} x_j^\pm v^\pm) \mathcal{I}_{\mathsf{S}}^{(N)}(\vec{x};\vec{y};t;cd) \prod_{j=1}^N \Gamma((pq)^\frac12 c^{-1} y_j^\pm v^\pm)  \,,
\end{align}
which was proven in Proposition 2.12 of \cite{2014arXiv1408.0305R}. It was later understood as a sort of generalization of Seiberg duality \cite{Seiberg:1994pq} and derived from the perspective of the $4d$ compactification of the $6d$ E-string theory on a torus with flux in \cite{Pasquetti:2019hxf}. It turns out that this duality can be derived by assuming only the Intriligator--Pouliot (IP) duality \cite{Intriligator:1995ne} and applying it iteratively, as it will be shown in \cite{BCP}.

 For the multiplication between $\mathsf S$ and $\mathsf T$, the operation is basically defined in the almost same way as that of $\mathsf S$ or $\mathsf T$ alone. We introduce an additional antisymmetric chiral $\Phi$ with the superpotential \eqref{eq:gluing sup} and gauge the diagonal $USp(2 N)$, but there is one additional ingredient: if a group element contains $\mathsf{S}$ sandwiched between two $\mathsf{T}$ elements, we introduce an extra superpotential
\begin{align}
\label{eq:triangle}
\mathrm{Tr} P_1 \Pi P_2 \,,
\end{align}
where $P_1$ and $P_2$ are two fundamental chirals in the $\mathsf{T}$-walls, and $\Pi$ is a bifundamental operator between two $USp(2 N)$ in the $\mathsf{S}$-wall (see Table \ref{eusptable} for its charges under the global symmetries). It will shortly become clear that this prescription ensures the group structure of $PSL(2,\mathbb Z)$.

\begin{figure}[!ht]
	\centering
	\includegraphics[scale=0.6]{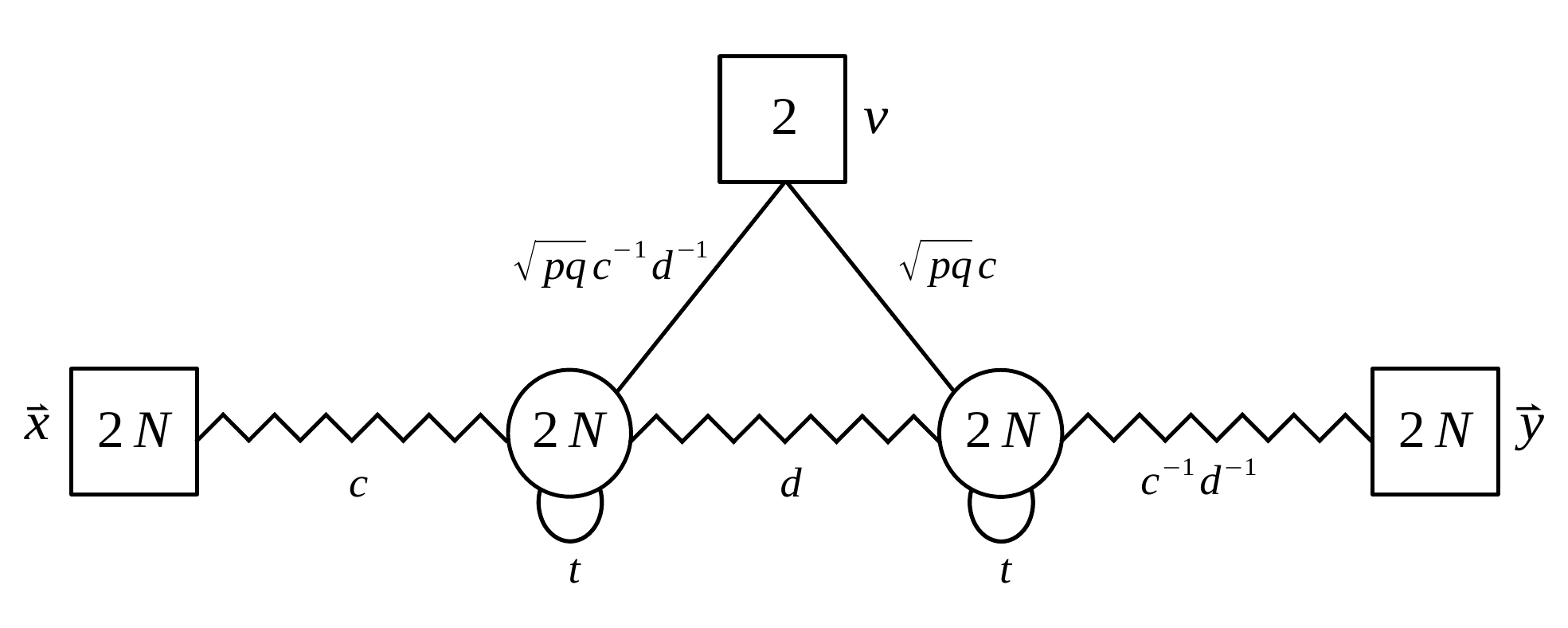}
	\caption{The $\mathsf{T}^{-1}$-wall given by $\mathsf{T}^{-1} = \mathsf{S} \mathsf{T} \mathsf{S} \mathsf{T} \mathsf{S}$ after getting rid of all the Identity-walls in the definition of each $\mathsf{T}$-wall.}
	\label{inverseT}
\end{figure}

As we have seen above, the inverse element of $\mathsf{S}$ is $\mathsf{S}$ itself. Namely, once two $\mathsf{S}$-walls are glued, they give rise to the Identity-wall as shown in \eqref{eq:deltaNN}. On the other hand, the inverse element of $\mathsf{T}$ is rather non-trivial and is given by $\mathsf{T}^{-1} = \mathsf{S} \mathsf{T} \mathsf{S} \mathsf{T} \mathsf{S}$, as shown in Figure \ref{inverseT}.

Note that all the abelian charges of the walls are fixed by the anomaly-free condition and the cubic superpotential \eqref{eq:triangle}; in this case, $P_1$ and $P_2$ in the superpotential are the two fundamental chirals in the two $\mathsf{T}$-walls, and $\Pi$ is the bifundamental operator of the $\mathsf{S}$-wall in the middle.
The index of the $\mathsf{T}^{-1}$-wall is expressed compactly in terms of those of the $\mathsf{S}$ and $\mathsf{T}$ components as follows:
\begin{align}
\mathcal{I}_{\mathsf{T}^{-1}}^{(N)}(\vec{x};\vec{y};v;t;d) &= \oint \left(\prod_{i = 1}^4 \udl{\vec{z}^{(i)}_N} \Gd_N(\vec{z}^{(i)};t)\right) \mathcal{I}_{\mathsf{S}}^{(N)}(\vec{x};\vec{z}^{(1)};t;c) \mathcal{I}_{\mathsf{T}}^{(N)}(\vec{z}^{(1)};\vec{z}^{(2)};v;t;c^{-1} d^{-1}) \nonumber \\
&\quad \times \mathcal{I}_{\mathsf{S}}^{(N)}(\vec{z}^{(2)};\vec{z}^{(3)};t;d) \mathcal{I}_{\mathsf{T}}^{(N)}(\vec{z}^{(3)};\vec{z}^{(4)};v;t;c) \mathcal{I}_{\mathsf{S}}^{(N)}(\vec{z}^{(4)};\vec{y};t;c^{-1} d^{-1}) \,.
\end{align}

\begin{figure}[!ht]
	\includegraphics[width=0.95\textwidth,center]{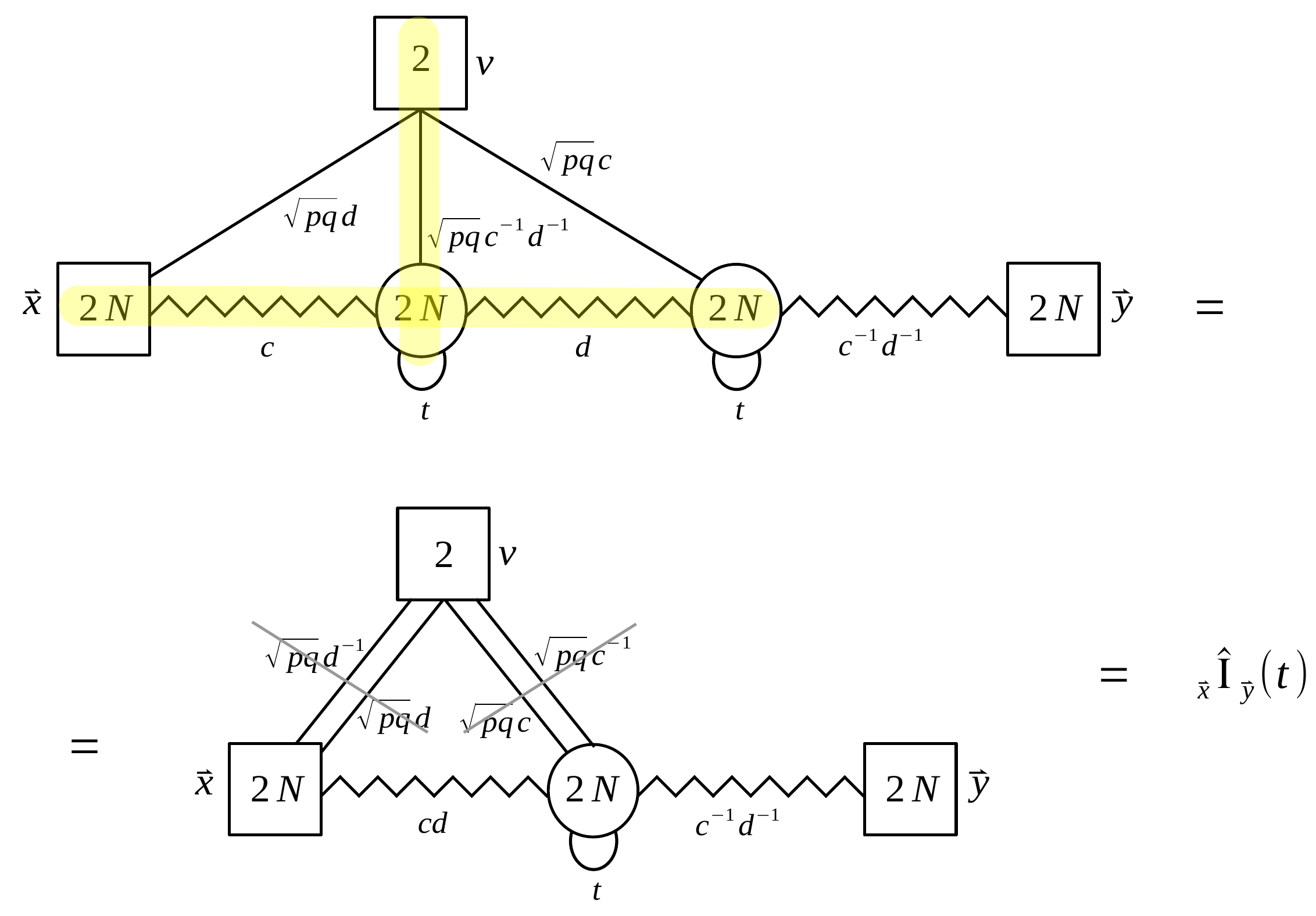}
	\caption{	The field theory realization of the relation $\mathsf{T}\mathsf{T}^{-1} = \mathsf{T}\mathsf{S} \mathsf{T} \mathsf{S} \mathsf{T} \mathsf{S}=I$, the other  $PSL(2,\mathbb{Z})$ defining relation.  Notice it  also encodes $\mathsf{T}^ T(\mathsf{T}^T)^{-1}=I$.
	 The grey lines are traced over  massive fields which can be integrated out.}
	\label{fig:TSTSTS}
\end{figure}

Now we want to check that $\mathsf{T}\mathsf{T}^{-1}=1$. According to the rule we gave above, we gauge the diagonal $USp(2N)$ in the presence of an extra antisymmetric chiral $\Phi$ and two types of the superpotential terms, one from \eqref{eq:gluing sup} and the other from \eqref{eq:triangle} as shown in the first diagram of Figure \ref{fig:TSTSTS}.

The corresponding index can then be written as follows:
\begin{align}
\label{eq:TSTSTS}
&\oint \udl{\vec{w}_N} \Gd_N(\vec{w};t)\mathcal{I}_{\mathsf{T}}^{(N)}(\vec{x};\vec{w};v;t;d) \mathcal{I}_{\mathsf{T}^{-1}}^{(N)}(\vec{w};\vec{y};v;t;d) \nonumber \\
&= \oint \udl{\vec{w}_N} \Gd_N(\vec{w};t) \left(\prod_{i = 1}^4 \udl{\vec{z}^{(i)}_N} \Gd_N(\vec{z}^{(i)};t)\right) \mathcal{I}_{\mathsf{T}}^{(N)}(\vec{x};\vec{w};v;t;d) \mathcal{I}_{\mathsf{S}}^{(N)}(\vec{w};\vec{z}^{(1)};t;c) \nonumber \\
&\quad \times \mathcal{I}_{\mathsf{T}}^{(N)}(\vec{z}^{(1)};\vec{z}^{(2)};v;t;c^{-1} d^{-1}) \mathcal{I}_{\mathsf{S}}^{(N)}(\vec{z}^{(2)};\vec{z}^{(3)};t;d) \mathcal{I}_{\mathsf{T}}^{(N)}(\vec{z}^{(3)};\vec{z}^{(4)};v;t;c) \mathcal{I}_{\mathsf{S}}^{(N)}(\vec{z}^{(4)};\vec{y};t;c^{-1} d^{-1}) \nonumber\\
&= \oint \left(\prod_{i = 2}^3 \udl{\vec{z}^{(i)}_N} \Gd_N(\vec{z}^{(i)};t)\right) \prod_{j=1}^N \Gamma((pq)^\frac12 d x_j^\pm v^\pm) \mathcal{I}_{\mathsf{S}}^{(N)}(\vec{x};\vec{z}^{(2)};t;c) \nonumber \\
&\quad \times \prod_{j=1}^N \Gamma((pq)^\frac12 c^{-1} d^{-1} z^{(2)}_j{}^\pm v^\pm) \mathcal{I}_{\mathsf{S}}^{(N)}(\vec{z}^{(2)};\vec{z}^{(3)};t;d) \prod_{j=1}^N \Gamma((pq)^\frac12 c z^{(3)}_j{}^\pm v^\pm) \mathcal{I}_{\mathsf{S}}^{(N)}(\vec{z}^{(3)};\vec{y};t;c^{-1} d^{-1}) \,.
\end{align}
In the last equality we wrote explicitly the contribution of the $\mathsf{T}$-walls in terms of doublets and Identity-walls using \eqref{eq:indTwall4d}, and we also used the delta functions inside the Identity-walls to get rid of some of the integrations. The final result is what is represented on the top of Figure \ref{fig:TSTSTS}.

Note that each triangle in the first quiver in Figure \ref{fig:TSTSTS} makes a superpotential of the form \eqref{eq:triangle}.
Because of the constraints on the fugacities imposed by this  superpotential, we can  apply the braid move of Figure \ref{braid} as highlighted in Figure \ref{fig:TSTSTS} to obtain the second quiver,  where, after removing the massive fields, we find two consecutive $\mathsf{S}$-walls  which are equivalent to a single Identity-wall. This gives us the index identity
\begin{align}
\label{eq:TT^{-1}}
&\oint \udl{\vec{w}_N} \Gd_N(\vec{w};t)\mathcal{I}_{\mathsf{T}}^{(N)}(\vec{x};\vec{w};v;t;d) \mathcal{I}_{\mathsf{T}^{-1}}^{(N)}(\vec{w};\vec{y};v;t;d) = {}_{\vec{x}}\hat{\mathbb{I}}_{\vec{y}}(t) \,.
\end{align}
We comment in passing that while on the l.h.s.~we have an explicit dependence on the fugacity $v$, the r.h.s.~doesn't depend on it. Nevertheless, the identity holds for any value of $v$. This tells us that also the l.h.s.~is actually independent of $v$. While here it is just a comment about the mathematical identity, we will see this phenomenon many times and later comment on its physical implications when discussing identities that are associated with genuine IR dualities. 

\begin{figure}[!ht]
	\centering
	\includegraphics[scale=0.65]{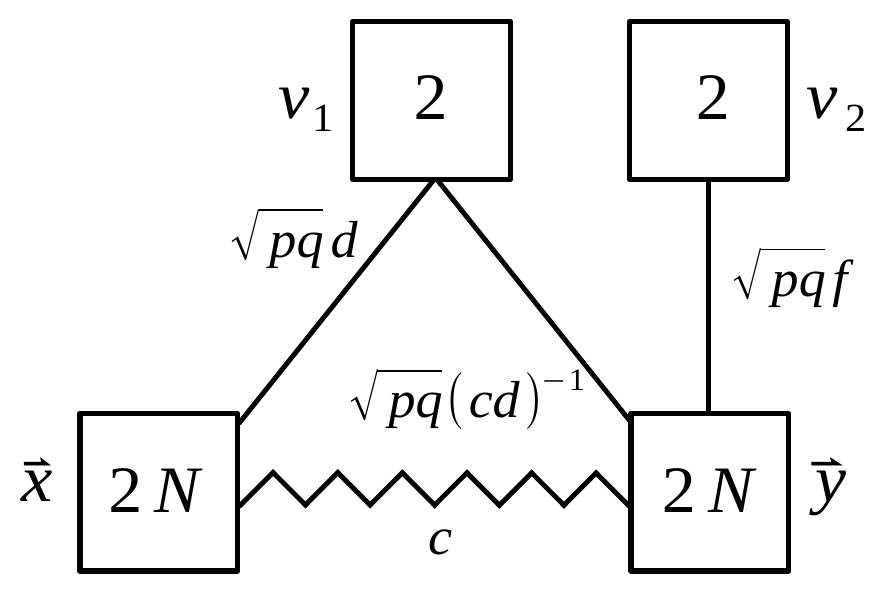}
	\caption{The field theory realization of the combination $\mathsf{T} \mathsf{S} \mathsf{T}^2$.}
	\label{tst2}
\end{figure}

As we have seen so far, a cubic superpotential \eqref{eq:triangle} must be turned on when an $\mathsf{S}$-wall sits between two $\mathsf{T}$-walls. What about if we have multiple $\mathsf{T}$-walls? For example, if we have a single $\mathsf{T}$-wall on the left and two $\mathsf{T}$-walls on the right of an $\mathsf{S}$-wall, two natural choices of the superpotential would be
\begin{align}
\mathrm{Tr} Q \Pi P_1\,, \qquad {\rm or} \qquad
\mathrm{Tr} Q \Pi (P_1+P_2) \,,
\end{align}
where $Q$ is the chiral from the left $\mathsf{T} $-wall and $P_1, \, P_2$ are the chirals from the two right $\mathsf{T}$-walls.
In fact, those two choices are equivalent up to a field redefinition. Therefore, a quiver associated with $\mathsf{T} \mathsf{S} \mathsf{T}^2$ would be as in Figure \ref{tst2}.

\begin{figure}[!ht]
	\includegraphics[width=\textwidth,center]{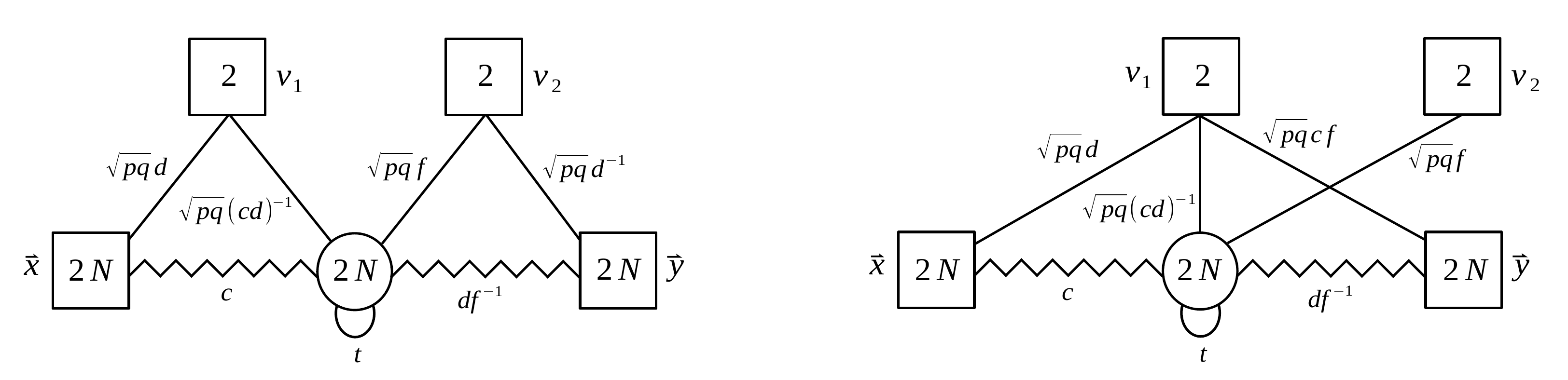}
	\caption{Two options for $\mathsf{TST}^2\mathsf{ST}$. As we explain in the text, only the first one is correct.}
	\label{twop}
\end{figure}

\begin{figure}[!ht]
	\includegraphics[width=\textwidth,center]{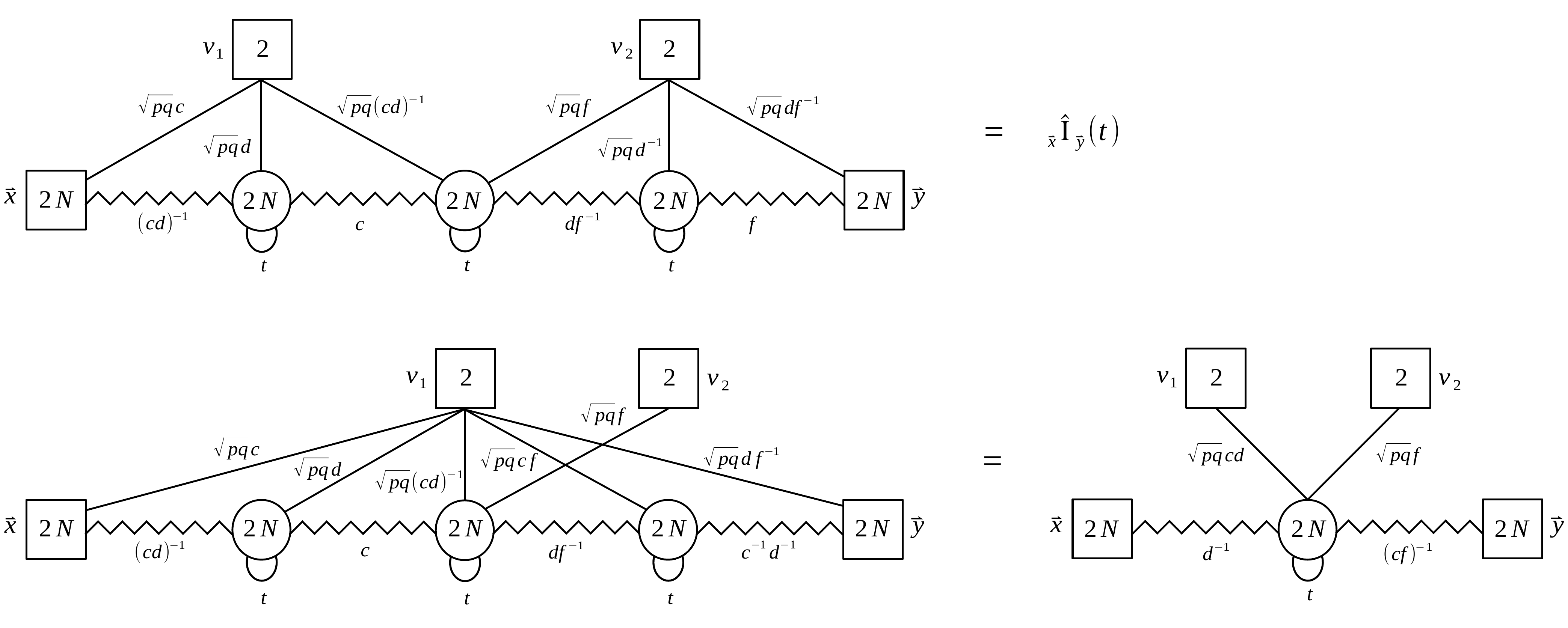}
	\caption{Two possible realizations of the combination $(\mathsf{TS})\mathsf{TST}^2\mathsf{ST}(\mathsf{ST})$. Only the first one is correct, since the second doesn't reproduce the Identity-wall expected from the relation $(\mathsf{TS})\mathsf{TST}^2\mathsf{ST}(\mathsf{ST})=I$.}
	\label{onlyone}
\end{figure}

\begin{figure}[!ht]
	\centering
	\includegraphics[scale=0.65]{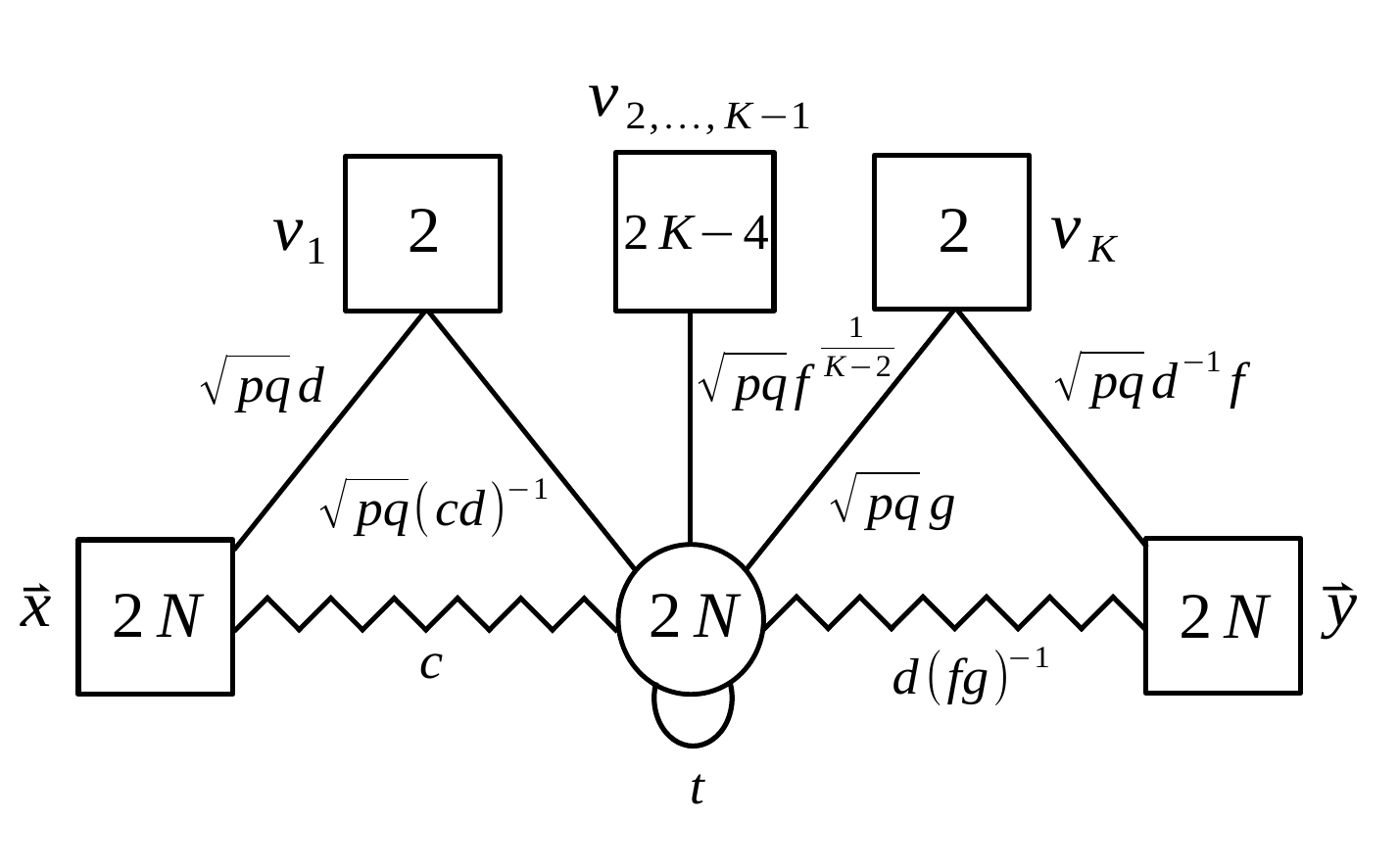}
	\caption{The field theory realization of the combination $\mathsf{TST}^k\mathsf{ST}$.}
	\label{tstkst}
\end{figure}

Moreover, we can also consider $\mathsf{TST}^2\mathsf{ST}$. In this case, we have the  two candidates depicted in Figure \ref{twop}.
One can find the correct choice by multiplying $\mathsf{TS}$ from the left and $\mathsf{ST}$ from the right, which is supposed to give 1.  As shown in Figure \ref{onlyone} only the first choice gives the expected Identity-wall. Thus, we conclude that the first quiver in Figure \ref{twop} is the correct implementation of $\mathsf{TST}^2\mathsf{ST}$. Generalizing this observation, we can also find the quiver corresponding to $\mathsf{TST}^k\mathsf{ST}$, which we give in Figure \ref{tstkst}.

In conclusion, with our gluing prescriptions, the  $\mathsf{S}$ and $\mathsf{T}$-walls  
satisfy two conditions $\mathsf{S}\mathsf{S}=1$ and $(\mathsf{ST})^3=1$, which generate the  $PSL(2,\mathbb Z)$ group.
Furthermore, as we will see in the next section, the $\mathsf{S}$ and $\mathsf{T}$-walls act on the QFT blocks consistently with the $PSL(2,\mathbb Z)$  structure.

\bigskip

\begin{figure}[!ht]
	\centering
	\includegraphics[scale=0.75]{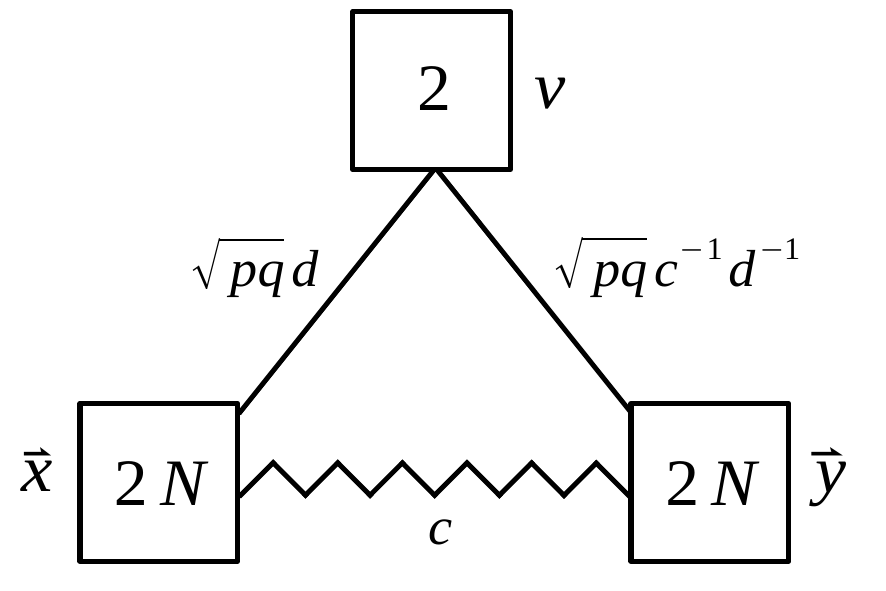}
	\caption{The $4d$ $\mathsf{T}^T$-wall.}
	\label{TTwall}
\end{figure}

\noindent\textbf{The $\mathsf{T}^T$-wall.} Finally, for later convenience, we would like to discuss another element of $PSL(2,\mathbb Z)$: $\mathsf{T}^T=\mathsf{TST}$.\footnote{The name $\mathsf{T}^T$ originates from the fact that the matrix representation of $\mathsf T^T = \mathsf{TST}$ is actually the transpose of the matrix associated with $\mathsf{T}$.} Although this is not an independent operation, it is interesting to consider it on its own, especially since we will consider the $\mathsf{T}^T$-dual of some quivers later on.
Its QFT realization consists of two sets of the $USp(2N)\times SU(2)$ bifundamental fields coupled to one copy of the $FE[USp(2N)]$ theory to form a triangle as depicted in Figure \ref{TTwall}, consistently with our previous discussion on the multiplication rules for the $PSL(2,\mathbb{Z})$ operators.
\\
\noindent
The index expression associated to the $\mathsf{T}^T$-wall is
\begin{align}
\label{eq:T^T}
\mathcal{I}_{\mathsf{T}^T}^{(N)} (\vec{x};\vec{y};v;t;c;d) &=\oint\udl{\vec{z}_N^{(1)}}\udl{\vec{z}_N^{(2)}}\Gd_N(\vec{z}^{(1)})\Gd_N(\vec{z}^{(2)})\mathcal{I}_{\mathsf{T}}^{(N)}(\vec{x};\vec{z}^{(1)};t;d)\nn\\
&\qquad\times\mathcal{I}_{\mathsf{S}}^{(N)}(\vec{z}^{(1)};\vec{z}^{(2)};t;c)\mathcal{I}_{\mathsf{T}}^{(N)}(\vec{z}^{(2)};\vec{y};t;(cd)^{-1})\nn\\
&= \mathcal{I}_{\mathsf{S}}^{(N)}(\vec{x};\vec{y};t;c) \prod_{j=1}^N \Gpq{(pq)^\half d x_j^\pm v^\pm} \Gpq{(pq)^\half (cd)^{-1} y_j^\pm v^\pm} \,,
\end{align}
where in the last equality we wrote explicitly the contribution of the $\mathsf{T}$-walls in terms of doublets and Identity-walls using \eqref{eq:indTwall4d} and we also used the delta functions inside the Identity-walls to get rid of some of the integrations.

\begin{figure}[!ht]
	\centering
	\includegraphics[scale=0.65]{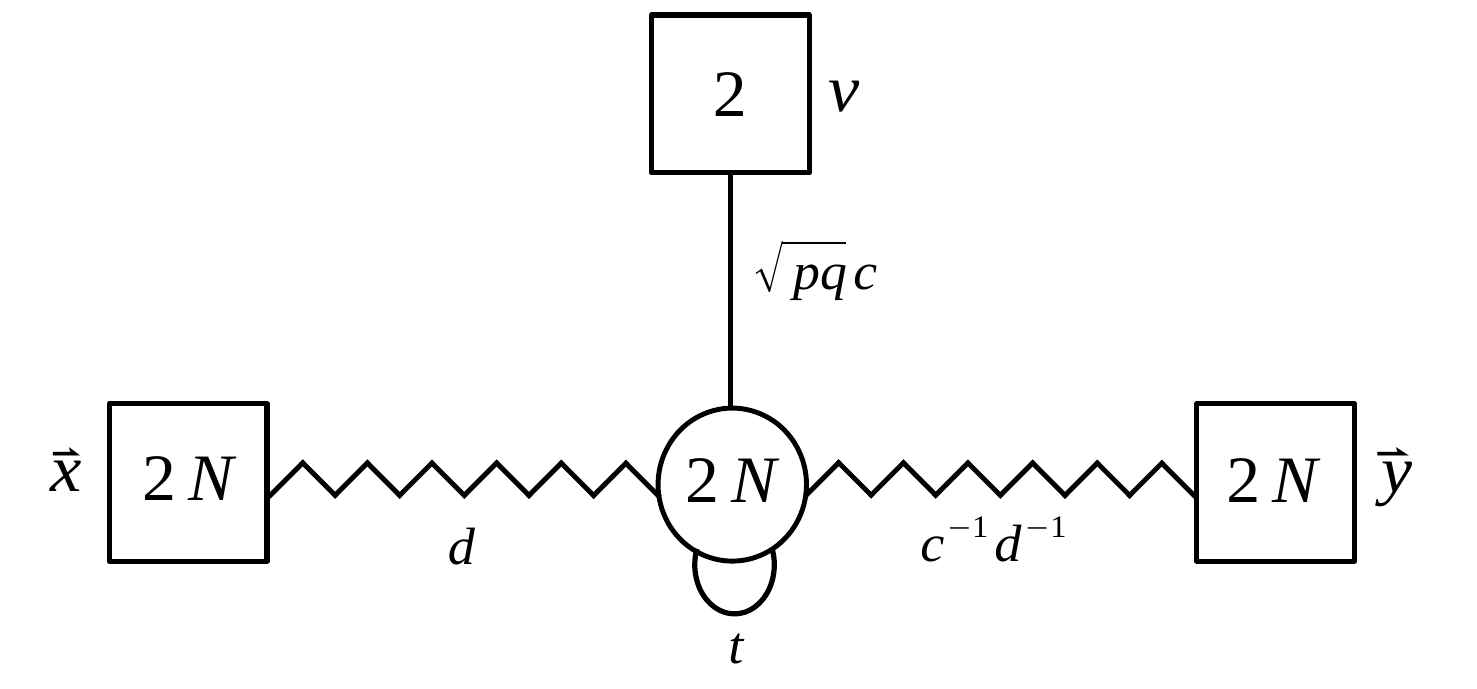}
	\caption{The $(\mathsf{T}^T)^{-1} = \mathsf{STS}$ wall. }\label{fig:TT-1wall4d}
\end{figure}

From the $PSL(2,\mathbb{Z})$ relation in Figure \ref{fig:TSTSTS} we can easily determine the inverse  $(\mathsf{T}^T)^{-1} = \mathsf{STS}$, whose associated quiver is given in Figure \ref{fig:TT-1wall4d}. 


\subsection{QFT building blocks}
In this section we introduce the three fundamental $4d$ QFT building blocks. By the way the $PSL(2,\mathbb{Z})$ operators in the previous subsection act on them, we can think of those QFT building blocks as being labelled by   vectors (1,0), (0,1) and (1,1) on which $PSL(2,\mathbb{Z})$ elements act as matrix multiplications. We will then see in Section \ref{sec:3d} that they reduce in $3d$ to the  QFT building blocks that are associated to the NS5, D5 and (1,1)-branes. For these reasons, we will call them $\mathsf{B}_{10}$, $\mathsf{B}_{01}$ and $\mathsf{B}_{11}$ blocks.

\bigskip

\begin{figure}[!ht]
	\centering
	\includegraphics[scale=0.75]{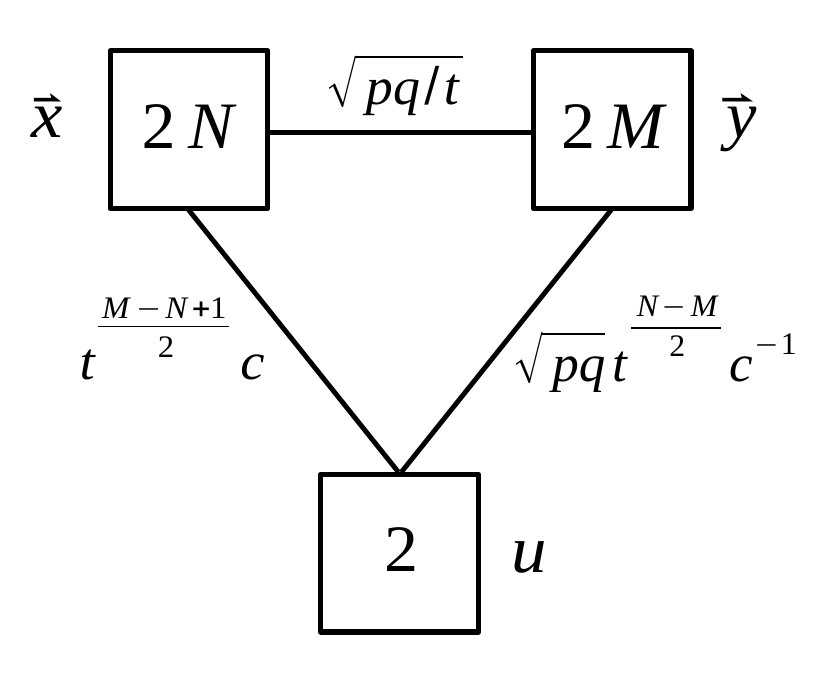}
	\caption{The $\mathsf{B}_{10}$ block.}
	\label{Figures/Section2/OperatorsAndBlocks/Triangle.pdf}
\end{figure}

\noindent\textbf{The $\mathsf{B}_{10}$ block.}
The first block is associated to an NS5 or (1,0)-brane. In $3d$ this would simply be a $U(N)\times U(M)$ bifundamental hypermultiplet. As proposed in \cite{Hwang:2021ulb}, its $4d$ uplift, the $\mathsf{B}_{10}$ block,  is a $USp(2N)\times USp(2M)$ bifundamental chiral multiplet plus four chirals, each two of which are in the fundamental representation of $USp(2N)$ or that of $USp(2M)$, respectively. These interact with a cubic superpotential, so they can be represented 
as in Figure \ref{Figures/Section2/OperatorsAndBlocks/Triangle.pdf}.
\\
The index of the $\mathsf{B}_{10}$  block is given by
\begin{align}\label{eq:indtriangle}
\mathcal{I}_{(1,0)}^{(N,M)}(\vec{x};\vec{y};u;t;c) & = \prod_{i=1}^N\prod_{j=1}^M \Gpq{(pq/t)^\half x_i^\pm y_j^\pm} \prod_{i=1}^N  \Gpq{t^{\frac{M-N+1}{2}} c x_j^\pm u^\pm} \nonumber\\
& \quad\times\prod_{j=1}^M\Gpq{(pq)^\half t^{\frac{N-M}{2}}c^{-1} y_j^\pm u^\pm} \,.
\end{align}

In the $3d$ limit we consider a real mass deformation that breaks the $USp$ groups down to $U$ and also a real mass deformation for $U(1)_c$ that gives mass to the two $SU(2)$ chiral doublets, thus leaving a $U(N)\times U(M)$ bifundamental hyper only. The process of integrating out these fields also produces background CS levels $\pm1$ and $\mp1$ for the $U(N)$ and $U(M)$ gauge nodes, respectively, where the signs depend on that of the real mass for $U(1)_c$. Nevertheless, these CS couplings always cancel out when gluing several building blocks together to form quivers. We will give more details on the $3d$ limit in Section \ref{sec:3d}.

\bigskip

\begin{figure}[!ht]
	\centering
	\includegraphics[scale=0.7]{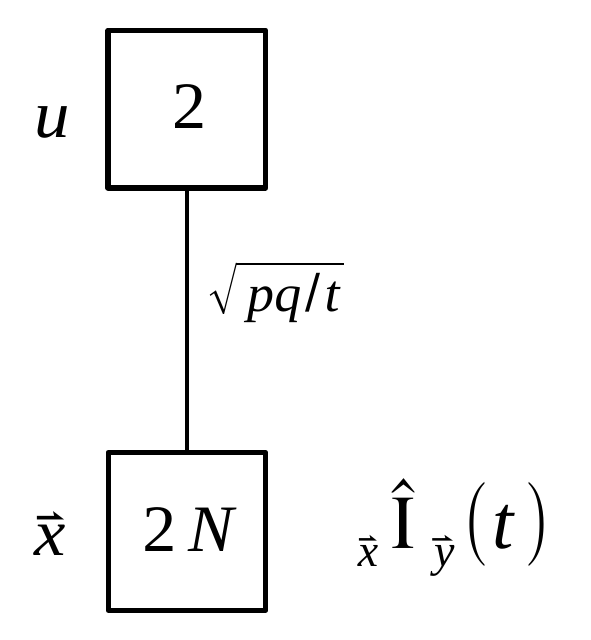}
	\caption{The $\mathsf{B}_{01}$ block.}
	\label{Figures/Section2/OperatorsAndBlocks/Flavor}
\end{figure}

\noindent\textbf{The $\mathsf{B}_{01}$ block.}
The second block is the one that we associate to a D5 or (0,1)-brane. In $3d$ this is just a single fundamental hypermultiplet, while its $4d$ uplift, the $\mathsf{B}_{01}$ block, is a pair of $USp(2N)$ fundamental chirals. Let us first consider the situation in which a D5 is suspended between the same number $N$ of D3-branes on the left and on the right so that these yield two $USp(2N)$ symmetries of the same rank. In order to give this QFT building block a structure with two $USp(2N)$ symmetries, we also attach to it an Identity-wall. Thus, we define the $\mathsf{B}_{01}$ block as in Figure \ref{Figures/Section2/OperatorsAndBlocks/Flavor}.
\\
To the $\mathsf{B}_{01}$ block we can associate the index expression
\begin{equation}
\mathcal{I}_{(0,1)}^{(N,N)} (\vec{x};\vec{y};u;t) = \prod_{j=1}^N \Gpq{(pq/t)^\half x_j^\pm u^\pm} {}_{\vec x}\hat{\mathbb{I}}_{\vec y}(t)\,.
\end{equation}

\begin{figure}[!ht]
	\includegraphics[width=\textwidth,center]{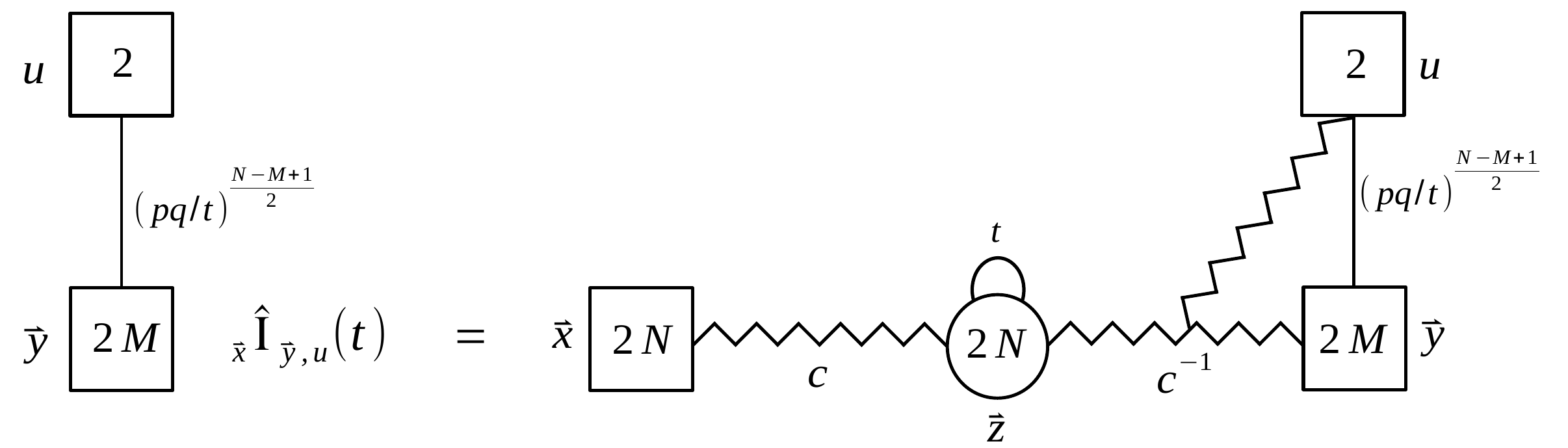}
	\caption{The asymmetric $\mathsf{B}_{01}$ block. }
	\label{Figures/Section2/OperatorsAndBlocks/Flavor_asymm}
\end{figure}

We can generalize this to the case corresponding to a  D5-brane suspended between different numbers of D3-branes, say $N$ D3's on the left and $M$ D3's on the right as shown in Figure \ref{Figures/Section2/OperatorsAndBlocks/Flavor_asymm}.

\noindent The associated index is given by
\begin{equation}
\label{eq:01NM}
\mathcal{I}_{(0,1)}^{(N,M)} (\vec{x};\vec{y};u;t) = \prod_{j=1}^M \Gpq{(pq/t)^\frac{N-M+1}{2} y_j^\pm u^\pm} {}_{\vec x}\hat{\mathbb{I}}_{\vec y,u}(t)\,,
\end{equation}
where ${}_{\vec x}\hat{\mathbb{I}}_{\vec y,v}(t)$ is the asymmetric Identity-wall defined in \eqref{eq:idop}.

\bigskip

\begin{figure}[!ht]
	\centering
	\includegraphics[scale=0.7]{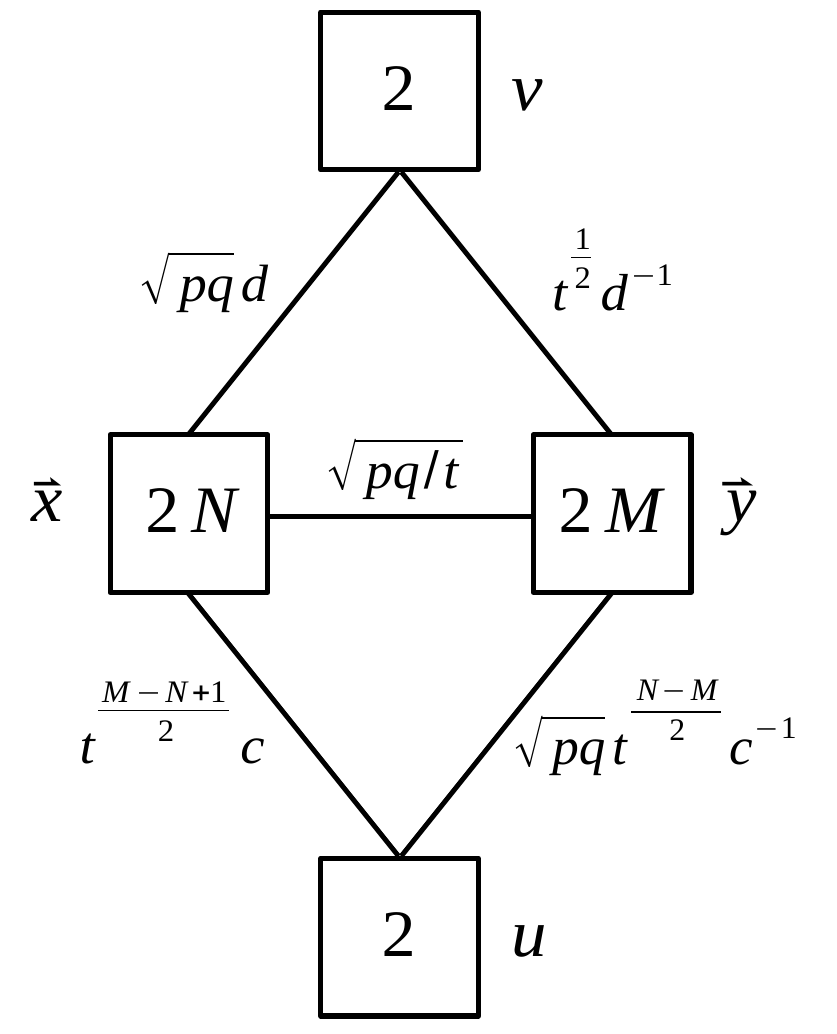}
	\caption{The $\mathsf{B}_{11}$ block.}\label{fig:B114d}
\end{figure}

\noindent\textbf{The $\mathsf{B}_{11}$ block.}
Finally, we consider the QFT building block associated to a $(1,1)$-brane. This wasn't considered in \cite{Hwang:2021ulb} and we propose it here to be a $USp(2N)\times USp(2M)$ bifundamental plus additional chiral fields that form a double triangle structure as in Figure \ref{fig:B114d}.\footnote{The two triangles can actually be combined to form a single triangle with an $SU(4)$ flavor node. Nevertheless, we will split the flavors into the two triangles so that only the $SU(2)^2\times U(1)$ subgroup is manifest because the two triangles and the associated $SU(2)$ symmetries will play different roles in the dualization algorithm.}

\noindent The corresponding index is
\begin{align}\label{eq:11NM}
\mathcal{I}_{(1,1)}^{(N,M)} (\vec{x};\vec{y};v;u;t;c;s) = & \prod_{i,j=1}^N \Gpq{(pq/t)^\half x_i^\pm y_j^\pm} \prod_{j=1}^N  \Gpq{(pq)^\half d x_j^\pm v^\pm} \Gpq{t^\half d^{-1} y_j^\pm v^\pm} \nonumber \\
&\times \Gpq{ t^{\tfrac{M-N+1}{2}} c x_j^\pm u^\pm} \Gpq{(pq)^\half  t^{\tfrac{N-M}{2}}  c^{-1} y_j^\pm u^\pm}\,.
\end{align}

When we consider the $3d$ limit, similarly to the case of the NS5-brane, the real mass deformation for $U(1)_c$ produces background CS levels for the two $U(N)$ nodes, which cancel out when considering a quiver. The real mass deformation for $U(1)_d$ also produces CS levels, but when considering a quiver these only cancel out if we have a stack of $(1,1)$-branes, while they remain if we have adjacent $(1,0)$ and $(1,1)$-branes. In such a case, they become dynamical CS couplings for the $U(N)$ gauge field arising from the D3's suspended between the $(1,0)$ and $(1,1)$-branes, as expected. One can also obtain the QFT building block corresponding to the $(1,-1)$-brane, which can be obtained as the $\mathcal{S}$-dualization of $(1,1)$. As we will see, both the $(1,1)$ and $(1,-1)$-branes correspond to double triangle building blocks with some relative charge difference, leading to the opposite CS levels in the $3d$ limit. Again, we postpone a more detailed discussion of the $3d$ counterpart to Section \ref{sec:3d}.

\subsection{Basic duality moves}
\label{subsec:4d_duality_moves}
In this section, we  present the basic duality moves involved in the dualization algorithm, which utilize all the ingredients we have introduced in the previous subsections. We will first review the $\mathsf{S}$-dualizations of the $\mathsf{B}_{10}$ and the $\mathsf{B}_{01}$ building blocks proposed in \cite{Hwang:2021ulb}. Then we will discuss new duality moves involving the other field theory ingredients we introduced.

\subsubsection{$\mathsf{S}$-dualization}

Let us first consider dualities for the QFT building blocks generated by the $\mathsf{S}$ operator.

\begin{figure}[!ht]
	\includegraphics[width=\textwidth,center]{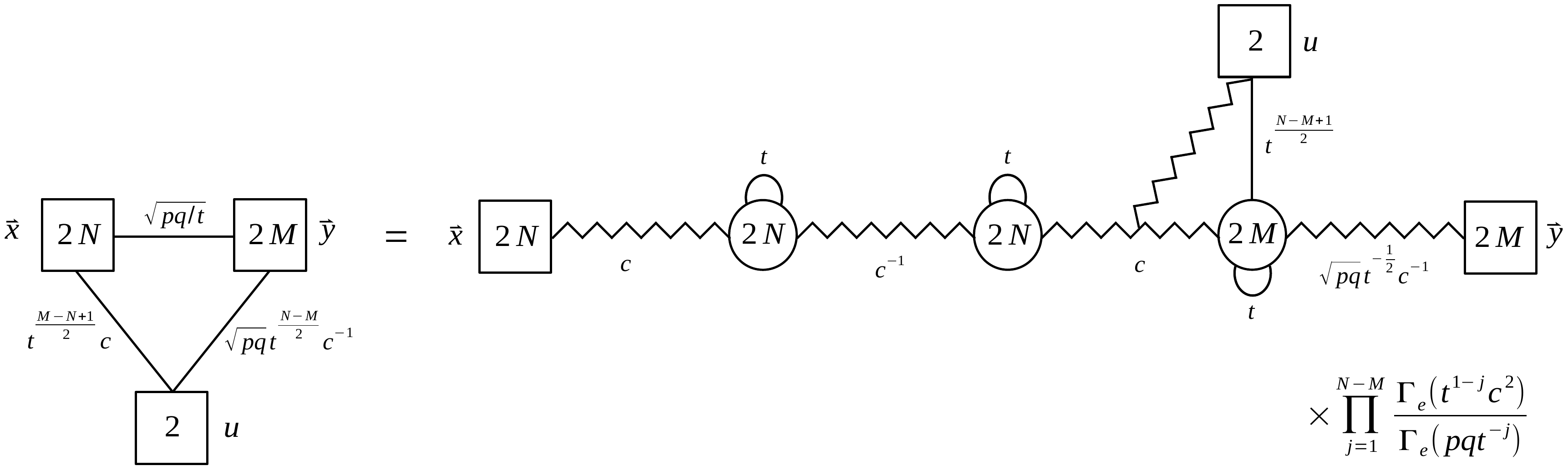}
	\caption{The $\mathsf{B}_{10}= \mathsf{S}\mathsf{B}_{01}  \mathsf{S}^{-1}$ duality move. Notice that for $N\neq M$ we have on the r.h.s.~the asymmetric $\mathsf{B}_{01}$ block, which is associated with a D5 with a different number $N$ and $M$ of D3's on each side. We represent the gauge singlet fields that are charged only under the abelian global symmetries and not the non-abelian ones by writing explicitly their index contribution.}
	\label{10S01}
\end{figure}

\bigskip
\noindent\textbf{The \boldmath$\mathsf{B}_{10}= \mathsf{S}\mathsf{B}_{01}  \mathsf{S}^{-1}$ duality move.}
We first consider the $4d$ QFT analogue of the $\mathsf{S}$-dualization of a D5-brane into an NS5-brane, which relates a $\mathsf{B}_{01}$ and a $\mathsf{B}_{10}$ block \cite{Hwang:2021ulb}. This can be associated with a genuine field theory duality relating the quiver theories in Figure \ref{10S01}.

This is a non-trivial IR duality that was derived by iteratively applying the Intriligator--Pouliot (IP) duality in \cite{Bottini:2021vms}.
At the level of the index, this dualization translates into the following integral identity:
\begin{eqnarray}
\label{eq:id1}
&&\mathcal{I}_{(1,0)}^{(N,M)}\left(\vec{x};\vec{y};u;t;ct^{\frac{M-N}{2}}\right) =\prod_{i=1}^{N-M}\frac{\Gpq{t^{1-i}c^2}}{\Gpq{pq\,t^{-i}}}\oint \udl{\vec{z}^{(1)}_N}\udl{\vec{z}^{(2)}_M}\Gd_N\left(\vec{z}^{(1)};t\right)\Gd_M\left(\vec{z}^{(2)};t\right)\nn\\
&&\quad\qquad\times\mathcal{I}_{\mathsf{S}}^{(N)}\left(\vec{x};\vec{z}^{(1)};t;c\right)
{\mathcal{I}_{(0,1)}^{(N,M)}\left(\vec{z}^{(1)};\vec{z}^{(2)};u;t\right)
\mathcal{I}_{\mathsf{S}}^{(M)}\left(\vec{z}^{(2)};\vec{y};t;(pq/t)^{\frac{1}{2}}c^{-1}\right)} \,,
\end{eqnarray}
where we recall that the index of the $\mathsf{B}_{10}$ and the $\mathsf{B}_{01}$ blocks were defined in \eqref{eq:indtriangle} and \eqref{eq:01NM} respectively.

\begin{figure}[!ht]
	\centering
	\includegraphics[width=\textwidth]{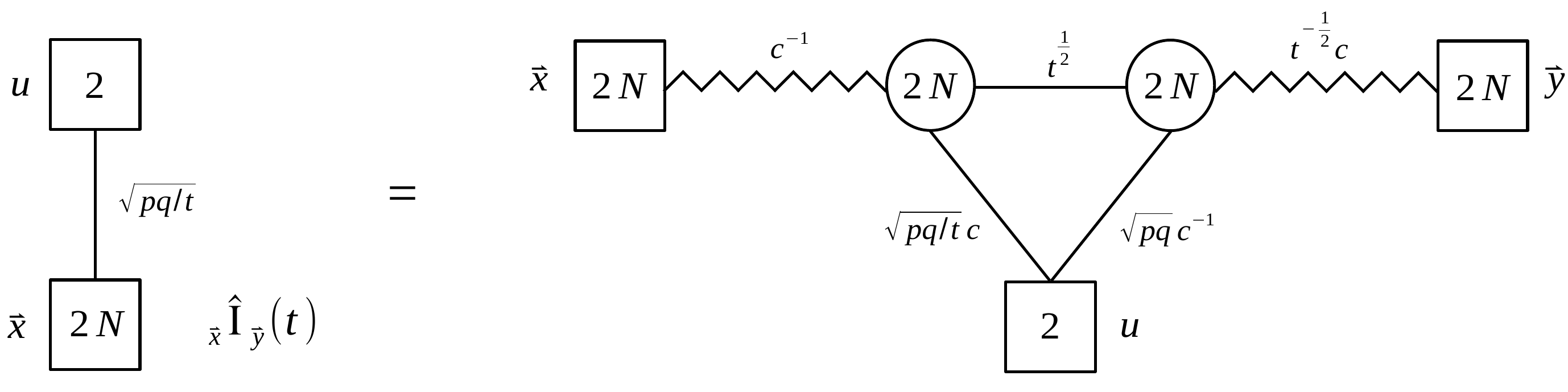}
	\caption{The $\mathsf{B}_{01}= \mathsf{S}\mathsf{B}_{10}  \mathsf{S}^{-1}$ duality move.}
	\label{01S10}
\end{figure}

\bigskip
\noindent\textbf{The \boldmath$\mathsf{B}_{01}= \mathsf{S}\mathsf{B}_{10}  \mathsf{S}^{-1}$ duality move.}
We next consider the QFT analogue of the $ \mathsf{S}$-dualization of an NS5-brane into a D5-brane. The quiver corresponding to this duality move is given in Figure  \ref{01S10}. 

\begin{figure}[!ht]
	\includegraphics[width=\textwidth,center]{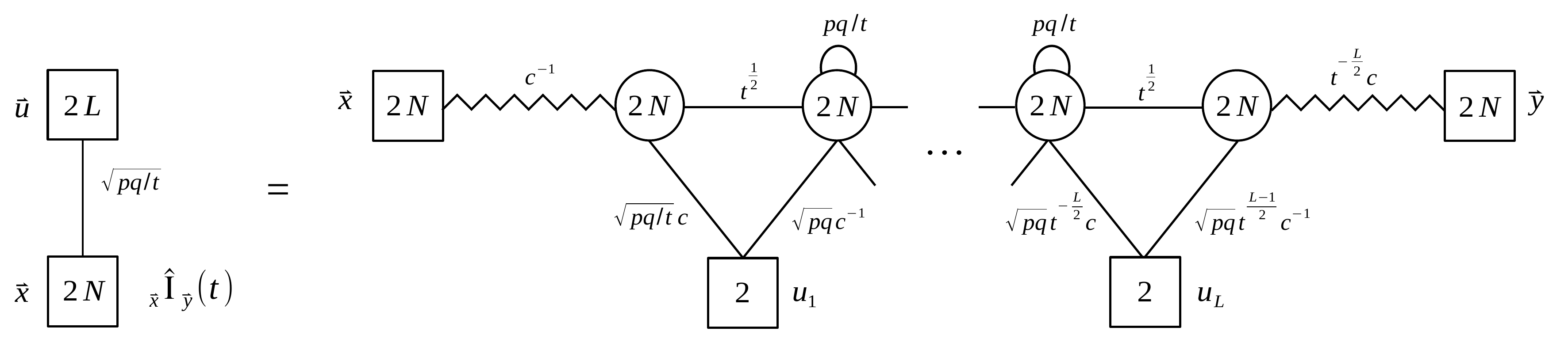}
	\caption{The $(\mathsf{B}_{01})^L= \mathsf{S}(\mathsf{B}_{10})^L  \mathsf{S}^{-1}$ duality move.}\label{L01S10}
\end{figure}

At the level of the index we have
\begin{eqnarray}
\label{eq:id2}
&&{\mathcal{I}_{(0,1)}^{(N,N)} (\vec{x};\vec{y};u;t)=\oint \udl{w^{(0)}_N}\udl{w^{(1)}_N} \Gd_N(\vec{w}^{(0)})  \Gd_N(\vec{w}^{(1)})} \nn\\
&&\quad\times{\mathcal{I}_{\mathsf{S}}^{(N)}(\vec{x};\vec{w}^{(0)};t;c^{-1}) \mathcal{I}_{(1,0)}^{(N,N)} \left(\vec{w}^{(0)};\vec{w}^{(1)};u;pq/t;c
\right)\mathcal{I}_{\mathsf{S}}^{(N)}(\vec{w}^{(1)};\vec{y};t;t^{-\frac{1}{2}}c) } \,.
\end{eqnarray}
For convenience, we also give the dualization of  $L$ D5-branes as $L$ NS5-branes between $\mathsf{S}$ and $\mathsf{S}^{-1}=\mathsf{S}$, shown in Figure \ref{L01S10}. 
\\
The index identity associated to this duality is explicitly
\begin{eqnarray}
&&{{}_{\vec{x}}\hat{\mathbb{I}}_{\vec{y}}(t)\prod_{i=1}^N\prod_{j=1}^L\Gpq{(pq/t)^{\frac{1}{2}}u_j^{\pm1}x_i^{\pm1}}=\oint\prod_{k=0}^L\udl{w^{(k)}_N} \Gd_N(\vec{w}^{(0)})} \nn\\
&&\quad\times{\mathcal{I}_{\mathsf{S}}^{(N)}(\vec{x};\vec{w}^{(0)};t;c^{-1}) \prod_{i=1}^L\mathcal{I}_{(1,0)}^{(N,N)} \left(\vec{w}^{(i-1)};\vec{w}^{(i)};u_i;pq/t;ct^{\frac{1-i}{2}}\right)}\nn\\
&&\quad\times{\prod_{k=1}^{L-1}\Gd_N(\vec{w}^{(k)};pq/t) \, \mathcal{I}_{\mathsf{S}}^{(N)}(\vec{w}^{(L)};\vec{y};t;t^{-\frac{L}{2}}c) \, \Gd_N(\vec{w}^{(L)}) \,.}
\end{eqnarray}

Also this result can be derived by iterating the IP duality as shown in \cite{Bottini:2021vms,Hwang:2021ulb}. Alternatively, it can be derived from the $\mathsf{S}$-dualization of a D5 into an NS5 by applying $\mathsf{S}$ on the left and $\mathsf{S}^{-1} = \mathsf{S}$ on the right and using the property we reviewed in the previous subsection that two concatenated $\mathsf{S}$-walls give an Identity-wall.

\begin{figure}[!ht]
	\includegraphics[width=\textwidth,center]{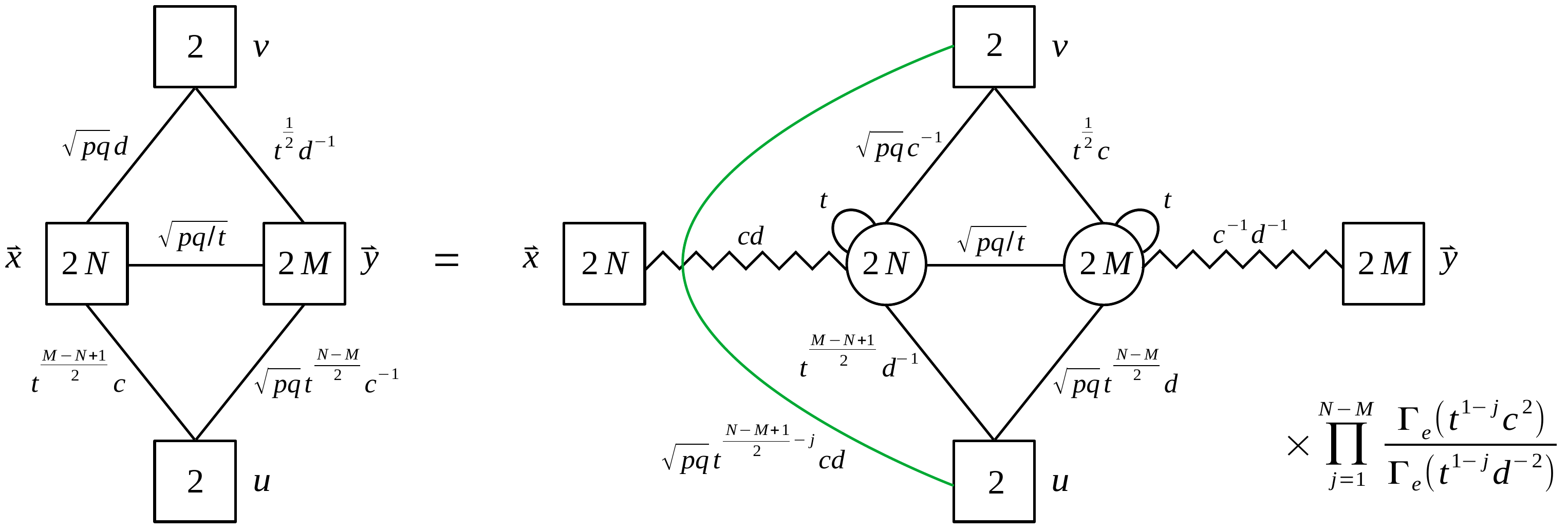}
	\caption{The $\mathsf{B}_{11}= \mathsf{S}\mathsf{B}_{1-1}  \mathsf{S}^{-1}$ duality move. The green line denotes a set of gauge singlet fields labelled $j=1,\dots,N-M$.}
	\label{11s1-1}
\end{figure}

\bigskip
\noindent\textbf{The \boldmath$\mathsf{B}_{11}= \mathsf{S}\mathsf{B}_{1-1}  \mathsf{S}^{-1}$ duality move.}
The last $\mathsf{S}$-dualization we consider is that of a (1,-1) into a (1,1)-brane, which can be represented as in Figure \ref{11s1-1}.

This duality move was not considered in \cite{Hwang:2021ulb}. We provide its derivation in Appendix \ref{proof_dualities}, where the main ingredient is the braid duality of Figure \ref{braid}. The corresponding index identity is 
\begin{align}\label{eq:11S1-1}
&\mathcal{I}_{(1,1)}^{(N,M)} (\vec{x};\vec{y};v;u;t;c;d) \nonumber \\
&= \prod_{j = 1}^{N-M} \frac{\Gpq{t^{1-j} c^2}}{\Gpq{t^{1-j} d^{-2}}} \Gpq{\sqrt{pq} t^{\frac{N-M+1}{2}-j} c d u^\pm v^\pm} \oint \udl{\vec{z}_N}\udl{\vec{w}_M} \Gd_N\left(\vec{z};t\right)  \Gd_M\left(\vec{w};t\right) \nonumber \\
&\quad \times \mathcal{I}_{\mathsf{S}}^{(N)}\left(\vec{x};\vec{z};t;c d\right) \mathcal{I}_{(1,-1)}^{(N,M)} (\vec{z};\vec{w};v;u;t;c;d) \mathcal{I}_{\mathsf{S}}^{(M)}\left(\vec{w};\vec{y};t;c^{-1} d^{-1}\right) ,
\end{align}
where the index contribution of the  $\mathsf{B}_{1-1}$ block is
\begin{equation}
\mathcal{I}_{(1,-1)}^{(N,M)} (\vec{z};\vec{w};v;u;t;c;d)=\mathcal{I}_{(1,1)}^{(N,M)} (\vec{z};\vec{w};v;u;t;d^{-1};c^{-1})\,,
\end{equation}
with $\mathcal{I}_{(1,1)}^{(N,M)}$ being defined in \eqref{eq:11NM}. 

Indeed, since the difference between the $(1,1)$-brane and the $(1,-1)$-brane is only a relative notion compared to other types of branes such as NS5 and D5, it is not surprising that they correspond to the same type of QFT building blocks with different charges. Assuming the $3d$ limit prescription involving the real mass deformation implemented by sending $\log c \rightarrow -\infty, \, \log d \rightarrow +\infty$ with hierarchy $\log(c d) \rightarrow -\infty$, one can distinguish the two branes by looking at the sign of the $U(1)_d$ charges of each block. For example, the left and right edges of the upper triangle of the $\mathsf{B}_{11}$ block that appears on the l.h.s.~of Figure \ref{11s1-1} have $U(1)_d$ charges +1 and -1, leading to CS levels +1 and -1 in the $3d$ limit, respectively. This is consistent with the relation
\begin{align}
\mathsf{B}_{11} = \mathsf{T} \mathsf{B}_{10} \mathsf{T}^{-1} \,,
\end{align}
which we will examine in the next subsubsection in detail. One the other hand, the edges of the lower triangle of the $\mathsf{B}_{1-1}$ block that appears on the r.h.s.~of Figure \ref{11s1-1} have the $U(1)_d$ charges -1 and +1, leading to CS levels -1 and +1 in the $3d$ limit, respectively. One should remember that there are extra CS level contributions from $U(1)_c$ charged fields, but, as mentioned before, they turn out to be canceled when the blocks are glued to form an entire quiver theory.

\subsubsection{$\mathsf{T}$-dualization}

We now consider dualities for the QFT building blocks generated by the $\mathsf{T}$ operator.

\begin{figure}[!ht]
	\includegraphics[width=\textwidth,center]{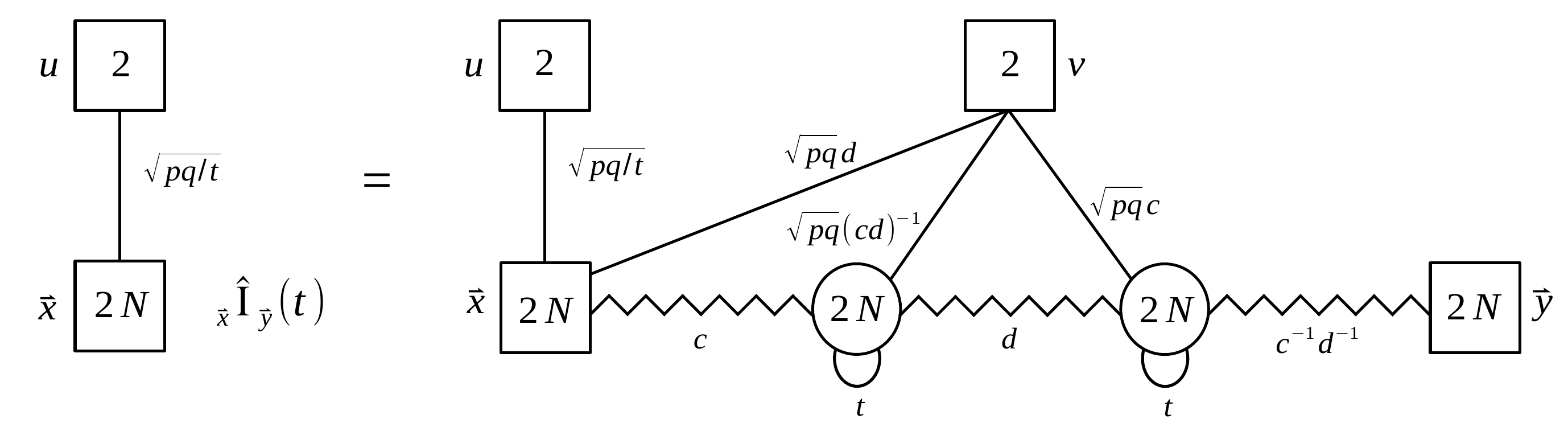}
	\caption{The $\mathsf{B}_{01}= \mathsf{T}\mathsf{B}_{01} \mathsf{T}^{-1}=
	\mathsf{T}\mathsf{B}_{01}  \mathsf{STSTS}$ duality move.}
	\label{01t01}
\end{figure}

\bigskip
\noindent\textbf{The \boldmath$\mathsf{B}_{01}= \mathsf{T}\mathsf{B}_{01}  \mathsf{T}^{-1}$ duality move.}
This move states that the $\mathsf{B}_{01}$ block is  transparent to the $\mathsf{T}$-dualization. 
To show this, we first recall that $\mathsf{T}^{-1} = \mathsf{STSTS}$. 
The quiver duality associated with this relation is then given 
in Figure \ref{01t01},  where according to  our gluing prescription we turn on a cubic superpotential for each triangle. This duality trivially follows from the relation in Figure \ref{fig:TSTSTS}.
\\

\noindent As an index identity we have
\begingroup\allowdisplaybreaks
\begin{align}\label{eq:T01}
&\mathcal{I}_{(0,1)}^{(N,N)} (\vec{x};\vec{y};u;t)\nonumber\\
&=\oint\left(\prod_{k=1}^6\udl{w^{(k)}_N}\Gd_N(\vec{z}^{(k)};t)\right)\mathcal{I}_{\mathsf{T}}^{(N)}(\vec{x};\vec{w}^{(1)};v;t;d)\mathcal{I}_{(0,1)}^{(N,N)} (\vec{w}^{(1)};\vec{w}^{(2)};u;t)\nn\\
&\qquad\times\mathcal{I}_{\mathsf{S}}^{(N)}(\vec{w}^{(2)};\vec{w}^{(3)};t;c)\mathcal{I}_{\mathsf{T}}^{(N)}(\vec{w}^{(3)};\vec{w}^{(4)};v;t;c^{-1} d^{-1})\mathcal{I}_{\mathsf{S}}^{(N)}(\vec{w}^{(4)};\vec{w}^{(5)};t;d)\nn\\
&\qquad\times\mathcal{I}_{\mathsf{T}}^{(N)}(\vec{w}^{(5)};\vec{w}^{(6)};v;t;c)\mathcal{I}_{\mathsf{S}}^{(N)}(\vec{w}^{(6)};\vec{y};t;c^{-1} d^{-1})\nonumber \\
&=\oint \left(\prod_{i = 1}^2 \udl{\vec{z}^{(i)}_N} \Gd_N(\vec{z}^{(i)};t)\right) \prod_{j=1}^N \Gpq{(pq/t)^\half x_j^\pm u^\pm} \prod_{j=1}^N \Gamma((pq)^\frac12 d x_j^\pm v^\pm) \mathcal{I}_{\mathsf{S}}^{(N)}(\vec{x};\vec{z}^{(1)};t;c) \nonumber \\
&\quad \times \prod_{j=1}^N \Gamma((pq)^\frac12 c^{-1} d^{-1} z^{(1)}_j{}^\pm v^\pm) \mathcal{I}_{\mathsf{S}}^{(N)}(\vec{z}^{(1)};\vec{z}^{(2)};t;d) \prod_{j=1}^N \Gamma((pq)^\frac12 c z^{(2)}_j{}^\pm v^\pm) \nonumber\\
&\quad\times\mathcal{I}_{\mathsf{S}}^{(N)}(\vec{z}^{(2)};\vec{y};t;c^{-1} d^{-1}) \,.
\end{align}
\endgroup 

\begin{figure}[!ht]
	\includegraphics[width=\textwidth,center]{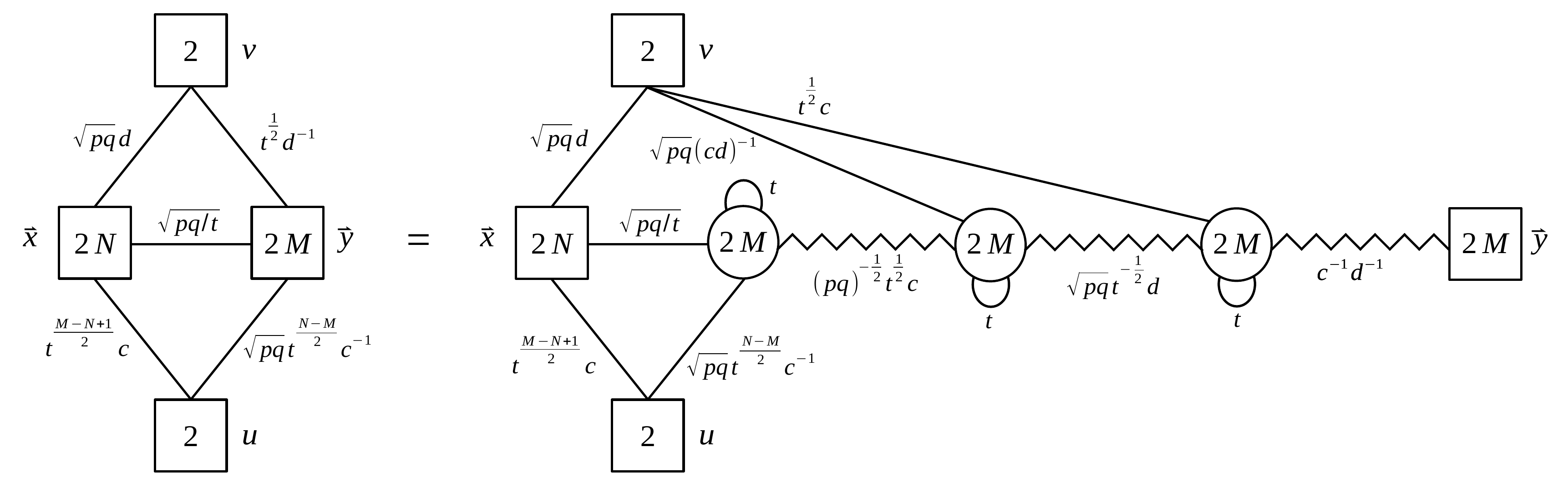}
	\caption{The $\mathsf{B}_{11}= \mathsf{T}\mathsf{B}_{10} \mathsf{T}^{-1} = \mathsf{T}\mathsf{B}_{10}\mathsf{STSTS}$ duality move.}
	\label{11t10}
\end{figure}

\begin{figure}[!ht]
	\includegraphics[width=.6\textwidth,center]{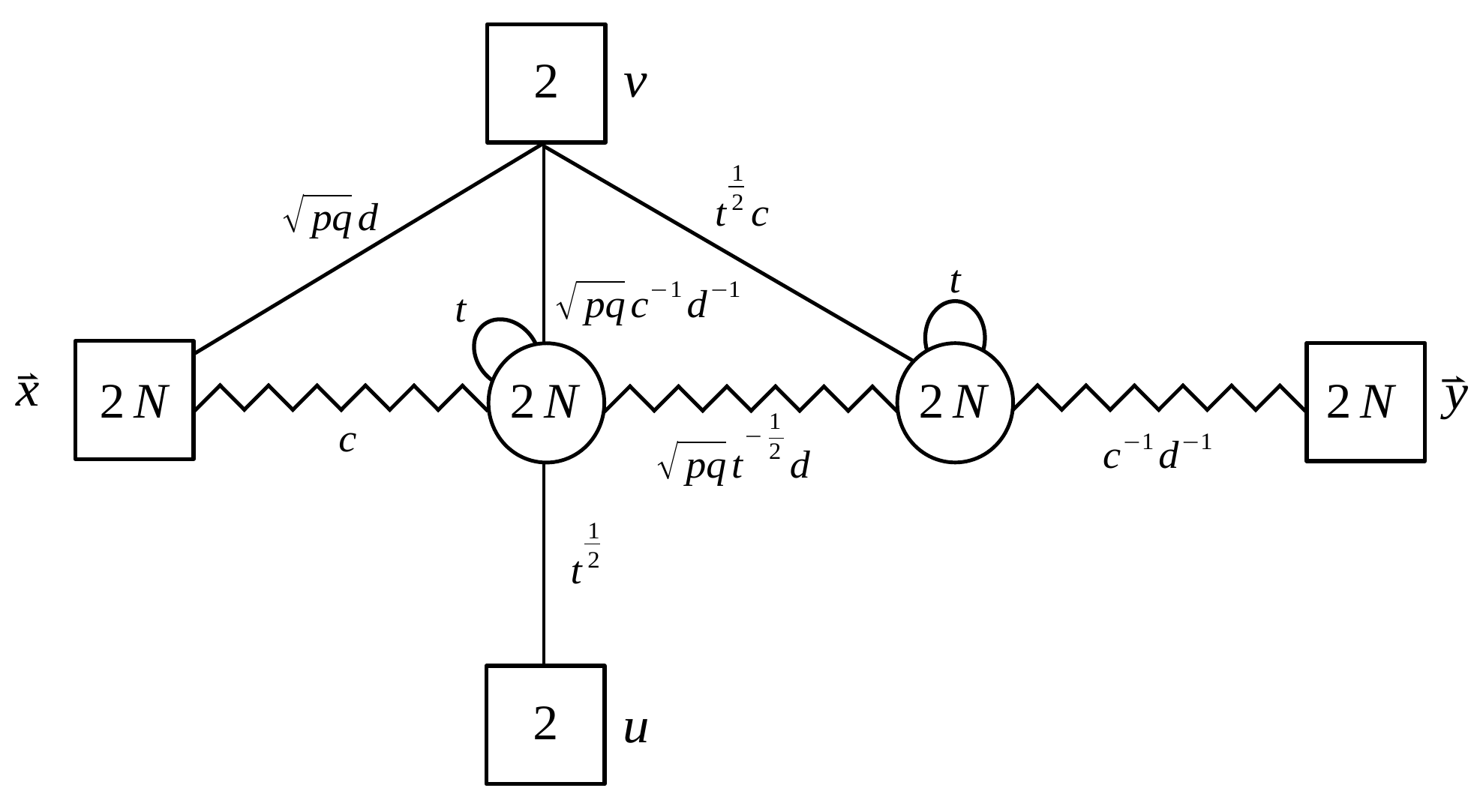}
	\caption{The quiver representation of $\mathsf{TS}\mathsf{B}_{01}\mathsf{TSTS}$, which is equal to $\mathsf{B}_{11}= \mathsf{T}\mathsf{B}_{10} \mathsf{STSTS}$. To obtain this quiver from the one on the right of Figure \ref{11t10}, we use the duality move $\mathsf{B}_{10}= \mathsf{S}\mathsf{B}_{01} \mathsf{S}$ of Figure \ref{10S01}.}
\label{ov}
\end{figure}

\bigskip
\noindent\textbf{The \boldmath$\mathsf{B}_{11}= \mathsf{T}\mathsf{B}_{10}  \mathsf{T}^{-1}$ duality move.}
The quiver representation of this duality move is given in Figure \ref{11t10}.
We can convince ourselves that the superpotentials are consistent with our gluing rules by observing that 
since $\mathsf{B}_{10}= \mathsf{S}\mathsf{B}_{01} \mathsf{S}$, we have
$\mathsf{B}_{11}= \mathsf{T}\mathsf{B}_{10} \mathsf{STSTS}=  \mathsf{TS}\mathsf{B}_{01}\mathsf{TSTS} $ 
which is shown (for simplicity for $N=M$) in Figure \ref{ov} and is clearly consistent with the rule of turning on the cubic 
superpotential in \eqref{eq:triangle}
whenever an $\mathsf{S}$-wall is sandwiched between two $\mathsf{T}$-walls.

\newpage
\noindent At the level of the index we have
\begingroup
\allowdisplaybreaks
\begin{align}\label{eq:T10}
&\mathcal{I}_{(1,1)}^{(N,M)}(\vec{x};\vec{y};v;u; t;c;d)
=\oint\left(\prod_{k=1}^6\udl{w^{(k)}_N}\Gd_N(\vec{z}^{(k)};t)\right)\mathcal{I}_{\mathsf{T}}^{(N)}(\vec{x};\vec{w}^{(1)};v;t;d)\mathcal{I}_{(1,0)}^{(N,M)} (\vec{w}^{(1)};\vec{w}^{(2)};u;t;c)\nn\\
&\qquad\times\mathcal{I}_{\mathsf{S}}^{(M)}(\vec{w}^{(2)};\vec{w}^{(3)};t;(pq/t)^{-\frac12} c)\mathcal{I}_{\mathsf{T}}^{(M)}(\vec{w}^{(3)};\vec{w}^{(4)};v;t;c^{-1} d^{-1})\mathcal{I}_{\mathsf{S}}^{(M)}(\vec{w}^{(4)};\vec{w}^{(5)};t;(pq/t)^\frac12 d) \nonumber \\
&\qquad\times \mathcal{I}_{\mathsf{T}}^{(M)}(\vec{w}^{(5)};\vec{w}^{(6)};v;t;c)\mathcal{I}_{\mathsf{S}}^{(M)}(\vec{w}^{(6)};\vec{y};t;c^{-1} d^{-1})\nn\\
&=\oint \left(\prod_{i = 1}^3 \udl{\vec{z}^{(i)}_M} \Gd_M(\vec{z}^{(i)};t)\right) \prod_{j=1}^N \Gamma((pq)^\frac12 d x_j{}^\pm v^\pm) \mathcal{I}_{(1,0)}^{(N,M)}(\vec{x};\vec{z}^{(1)};u;t;c) \nonumber \\
&\quad \times \mathcal{I}_{\mathsf{S}}^{(M)}(\vec{z}^{(1)};\vec{z}^{(2)};t;(pq/t)^{-\frac12} c) \prod_{j=1}^M \Gamma((pq)^\frac12 c^{-1} d^{-1} z^{(2)}_j{}^\pm v^\pm) \mathcal{I}_{\mathsf{S}}^{(M)}(\vec{z}^{(2)};\vec{z}^{(3)};t;(pq/t)^\frac12 d) \nonumber \\
&\quad \times \prod_{j=1}^M \Gamma(t^\frac12 c z^{(3)}_j{}^\pm v^\pm) \mathcal{I}_{\mathsf{S}}^{(M)}(\vec{z}^{(3)};\vec{y};t;c^{-1} d^{-1}) \,.
\end{align}
\endgroup

The duality move of Figure \ref{11t10}  can be easily proven by applying the braid duality move of Figure \ref{braid} to the second gauge node
on the r.h.s., removing the massive fields and reconstructing an Identity-wall from the fusion of the last two  $\mathsf{S}$-walls.

\begin{figure}[!ht]
	\includegraphics[width=\textwidth,center]{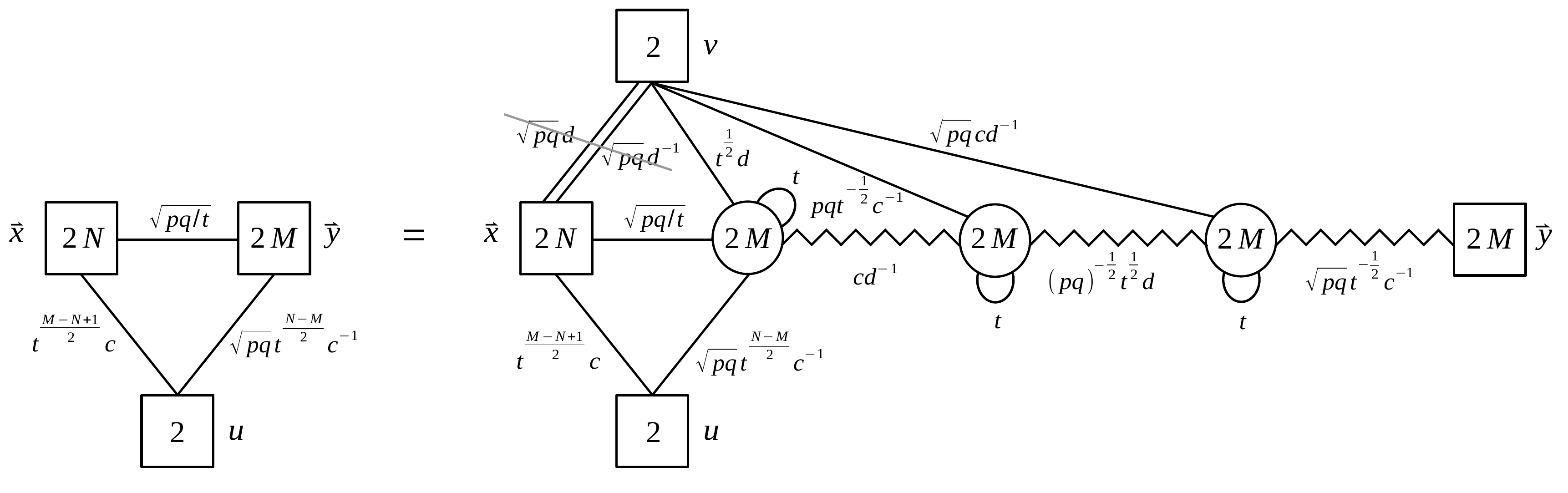}
	\caption{The 
	$\mathsf{B}_{10}= \mathsf{T}\mathsf{B}_{1-1}  \mathsf{T}^{-1}=\mathsf{T}\mathsf{B}_{1-1}  \mathsf{STSTS}$ duality move. On the r.h.s.~we included   two massive fields (barred by the grey line)  to explicitly display the $\mathsf{B}_{1-1}$ block between the $\mathsf{T}$ and $\mathsf{T}^{-1}$ operators. }
	\label{10t1-1}
\end{figure}

\bigskip
\noindent\textbf{The \boldmath$\mathsf{B}_{10}= \mathsf{T}\mathsf{B}_{1-1}  \mathsf{T}^{-1}$ duality move.}
The quiver associated with this duality move is given in Figure \ref{10t1-1}, and the corresponding index identity is given by
\begin{align}\label{eq:T11}
&\mathcal{I}_{(1,0)}^{(N,M)}(\vec{x};\vec{y};u;t;c) \nonumber \\
&=\oint \left(\prod_{i = 1}^3 \udl{\vec{z}^{(i)}_M} \Gd_M(\vec{z}^{(i)};t)\right) \prod_{j=1}^N \Gamma((pq)^\frac12 d x_j{}^\pm v^\pm) \mathcal{I}_{(1,1)}^{(N,M)}(\vec{x};\vec{z}^{(1)};v;u;t;c;d^{-1}) \nonumber \\
&\quad \times \mathcal{I}_{\mathsf{S}}^{(M)}(\vec{z}^{(1)};\vec{z}^{(2)};t;c d^{-1}) \prod_{j=1}^M \Gamma(pq t^{-\frac12} c^{-1} z^{(2)}_j{}^\pm v^\pm) \mathcal{I}_{\mathsf{S}}^{(M)}(\vec{z}^{(2)};\vec{z}^{(3)};t;(pq/t)^{-\frac12} d) \nonumber \\
&\quad \times \prod_{j=1}^M \Gamma((pq)^\frac12 c d^{-1} z^{(3)}_j{}^\pm v^\pm) \mathcal{I}_{\mathsf{S}}^{(M)}(\vec{z}^{(3)};\vec{y};t;(pq/t)^\frac12 c^{-1}) \,,
\end{align}
where from now on we will only give the expressions where we have already replaced the contribution of the $\mathsf{T}$-walls and simplified all Identity-walls.

Notice that on the r.h.s.~there is an $SU(2)_v$ symmetry that instead doesn't appear on the l.h.s.. Related to this, we can see that in the index identity \eqref{eq:T11} the fugacity $v$ appears on the r.h.s.~but not on the l.h.s.. Nevertheless, the identity holds for any value of $v$, which tells us that also the r.h.s.~is secretly independent of it. Remember that the index is counting operators in our theory graded by their quantum numbers, as one can see by expanding it as a power series. Each term of the expansion indeed corresponds to an operator in the theory whose quantum numbers are encoded in the powers of the corresponding fugacity that appear. We can then understand our observation as the fact that there is no operator in the spectrum of our theory that transforms under $SU(2)_v$, or in other words, the symmetry acts trivially on the spectrum of the low energy theory. Consistently with this, we find that all the anomalies involving $SU(2)_v$ vanish. We will see many examples of this phenomenon in the rest of the paper.

Again this duality move can be easily proven by applying the braid duality move of Figure \ref{braid} to the second gauge node on the r.h.s., removing the massive fields and reconstructing an Identity-wall from the fusion of the last two  $\mathsf{S}$-walls.

\subsubsection{$\mathsf{T}^T$-dualization}

Finally, we consider dualities for the QFT building blocks generated by the $\mathsf{T}^T$ operator.

\begin{figure}[!ht]
	\includegraphics[width=\textwidth,center]{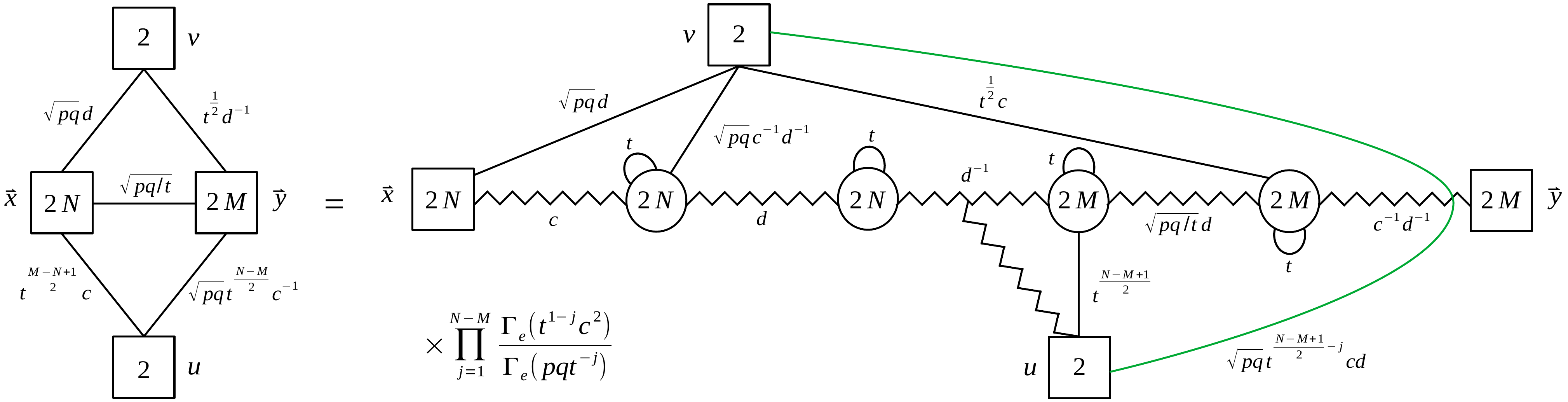}
	\caption{The $\mathsf{B}_{11}= \mathsf{T}^T\mathsf{B}_{01}  (\mathsf{T}^T)^{-1}$ duality move. The green line denotes gauge singlet fields labelled by $j=1,\dots,N-M$.}
	\label{11tt01}
\end{figure}

\bigskip
\noindent\textbf{The \boldmath$\mathsf{B}_{11}= \mathsf{T}^T\mathsf{B}_{01}  (\mathsf{T}^T)^{-1}$ duality move.}
Remembering that  $(\mathsf{T}^T)^{-1}= \mathsf{STS}$ we find that the duality move associated with this relation is as given in Figure \ref{11tt01}.

\noindent At the level of the index this translates into the identity
\begingroup\allowdisplaybreaks
\begin{align}\label{eq:TT01}
& \mathcal{I}_{(1,1)}^{(N,M)}(\vec{x};\vec{y};v;u; t;c;d) \nonumber \\
&=\prod_{j = 1}^{N-M} \frac{\Gpq{t^{1-j} c^2}}{\Gpq{pq t^{-j}}} \Gpq{(pq)^\frac12 t^{\frac{N-M+1}{2}-j} c d v^\pm u^\pm} \nonumber \\
&\quad\times\oint \udl{\vec{z}_N} \Gd_N(\vec{z};t) \left(\prod_{i = 1}^2 \udl{\vec{w}^{(i)}_M} \Gd_M(\vec{w}^{(i)};t)\right) \prod_{j=1}^N \Gamma((pq)^\frac12 d x_j{}^\pm v^\pm)  \nonumber \\
&\quad \times \mathcal{I}_{\mathsf{S}}^{(N)}(\vec{x};\vec{z};t;c)\prod_{j=1}^N \Gamma((pq)^\frac12 c^{-1} d^{-1} z_j{}^\pm v^\pm) \mathcal{I}_{(0,1)}^{(N,M)}(\vec{z};\vec{w}^{(1)};u;pq/t) \nonumber \\
&\quad \times \mathcal{I}_{\mathsf{S}}^{(M)}(\vec{w}^{(1)};\vec{w}^{(2)};t;(pq/t)^\frac12 d) \prod_{j=1}^M \Gamma(t^\frac12 c w^{(2)}_j{}^\pm v^\pm) \mathcal{I}_{\mathsf{S}}^{(M)}(\vec{w}^{(2)};\vec{y};t;c^{-1} d^{-1}) \,,
\end{align}
\endgroup
where $\mathcal{I}_{(0,1)}^{(N,M)}(\vec{x};\vec{y};u;t) $ is defined in \eqref{eq:01NM}. We give a proof of this duality in Appendix \ref{proof_dualities}.

\begin{figure}[!ht]
	\includegraphics[width=\textwidth,center]{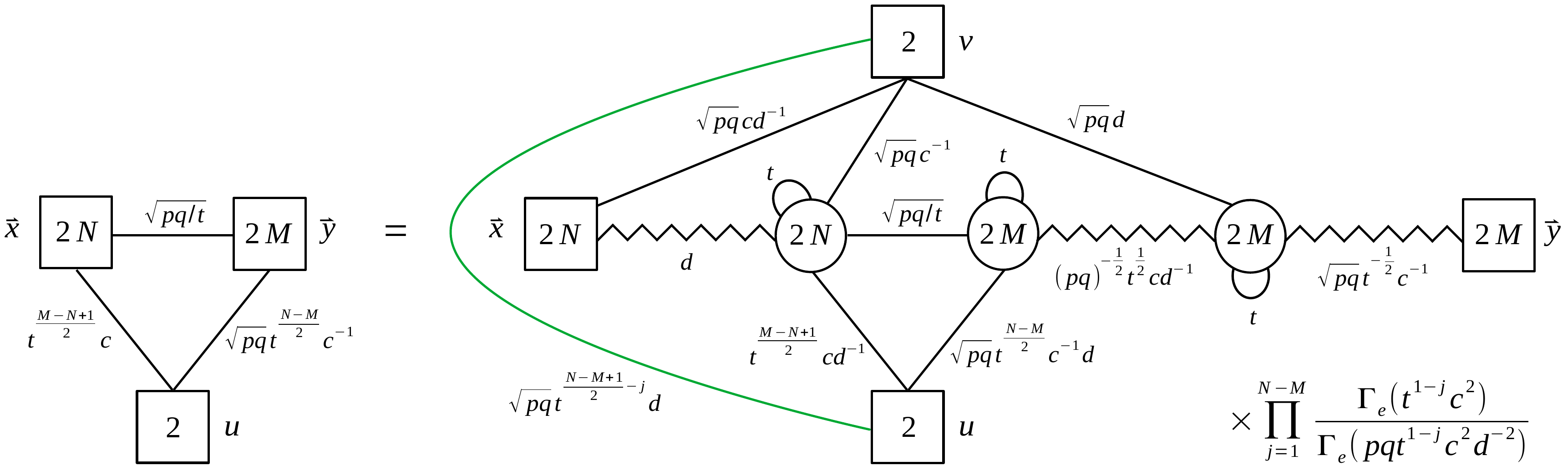}
	\caption{The $\mathsf{B}_{10}= \mathsf{T}^T\mathsf{B}_{10}  (\mathsf{T}^T)^{-1}$ duality move. The green line denotes gauge singlet fields labelled by $j=1,\dots,N-M$.}
	\label{10tt10}
\end{figure}

\bigskip
\noindent\textbf{The \boldmath$\mathsf{B}_{10}= \mathsf{T}^T\mathsf{B}_{10}  (\mathsf{T}^T)^{-1}$ duality move.}
This move states that the $\mathsf{B}_{10}$ block is  transparent to the $\mathsf{T}^T$-dualization. The corresponding duality move is represented in Figure \ref{10tt10}.

\noindent The index identity associated to this duality is
\begin{align}\label{eq:TT10}
&\mathcal{I}_{(1,0)}^{(N,M)}(\vec{x};\vec{y};u;t;c)\nonumber \\
&=\prod_{j = 1}^{N-M} \frac{\Gpq{t^{1-j} c^2}}{\Gpq{t^{1-j} c^2 d^{-2}}} \Gpq{(pq)^\frac12 t^{\frac{N-M+1}{2}-j} d} \nonumber \\
&\quad\times \oint \udl{\vec{z}_N} \Gd_N(\vec{z};t) \left(\prod_{i = 1}^2 \udl{\vec{w}^{(i)}_M} \Gd_M(\vec{w}^{(i)};t)\right) \prod_{j=1}^N \Gamma((pq)^\frac12 c d^{-1} x_j{}^\pm v^\pm) \mathcal{I}_{\mathsf{S}}^{(N)}(\vec{x};\vec{z};t;d) \nonumber \\
&\quad \times \prod_{j=1}^N \Gamma((pq)^{\frac12} c^{-1} z_j{}^\pm v^\pm) \mathcal{I}_{(1,0)}^{(N,M)}(\vec{z};\vec{w}^{(1)};u;t;c d^{-1}) \mathcal{I}_{\mathsf{S}}^{(M)}(\vec{w}^{(1)};\vec{w}^{(2)};t;(pq/t)^{-\frac12} c d^{-1}) \nonumber \\
&\quad \times \prod_{j=1}^M \Gamma((pq)^\frac12 d w^{(2)}_j{}^\pm v^\pm) \mathcal{I}_{\mathsf{S}}^{(M)}(\vec{w}^{(2)};\vec{y};t;(pq/t)^\frac12 c^{-1}) \,.
\end{align}
We give a proof of this duality move in Appendix \ref{proof_dualities}.

\begin{figure}[!ht]
	\includegraphics[width=\textwidth,center]{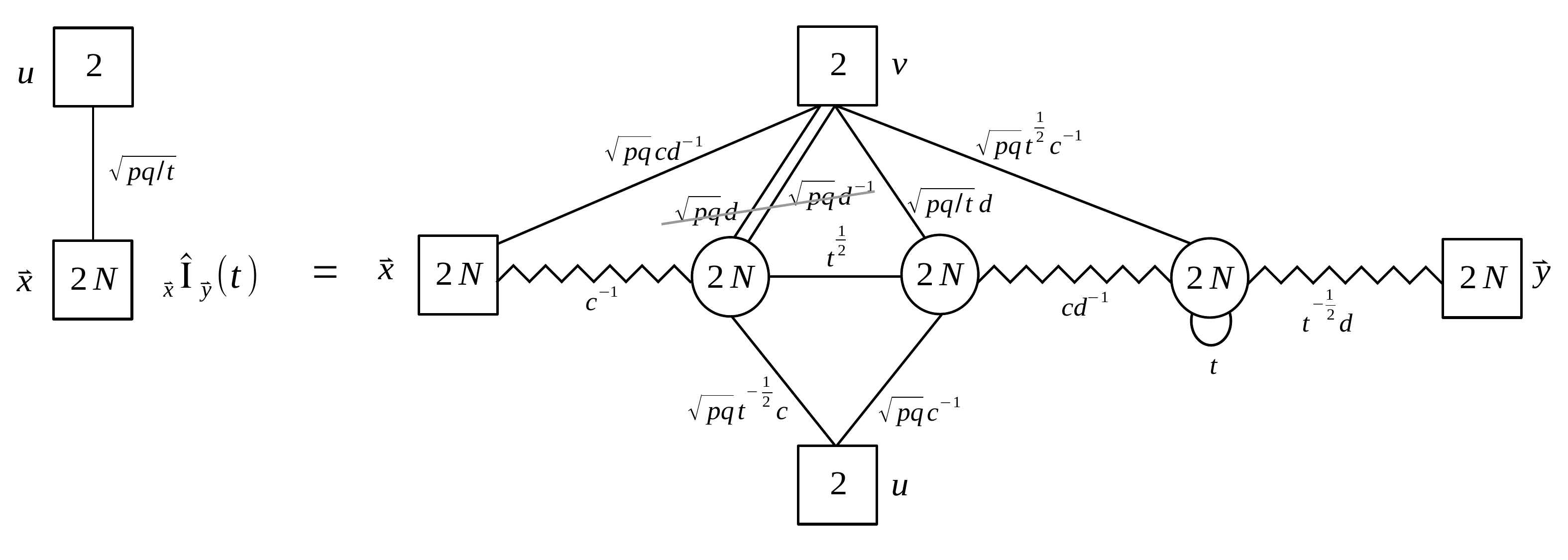}
	\caption{The $\mathsf{B}_{01}= \mathsf{T}^T\mathsf{B}_{1-1}  (\mathsf{T}^T)^{-1}$ duality move.
	On the r.h.s.~two massive fields are included to explicitly display the $\mathsf{B}_{1-1}$ block between the $\mathsf{T}^T$ and $(\mathsf{T}^T)^{-1}$ operators.}
	\label{10tt1-1}
\end{figure}

\bigskip
\noindent\textbf{The \boldmath$\mathsf{B}_{01}= \mathsf{T}^T\mathsf{B}_{1-1}  (\mathsf{T}^T)^{-1}$ duality move.} Finally, we have the $\mathsf{T}^T$ dual of a $\mathsf{B}_{1-1}$ block giving a $\mathsf{B}_{01}$ block, which we represent in Figure \ref{10tt1-1}.
The resulting index identity is given by
\begin{align}\label{eq:TT11}
& \prod_{j=1}^N \Gpq{(pq/t)^\half x_j^\pm u^\pm} {}_{\vec x}\hat{\mathbb{I}}_{\vec y}(t) \nonumber \\
&=\oint \left(\prod_{i = 1}^3 \udl{\vec{z}^{(i)}_N} \Gd_N(\vec{z}^{(i)};t)\right) \prod_{j=1}^N \Gamma((pq)^\frac12 c d^{-1} x_j{}^\pm v^\pm) \mathcal{I}_{\mathsf{S}}^{(M)}(\vec{x};\vec{z}^{(1)};t;c^{-1})  \nonumber \\
&\quad \times \prod_{j=1}^N \Gamma((pq)^{\frac12} d z^{(1)}_j{}^\pm v^\pm)\mathcal{I}_{(1,1)}^{(N)}(\vec{z}^{(1)};\vec{z}^{(2)};v;u;pq/t;c;d^{-1}) \mathcal{I}_{\mathsf{S}}^{(N)}(\vec{z}^{(2)};\vec{z}^{(3)};t;c d^{-1}) \nonumber \\
&\quad \times \prod_{j=1}^N \Gamma((pq t)^\frac12 c^{-1} z^{(3)}_j{}^\pm v^\pm) \mathcal{I}_{\mathsf{S}}^{(M)}(\vec{z}^{(3)};\vec{y};t;t^{-\frac12} d) \,.
\end{align}
We give a proof of this duality move in Appendix \ref{proof_dualities}.

\section{Dualization algorithm, VEVs, RG flows and HW moves}
\label{sec:algorithm}

In this section, we describe the local dualization algorithm using the ingredients we introduced in the previous section. Specifically, we illustrate the implementation of each step in the mirror-dualization of  the SQCD.

The first step of the dualization algorithm consist in chopping a theory into QFT blocks by un-gauging each gauge node. This step produces several QFT blocks as well as non-dynamical vector multiplets and chirals fields in the antisymmetric representations of the frozen gauge symmetries. 
It is important to keep track of these  fields to restore the correct gauging after the dualization.

In the second step, we dualize each QFT block by means of the basic  duality moves.

In the third step, we restore the gauging of the original gauge nodes, which generates several Identity-walls.  In case only symmetric Identity-walls are generated we just need to implement  them to read out the dual theory. If asymmetric walls are generated instead, we need to take care of their effect since fields generically acquire VEVs when we implement the identifications that the asymmetric walls prescribe.

In the fourth step, when necessary, we  study the RG flow triggered by these VEVs to obtain the final IR dual theory. There are various strategies we can follow at this point. We will see below how we can efficiently study the sequential Higgsing generated by these VEVs at the level of the index
or alternatively we can  reach the final IR configuration by a sequential applications of  the IP duality, which basically trades VEVs for massive deformations.

We will also see a third strategy where the effect of this sequential Higgsing or the iteration of the IP duality is modularized into a new duality move shown in Figure \ref{nHWmove}, which we prove in Appendix \ref{HW_proof_IP}.
\begin{figure}[!ht]
	\includegraphics[width=\textwidth,center]{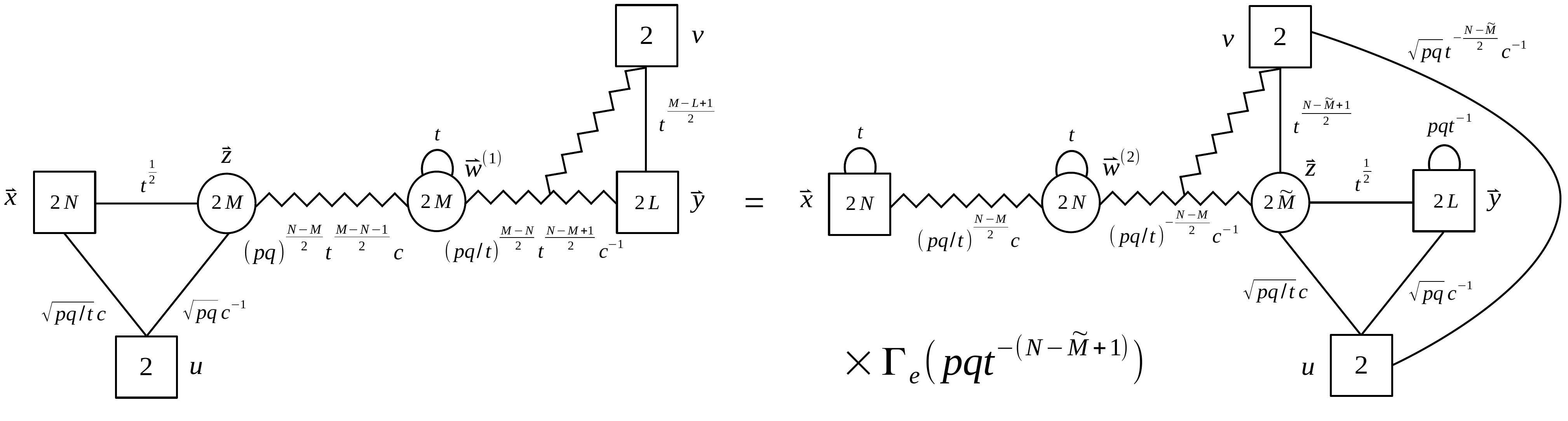}
	\caption{The Hanany--Witten duality move. Notice that  on the r.h.s.~the rank of the second gauge node is now  $\widetilde{M}=N+L-M+1$.
	 }
\label{nHWmove}
\end{figure}

We call this duality move, which swaps a  $\mathsf{B}_{01}$ with a $\mathsf{B}_{10}$ block,
the Hanany--Witten (HW) duality move because, as we will see in Section \ref{sec:3d}, its $3d$ version exactly corresponds 
to the local effect of an  HW brane move in the brane set-up realizing the $3d$ quiver theory. 
For $N\geq\widetilde{M}\geq 0$, 
the associated index identity is given by 
\begin{align}\label{eq:HWid}
    & \oint d\Vec{z}_M \,\, \Delta_M\left(\Vec{z}\right) 
    \mathcal{I}^{(N,M)}_{(1,0)}\left( \vec{x};\vec{z};u;pqt^{-1};c (pqt^{-1})^\frac{ N-M}{2}\right)
    \,_{\vec{z}}\hat{\mathbb{I}}_{\Vec{y},\,v}(t)
    \prod_{k=1}^L \Gamma_e\left( t^{\frac{M-L+1}{2}} y_k^{\pm} v^{\pm} \right) \nonumber\\
    & \quad = 
    \Gamma_e\left( pqt^{-(N-\widetilde{M}+1)} \right) \Gamma_e\left( (pq)^{\frac{1}{2}} t^{-\frac{N-\widetilde{M}}{2}} c^{-1} u^{\pm} v^{\pm} \right)  A_N\left(\vec{x};t\right) A_L\left(\vec{y};pq/t\right) \nonumber\\
    & \qquad \times 
    \oint d\Vec{z}_{\widetilde{M}}\,\, \Delta_{\widetilde{M}}\left( \Vec{z} \right) \,_{\vec{x}}\hat{\mathbb{I}}_{\Vec{z},\,v}(t) 
    \prod_{j=1}^{\widetilde{M}} \Gamma_e\left( t^{\frac{N-\widetilde{M}+1}{2}} z_j^{\pm} v^{\pm} \right)  
    \mathcal{I}^{(\widetilde{M},L)}_{(1,0)}\left( \vec{z};\vec{y};u;pqt^{-1};c (pqt^{-1})^\frac{ \widetilde{M}-L}{2}\right)\,.
\end{align}
With few manipulations, we can recombine the terms
$    \,_{\vec{z}}\hat{\mathbb{I}}_{\Vec{y},\,v}(t)
    \prod_{k=1}^L \Gamma_e\left( t^{\frac{M-L+1}{2}} y_k^{\pm} v^{\pm} \right)$
    in the first line into $\mathcal{I}^{(M,L)}_{(0,1)}(\vec{z};\vec{y};v;pq/t)$
 (up to some factors) and similarly on the third line, to explicitly display the swap of a $\mathsf{B}_{(0,1)}$  with a $\mathsf{B}_{(1,0)}$ block.
    The manipulations involve performing flip-flip duality (see for example eq. (2.14) in   \cite{Bottini:2021vms}) 
 on all the $\mathsf{S}$-walls so to produce Identity-walls
    with fugacity $pq/t$. However we prefer to keep the formula as it is, since this is the most convenient form to implement the dualizations.   

As we will see in Subsection \ref{sec:HW}, we can obtain the final configuration of the mirror dual theory by repeating this HW duality move, performing the field theory counterpart of the brane rearrangement that is needed to reach a configuration where there is a zero net number of D3-branes ending on the the D5-branes and so it is possible to read out the QFT.

Summarizing, if  asymmetric Identity-walls are generated after the dualization of the QFT blocks, we can study their effect by using 
the sequential Higgsing, the IP iteration, or the HW duality move to obtain the final IR $PSL(2,\mathbb{Z})$ dual theory.
We will illustrate these three strategies in the case of the SQCD in the following subsections.

\subsection{$\mathsf{S}$-dualization of SQCD}
\label{subsec:Sdualisation_SQCD}
\begin{figure}[!ht]
	\includegraphics[width=\textwidth,center]{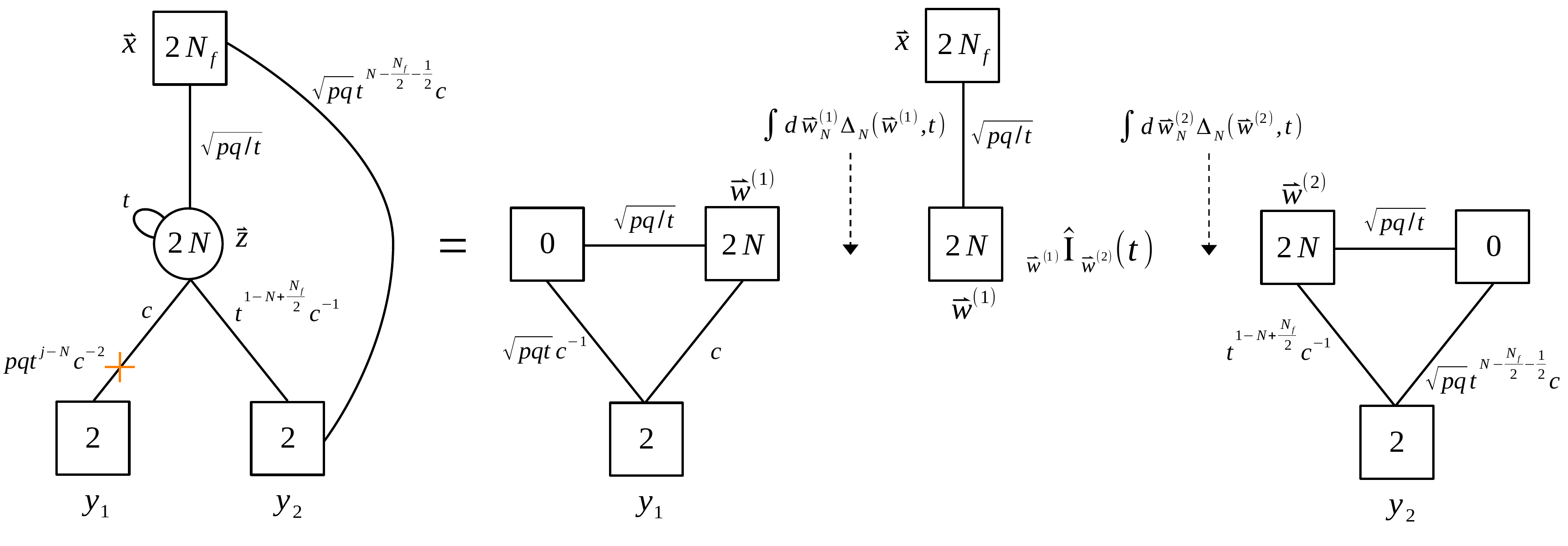}
	\caption{The SQCD and its block decomposition. The orange cross denotes a tower of singlets flipping the meson dressed with the $j$-th power of the antisymmetric  with $j=0,\dots,N-1$. We also explicitly indicate the frozen integration measure containing vector and antisymmetric chiral fields. Notice that we add trivial lines to reconstruct the triangle structure that identifies the $\mathsf{B}_{10}$ blocks.}
		 \label{SQCDzero}
\end{figure}

We will focus on the example of the $4d$ $\mathcal{N}=1$ $USp(2 N)$ SQCD with $2N_f+4$ fundamental chirals and $N_f\ge 2N$, represented on the left of Figure \ref{SQCDzero}. 
This theory corresponds to the element $\rho=[N_f-N,N]$, $\sigma=[1^{N_f}]$ of the $E_\rho^\sigma[USp(2N)]$ family studied in  \cite{Hwang:2020wpd} and  reduces to the $U(N)$ SQCD with  $N_f$ flavors in the $3d$ limit.

This theory can be decomposed into  QFT blocks as shown on the right of Figure \ref{SQCDzero}. When we  chop it into the blocks, to avoid clutter, we remove the singlets of the original SQCD, the orange cross indicating a tower of singlets flipping the $y_1$ meson dressed up to the $(N-1)$th power of the antisymmetric field and the $SU(2)_{y_2}\times USp(2N_f)$ bifundamental. We will always adopt this convention of removing singlets before chopping theories into QFT blocks, then perform all the duality moves, and restore the original singlets after reaching the dual theory. 

\begin{figure}[!ht]
	\includegraphics[width=\textwidth,center]{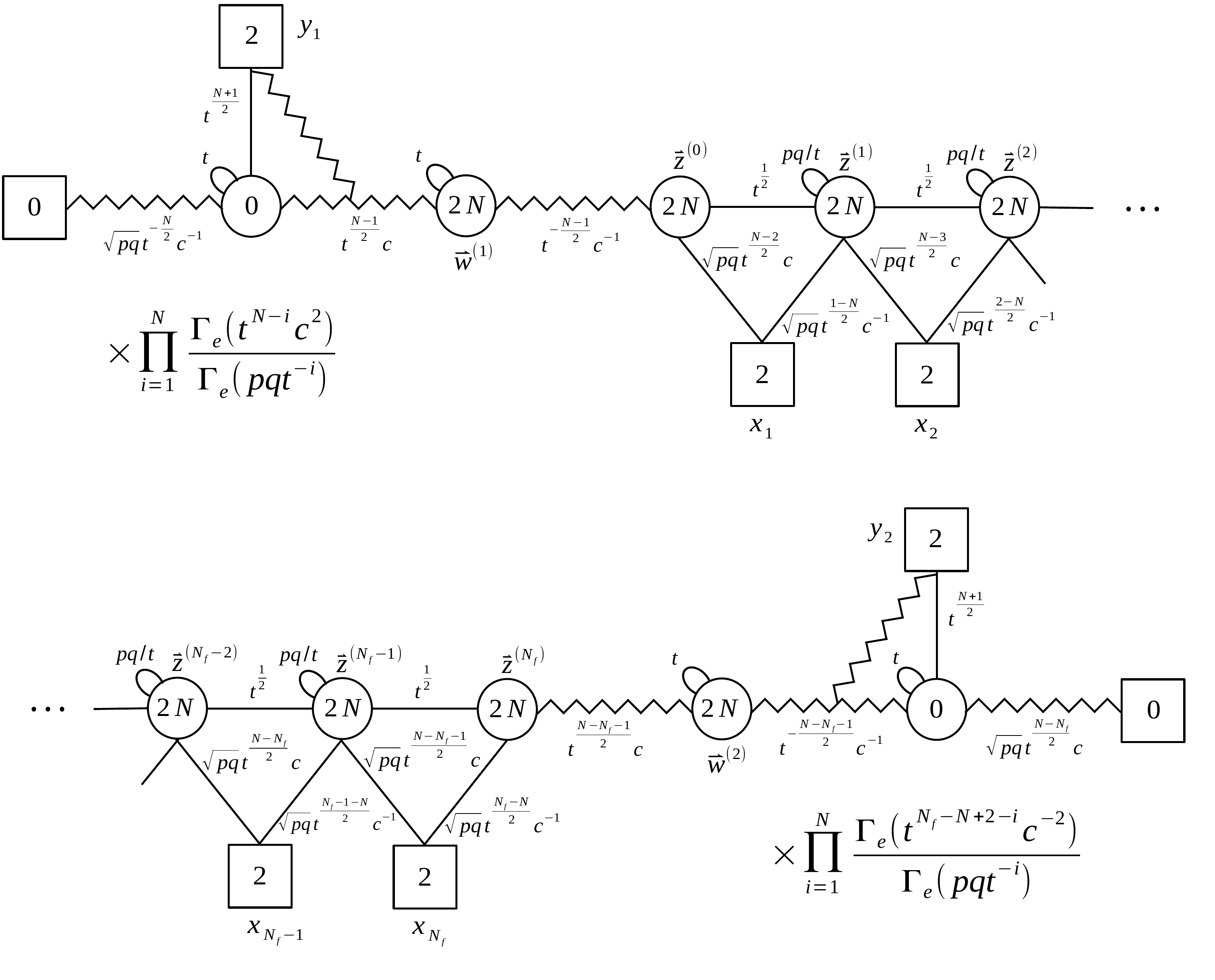}
	\caption{Result of the block dualization of the SQCD.}
	 \label{SQCDone}
\end{figure}

We now dualize each QFT block using the basic $\mathsf{S}$-duality moves of Figures
 \ref{10S01} and \ref{01S10} and glue them back by restoring the original gauge symmetries. This results in the dualized quiver shown in Figure \ref{SQCDone}. The extra singlets we get from the dualization are denoted in the figure by their index contributions.

\begin{figure}[!ht]
	\includegraphics[width=\textwidth,center]{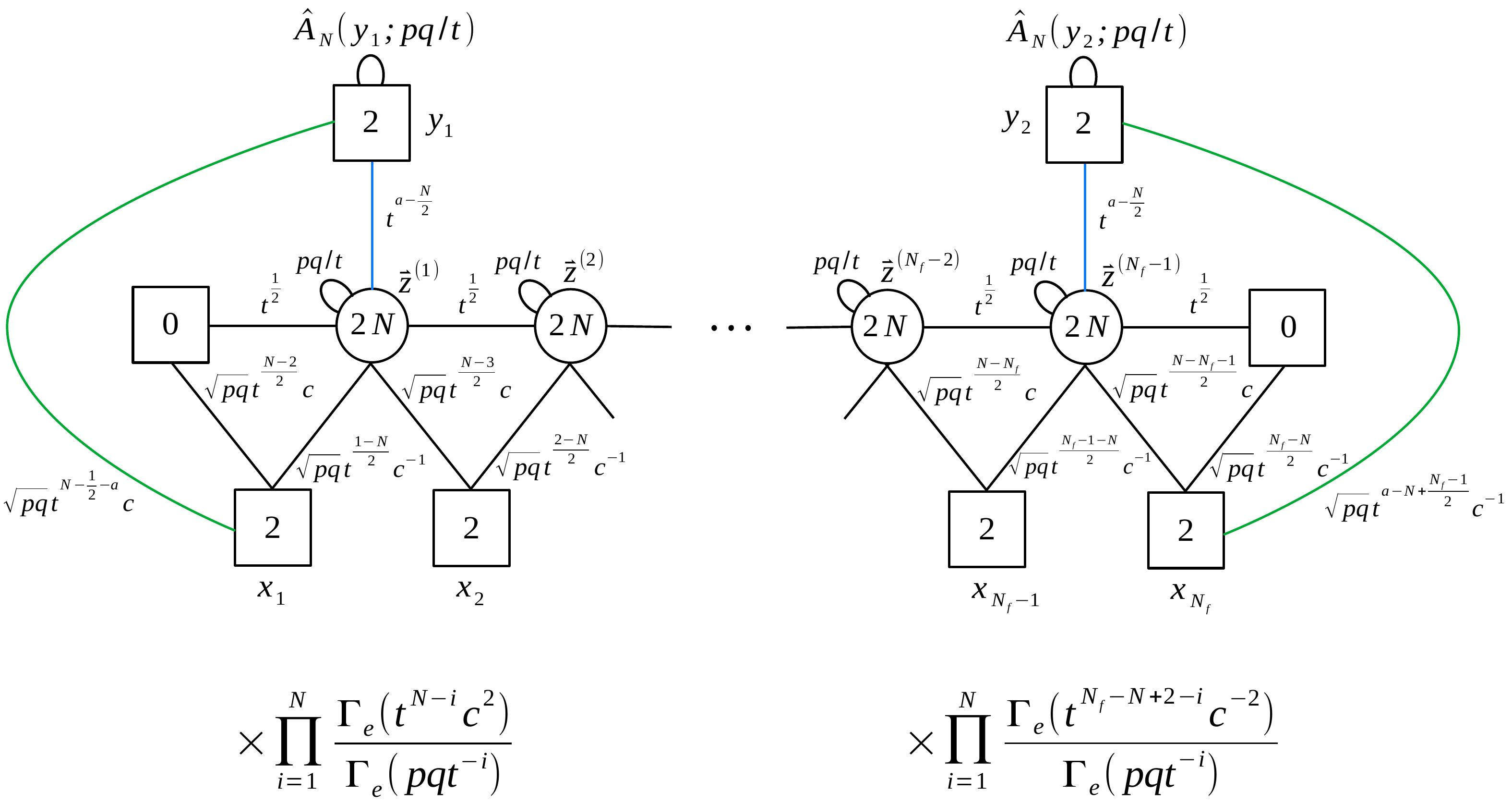}
	\caption{Result of the $\mathsf{S}$-dualization of the SQCD. The blue and green legs denote sets of chirals labelled by  $a=1,\dots,N$. Mesons constructed with blue chirals acquire a VEV.	}
	\label{SQCDtwo}
\end{figure}

We can now implement the identifications of the asymmetric Identity-walls at the two sides of the quiver,
which specialize the gauge fugacities $\vec{z}^{(0)}$ and $\vec{z}^{(N_f)}$ of the leftmost and rightmost nodes in a geometric progression as 
\begin{align}
z^{(0)}_j=t^{\tfrac{N+1-2j}{2}} y_1\,, \qquad z^{(N_f)}_j=t^{\tfrac{N+1-2j}{2}} y_2\,, \qquad j=1,\cdots,N\,,
\end{align}
to arrive at the quiver in Figure \ref{SQCDtwo}. 
The blue lines indicate two sets of $N$ chirals in the bifundamental of $SU(2)_{y_1} \times USp(2N)_{z^{(1)}}$ and that of $SU(2)_{y_2} \times USp(2N)_{z^{(N_f-1)}}$, respectively. They contribute to the index by
\begin{align}
  \prod_{i=1}^{N}\Gpq{t^{a-\tfrac{N}{2}}z^{(1)\pm}_i y_1^\pm  }, \qquad 
  \prod_{i=1}^{N}\Gpq{t^{a-\tfrac{N}{2}}z^{(N_f-1)\pm}_i y_2^\pm  }
 \qquad a=0,\, 1, \, \cdots, \, N \,.
 \label{blue}
 \end{align}
In addition, we have the following singlets  coming from the dualization:
\begin{align}
\nonumber& \widehat{A}_N(y_1;pq/t )\prod_{a=1}^N\Gpq{\sqrt{pq} t^{N-1/2-a} c x_1^\pm y_1^\pm } 
 \frac{ \Gpq{t^{N_c-a}c^2 }}{\Gpq{p q t^{-a}}} \\
&\times \widehat{A}_N(y_2;pq/t ) \prod_{a=1}^N \Gpq{\sqrt{pq} t^{a-N+\frac{N_f-1}{2}} c^{-1} x_{N_f}^\pm y_2^\pm }
  \frac{  \Gpq{t^{N_f-N_c+2-a} c^{-2}}}{\Gpq{p q t^{-a}}} \,,
\end{align}
where we defined
\begin{align}
\label{eq:spec_antisymm}
\widehat{A}_N(v;pq/t)={A}_N\left(t^{\tfrac{N-1}{2}} v,\cdots,t^{\tfrac{1-N}{2}} v;pq/t\right)  \,.
\end{align}

Now we have reached a quiver theory with no duality-walls left. As we are going to see shortly, there are some VEVs which trigger an RG flow, which we study in several ways in the following subsections.

\subsection{Method I:  VEV propagation via sequential Higgsing}

%
%

In the quiver in Figure \ref{SQCDtwo}, some of the mesons constructed with the chirals in blue acquire VEVs Higgsing the first and last node down to $SU(2)$. At the level of the index, following  the strategy of \cite{Gaiotto:2012xa}, this can be seen by noticing that, for example, the $a=1,N-1$ chrials in the first set of eq.~\eqref{blue} can be paired up as
\begin{align}
  \prod_{i=1}^{N}\Gpq{t^{1-\tfrac{N}{2}}z^{(1)\pm}_i y_1^\pm}  \Gpq{t^{\tfrac{N}{2} -1}z^{(1)\pm}_i y_1^\pm}=   \prod_{i=1}^{N}\Gpq{z^{(1)\pm}_i (t^{1-\tfrac{N}{2}} y_1)^\pm}   \Gpq{z^{(1)\pm}_i (t^{\tfrac{N}{2}-1} y_1)^\pm} .
\end{align}
On the r.h.s., we have two gamma functions such that the product of their arguments is 1, signaling that the meson constructed from the corresponding chirals is uncharged under all the abelian symmetries including the R-symmetry and is thus taking a VEV.
In that case, the location of the poles of these gamma functions collide and
pinch the integration contour at two points, say at $ z^{(1)}_1 = t^{1-\frac{N}{2}} y_1$ and  $z^{(1)}_{N-1} = t^{\frac{N}{2}-1} y_1$.
Similarly, all the other chirals in  \eqref{blue}  but $a=0, \, N$ pair up producing colliding poles that pinch the integration contour at
\begin{equation}
    \begin{cases}
    z^{(1)}_1 = t^{1-\frac{N}{2}} y_1 \,, \\[5pt]
    z^{(1)}_2 = t^{2-\frac{N}{2}} y_1 \,, \\[5pt]
    \vdots \\
    z^{(1)}_{N-2} = t^{(N-2)-\frac{N}{2}} y_1 \,, \\[5pt]
    z^{(1)}_{N-1} = t^{(N-1)-\frac{N}{2}} y_1 \,, \\[5pt]
    z^{(1)}_N = \tilde{z}^{(1)} \,,
    \end{cases}
    \end{equation}
and an analogous freezing condition is imposed on the last $\vec{z}^{(N_f-1)}$ node. The index after Higgsing is obtained by evaluating the residues at such points. Notice that when we evaluate the residues, there are pairs of poles and zeros canceling each other, coming from $\Gpq{1}$ contained in \eqref{blue} and $\Gpq{pq}$ contained in $A_N(y_1, pq/t)$ and $A_N(y_2, pq/t)$, respectively.

\begin{figure}[!ht]
	\includegraphics[width=.7\textwidth,center]{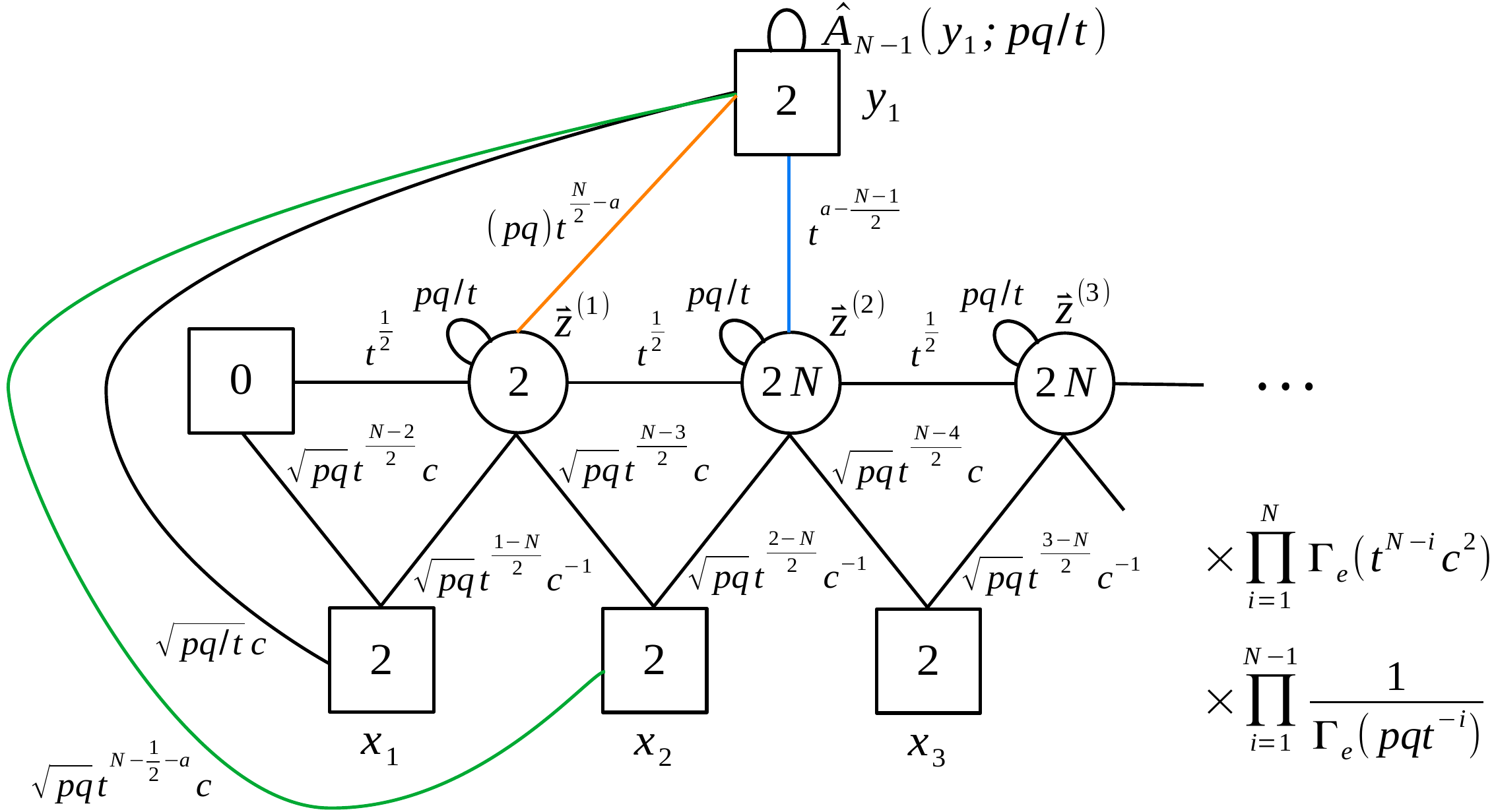}
	\caption{The result of the Higgsing of the leftmost $USp(2N)$ gauge node. The blue  line denotes a set of chirals with $a=1,\dots,N-1$, the orange  line denotes a set of chirals with $a=2,\dots,N-1$, 
while the green leg denotes a set of chirals with  $a=2,\dots,N$. Again the mesons constructed with the blue legs acquire a VEV. }
	    \label{pinch2}
\end{figure}

Let's now focus on the left part of the quiver.  Taking the residues at these poles, we obtain the theory after the Higgsing induced by the VEV shown in  Figure \ref{pinch2}.
We now  have $N-1$ chirals (in blue) in the bifundamental of $SU(2)_{y_1}\times USp(2N)_{z^{(2)}}$  contributing to the index as
\begin{align}
  \prod_{i=1}^{N}\Gpq{t^{a-\tfrac{N-1}{2}}z^{(2)\pm}_i y_1^\pm  }
 \qquad a=0, \, 1, \, \cdots, \, N-1 \,,
  \end{align}
which now Higgs the leftmost $USp(2N)$ node, next to the $USp(2)$ node, down to $USp(4)$. Indeed we have $N-2$ colliding poles and $N-2$ zeros  from the singlets $A_{N-1}(y_1, pq/t)$.

\begin{figure}[!ht]
	\includegraphics[width=\textwidth,center]{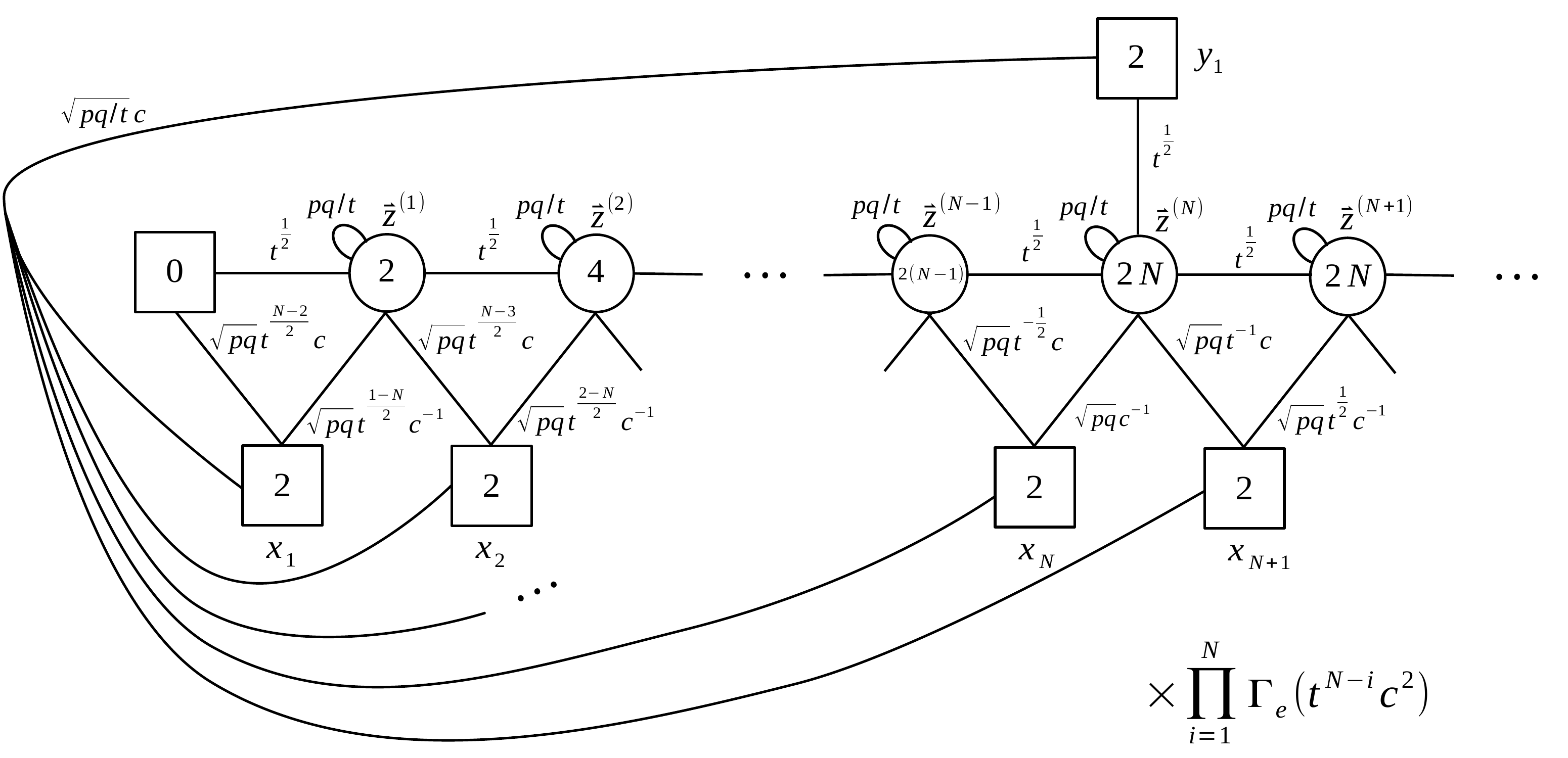}
	\caption{The final result of the sequential Higgsing. In this configuration there are no VEVs turned on.}
	   \label{pinch3}
\end{figure}

\begin{landscape}
\begin{figure}[!ht]
	\includegraphics[scale=.4,center]{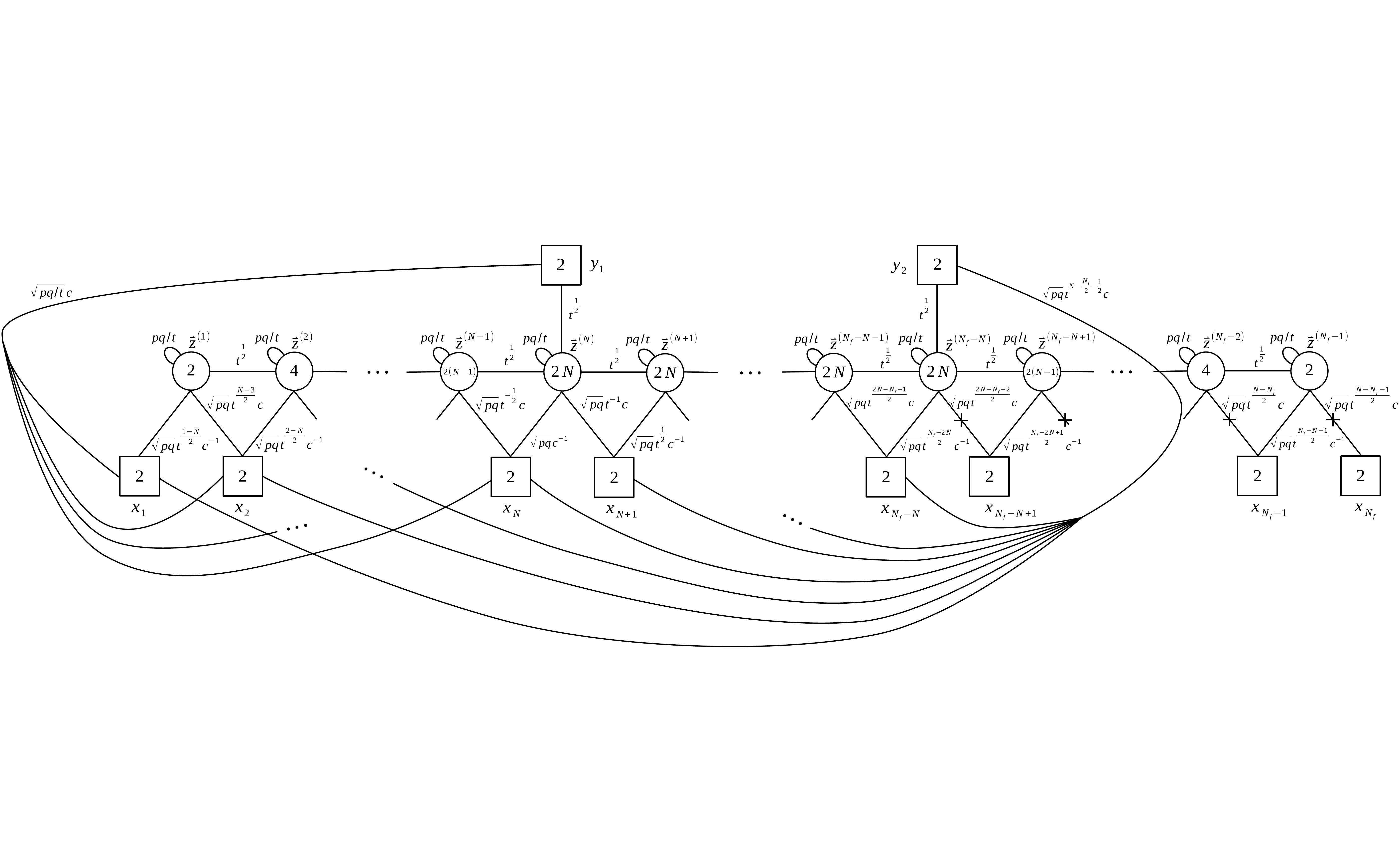}
	\caption{Mirror of the SQCD.}
	 \label{finaldualSQCD}
\end{figure}
\end{landscape}

We can iterate this procedure $N$ times creating a tail of nodes with increasing ranks as in Figure \ref{pinch3}.
Now there are no blue legs, the vertical chirals interact with a cubic superpotential with the antisymmetric
\begin{align}
  \prod_{i=1}^{N}\Gamma\left(t^{\tfrac{1}{2}}z^{(N)\pm}_i y_1^\pm  \right),\qquad
  \prod_{i=1}^{N}\Gamma\left(t^{\tfrac{1}{2}}z^{(N_f-N)\pm}_i y_2^\pm  \right) \,,
\end{align}
and there are no $A_k(y_1, pq/t)$ singlets, so no zeros. Therefore, there is no further VEV to implement.
Repeating the procedure on the right part of the quiver and restoring the singlets of the original SQCD that we dropped when we chopped it into QFT blocks, we obtain the mirror dual shown in Figure \ref{finaldualSQCD}.
Notice that some of the singlets produced by the Higgsing cancel out with the original singlets, and we are left with the correct mirror dual of the $4d$ SQCD found in \cite{Hwang:2020wpd}.

\subsection{Method II:  VEV propagation  via the IP duality}
\label{sec:Higgsing via IP}

The sequential Higgs mechanism due to the VEV propagation explained in the previous subsection can alternatively  be implemented by iteratively applying the IP duality,  basically trading VEVs for massive deformations.
The idea is similar to the correspondence between the mass deformation and the Higgsing  under the Seiberg duality \cite{Seiberg:1994pq}.

Let's go back to the quiver in Figure \ref{SQCDtwo}.
We focus again on the l.h.s.~of the quiver and use the  relation shown in Figure \ref{fig:N_to_1_chir} and proved in Appendix 
\ref{proof_N_to_1_chir} to trade the $N$ bifundamentals chirals (in blue) for a single bifundamental. 
\begin{figure}[!ht]
	\centering
	\includegraphics[width=\textwidth,center]{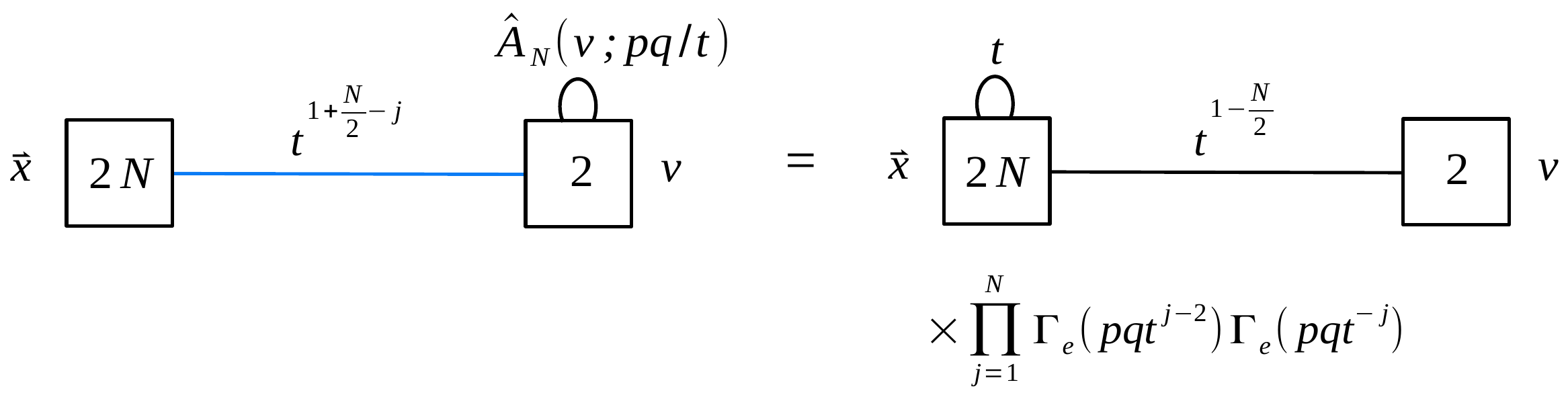}
	\caption{A relation between two WZ models. On the l.h.s.~the blue leg denotes a set of chirals with  $j=1,\dots,N$.}
	\label{fig:N_to_1_chir}
\end{figure}

%

The result is given by the first quiver in  Figure \ref{fig:IP_first_run}.
Notice that, since the first node $USp(2N)_{z^{(1)}}$ (in orange) has no antisymmetric field, we can apply the IP duality to it, so to obtain the second quiver  theory  in Figure \ref{fig:IP_first_run} where the dualized gauge node has now become $USp(2)_{z^{(1)}}$. As an effect of the dualization, the $y_1$ flavor node has now {\it moved}: we have a $y_1$ flavor connected to  the $USp(2N)_{z^{(2)}}$ node with $U(1)_t$ charge increased by $1/2$, and a $y_1$ flavor connected to  the $USp(2N)_{z^{(1)}}$ node  with charge $\sqrt{pq}t^{\frac{N}{2}-1}$.
 Also, some singlets are produced and some are cancelled. Consider, for example, the tower of $N$ $SU(2)_{y_1}\times SU(2)_{x_1}$ bifundamental singlets: one of them is cancelled and instead one $SU(2)_{y_1}\times SU(2)_{x_2}$ bifundamental singlet is produced. Moreover, the legs of the saw are reorganized such that now $x_1$ is {\it swapped} with $x_2$.
 
Another effect of the dualization was to give mass to the antisymmetric chiral of the adjacent $USp(2N)_{z^{(2)}}$ (in red in the second quiver of Figure \ref{fig:IP_first_run}) so that we can apply again the IP duality to it and obtain the third quiver where the dualized gauge node is now $USp(4)_{z^{(2)}}$.
Notice that the $y_1$ flavor node keeps moving to the right. Also, $x_1$ keeps moving to the right.

It is clear that we can keep iterating this application of the IP duality, since at each step we remove the antisymmetric at the node on the right of the dualized one and restore the one at the node on the left. After $N-2$ steps, we reach the first quiver in the second line of Figure \ref{fig:IP_first_run}, where we have created
 a tail of gauge nodes with increasing ranks: $USp(2)$, $USp(4)$, and so on up to $USp(2N-4)$. Now the $y_1$ flavor node is attached to the $USp(2N)_{z^{(N-1)}}$ gauge node with a chiral of charge $t^0$, that is uncharged under all abelian symmetries. This indicates that the associated meson is taking a VEV, which triggers a Higgsing that we could implement by taking the residues at collinding poles as we did in the previous subsection. This is supported by the fact that one of the singlets leads to a zero in the index since its index conbribution is $\Gamma_e(pq)$. We then expect the rank of this node to be Higgsed by one unit. 
 
\begin{figure}[!ht]
	\includegraphics[scale=.39,center]{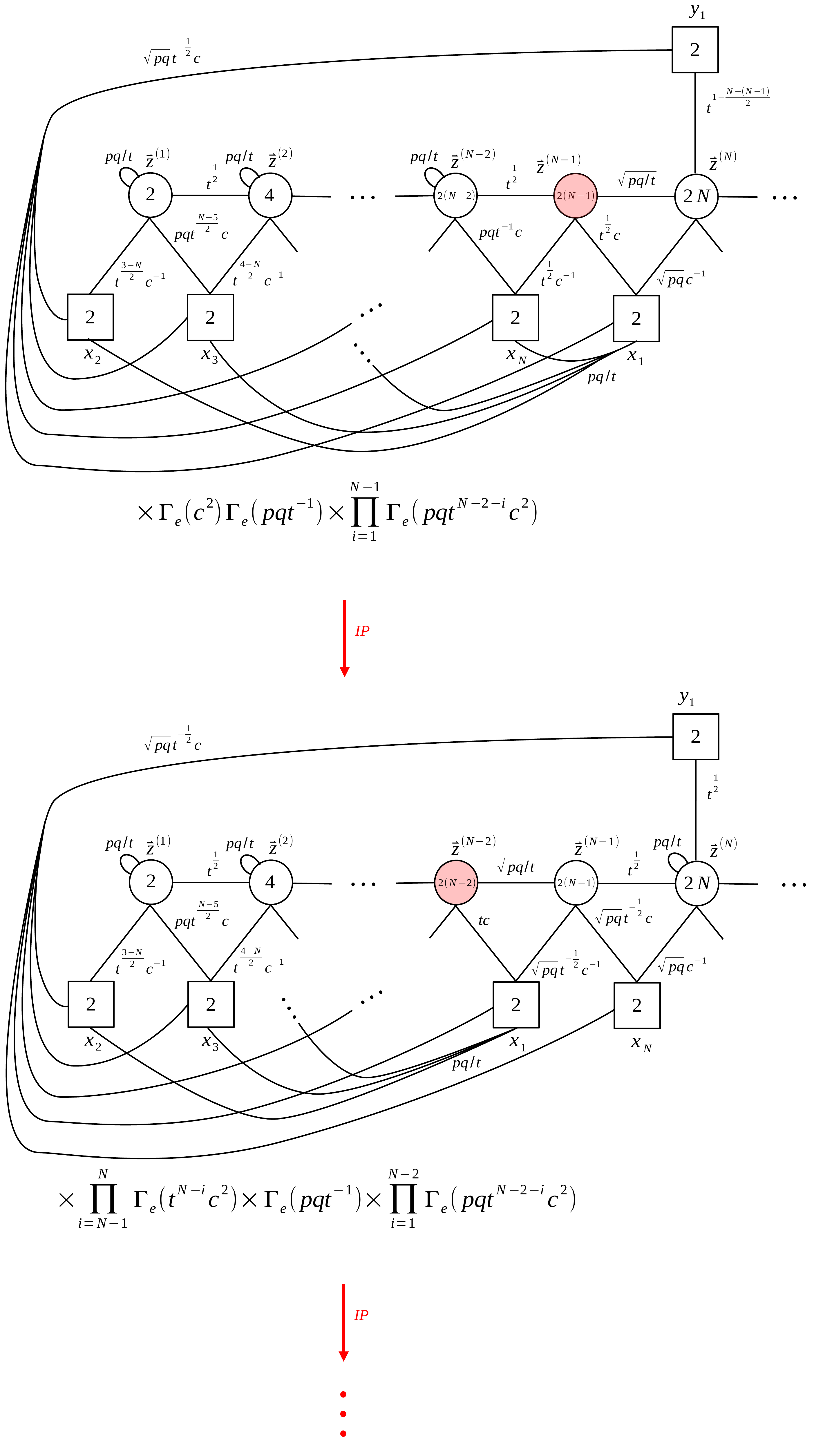}
	 \caption{Second round of applications of the IP duality.}
    \label{fig:IP_second_run}
\end{figure}

Instead, we shall now employ a slightly different strategy which allows us to see the overall effect of the Higgsing using the IP duality twice. We hence apply the IP duality on the $USp(2N)_{z^{(N-1)}}$  node  which becomes $USp(2N-2)_{z^{(N-1)}}$ to obtain the last quiver in Figure  \ref{fig:IP_first_run}. Notice that the chirals connecting the $USp(2N-2)_{z^{(N-1)}}$ node to the $y_1$ flavor node have the charge of $\sqrt{pq}$, which means that they are massive and can be integrated out. This is due to the fact that this flavor had a VEV before the dualization, which turned the VEV into a complex mass deformation as usual in Seiberg-like dualities. After having integrated out the massive fields, we obtain the first quiver in Figure \ref{fig:IP_second_run}.

Now we have a tail with increasing ranks from $USp(2)$ to $USp(2N)$. Notice that the flavor node $x_1$ now sits between the $USp(2N-2)_{z^{(N-1)}}$ and $USp(2N)_{z^{(N)}}$ gauge nodes.
Since the red $USp(2N-2)_{z^{(N-1)}}$   node has no antisymmetric, we can apply the IP duality on it. Notice that this doesn't simply revert the previous dualization that we did of the same node, since in between we integrated out the massive fields. We indeed obtain the second quiver in Figure \ref{fig:IP_second_run}. We then continue applying the IP duality moving towards the left until we reach the first $USp(2)_{z^{(1)}}$ node. At each step, we restore the antisymmetric on the previous node and remove the one at the subsequent node, which is what allows us to keep moving to the left with the application of the IP duality. Moreover, at each dualization the rank of the gauge node doesn't change, so that eventually we arrive at the same result as the end point of the VEV propagation studied in the previous subsection, which we show in Figure \ref{pinch3}.

By repeating the same procedure also on the right side of the $\mathsf{S}$-dualized SQCD quiver in Figure \ref{SQCDone}, we eventually  arrive at the mirror dual drawn in Figure \ref{finaldualSQCD}.

\begin{landscape}
\null
\vfill
	\begin{figure}[!ht]
	\includegraphics[scale=.3,center]{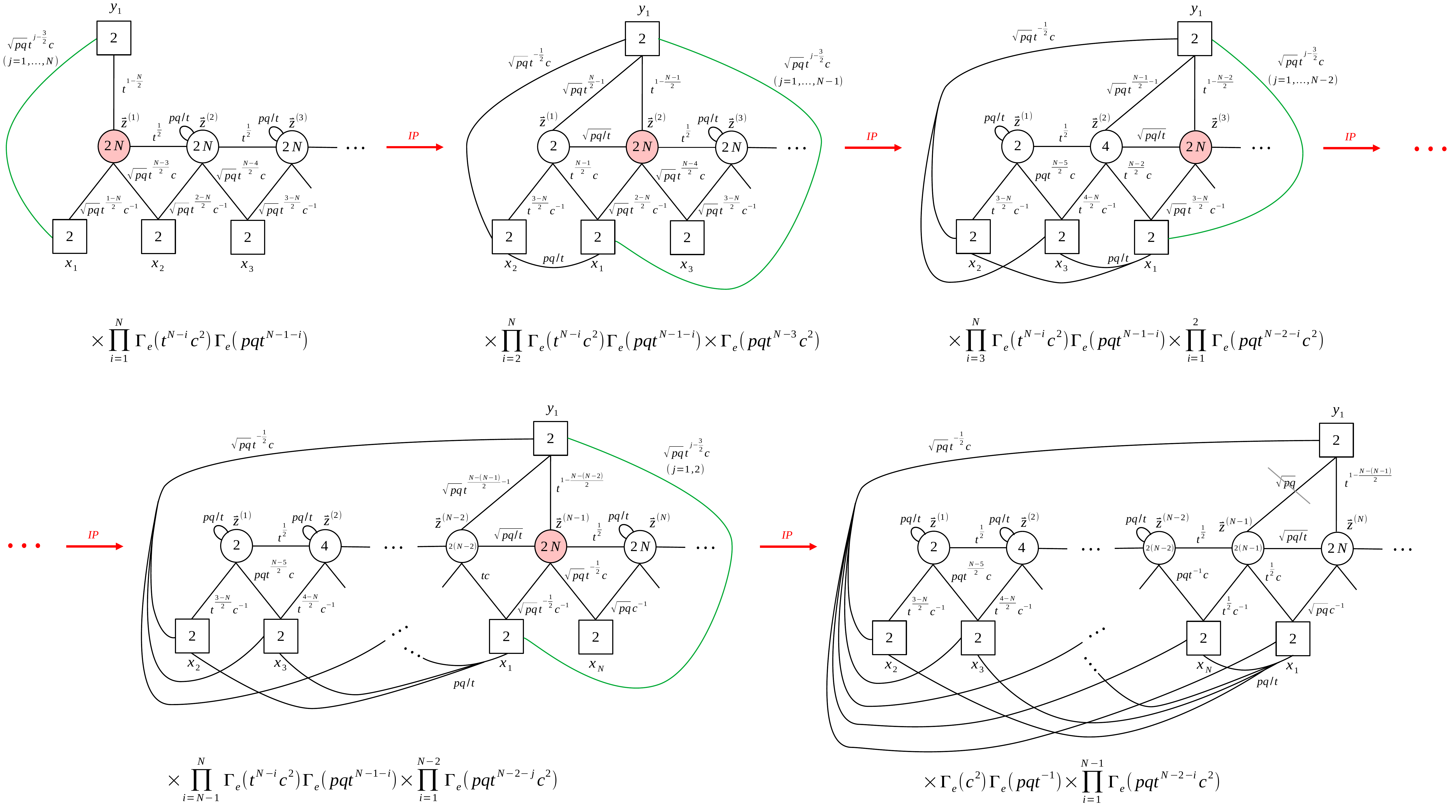}
	 \caption{First round of applications of the IP duality.}
    \label{fig:IP_first_run}
\end{figure}
\end{landscape}

%

\subsection{Method III: VEV propagation via the HW duality move}
\label{sec:HW}

In this subsection we will illustrate  the third and final strategy to implement the VEVs generated in the dualization algorithm, which uses the HW moves in Figure \ref{nHWmove} to reach the final IR frame.

\begin{figure}[h]
\includegraphics[scale=.36,center]{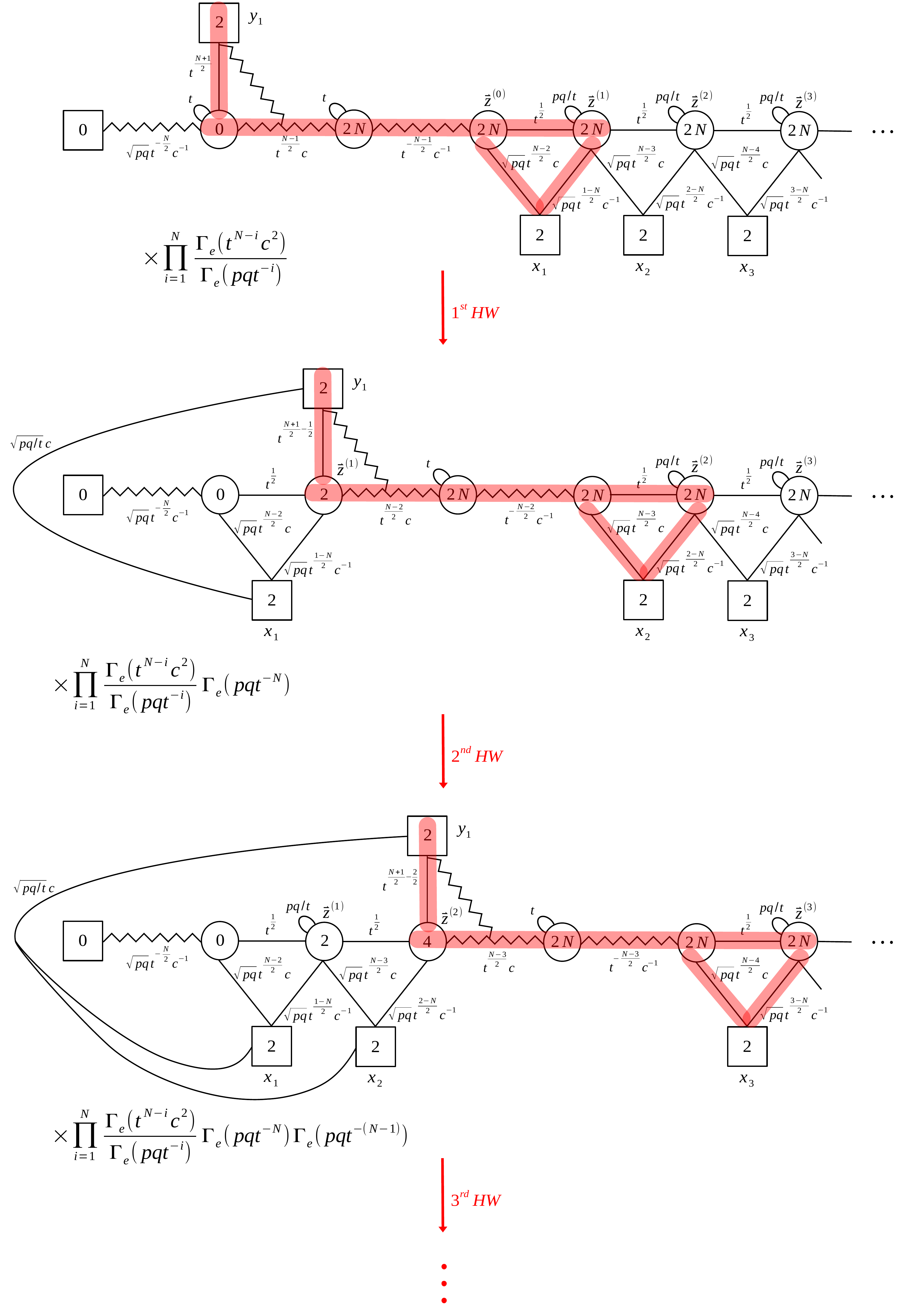}
\caption{Initial applications of the HW duality move.}
\label{fig:HW_0_1_2}
\end{figure}

\begin{figure}[h]
\includegraphics[scale=.31,center]{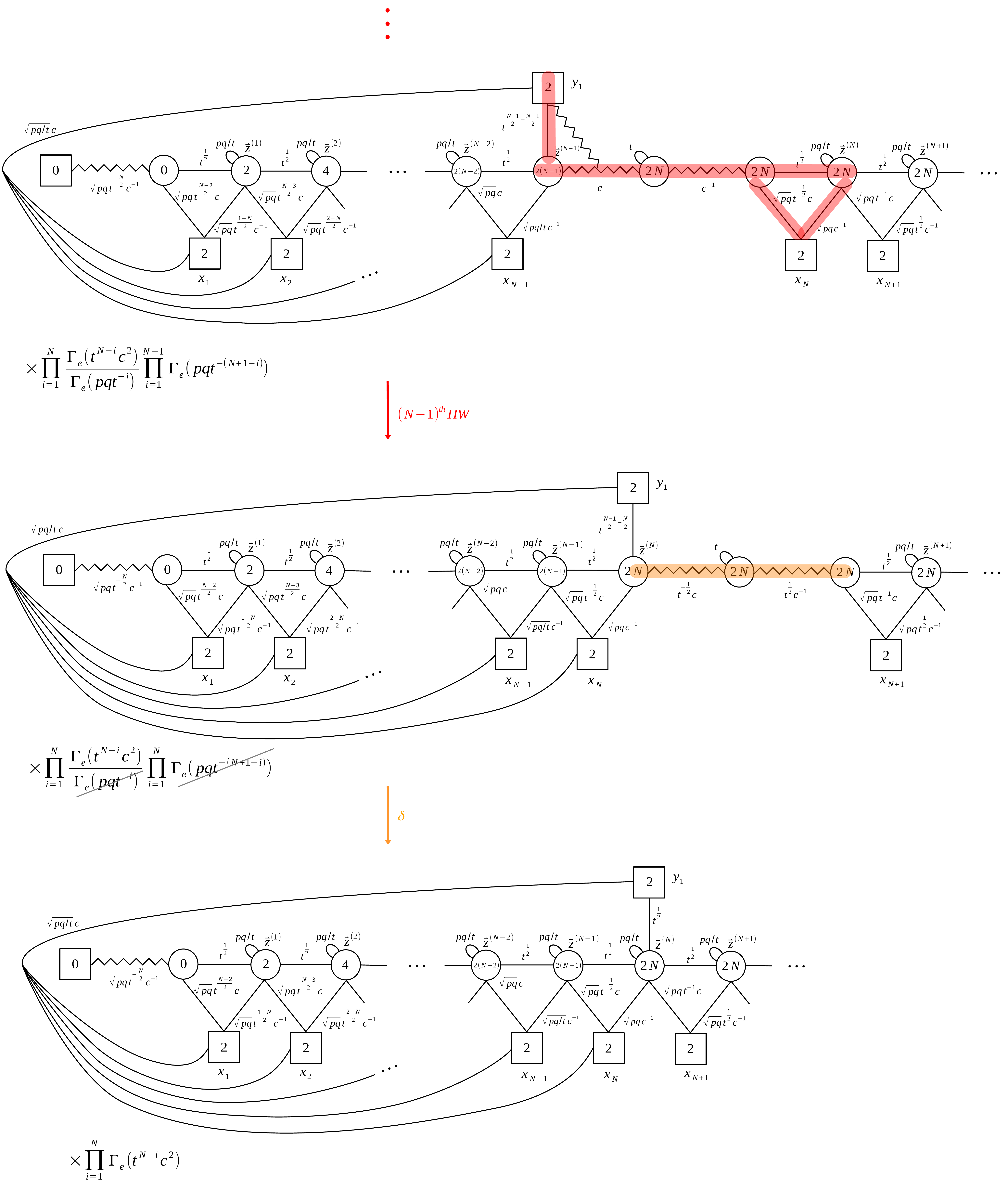}
\caption{Last application of the HW duality move and the final manipulations leading to the IR frame.}
\label{fig:HW_Nmin1_N_final}
\end{figure}

Consider the $\mathsf{S}$-dualized SQCD in Figure \ref{SQCDone}, again focusing on the left part of the quiver to begin with.
We apply the HW move to the sequence of $\mathsf{B}_{01}$-$\mathsf{B}_{10}$ blocks highlighted in red in the first quiver in Figure
\ref{fig:HW_0_1_2} and reach the  second quiver.
Notice that  the  $USp(2N)_{z^{(1)}}$ gauge node becomes a $USp(2)_{z^{(1)}}$ node and two singlets have been produced:  the $SU(2)_{y_1}\times SU(2)_{x_1}$ bifundamental and the one corresponding to $\Gamma_e(pqt^{-N})$, which actually cancels one of the singlets of Figure \ref{SQCDone}. We then apply again the HW move to the highlighted sequence of $\mathsf{B}_{01}$-$\mathsf{B}_{10}$ blocks in the second quiver of Figure
\ref{fig:HW_0_1_2} to reach the third quiver and continue moving to the right.

After $N-1$ dualizations we arrive at the first quiver in Figure \ref{fig:HW_Nmin1_N_final}.  If we apply the HW duality move once again, we obtain the second theory in Figure \ref{fig:HW_Nmin1_N_final}  where the  $USp(2N)_{z^{(N)}}$ node remains $USp(2N)$ without any rank change and the vertical flavor has exactly charge $t^{1/2}$ and forms the standard cubic superpotential with the antisymmetric field. In particular, the VEV has been completely extinguished by this sequence of applications of the HW move. This implies that now the Identity-wall is actually symmetric and so it can be easily implemented by identifying the two adjacent nodes, which results in the third quiver in Figure \ref{fig:HW_Nmin1_N_final}. As expected, the result exactly coincides with the end point of the Higgsing in Figure \ref{pinch3}. 

Again we can perform an analogous sequence of the HW duality moves on the right part of the quiver in Figure \ref{SQCDone} and obtain the mirror dual SQCD shown in Figure \ref{finaldualSQCD} after restoring the singlets of the original SQCD.

\clearpage
\section{Algorithm in \boldmath$3d$}
\label{sec:3d}

In this section, we discuss the $3d$ version of the dualization algorithm. We will, in particular, present all of its basic ingredients, which are obtained as a limit of those in $4d$ that we saw in the previous two sections. We will not explicitly discuss any example of application of the algorithm in $3d$, since this works exactly as in the $4d$ case.

\subsection{$SL(2,\mathbb{Z})$ operators}

We begin by introducing the  $SL(2,\mathbb{Z})$ duality-walls. 
\bigskip

\begin{figure}[!ht]
	\centering
	\includegraphics[scale=0.6]{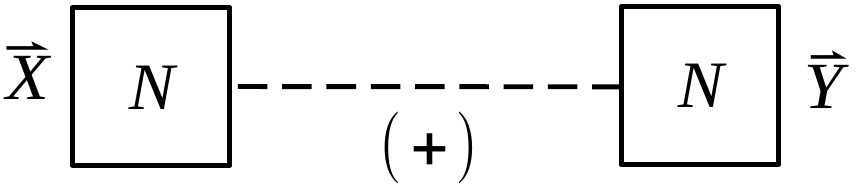}
	\hspace{1.5cm}
	\includegraphics[scale=0.6]{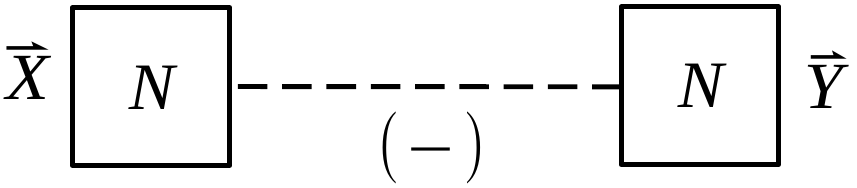}
	\caption{The $3d$ $\mathcal{S}$-wall and $\mathcal{S}^{-1}$ walls. }
	\label{Figures/Section3d/OperatorsAndBlocks/Sop3d.pdf}
\end{figure}

\noindent\textbf{$\mathcal{S}$-wall}. As we already mentioned, in $3d$ we associate the $\mathcal{S}$ generator of $SL(2,\mathbb{Z})$ with the $FT[U(N)]$ theory \cite{Aprile:2018oau}, which is the $T[U(N)]$ theory \cite{Gaiotto:2008ak} modified by adding a set of singlet chiral fields in the adjoint representation of the manifest $U(N)_X$ flavor symmetry:\footnote{Notice that w.r.t.~to   \cite{Bottini:2021vms,Hwang:2021ulb} here
we include an extra minus sign in front of either the  $\vec{X}$ or $\vec{Y}$ mass parameters. }
\be\label{indexSwall3d}
&&\mathcal{Z}_{\mathcal{\mathcal{S}}}^{(N)}(\vec{X};\vec{Y};m_A)\equiv\mathcal{Z}_{FT[U(N)]}(\vec{X};-\vec{Y};m_A)=\mathcal{Z}_{FT[U(N)]}(-\vec{X};\vec{Y};m_A)
\ee
and
\be\label{indexSwall3d}
&&\mathcal{Z}_{\mathcal{\mathcal{S}}^{-1}}^{(N)}(\vec{X};\vec{Y};m_A)=\mathcal{Z}_{\mathcal{\mathcal{S}}}^{(N)}(\vec{X};-\vec{Y};m_A)=\mathcal{Z}_{\mathcal{\mathcal{S}}}^{(N)}(-\vec{X};\vec{Y};m_A) \,.
\ee
We will schematically represent the $3d$ $\mathcal{S}$-wall as in Figure \ref{Figures/Section3d/OperatorsAndBlocks/Sop3d.pdf} to display both of its $U(N)$ global symmetries. We will distinguish between $\mathcal{S}$ and $\mathcal{S}^{-1}$ by a $(+)$ and $(-)$ label respectively.

Similarly to its $4d$ counterpart, the $FE[USp(2N)]$ theory, the $S^3_b$ partition function 
\cite{Kapustin:2009kz,Jafferis:2010un,Hama:2010av,Hama:2011ea}
of $FT[U(N)]$ admits  an integral form given by the following recursive definition:
\be\label{indexSwall3d}
&&\mathcal{Z}_{FT[U(N)]}(\vec{X};\vec{Y};m_A)=\e^{2\pi iY_N\sum_{i=1}^NX_i}\prod_{i,j=1}^N\sbfunc{i\frac{Q}{2}\pm(X_i-X_j)-2m_A}\nn\\
&&\quad\times\int\udl{\vec{Z}^{(N-1)}_{N-1}}  \Delta^{(3d)}_{N-1}\left(\vec Z^{(N-1)}\right) \e^{-2\pi i Y_N\sum_{a=1}^{N-1}Z_a^{(N-1)}} \prod_{a=1}^{N-1}\prod_{i=1}^Ns_b\left(\pm(Z_a^{(N-1)}-X_i)+m_A\right)\nn\\
&&\quad\times\mathcal{Z}_{FT[U(N-1)]}\left(Z_1^{(N-1)},\cdots,Z_{N-1}^{(N-1)};Y_1,\cdots,Y_{N-1};m_A\right)\,,
\label{pftsun}
\ee
where in the $3d$ partition function the integration measure is defined as
\be
\udl{\vec{Z}_{n}}=\frac{\prod_{i=1}^{n}\udl{Z_a}}{n!}\, ,\quad \Delta^{(3d)}_N\left(\vec Z\right)=\frac{1 }{\prod_{a<b}^{N}s_b\left(i\frac{Q}{2}\pm(Z_a-Z_b)\right)} \,.
\ee

Here $X_i$, $Y_j$ and $m_A$ are the real mass parameters respectively for the $U(N)_X$, $U(N)_Y$, $U(1)_A$ global symmetries, where $U(1)_A$ is the axial symmetry which together with the trial R-symmetry $U(1)_R$ is obtained from a suitable linear combination of the Cartans of the $SO(4)_R\cong SU(2)_H\times SU(2)_C$ R-symmetry of $3d$ $\mathcal{N}=4$ theories. The deformation by the real mass $m_A$ then breaks supersymmetry to $\mathcal{N}=2^*$ \cite{Tong:2000ky}. In particular our parametrization of $U(1)_A$ and $U(1)_R$ is such that the canonical R-symmetry under which all hypermultiplets have charge $\frac{1}{2}$ is obtained by setting the mixing coefficient $R_A$ between $U(1)_A$ and $U(1)_R$ to $R_A=\frac{1}{2}$. Equivalently at the level of the sphere partition function to recover $\mathcal{N}=4$ we should turn off $m_A$ setting $m_A=i\frac{Q}{4}$. 

The partition function \eqref{pftsun} can be obtained as a limit of the $4d$ index of the $FE[USp(2N)]$ theory \eqref{indexSwall}  as it was shown in \cite{Pasquetti:2019tix,Pasquetti:2019hxf}. This consists of three main steps \cite{Pasquetti:2019tix,Pasquetti:2019hxf,Bottini:2021vms}. The first one is a circle reduction from $4d$ to $3d$, which in particular relates the $S^3\times S^1$ index to the $S^3_b$ partition function. This is done by using the following property:
\be\label{eq:gammatosb}
\lim_{\beta\to0}\Gc_e\left(\e^{2\pi i\beta z};p=\e^{-2\pi \beta b},q=\e^{-2\pi \beta b^{-1}}\right)=\e^{-\frac{i\pi}{6\beta}\left(i\frac{Q}{2}-z\right)}\sbfunc{i\frac{Q}{2}-z}\,,
\ee
where $\beta$ is the $S^1$ radius that we send to zero. The $4d$ fugacities and the $3d$ real masses are related by
\be
\label{eq:var}
x_i=\e^{2\pi i\beta X_i},\quad y_j=\e^{2\pi i\beta Y_j},\quad t=\e^{2\pi i\beta (iQ-2m_A)},\quad c=\e^{2\pi i\beta \Delta}\,.
\ee

The second step is a real mass deformation that breaks the $USp(2N)$ symmetry to $U(N)$
\be
X_i\to X_i+s,\quad Y_j\to Y_j+s\, \quad s\to+\infty\,,
\ee
which is done using the following property:
\be\label{eq:sbasymp}
\lim_{z\to\pm\infty}\sbfunc{z}=\e^{\pm i\frac{\pi}{2}z^2}\,.
\ee
This leads to the $FM[U(N)](\vec{X},\vec{Y},\Delta)$ theory, introduced in \cite{Pasquetti:2019tix}, which contains a  monopole superpotential.

Finally, we consider the real mass deformation
\be
\Delta\to\pm\infty\,,
\ee
which leads to 
\be
\lim_{\Delta\to\pm\infty} FM[U(N)](\vec{X},\vec{Y},\Delta)= \mathcal A \, FT[U(N)](\vec{X},\pm\vec{Y}) \,,
\ee
where the prefactor $\mathcal A$ can be determined by \eqref{eq:gammatosb}; see (5.21) in \cite{Bottini:2021vms} for the result.

\begin{figure}[!ht]
	\centering
	\includegraphics[scale=0.7]{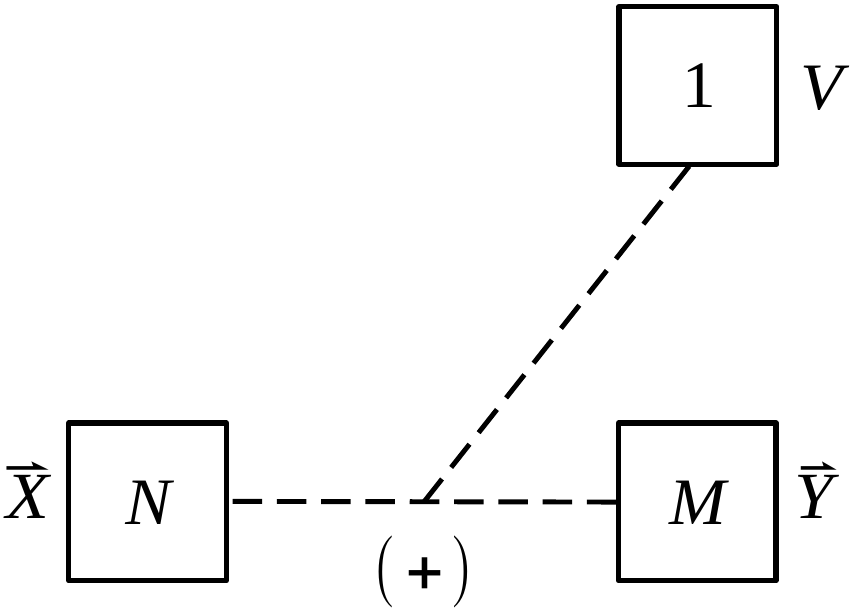}
	\caption{The asymmetric $3d$ $\mathcal{S}$-wall.}
	\label{Figures/Section3d/OperatorsAndBlocks/AsymmS3d.pdf}
\end{figure}

We also define an asymmetric $\mathcal{S}$-wall, where one of the $U(N)$ symmetries is broken to $U(M)\times U(1)$ with $M<N$. This is obtained by a deformation of the ordinary $\mathcal{S}$-wall that is completely analogous to the one considered in $4d$ in \cite{Hwang:2020wpd,Bottini:2021vms}, and it implies the following specialization of the real mass parameters for the broken symmetry:
\begin{equation}
Y_{M+j}=\frac{N-M+1-2j}{2}(iQ-2m_A)+V,\qquad j=1,\cdots,N-M\,.
\end{equation}
We will schematically represent the asymmetric $\mathcal{S}$-wall as in Figure \ref{Figures/Section3d/OperatorsAndBlocks/AsymmS3d.pdf}.

\bigskip

\begin{figure}[!ht]
	\centering
	\includegraphics[scale=0.42]{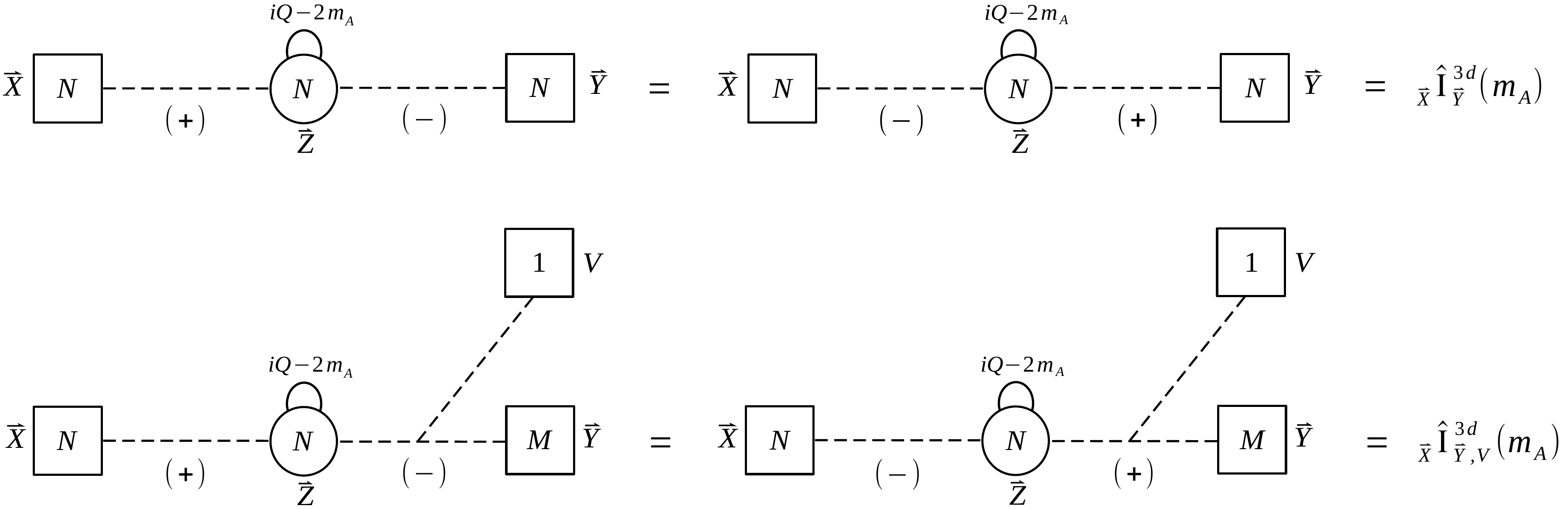}
	\caption{Gluing two $\mathcal{S}$-walls to gives an Identity-wall,  in the symmetric case first line and in the asymmetric case second line. }
	\label{Figures/Section3d/OperatorsAndBlocks/SSid3d.pdf}
\end{figure}

\noindent\textbf{Identity-wall}. As we discussed extensively in $4d$, also in $3d$ we expect the $SL(2,\mathbb{Z})$ operators to obey the multiplication rules of the $SL(2,\mathbb Z)$ group. In particular, we expect the $\mathcal{S}$-wall to satisfy the relations $\mathcal{S}\mathcal{S}^{-1}=1$ and $\mathcal{S}^2=-1$. Unlike in the $4d$ case, these two correspond to distinct identities satisfied by the partition function of the $3d$ $\mathcal{S}$-wall. Indeed, as it was shown in \cite{Bottini:2021vms}, from the $4d$ identity \eqref{eq:deltaNN} one can obtain two identities, namely\footnote{
These identities  were already known to follow  from the closed-form expression of the round sphere partition function of the $T[SU(N)]$ found in \cite{Benvenuti:2011ga,Nishioka:2011dq}.}

\begin{align}
\label{3ddeltaTSUN}
&\int\udl{\vec{Z}_N}\Delta^{3d}_N(\vec{Z};m_A)\mathcal{Z}_{\mathcal{S}}^{(N)}(\vec{Z};\vec{X};m_A)\mathcal{Z}_{\mathcal{S}}^{(N)}(\vec{Z};-\vec{Y};m_A)={}_{\vec X}\hat{\mathbb{I}}^{3d}_{\vec Y}(m_A)
\end{align}
and
\begin{align}
&\int\udl{\vec{Z}_N}\Delta^{3d}_N(\vec{Z};m_A)\mathcal{Z}_{\mathcal{S}}^{(N)}(\vec{Z};\vec{X};m_A)\mathcal{Z}_{\mathcal{S}}^{(N)}(\vec{Z};\vec{Y};m_A)={}_{\vec X}\hat{\mathbb{I}}^{3d}_{-\vec Y}(m_A)\,,
\end{align}
depending on which limit we take between
\be
X_i\to X_i+s,\quad Y_i\to Y_i+s\, \quad s\to+\infty\,,
\ee
and
\be
X_i\to X_i+s,\quad Y_i\to Y_i-s\, \quad s\to+\infty\,.
\ee
In both cases, we also take $\Delta \to -\infty$. We defined the three-dimensional Identity operator
\begin{equation}\label{eq:idopNN3d}
{}_{\vec X}\hat{\mathbb{I}}^{3d}_{\vec Y}(m_A)=\frac{\sum_{\gs\in S_N}\prod_{j=1}^N\gd\left(X_j-Y_{\gs(j)}\right)}{\Gd^{3d}_N(\vec{X};m_A)}
\end{equation}
with
\begin{equation}
\Gd_N^{3d}(\vec{X};m_A)=\frac{\prod_{i,j=1}^N\sbfunc{-i\frac{Q}{2}+(X_i-X_j)+2m_A}}{\prod_{i<j}^N\sbfunc{i\frac{Q}{2}\pm(X_i-X_j)}}\,.
\end{equation}
We also define the three-dimensional asymmetric Identity-wall ${}_{\vec X}\hat{\mathbb{I}}^{3d}_{\vec Y,V}(m_A)$  as
\begin{equation}
\label{eq:idop3d}
{}_{\vec X}\hat{\mathbb{I}}^{3d}_{\vec Y,V}(m_A)=\frac{1}{\Gd^{3d}_N(\vec{X};m_A)}\left.\sum_{\sigma \in S_N}\prod_{i=1}^{N}\delta\left(X_i-Y_{\sigma(i)}\right)\right|_{Y_{M+j}=\frac{N-M+1-2j}{2}(iQ-2m_A)+V}\,.
\end{equation}
Both the symmetric and the asymmetric Identity relations are represented in Figure \ref{Figures/Section3d/OperatorsAndBlocks/SSid3d.pdf}.

\bigskip

\begin{figure}[!ht]
	\centering
	\includegraphics[scale=0.8]{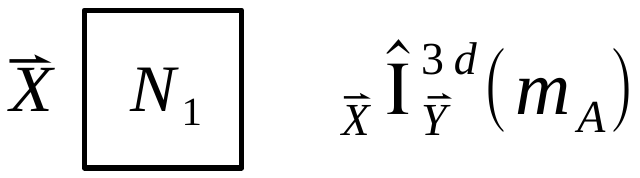}
	\caption{The $3d$ $\mathcal{T}$-wall.}
	\label{Figures/Section3d/OperatorsAndBlocks/Top3d.pdf}
\end{figure}

\noindent\textbf{$\mathcal{T}$-wall}. The $\mathcal{T}$ generator of $SL(2,\mathbb{Z})$ in $3d$ is simply associated with the addition of a CS level (see Figure \ref{Figures/Section3d/OperatorsAndBlocks/Top3d.pdf}).
\\
We then define its contribution to the $S^3_b$ partition function as
\begin{equation}\label{eq:Tgen3d}
\mathcal{Z}_{\mathcal{T}}^{(N)}(\vec{X};\vec{Y};m_A) =\e^{\frac{i\pi N}{24}(8m_A^2(N-1)-4im_AQ(N-1)+Q^2)} \e^{-i\pi\sum_{i=1}^NX_i^2} {}_{\vec x}\hat{\mathbb{I}}^{3d}_{\vec y}(m_A)\,,
\end{equation}
which can be obtained as a limit of \eqref{eq:indTwall4d} by taking
\begin{align}
X_i\to X_i+s,\quad Y_j\to Y_j+s, \quad V \to V+s, \quad s\to+\infty\,,
\end{align}
and then
\begin{align}
\Delta \to \infty, \qquad D \to \infty
\end{align}
with $v = \e^{2 \pi i \beta V}$ and $d=\e^{2\pi i\beta D}$; the other variables are defined in \eqref{eq:var}.
The first prefactor is included so to simplify the identities for the basic duality moves and to recover  the relation $(\mathcal{S}\mathcal{T})^3=1$ as will see momentarily.

\bigskip

\begin{figure}[!ht]
	\centering
	\includegraphics[scale=0.8]{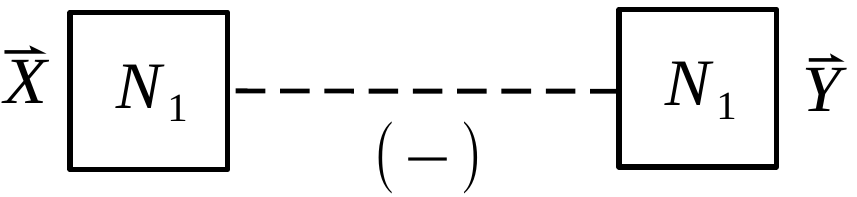}
	\caption{The $3d$ $\mathcal{T}^T$-wall.}
	\label{Figures/Section3d/OperatorsAndBlocks/TTop3d.pdf}
\end{figure}

\noindent\textbf{$\mathcal{T}^T$-wall}. Finally, for completeness we also discuss the $\mathcal{T}^T$ operator. This is not an independent operator, since it can be written in terms of the $\mathcal S$ and $\mathcal{T}$ generators as $\mathcal{T}^T=-\mathcal{T}\mathcal{S}\mathcal{T}$. Accordingly, we associate to it one copy of the $FT[U(N)]$ $\mathcal{S}$-wall theory with background CS level $+1$ for both of its $U(N)$ global symmetries (see Figure \ref{Figures/Section3d/OperatorsAndBlocks/TTop3d.pdf}). 

The $\mathcal T^T$ operator can also be obtained from the $4d$ $\mathsf T^T$ operator 
like the $\mathcal S$ and $\mathcal T$ are obtained from $\mathsf S$ and $\mathsf T$ in $4d$ by taking some limits and discarding unwanted background CS couplings.
Specifically, at the level of the sphere partition function, we can take the $3d$ limit of eq.~\eqref{eq:T^T} with $\Delta\to\infty$ and $D\to\infty$, which leads to
\begin{align}
\label{eq:T^T3d}
\mathcal{Z}_{\mathcal{T}^T}^{(N)} (\vec{X};\vec{Y};m_A) &=\int\udl{\vec{Z}^{(1)}_N}\udl{\vec{Z}^{(2)}_N}\Gd_N^{3d}(\vec{Z}^{(1)};m_A)\Gd_N^{3d}(\vec{Z}^{(2)};m_A)\mathcal{Z}_{\mathcal{T}}^{(N)}(\vec{X};\vec{Z}^{(1)};m_A)\nn\\
&\qquad\times \mathcal{Z}_{\mathcal{S}^{-1}}^{(N)}(\vec{Z}^{(1)};\vec{Z}^{(2)};m_A)\mathcal{Z}_{\mathcal{T}}^{(N)}(\vec{Z}^{(2)};\vec{Y};m_A)\nn\\
&=\e^{\frac{i\pi N}{12}(8m_A^2(N-1)-4im_AQ(N-1)+Q^2)} \e^{-i\pi(\sum_{i=1}^NX_i^2+Y_i^2)} \mathcal{Z}_{\mathcal{S}^{-1}}^{(N)}(\vec{X};\vec{Y};m_A)
\end{align}
after subtracting the flavor CS levels by $+1$ and $-1$ for $U(N)_X$ and $U(N)_Y$, respectively.
If we replace $\mathcal{Z}_{\mathcal{S}^{-1}}^{(N)}(\vec{X};\vec{Y};m_A)$ by
\begin{align}
\label{eq:Sinv}
\mathcal{Z}_{\mathcal{S}^{-1}}^{(N)}(\vec{X};\vec{Y};m_A) = \int \ud \vec Z_N \Gd^{3d}_N(\vec{Z};m_A) \mathcal{Z}_{\mathcal{S}}^{(N)}(\vec{X};\vec{Z};m_A) {}_{\vec Z}\hat{\mathbb{I}}^{3d}_{-\vec Y}(m_A) \,,
\end{align}
which corresponds to $\mathcal S^{-1} = -\mathcal S$, we can see this is exactly the relation $\mathcal{T}^T=-\mathcal{T}\mathcal{S}\mathcal{T}$.

\bigskip

\noindent\textbf{$SL(2,\mathbb{Z})$ relations}. We have already discussed the relation $\mathcal{S}^2=-1$ .
The relation $(\mathcal{S}\mathcal{T})^3=1$ can be derived as shown in \cite{Bottini:2021vms} as a limit of the braid relation \eqref{eq:braid}.
 
Implementing the $3d$ limit following the steps above and turning on real masses for the $\Delta\to\infty$ and $D\to\infty$ we obtain the following relation 
\be
 \mathcal{T} \mathcal{S}^{-1}\mathcal{TS}^{-1}\mathcal{T}= \mathcal{S}^{-1} \,,
\ee
or equivalently, $(\mathcal{ST})^3=1$ once we use \eqref{eq:Sinv} and implement the delta functions.

\subsection{QFT building blocks}

The fundamental QFT building blocks in $3d$ are those naturally associated with the possible types of 5-branes that we can have in the Hanany--Witten brane setup. We focus in particular on NS5, D5 and (1,1)-branes.

\bigskip

\begin{figure}[!ht]
	\centering
	\includegraphics[scale=0.8]{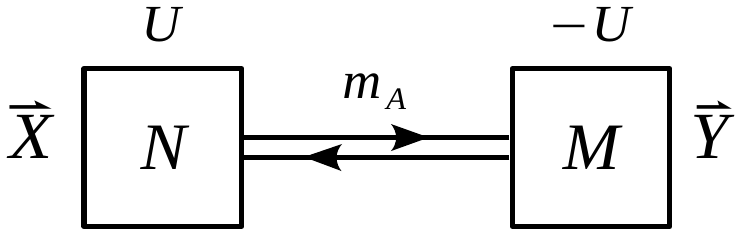}
	\caption{The $\mathcal{B}_{10}$ block.}
	\label{Figures/Section3d/OperatorsAndBlocks/B103d.pdf}
\end{figure}

\noindent\textbf{The $\mathcal{B}_{10}$ block.}
To an NS5-brane with $N$ D3-branes ending on its left and $M$ on its right is associated a $U(N)\times U(M)$ bifundamental hypermultiplet, which we will also call $\mathcal{B}_{10}$ block (see Figure \ref{Figures/Section3d/OperatorsAndBlocks/B103d.pdf}). 

\noindent Its contribution to the $S^3_b$ partition function  is given by:
\begin{equation}\label{eq:ZNS5}
\mathcal{Z}_{(1,0)}^{(N,M)}(\vec{X};\vec{Y};U;m_A) =\e^{2\pi iU\left(\sum_{i=1}^NX_i-\sum_{j=1}^MY_j\right)} \prod_{i=1}^N\prod_{j=1}^M \sbfunc{\frac{i Q}{2}-m_A\pm(X_i-Y_j)}\,.
\end{equation}

Here $U$ appears as an FI parameter.\footnote{More precisely, at this level since the $U(N)$ and $U(M)$ symmetries are not gauged it is actually a background BF coupling. Inside quivers in which the $U(N)$ and $U(M)$ symmetries are gauged it really becomes an FI.}
This expression can be obtained from  the $4d$  $\mathsf{B}_{10}$ block in \eqref{eq:indtriangle} by taking the $3d$ limit with the usual three steps.
The first involves the $3d$ reduction where the $4d$ fugacities and the $3d$ real masses are related by
\be
x_i=\e^{2\pi i\beta X_i},\quad y_j=\e^{2\pi i\beta Y_j},\quad t=\e^{2\pi i\beta (iQ-2m_A)},\quad c=\e^{2\pi i\beta \Delta},\quad u=\e^{2\pi i\beta U}\,.
\ee
The second step is a real mass deformation that breaks the $USp(2N)\times USp(2M)$ symmetry to $U(N)\times U(M)$
\be
\label{eq:10 limit}
X_i\to X_i+s\,,\quad Y_j\to Y_j+s\,, \quad U \to U-s\,, \quad s\to+\infty\,,
\ee
Finally, we consider the real mass deformation $\Delta\to-\infty$,
which in particular has the effect of removing the lower part of the triangle in Figure \ref{Figures/Section2/OperatorsAndBlocks/Triangle.pdf}. This also makes the parameter $\pm U$ appear as an FI parameter and produces a background CS level $\mp1$ for the $U(N)$ and $U(M)$ nodes respectively. Removing all the exponential prefactors that are produced in the limit except for those encoding the FI's, from the index \eqref{eq:indtriangle} of the $\mathsf{B}_{10}$ block we recover the $S^3_b$ partition function \eqref{eq:ZNS5} of the $\mathcal{B}_{10}$ block.

Note that we can also obtain the block $\mathcal B_{-10}$ by acting with $-I$ on each side of $\mathcal B_{10}$, which can be expressed by the following identity:
\begin{align}
\mathcal{Z}_{(-1,0)}^{(N,M)}(\vec{X};\vec{Y};U;m_A) &= \oint \udl{\vec{Z}_{N}} \udl{\vec{W}_{M}} \Gd^{3d}_N(\vec{Z};m_A) \Gd^{3d}_M(\vec{W};m_A) {}_{\vec Z}\hat{\mathbb{I}}^{3d}_{-\vec X}(m_A) {}_{\vec W}\hat{\mathbb{I}}^{3d}_{-\vec Y}(m_A) \nonumber \\
&\quad \times \e^{2\pi iU\left(\sum_{i=1}^NZ_i-\sum_{j=1}^MW_j\right)} \prod_{i=1}^N\prod_{j=1}^M \sbfunc{\frac{i Q}{2}-m_A\pm(Z_i-W_j)} \nonumber\\
&=\e^{-2\pi iU\left(\sum_{i=1}^NX_i-\sum_{j=1}^MY_j\right)} \prod_{i=1}^N\prod_{j=1}^M \sbfunc{\frac{i Q}{2}-m_A\pm(X_i-Y_j)} , \label{eq:B-10}
\end{align}
which is simply $\mathcal{Z}_{(1,0)}^{(N,M)}(\vec{X};\vec{Y};-U;m_A)$ where the sign of $U$ is flipped.

\bigskip

\begin{figure}[!ht]
	\centering
	\includegraphics[scale=0.8]{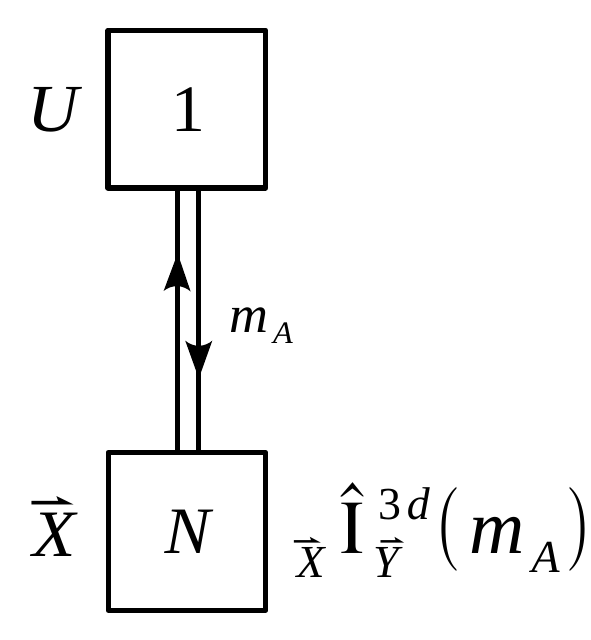}
	\caption{The symmetric $\mathcal{B}_{01}$ block. Here and in the following figures, a parameter $F$ close to a chiral corresponds to part of the argument of its contribution to the $S^3_b$ partition function $\sbfunc{i\frac{Q}{2}-F+\text{(non-Abelian)}}$.}
	\label{Figures/Section3d/OperatorsAndBlocks/B013d.pdf}
\end{figure}

\noindent\textbf{The $\mathcal{B}_{01}$ block}.
To a (0,1)-brane  with $N$ D3-branes ending both on its left and on its right we associated a single $U(N)$ fundamental hypermultiplet, the $\mathcal{B}_{01}$ block depicted in Figure \ref{Figures/Section3d/OperatorsAndBlocks/B013d.pdf}.
Similarly to the $4d$ $\mathsf{B}_{01}$ block, we equip this with one copy of the Identity-wall to give it a structure with two sets of $U(N)$ indices. 
\\
Its contribution to the $S^3_b$ partition function is given by
\begin{equation}
\mathcal{Z}_{(0,1)}^{(N,N)} (\vec{X};\vec{Y};U;m_A) = \prod_{j=1}^N \sbfunc{\frac{iQ}{2}-m_A\pm(X_i-U)} {}_{\vec X}\hat{\mathbb{I}}^{3d}_{\vec Y}(m_A)\,.
\end{equation}
In analogy with what we did in $4d$, it is useful to generalize this to a D5 suspended between different numbers of D3-branes, say $N$ D3's on the left and $M$ D3's on the right. The associated partition function is given by
\begin{equation}
\label{eq:01NM3d}
\mathcal{Z}_{(0,1)}^{(N,M)} (\vec{X};\vec{Y};U;m_A) = \prod_{j=1}^M \sbfunc{\frac{iQ}{2} - m_A(N-M+1) \pm(Y_j-U)} {}_{\vec X}\hat{\mathbb{I}}^{3d}_{\vec Y,U}(m_A)\,.
\end{equation}

Similarly to our previous discussion for the NS5 basic block, one can obtain the expression \eqref{eq:01NM3d} for the $S^3_b$ partition function of the $\mathcal{B}_{01}$ block as a limit of that \eqref{eq:01NM} for the $4d$ index of the $\mathsf{B}_{01}$ block with the identifications 
\be
x_i=\e^{2\pi i\beta X_i},\quad y_j=\e^{2\pi i\beta Y_j},\quad t=\e^{2\pi i\beta (iQ-2m_A)},\quad u=\e^{2\pi i\beta U}\,,
\ee
followed by the real mass $X_i\to X_i+s,\quad Y_j\to Y_j+s, \quad U \to U+s, \quad s\to+\infty$.
We can also introduce the block $\mathcal B_{0-1}$ by acting with $-I$ on each side of $\mathcal B_{01}$
which results into
 $\mathcal{Z}_{(0,-1)}^{(N,M)} (\vec{X};\vec{Y};U;m_A)=\mathcal{Z}_{(0,1)}^{(N,M)} (\vec{X};\vec{Y};-U;m_A)$.

\bigskip

\begin{figure}[!ht]
	\centering
	\includegraphics[scale=0.8]{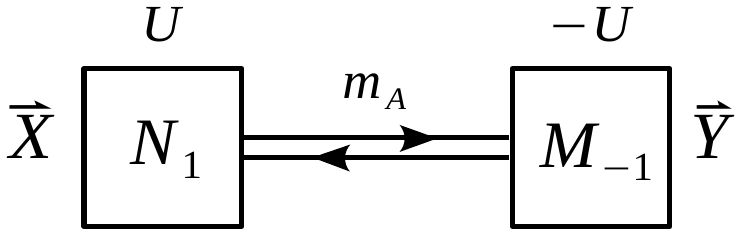}
	\caption{The $\mathcal{B}_{11}$ block.}
	\label{Figures/Section3d/OperatorsAndBlocks/B113d.pdf}
\end{figure}

\noindent\textbf{The $\mathcal{B}_{11}$ block.}
Finally, the QFT building block associated to a $(1,1)$-brane with $N$ D3 branes ending on its left and $M$ on its right is a $U(N)\times U(M)$ bifundamental hypermultiplet with the addition of background CS levels $+1$ and $-1$ respectively for the $U(N)$ and the $U(M)$ global symmetries, which we will also call $\mathcal{B}_{11}$ block (see Figure \ref{Figures/Section3d/OperatorsAndBlocks/B113d.pdf}).

\noindent At the level of the sphere partition function we have
\begin{align}\label{eq:Z11brane}
\mathcal{Z}_{(1,1)}^{(N,M)}(\vec{X};\vec{Y};U;m_A) &=\e^{-i\pi\left(\sum_{i=1}^NX_i^2-\sum_{j=1}^MY_j^2\right)}\e^{2\pi iU\left(\sum_{i=1}^NX_i-\sum_{j=1}^MY_j\right)} \nn\\
&\times\prod_{i=1}^N\prod_{j=1}^M \sbfunc{\frac{iQ}{2}-m_A\pm(X_i-Y_j)}\,,
\end{align}
which again can be obtained as the $3d$ limit of the $4d$ index \eqref{eq:11NM} (with $D \to +\infty$ in addition to \eqref{eq:10 limit}) of the $\mathsf{B}_{11}$ block after removing some of the exponential prefactors.

\subsection{Basic duality moves}

Having defined the $3d$ QFT blocks and the $SL(2,\mathbb{Z})$ operators, we are ready to present the $3d$ version of the basic duality moves that play a central role in the dualization algorithm. We will get them as a limit of the $4d$ ones, but we stress that they can be independently derived by iterative application of the Aharony duality \cite{Aharony:1997gp}.


\subsubsection{$\mathcal{S}$-dualization}

\begin{figure}[!ht]
\includegraphics[scale=0.4,center]{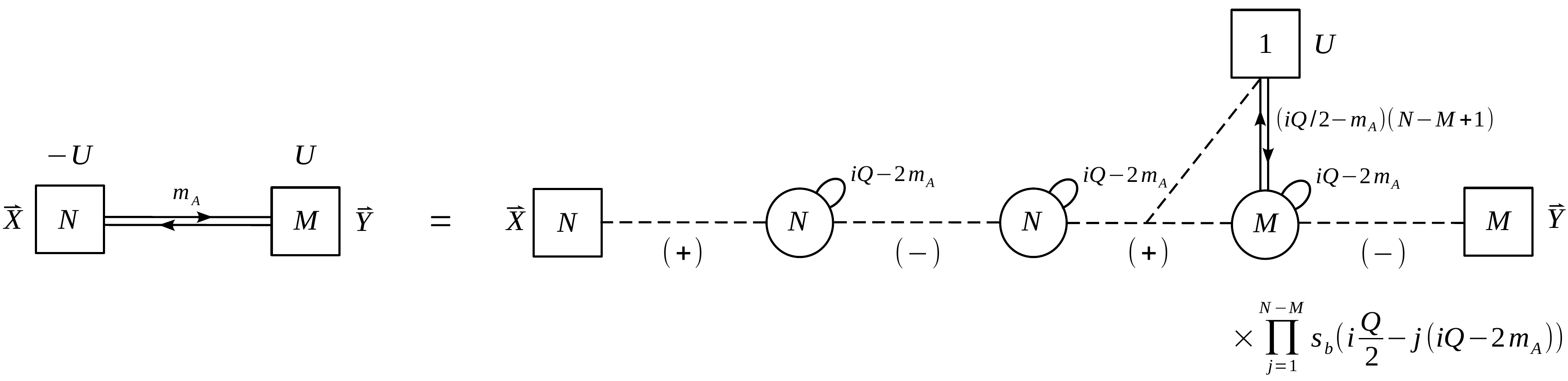}
\caption{The $\mathcal{B}_{-10}=\mathcal{S}\mathcal{B}_{01} \mathcal{S}^{-1}$ duality move.}
\label{fig:10_Sdual_3d}
\end{figure}

\noindent\textbf{\boldmath$\mathcal{B}_{-10}=\mathcal{S}\mathcal{B}_{01} \mathcal{S}^{-1}$}. The first duality move is the $\mathcal{S}$-dualization of a $(0,1)$-brane into a $(-1,0)$-brane, a $\overline{\text{NS5}}$, which we interpret as the  field theory duality in Figure \ref{fig:10_Sdual_3d}.

\noindent This corresponds to the  following identity of $S^3_b$ partition functions:
%
\begin{align}\label{eq:id13d}
&\mathcal{Z}_{(-1,0)}^{(N,M)} (\vec{X};\vec{Y};U;m_A)= \prod_{j=1}^{N-M} s_b\left(i\frac{Q}{2}-j(iQ-2m_A) \right)
\int\left(\prod_{k=1}^2\udl{\vec{Z}^{(k)}_M}\Gd_M^{3d}(\vec{Z}^{(k)};m_A)\right)\nn\\
&\qquad\qquad\times
\mathcal{Z}_{\mathcal{S}}^{(N)}\left(\vec{X};\vec{Z}^{(1)}\right)
\mathcal{Z}_{(0,1)}^{(N,M)} \left(\vec{Z}^{(1)};\vec{Z}^{(2)};U;i\frac{Q}{2}-m_A\right)
\mathcal{Z}_{\mathcal{S}^{-1}}^{(M)}(\vec{Z}^{(2)};\vec{Y};m_A) \,,
\end{align}
where the partition function for the $\mathcal B_{-10}$ block on the left hand side is given by that of $\mathcal B_{10}$ with the FI parameter $-U$ as shown in \eqref{eq:B-10}.
This identity can be obtained as a limit of the $4d$ one in \eqref{eq:id1} following the usual three steps. In the process we define the $3d$ parameters as
\be
&&x_i=\e^{2\pi i\beta X_i},\quad y_j=\e^{2\pi i\beta Y_j},\quad Z_a^{(k)}=\e^{2\pi i\beta z^{(k)}_a},\nn\\
&&t=\e^{2\pi i\beta (iQ-2m_A)},\quad c=\e^{2\pi i\beta \Delta},\quad u=\e^{2\pi i\beta U}\,.
\ee
 Then we consider the real mass deformation breaking all the symplectic groups down to unitary:
\begin{align}
X_i\to X_i+s,\quad Y_j\to Y_j+s,\quad Z^{(k)}_a\to Z^{(k)}_a+s, \quad U \to U+s, \quad s\to+\infty\,.
\end{align}
Finally, the last step is the real mass deformation
\begin{equation}
\Delta\to-\infty\,.
\end{equation}


It turns out that all the exponential prefactors produced in the first two steps perfectly cancel between the two sides of the identity, while the only prefactor surviving in the last step is the FI containing $U$ that we included in our definition \eqref{eq:B-10} of the $\mathsf{B}_{-10}$ block, thus providing a justification for it. In particular, it is crucial that all the factors involving the parameters that we sent to infinity cancel, so to give a well-defined identity also in $3d$.

\bigskip

\begin{figure}[!ht]
\includegraphics[scale=0.4,center]{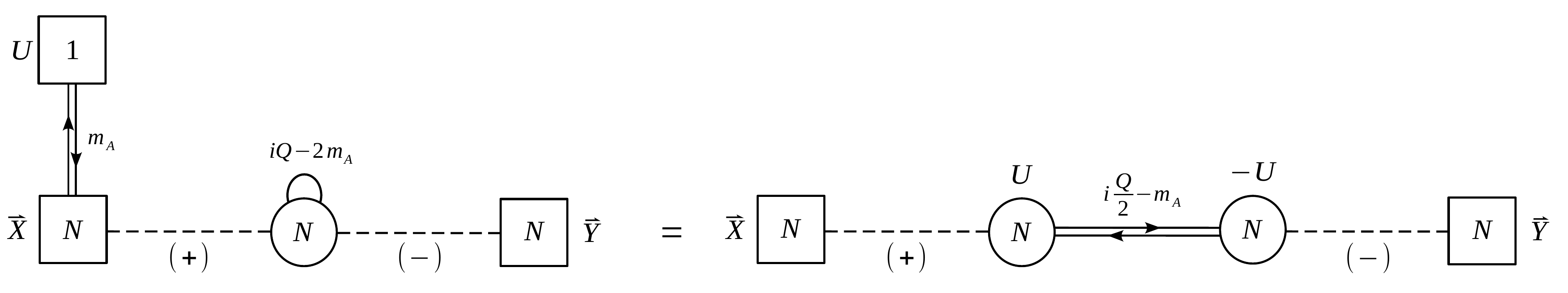}
\caption{The $\mathcal{B}_{01}=\mathcal{S}\mathcal{B}_{10} \mathcal{S}^{-1}$ duality move.}
\label{fig:01_Sdual_3d}
\end{figure}

\noindent\textbf{\boldmath$\mathcal{B}_{01}=\mathcal{S}\mathcal{B}_{10} \mathcal{S}^{-1}$}. We next consider the QFT analogue of the $\mathcal{S}$-dualization of a $(1,0)$-brane into a $(0,1)$-brane  shown in Figure \ref{fig:01_Sdual_3d}.

\noindent In this case, the associated partition function identity is
\begin{align}\label{eq:id23d}
\mathcal{Z}_{(0,1)}^{(N)} (\vec{X};\vec{Y};U;m_A)&=\int\udl{\vec{Z}^{(1)}_N}\udl{\vec{Z}^{(2)}_N}\Gd_N^{3d}(\vec{Z}^{(1)})\Gd_N^{3d}(\vec{Z}^{(2)})\mathcal{Z}_{\mathcal{S}}^{(N)}(\vec{X};\vec{Z}^{(1)};m_A)\nonumber\\
&\quad\times\mathcal{Z}_{(1,0)}^{(N,N)}\left(\vec{Z}^{(1)};\vec{Z}^{(2)};U;i\frac{Q}{2}-m_A\right)\mathcal{Z}_{\mathcal{S}^{-1}}^{(N)}(\vec{Z}^{(2)};\vec{Y};m_A)\,.
\end{align}

\noindent This can be obtained as a limit of the $4d$ identity \eqref{eq:id2} that is completely analogous to the 
previous duality move, now with $\Delta \to +\infty$.

\bigskip

\begin{figure}[!ht]
\includegraphics[scale=0.5,center]{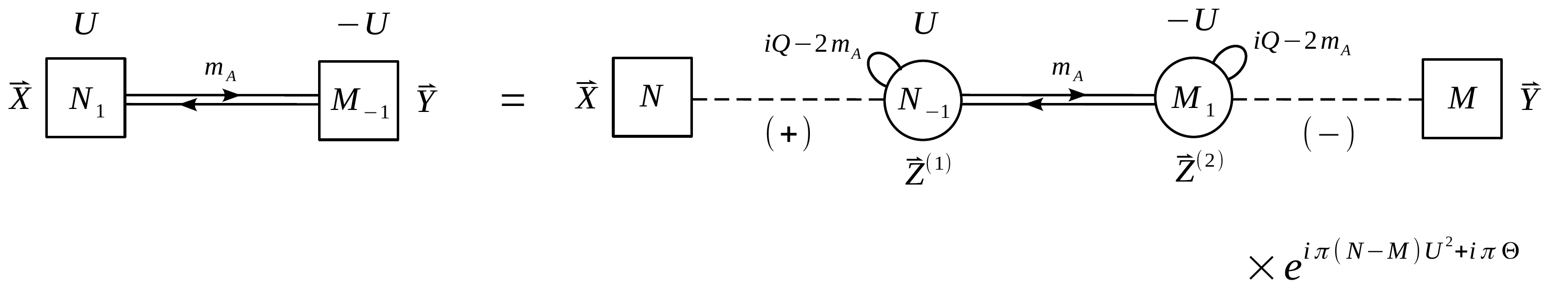}
\caption{The $\mathcal{B}_{11}=\mathcal{S}\mathcal{B}_{1-1} \mathcal{S}^{-1}$ duality move.}
\label{fig:11_Sdual_3d}
\end{figure}

\noindent\textbf{\boldmath$\mathcal{B}_{11}=\mathcal{S}\mathcal{B}_{1-1} \mathcal{S}^{-1}$}. The last $\mathcal{S}$-dualization we consider is the one relating a $(1,1)$-brane and a (1,-1)-brane shown in Figure \ref{fig:11_Sdual_3d}.

\noindent At the level of the $S^3_b$ partition function, this duality move is encoded in the following identity:
\begin{align}\label{eq:S113d}
\mathcal{Z}_{(1,1)}^{(N,M)} (\vec{X};\vec{Y};U;m_A)&=\e^{i\pi(N-M)U^2+i\pi \Theta}\int\udl{\vec{Z}^{(1)}_N}\udl{\vec{Z}^{(2)}_M}\Gd_N^{3d}(\vec{Z}^{(1)};m_A)\Gd_M^{3d}(\vec{Z}^{(2)};m_A)\nonumber\\
&\quad\times\mathcal{Z}_{\mathcal{S}}^{(N)}(\vec{X};\vec{Z}^{(1)};m_A)\mathcal{Z}_{(1,-1)}^{(N,M)}(\vec{Z}^{(1)};\vec{Z}^{(2)};U;m_A)\mathcal{Z}_{\mathcal{S}^{-1}}^{(M)}(\vec{Z}^{(2)};\vec{Y};m_A)\,,
\end{align}
where 
\begin{align}\label{eq:thetaschif}
\Theta&=\frac{N-M}{24}\left[8m_a^2\left((N-M)^2-3(N+M)+2\right)\right.\nn\\
&\quad\left.-4im_AQ\left(2(N-M)^2-3(N+M)+1\right)-Q^2\left(2(N-M)^2+1\right)\right] \,.
\end{align}
We have defined the contribution of the $(1,-1)$-brane, that is the $\mathsf{B}_{\text{1-1}}$ block, as
\begin{align}\label{eq:1m13d}
\mathcal{Z}_{\text{(1,-1)}}^{(N,M)}(\vec{X};\vec{Y};U;m_A)&=\e^{i\pi\left(\sum_{i=1}^NX_i^2-\sum_{j=1}^MY_j^2\right)}\e^{2\pi iU\left(\sum_{i=1}^NX_i-\sum_{j=1}^MY_j\right)} \nn\\
&\quad\times\prod_{i=1}^N\prod_{j=1}^M   \sbfunc{\frac{i Q}{2}-m_A\pm(X_i-Y_j)}.
\end{align}
Notice that this differs from the contribution of a $(1,1)$ for the fact that the CS levels are inverted.
The identity \eqref{eq:S113d} again can be obtained as a limit of \eqref{eq:11S1-1}
with 
\begin{align}
\label{eq:3d limit}
X_i\to X_i+s,\quad Y_j\to Y_j+s,\quad Z^{(k)}_a\to Z^{(k)}_a+s, \quad U \to U-s, \quad V \to V+s, \quad s\to+\infty
\end{align}
followed by $\Delta \to -\infty$ and $D \to +\infty$ satisfying $\Delta+D \to -\infty$, where $V$ here and below is defined such that $v = \e^{2 \pi i \beta V}$. 
By looking at the result of the limit  it is easy to check that  one of the two parameters  $U$ and $V$ is redundant, and we can take $V = 0$ to obtain \eqref{eq:1m13d}. In other words, we can simply take $V \to s\to +\infty$ in \eqref{eq:3d limit} to obtain \eqref{eq:1m13d}.

\subsubsection{$\mathcal{T}$-dualization}

\begin{figure}[!ht]
\includegraphics[scale=0.4,center]{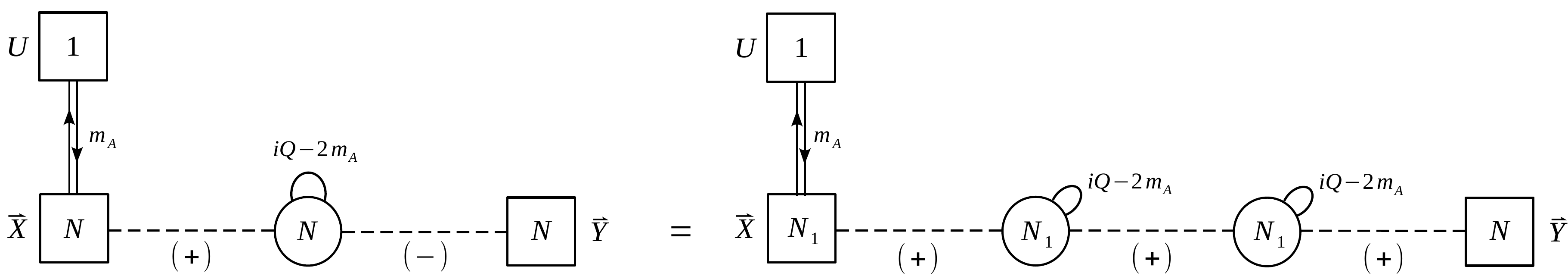}
\caption{The $\mathcal{B}_{01}=\mathcal{T}\mathcal{B}_{01} \mathcal{T}^{-1}$ duality move.}
\label{fig:01_Tdual_3d}
\end{figure}

\noindent\textbf{\boldmath$\mathcal{B}_{01}=\mathcal{T}\mathcal{B}_{01} \mathcal{T}^{-1}$.} A D5-brane is invariant under the $\mathcal{T}$-dualization as shown in Figure \ref{fig:01_Tdual_3d}.

\noindent Recalling that the inverse of $\mathcal{T}$ is given by $\mathcal{T}^{-1} = \mathcal{S}\mathcal{T}\mathcal{S}\mathcal{T}\mathcal{S}$, we find that this dualization implies the following identity of $S^3_b$ partition functions:
\begin{align}
\mathcal{Z}_{(0,1)}^{(N)} (\vec{X};\vec{Y};U;m_A)&=\int\left(\prod_{k=1}^6\udl{\vec{Z}^{(k)}_N}\Gd_N^{3d}(\vec{Z}^{(k)};m_A)\right)\mathcal{Z}_{\mathcal{T}}^{(N)}(\vec{X};\vec{Z}^{(1)};m_A)\nn\\
&\times\mathcal{Z}_{(0,1)}^{(N)} (\vec{Z}^{(1)};\vec{Z}^{(2)};U;m_A)\mathcal{Z}_{\mathcal{S}}^{(N)}(\vec{Z}^{(2)};\vec{Z}^{(3)};m_A)\mathcal{Z}_{\mathcal{T}}^{(N)}(\vec{Z}^{(3)};\vec{Z}^{(4)};m_A)\nn\\
&\times\mathcal{Z}_{\mathcal{S}}^{(N)}(\vec{Z}^{(4)};\vec{Z}^{(5)};m_A)\mathcal{Z}_{\mathcal{T}}^{(N)}(\vec{Z}^{(5)};\vec{Z}^{(6)};m_A)\mathcal{Z}_{\mathcal{S}}^{(N)}(\vec{Z}^{(6)};\vec{Y};m_A)\,.
\end{align}

\noindent We point out that this identity doesn't have any additional prefactor thanks to the fact that we defined the $\mathcal{T}$ generator in \eqref{eq:Tgen3d} with an extra prefactor. Again this can be obtained as a limit of the corresponding $4d$ identity \eqref{eq:T01}
with 
\begin{gather}
\begin{gathered}
X_i\to X_i+s,\quad Y_j\to Y_j+s,\quad Z^{(1,\dots,4)}_a\to Z^{(1,\dots,4)}_a+s, \\
Z^{(5,6)}_a\to Z^{(5,6)}_a-s, \quad U \to U+s, \quad V \to s \,, \quad s\to+\infty
\end{gathered}
\end{gather}
followed by $\Delta \to -\infty$ and $D \to +\infty$ satisfying $\Delta+D \to -\infty$. Note that the sign of $s$ for $Z^{(5,6)}$ is flipped to obtain $\mathcal S$ rather than $\mathcal S^{-1}$ directly. In addition, we simply take $V \to s \to +\infty$ because $V$ becomes redundant after the limit, as mentioned before.

\bigskip

\begin{figure}[!ht]
\includegraphics[scale=0.4,center]{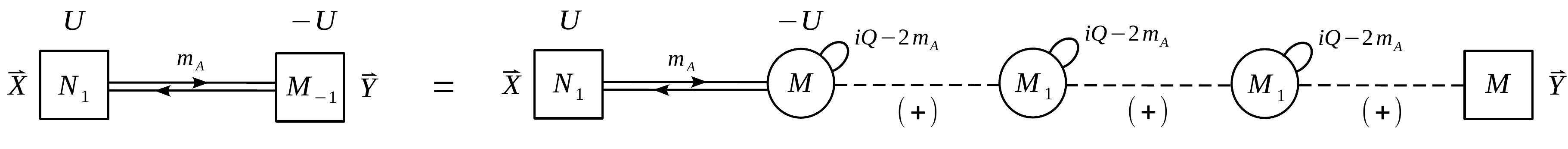}
\caption{The $\mathcal{B}_{11}=\mathcal{T}\mathcal{B}_{10} \mathcal{T}^{-1}=\mathcal{T}\mathcal{B}_{10} \mathcal{S}\mathcal{T}\mathcal{S}\mathcal{T}\mathcal{S}$ duality move.}
\label{fig:11_Tdual_3d}
\end{figure}

\noindent\textbf{\boldmath$\mathcal{B}_{11}=\mathcal{T}\mathcal{B}_{10} \mathcal{T}^{-1}$.} We now consider the  $\mathcal{T}$-dualization  of a $(1,0)$-brane into a $(1,1)$-brane. Using again that $\mathcal{T}^{-1} = \mathcal{STSTS}$ this corresponds to the duality
in Figure \ref{fig:11_Tdual_3d}.

\noindent At the level of partition functions we have the following identity:
\begin{align}
\mathcal{Z}_{(1,1)}^{(N,M)} (\vec{X};\vec{Y};U;m_A)&=\int\udl{\vec{Z}^{(1)}_N}\Gd_N^{3d}(\vec{Z}^{(1)};m_A)\left(\prod_{k=2}^6\udl{\vec{Z}^{(k)}_M}\Gd_M^{3d}(\vec{Z}^{(k)};m_A)\right)\mathcal{Z}_{\mathcal{T}}^{(N)}(\vec{X};\vec{Z}^{(1)};m_A)\nn\\
&\times\mathcal{Z}_{(1,0)}^{(N,M)} (\vec{Z}^{(1)};\vec{Z}^{(2)};U;m_A)\mathcal{Z}_{\mathcal{S}}^{(M)}(\vec{Z}^{(2)};\vec{Z}^{(3)};m_A)\mathcal{Z}_{\mathcal{T}}^{(M)}(\vec{Z}^{(3)};\vec{Z}^{(4)};m_A)\nn\\
&\times\mathcal{Z}_{\mathcal{S}}^{(M)}(\vec{Z}^{(4)};\vec{Z}^{(5)};m_A)\mathcal{Z}_{\mathcal{T}}^{(M)}(\vec{Z}^{(5)};\vec{Z}^{(6)};m_A)\mathcal{Z}_{\mathcal{S}}^{(M)}(\vec{Z}^{(6)};\vec{Y};m_A)\,,
\end{align}
which again can be obtained as a limit of the corresponding $4d$ identity \eqref{eq:T11}
with 
\begin{gather}
\begin{gathered}
\label{eq:limit}
X_i\to X_i+s,\quad Y_j\to Y_j+s,\quad Z^{(1,\dots,4)}_a\to Z^{(1,\dots,4)}_a+s, \\
Z^{(5,6)}_a\to Z^{(5,6)}_a-s, \quad U \to U-s, \quad V \to s, \quad s\to+\infty
\end{gathered}
\end{gather}
followed by $\Delta \to -\infty$ and $D \to +\infty$ satisfying $\Delta+D \to -\infty$. Again, we simply send $V \to s \to +\infty$ to obtain the identity independent of $V$.

\bigskip

\begin{figure}[!ht]
\includegraphics[scale=0.4,center]{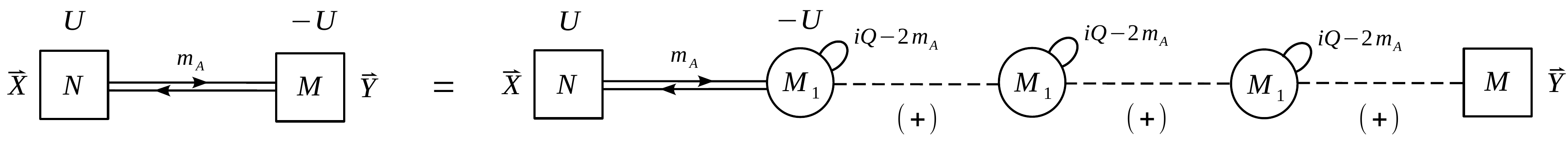}
\caption{The $\mathcal{B}_{10}=\mathcal{T}\mathcal{B}_{1-1} \mathcal{T}^{-1}=\mathcal{T}\mathcal{B}_{1-1} \mathcal{S}\mathcal{T}\mathcal{S}\mathcal{T}\mathcal{S}$ duality move. 
}
\label{fig:10_Tdual_3d}
\end{figure}

\noindent\textbf{\boldmath$\mathcal{B}_{10}=\mathcal{T}\mathcal{B}_{1-1} \mathcal{T}^{-1}$.} Lastly, we consider the $\mathcal{T}$-dualization of a $(1,-1)$-brane into a $(1,0)$-brane, which corresponds to the duality in Figure \ref{fig:10_Tdual_3d}.

\noindent This translates into the identity
\begin{align}
\mathcal{Z}_{(1,0)}^{(N,M)} (\vec{X};\vec{Y};U;m_A)&=\int\udl{\vec{Z}^{(1)}_N}\Gd_N^{3d}(\vec{Z}^{(1)};m_A)\left(\prod_{k=2}^6\udl{\vec{Z}^{(k)}_M}\Gd_M^{3d}(\vec{Z}^{(k)};m_A)\right)\mathcal{Z}_{\mathcal{T}}^{(N)}(\vec{X};\vec{Z}^{(1)};m_A)\nn\\
&\times\mathcal{Z}_{\text{(1,-1)}}^{(N,M)} (\vec{Z}^{(1)};\vec{Z}^{(2)};U;m_A)\mathcal{Z}_{\mathcal{S}}^{(M)}(\vec{Z}^{(2)};\vec{Z}^{(3)};m_A)\mathcal{Z}_{\mathcal{T}}^{(M)}(\vec{Z}^{(3)};\vec{Z}^{(4)};m_A)\nn\\
&\times\mathcal{Z}_{\mathcal{S}}^{(M)}(\vec{Z}^{(4)};\vec{Z}^{(5)};m_A)\mathcal{Z}_{\mathcal{T}}^{(M)}(\vec{Z}^{(5)};\vec{Z}^{(6)};m_A)\mathcal{Z}_{\mathcal{S}}^{(M)}(\vec{Z}^{(6)};\vec{Y};m_A)\,,
\end{align}
where we recall that the contribution of a $(1,-1)$-brane was defined in \eqref{eq:1m13d}. Once again, this identity can be obtained from the corresponding $4d$ identity \eqref{eq:T11} with the limit \eqref{eq:limit}.
%
%

\bigskip

\subsubsection{$\mathcal{T}^T$-dualization}

We also discuss below the basic moves involving the  $SL(2,\mathbb{Z})$ element $\mathcal{T}^T = -\mathcal{T}\mathcal{S}\mathcal{T}$ with inverse $\left(\mathcal{T}^T\right)^{-1} = -\mathcal{S}\mathcal{T}\mathcal{S}$. 

\bigskip

\begin{figure}[!ht]
\includegraphics[scale=0.35,center]{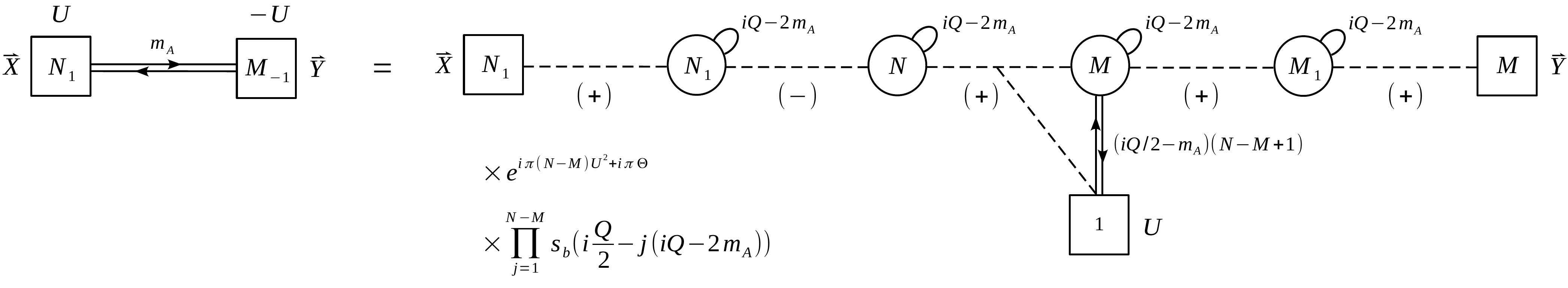}
\caption{The $\mathcal{B}_{11}=\mathcal{T}^T\mathcal{B}_{01} \left(\mathcal{T}^T\right)^{-1}=
\mathcal{T}\mathcal{S}\mathcal{T}\mathcal{B}_{01} \mathcal{S}\mathcal{T}\mathcal{S}$ duality move.
}
\label{fig:11_Ttrdual_3d}
\end{figure}

\noindent\textbf{\boldmath$\mathcal{B}_{11}=\mathcal{T}^T\mathcal{B}_{01} \left(\mathcal{T}^T\right)^{-1}$.} 
The $\mathcal{T}^T$-dualization of a $(0,1)$-brane into a $(1,1)$-brane corresponds to the duality in Figure \ref{fig:11_Ttrdual_3d}.

\noindent 
We then have the following identity of $S^3_b$ partition functions:
\begin{align}
&\mathcal{Z}_{(1,1)}^{(N,M)} (\vec{X};\vec{Y};U;m_A)=\e^{i\pi(N-M)U^2-i\pi\Theta}
\prod_{j=1}^{N-M} s_b\left(i\frac{Q}{2}-j(iQ-2m_A) \right)\nn\\
&\quad\times\int\left(\prod_{k=1}^3\udl{\vec{Z}^{(k)}_N}\Gd_N^{3d}(\vec{Z}^{(k)};m_A)\right)\left(\prod_{k=4}^6\udl{\vec{Z}^{(k)}_M}\Gd_N^{3d}(\vec{Z}^{(M)};m_A)\right)\nn\\
&\quad\times \mathcal{Z}_{\mathcal{T}}^{(N)}(-\vec{X};\vec{Z}^{(1)};m_A)\mathcal{Z}_{\mathcal{S}}^{(N)}(\vec{Z}^{(1)};\vec{Z}^{(2)};m_A)\mathcal{Z}_{\mathcal{T}}^{(N)}(\vec{Z}^{(2)};\vec{Z}^{(3)};m_A)\nn\\
&\quad\times\mathcal{Z}_{(0,1)}^{(N,M)} \left(\vec{Z}^{(3)};\vec{Z}^{(4)};U;i\frac{Q}{2}-m_A\right)
\mathcal{Z}_{\mathcal{S}}^{(M)}(\vec{Z}^{(4)};\vec{Z}^{(5)};m_A)\mathcal{Z}_{\mathcal{T}}^{(M)}(\vec{Z}^{(5)};\vec{Z}^{(6)};m_A)\nn\\
&\quad\times\mathcal{Z}_{\mathcal{S}}^{(M)}(\vec{Z}^{(6)};-\vec{Y};m_A)\,,
\end{align}
where we recall that $\Theta$ was defined in \eqref{eq:thetaschif}. Note that we have inserted the extra signs of $-\vec X$ and $-\vec Y$ on the right hand side, which reflect the minus sign of $\mathcal T^T = -\mathcal{TST}$ and its inverse. This identity can be obtained, as usual, as a limit of \eqref{eq:TT01}
with 
\begin{gather}
\begin{gathered}
X_i\to X_i-s,\quad Y_j\to Y_j-s,\quad Z^{(1,\dots,4)}_a\to Z^{(1,\dots,4)}_a+s, \\
Z^{(5,6)}_a\to Z^{(5,6)}_a-s, \quad U \to U+s, \quad V \to s, \quad s\to+\infty
\end{gathered}
\end{gather}
followed by $\Delta \to -\infty$ and $D \to +\infty$ satisfying $\Delta+D \to -\infty$.

\bigskip

\begin{figure}[!ht]
\includegraphics[scale=0.4,center]{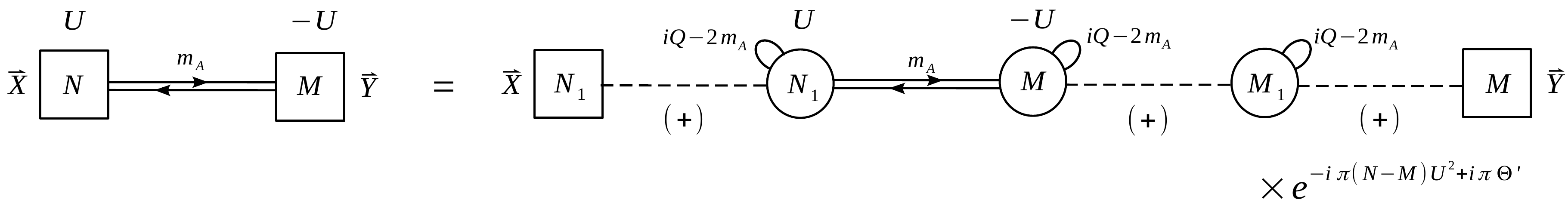}
\caption{The $\mathcal{B}_{10}=\mathcal{T}^T\mathcal{B}_{10} \left(\mathcal{T}^T\right)^{-1}=
\mathcal{T}\mathcal{S}\mathcal{T}\mathcal{B}_{10} \mathcal{S}\mathcal{T}\mathcal{S}$
 duality move.}
\label{fig:10_Ttrdual_3d}
\end{figure}

\noindent\textbf{\boldmath$\mathcal{B}_{10}=\mathcal{T}^T\mathcal{B}_{10} \left(\mathcal{T}^T\right)^{-1}$.} A $(1,0)$-brane is transparent under the $T^T$-dualization, which is translated into the non-trivial duality in Figure \ref{fig:10_Ttrdual_3d}.

\noindent The associated $S^3_b$ partition function identity is
\begin{align}
&\mathcal{Z}_{(1,0)}^{(N,M)} (\vec{X};\vec{Y};U;m_A)=\e^{-i\pi(N-M)U^2-i\pi\Theta'}\nn\\
&\qquad\times\int\left(\prod_{k=1}^3\udl{\vec{Z}^{(k)}_N}\Gd_N^{3d}(\vec{Z}^{(k)};m_A)\right)\left(\prod_{k=4}^6\udl{\vec{Z}^{(k)}_M}\Gd_N^{3d}(\vec{Z}^{(M)};m_A)\right)\nn\\
&\qquad\times \mathcal{Z}_{\mathcal{T}}^{(N)}(-\vec{X};\vec{Z}^{(1)};m_A)\mathcal{Z}_{\mathcal{S}}^{(N)}(\vec{Z}^{(1)};\vec{Z}^{(2)};m_A)\mathcal{Z}_{\mathcal{T}}^{(N)}(\vec{Z}^{(2)};\vec{Z}^{(3)};m_A)\nn\\
&\qquad\times\mathcal{Z}_{(1,0)}^{(N,M)} (\vec{Z}^{(3)};\vec{Z}^{(4)};U;m_A)\mathcal{Z}_{\mathcal{S}}^{(M)}(\vec{Z}^{(4)};\vec{Z}^{(5)};m_A)\mathcal{Z}_{\mathcal{T}}^{(M)}(\vec{Z}^{(5)};\vec{Z}^{(6)};m_A)\nn\\
&\qquad\times\mathcal{Z}_{\mathcal{S}}^{(M)}(\vec{Z}^{(6)};-\vec{Y};m_A)\,,
\end{align}
where we defined
\begin{align}
\Theta'=\frac{N-M}{12}(iQ-2m_A)^2\left[(N-M)^2-1\right]\,.
\end{align}
This identity can be obtained as a limit of \eqref{eq:TT10}
with 
\begin{gather}
\begin{gathered}
\label{eq:limit2}
X_i\to X_i+s,\quad Y_j\to Y_j+s, \quad Z^{(1)}_a\to Z^{(1)}_a-s, \\
Z^{(k>1)}_a\to Z^{(k>1)}_a+s, \quad U \to U-s, \quad V \to s, \quad s\to+\infty
\end{gathered}
\end{gather}
followed by $\Delta \to -\infty$ and $D \to +\infty$ satisfying $\Delta+D \to -\infty$.

\bigskip

\begin{figure}[!ht]
\includegraphics[scale=0.35,center]{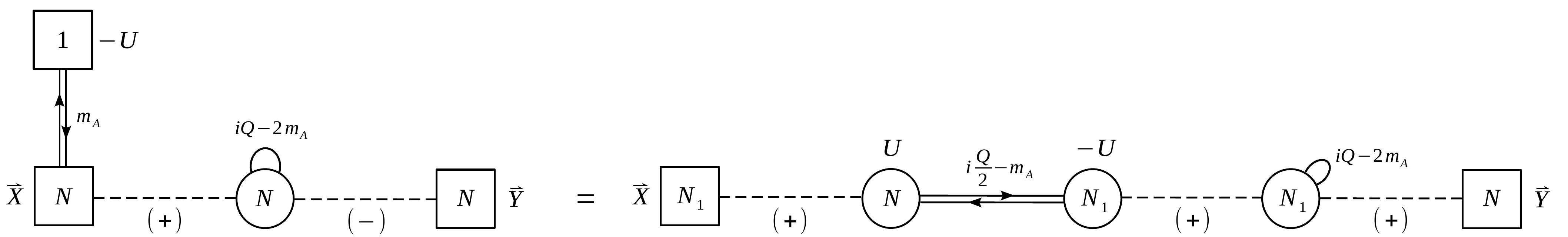}
\caption{The $\mathcal{B}_{0-1}=\mathcal{T}^T\mathcal{B}_{1-1} \left(\mathcal{T}^T\right)^{-1}$ duality move.}
\label{fig:01_Ttrdual_3d}
\end{figure}

\noindent\textbf{\boldmath$\mathcal{B}_{0-1}=\mathcal{T}^T\mathcal{B}_{1-1} \left(\mathcal{T}^T\right)^{-1}$.}  Lastly, we consider the $\mathcal{T}^T$-dualization of a $(1,-1)$-brane into a $(0,-1)$-brane, or $\overline{\text{D5}}$, which corresponds to the duality in Figure \ref{fig:01_Ttrdual_3d}

\begingroup
\allowdisplaybreaks
\noindent At the level of  $S^3_b$ partition function we have the following identity:
\begin{align}
\mathcal{Z}_{(0,-1)}^{(N)} (\vec{X};\vec{Y};U;m_A)&=\int\left(\prod_{k=1}^6\udl{\vec{Z}^{(k)}_N}\Gd_N^{3d}(\vec{Z}^{(k)};m_A)\right)\mathcal{Z}_T^{(N)}(-\vec{X};\vec{Z}^{(1)};m_A)\nn\\
&\times \mathcal{Z}_{\mathcal{S}}^{(N)}(\vec{Z}^{(1)};\vec{Z}^{(2)};m_A)\mathcal{Z}_{\mathcal{T}}^{(N)}(\vec{Z}^{(2)};\vec{Z}^{(3)};m_A)\nn\\
&\times 
	\mathcal{Z}_{\text{(1,-1)}}^{(N,N)}\left(\vec{Z}^{(3)};\vec{Z}^{(4)};U;i\frac{Q}{2}-m_A\right) \mathcal{Z}_{\mathcal{S}}^{(N)}(\vec{Z}^{(4)};\vec{Z}^{(5)};m_A)\nn\\
&\times 
	\mathcal{Z}_{\mathcal{T}}^{(N)}(\vec{Z}^{(5)};\vec{Z}^{(6)};m_A)
	\mathcal{Z}_{\mathcal{S}}^{(N)}(\vec{Z}^{(6)};-\vec{Y};m_A)\,,
\end{align}
\endgroup
where $\mathcal{Z}_{(0,-1)}^{(N,M)}(\vec{X};\vec{Y};U;m_A)=\mathcal{Z}_{(0,1)}^{(N,M)}(\vec{X};\vec{Y};-U;m_A)$,
which we obtain from the corresponding $4d$ identity \eqref{eq:TT11} with the limit \eqref{eq:limit2}.

\subsection{The Hanany--Witten duality move}

\begin{figure}[!ht]
\includegraphics[width=\textwidth,center]{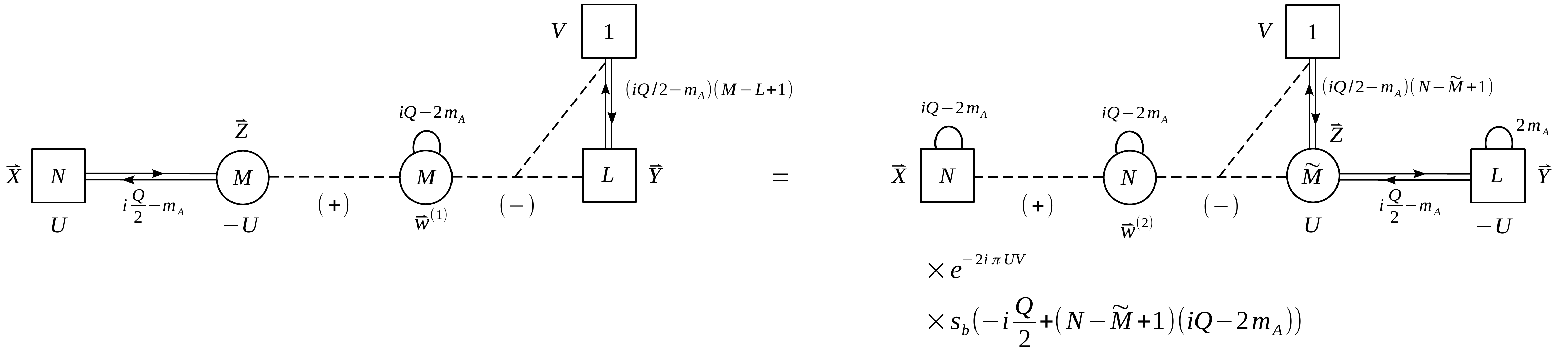}
\caption{The HW duality move in $3d$.}
\label{fig:HW_move_3d}
\end{figure}

As we stressed many times, the algorithm to dualize a quiver into its mirror dual works exactly in the same way in $4d$ and in $3d$. In particular, also in $3d$ we might end up with a theory in which some operator has a non-trivial VEV after we dualize the blocks of the chopped quiver via the basic duality moves. This will happen every time we start from a quiver with gauge nodes of unequal ranks, and it is the manifestation of the fact that in the brane setup after applying $\mathcal{S}$-duality we have D5-branes with different numbers of D3's on the left and on the right. One then has to reorder the branes using Hanany--Witten (HW) moves until we reach the situation in which all D5-branes have an equal number of D3's on the left and on the right. In such a situation we are able to read off the mirror dual quiver gauge theory. In field theory this operation amounts to studying the aforementioned VEVs. In Section \ref{sec:algorithm}, we have seen three equivalent ways of studying these VEVs. One of these was a duality that realizes the HW move in the field theory language. In order to complete the algorithm we then need an analogous duality in $3d$.
The $3d$ HW duality move is given in Figure \ref{fig:HW_move_3d}.
 
We can obtain the corresponding partition function identity by studying a limit of the  index identity \eqref{eq:HWid} associated to the $4d$ HW move duality, which is realized by
\begin{align}
X_i\to X_i+s,\quad Y_j\to Y_j+s, \quad Z_a\to Z_a+s, \quad U \to U-s, \quad V \to V+s, \quad s\to+\infty
\end{align}
followed by $\Delta \to -\infty$. Then this limit results in the following $3d$ partition function identity for $N\geq\widetilde{M}\geq 0$ as consistent with the S-rule:
\bigskip\bigskip\bigskip\bigskip
\begingroup\allowdisplaybreaks
\begin{align}
&\int\udl{\vec{Z}_M}\Gd^{3d}_M(\vec{Z})
\mathcal{Z}_{(1,0)}^{(N,M)} \left(\vec{X};\vec{Z};U;i\frac{Q}{2}-m_A \right)
 {}_{\vec Z}\hat{\mathbb{I}}^{3d}_{\vec Y,V}(m_A)\
\nn\\
&\qquad \times
\prod_{j=1}^L \sbfunc{\frac{iQ}{2} -\left(\frac{iQ}{2}-m_A\right) (M-L+1) \pm(Y_j-V)}
\nn\\
&\qquad 
=
\e^{-2i\pi UV} s_b\left(-i\frac{Q}{2}+(N-\widetilde{M}+1)(iQ-2m_A) \right) \nn\\
& \qquad\quad\times \prod_{i,j=1}^N s_b\left( -i\frac{Q}{2}+2m_A  \pm (X_i-X_j) \right) \prod_{i,j=1}^L s_b\left (i\frac{Q}{2} -2m_A \pm (Y_i-Y_j)\right) \nn\\
&\qquad\quad\times \int\udl{\vec{Z}_{\widetilde{M}}}\Gd^{3d}_{\widetilde{M}}(\vec{Z})
 \prod_{j=1}^{\widetilde{M}} \sbfunc{\frac{iQ}{2} -\left(\frac{iQ}{2}-m_A\right) (N-\widetilde{M}+1) \pm(Z_j-V)}
 \nn\\
&\qquad\quad \times
   {}_{\vec X}\hat{\mathbb{I}}^{3d}_{\vec Z,V}(m_A)
 \mathcal{Z}_{(1,0)}^{(\widetilde{M},L)}\left(\vec{Z};\vec{Y};U;i\frac{Q}{2}-m_A\right)\,.
\end{align}
\endgroup

\begin{figure}[!ht]
\includegraphics[scale=0.5,center]{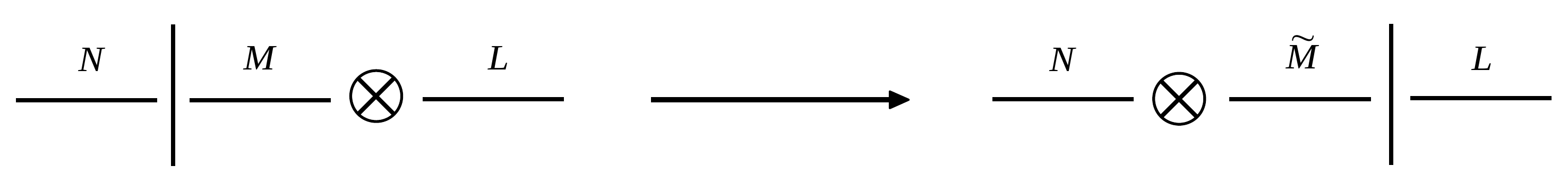}
\caption{The HW brane move swapping NS5 (a solid vertical line) and D5 (a crossed circle), where $\widetilde{M}=N+L-M+1$.}
\label{fig:HWmove_3d_brane}
\end{figure}

\begin{figure}[!ht]
    \centering
    \includegraphics[scale=0.5]{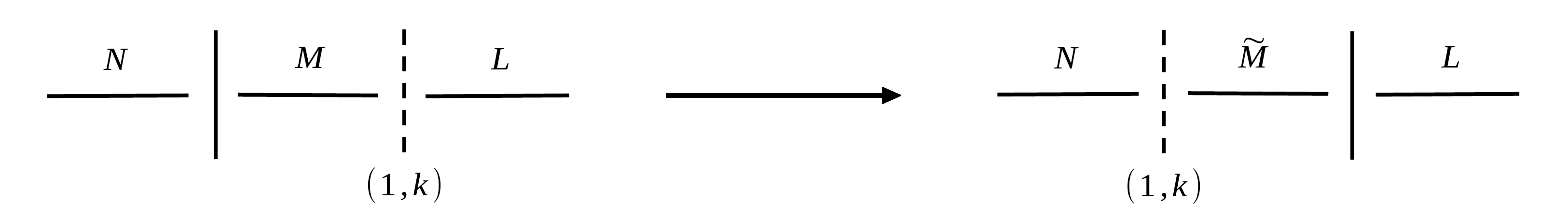}
    \caption{The HW brane move swapping NS5 (a solid vertical line) and $(1,k)$ (a dashed vertical line), where $\widetilde{M}=N+L-M+| k |$, which in field theory is realized by a local application of the Giveon--Kutasov duality}
    \label{GKbrane}
\end{figure}

This encodes in field theory the Hanany--Witten move that swaps a $(1,0)$ with a $(0,1)$-brane as shown in Figure \ref{fig:HWmove_3d_brane}, where before the transition we have $N$-D3 branes on the left of the $(1,0)$, $M$ D3-branes between the $(1,0)$ and the $(0,1)$-brane, and $L$ D3-branes on the right of the $(0,1)$, while after the transition we have $N$ D3-branes on the left of the $(0,1)$, $\widetilde{M}$ D3-branes between the $(0,1)$ and the $(1,0)$, and $L$ D3-branes on the right of the $(1,0)$.
One can also consider HW moves swapping a $(1,0)$ and a $(1,1)$-brane as shown in Figure \ref{GKbrane}.
These moves are not strictly necessary, since we can read off a gauge theory from the brane setup even if we have $(1,1)$-branes with a different number of D3's on each side. Nevertheless, this is still interesting since the two equivalent brane configurations before and after the HW move are usually associated with distinct gauge theories, leading to a non-trivial IR duality. In field theory, this amounts to using the Giveon--Kutasov duality \cite{Giveon:2008zn} to dualize the gauge node associated with the D3-branes suspended between the (1,0) and the $(1,1)$-brane involved in the HW move.

The Giveon--Kutasov duality relates a $3d$ $\mathcal{N}=2$ $U(N_c)_k$ theory with $N_f$ fundamental flavors $Q^i$, $\tilde{Q}_i$, $i=1,\cdots N_f$ and no superpotential, $\mathcal{W}=0$, to a $U(N_f+|k|-N_c)_{-k}$ theory with $N_f$ fundamental flavors $q_i$, $\tilde{q}^i$, $N_f^2$ gauge singlets $M^i_j$ and superpotential $\hat{\mathcal{W}}=M^i_jq_i\tilde{q}^j$. The reason why we can apply this duality is because the node of the quiver that we want to dualize has a non-zero CS level due to the presence of the $(1,1)$-brane, which makes the $\mathcal{N}=2$ adjoint chiral in the $\mathcal N=4$ vector multiplet massive so that at low energies we have an effective $\mathcal{N}=2$ theory with only fundamental flavors and a quartic superpotential. The identity of the $S^3_b$ partition functions associated with the Giveon--Kutasov duality is \cite{Kapustin:2010mh,Benini:2011mf}

\begin{align}
&\int\udl{\vec{Z}_{N_c}}\Gd_{N_c}^{3d}(\vec{Z})\e^{-ik\pi\sum_{i=1}^{N_c}Z_i^2-\pi i\eta\sum_{i=1}^NZ_i}\prod_{i=1}^{N_c}\prod_{a=1}^{N_f}\sbfunc{i\frac{Q}{2}\mp Z_i-\mu^\pm_a)}\nn\\
&=e^{\frac{i \pi}{24} \left(k^2+2\right)-\frac{i \pi}{2} \phi} \prod_{a,b=1}^{N_f}\sbfunc{i\frac{Q}{2}-\mu^+_a-\mu^-_b}\nn\\
&\quad\times\int\udl{\vec{Z}_{N_c'}}\Gd_{N_c'}^{3d}(\vec{Z})\e^{ik\pi\sum_{i=1}^{N_c'}Z_i^2+\pi i\eta\sum_{i=1}^{N_c'}Z_i}\prod_{i=1}^{N_c'}\prod_{a=1}^{N_f}\sbfunc{\mp Z_i+\mu^\pm_a}\,.
\end{align}
where $k > 0$ and
\begin{align}
\phi &= k \left(\sum_{a=1}^{N_f} \left(\left(\mu^+_a\right)^2+\left(\mu^-_a\right)^2\right)-\frac{Q^2}{4} k \left(k-2 N_c'\right)+\frac12 \eta^2-i Q k \sum_{a = 1}^{N_f} \left(\mu^+_a+\mu^-_a\right)\right. \nonumber \\
&\quad \left.+\eta \sum_{a = 1}^{N_f} \left(\mu^+_a-\mu^-_a\right)+\frac12 \left(i Q N_c'-\sum_{a = 1}^{N_f} \left(\mu^+_a+\mu^-_a\right)\right)^2\right)
\end{align}
where $\mu_a^\pm$ are the real masses for the $U(N_f)\times U(N_f)/U(1)$ flavor symmetry.

Notice the exponential prefactors that can be interpreted as contact terms for the global symmetries \cite{Closset:2012vg,Closset:2012vp}. These are crucial since inside a quiver such symmetries are actually gauged, and the contact terms induce CS couplings for the gauge nodes that are adjacent to the quiver we dualized. This is expected from the brane perspective, since moving a $(1,1)$ through a (1,0) not only affects the CS level of the gauge node associated to the D3's suspended between them by changing its sign, but also those of the nodes associated to the adjacent intervals. Moreover, the gauge singlets produced with the duality might give mass to/produce adjoint chiral fields for the adjacent nodes, again compatibly with the fact that their CS levels change after the dualization.

In the next section, we will encounter the $4d$ analogue of the Hanany--Witten move between a $(1,0)$ and a $(1,1)$-brane. Also in that case the theories before and after the transition can be related by a simple dualization of the relevant gauge node in the quiver, but in $4d$ the basic duality will be once again the Intriligator--Pouliot duality. This is compatible with the fact that the Giveon--Kutasov duality in $3d$ can also be obtained from the Intriligator--Pouliot duality in $4d$.

\section{Duality webs}
\label{sec:dualityweb}

In this section we apply the algorithm to construct a web of dualities for an SQCD example both in $4d$  and in $3d$.

\subsection{$4d$ $PSL(2,\mathbb{Z})$ duality web}

\begin{figure}[!ht]
\includegraphics[scale=0.34,center]{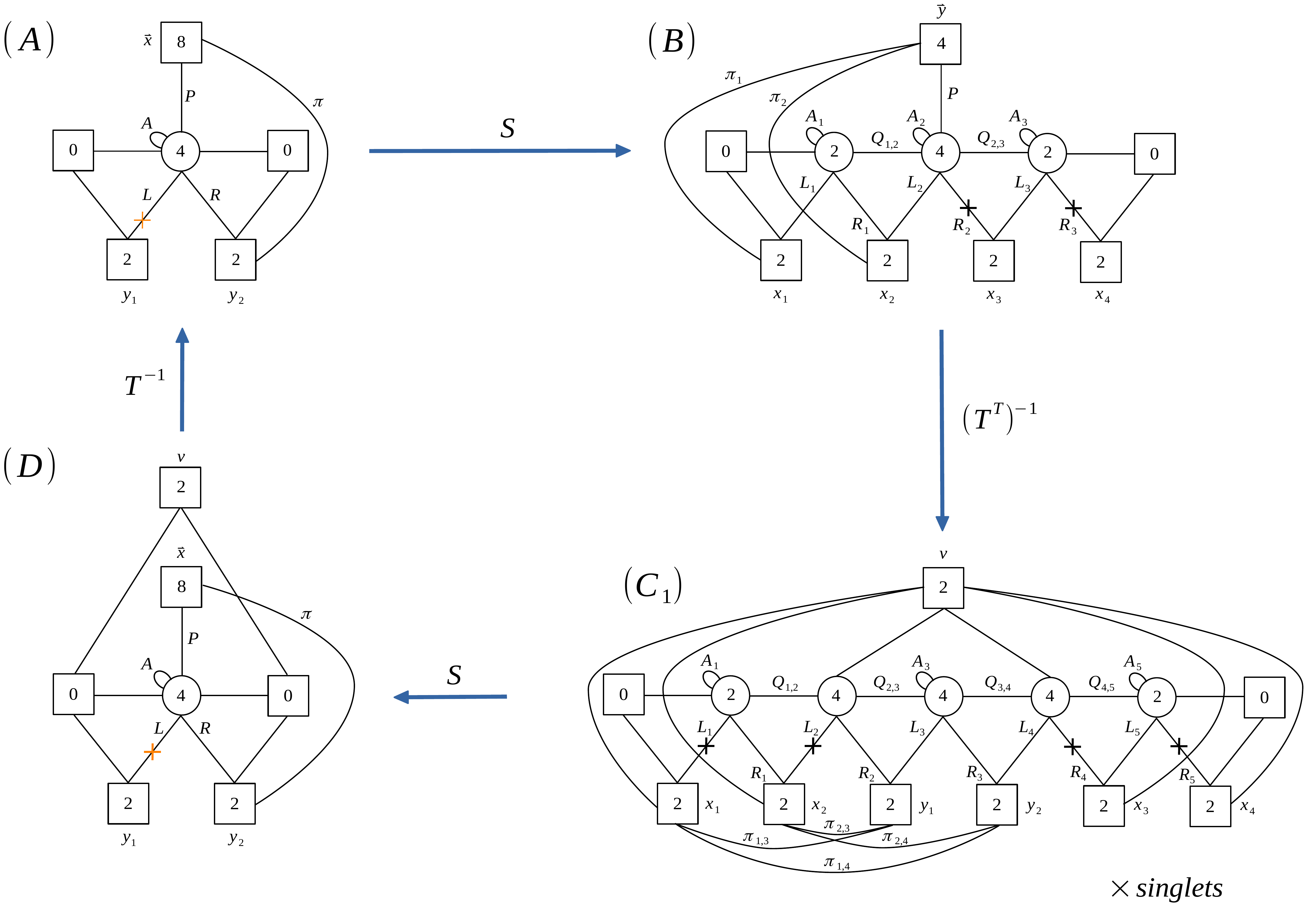}
\caption{A $PSL(2,\mathbb{Z})$ duality web for the $4d$ SQCD with $N_c=2$ and $N_f=4$. The theory $C_1$ has extra singlets that we specify in Figures \ref{fig:B_dualized_to_C1} and \ref{C123}.}
\label{sl2zweb}
\end{figure}

\begin{figure}[!ht]
\includegraphics[scale=0.34,center]{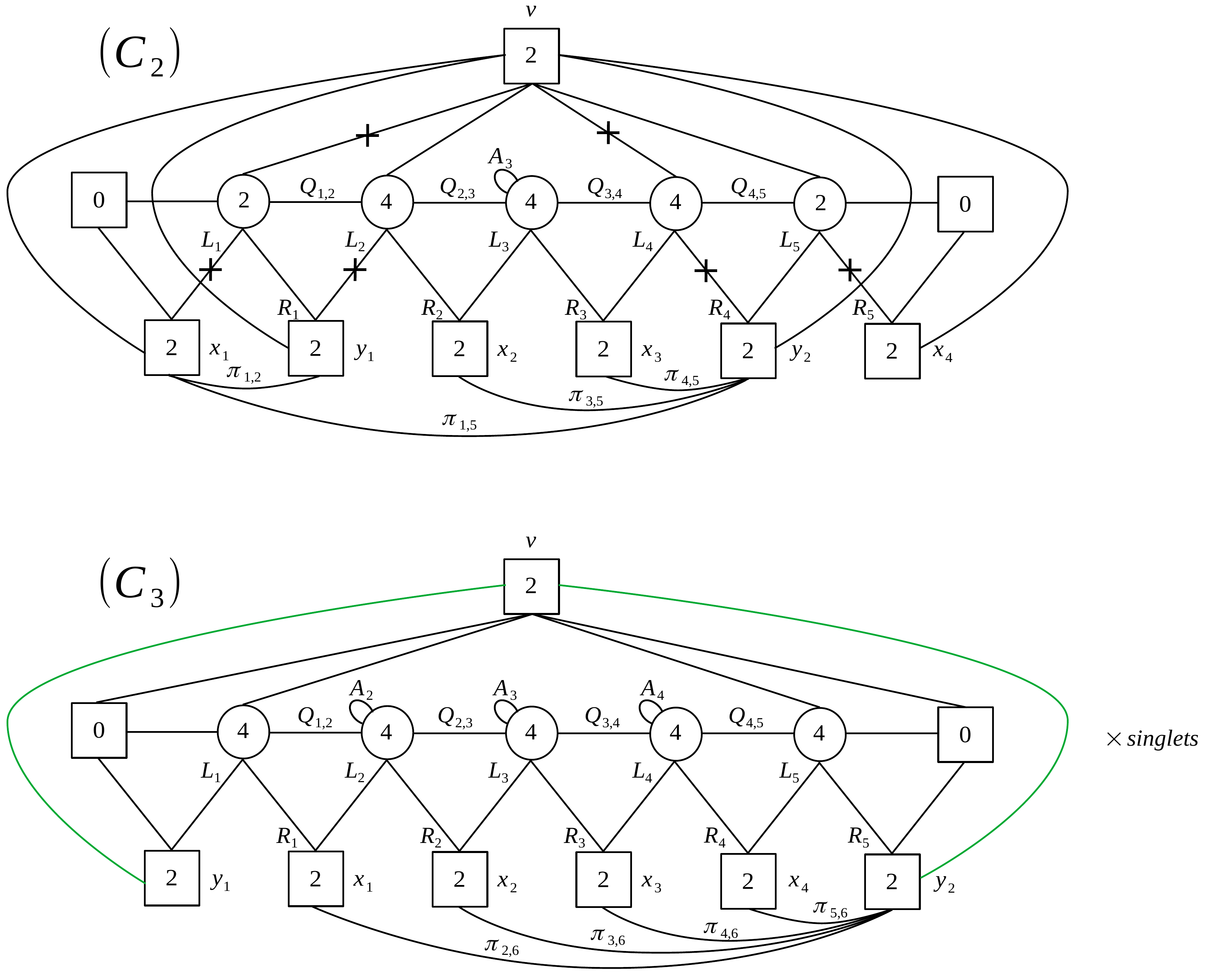}
\caption{The IP dual frames $C_2$ and $C_3$. Theory $C_3$ has extra singlets that we specify in Figure \ref{C123}.}
 \label{fig:C}
\end{figure}

Using the  $PSL(2,\mathbb{Z})$ duality moves we introduced in Section \ref{sec:QFTblocks} we construct the web of dualities in Figure \ref{sl2zweb}. We start from the SQCD  with $N_c=2$ and $N_f=4$, which we call Theory $A$ and depict in the upper left corner of the figure, and sequentially apply different duality moves corresponding to $\mathsf S$, $(\mathsf T^ T)^{-1}$, $\mathsf S$, and $\mathsf T^{-1}$
to reach the other corners of the diagram.
Theory $B$ is the mirror dual of Theory $A$, i.e.~its $\mathsf S$-dual.
Theory $C_1$ is obtained by acting with $(\mathsf T^ T)^{-1}$ on Theory $B$.
In addition to $C_1$ we have two extra frames, $C_2$ and $C_3$ shown in Figure \ref{fig:C}, which  are obtained by IP dualizations. Theory $D$ is  obtained by acting with $\mathsf S$ on $C_1$ (or  $C_2$, or $C_3$) and the web closes by acting with $\mathsf T^{-1}$ on Theory $D$ to go back to Theory $A$, since $\mathsf S (\mathsf T^ T)^{-1} \mathsf S \mathsf T^{-1}=\mathsf S (\mathsf S\mathsf T\mathsf S) \mathsf S \mathsf T^{-1}=1$.

\begin{figure}[!ht]
	\includegraphics[scale=0.34,center]{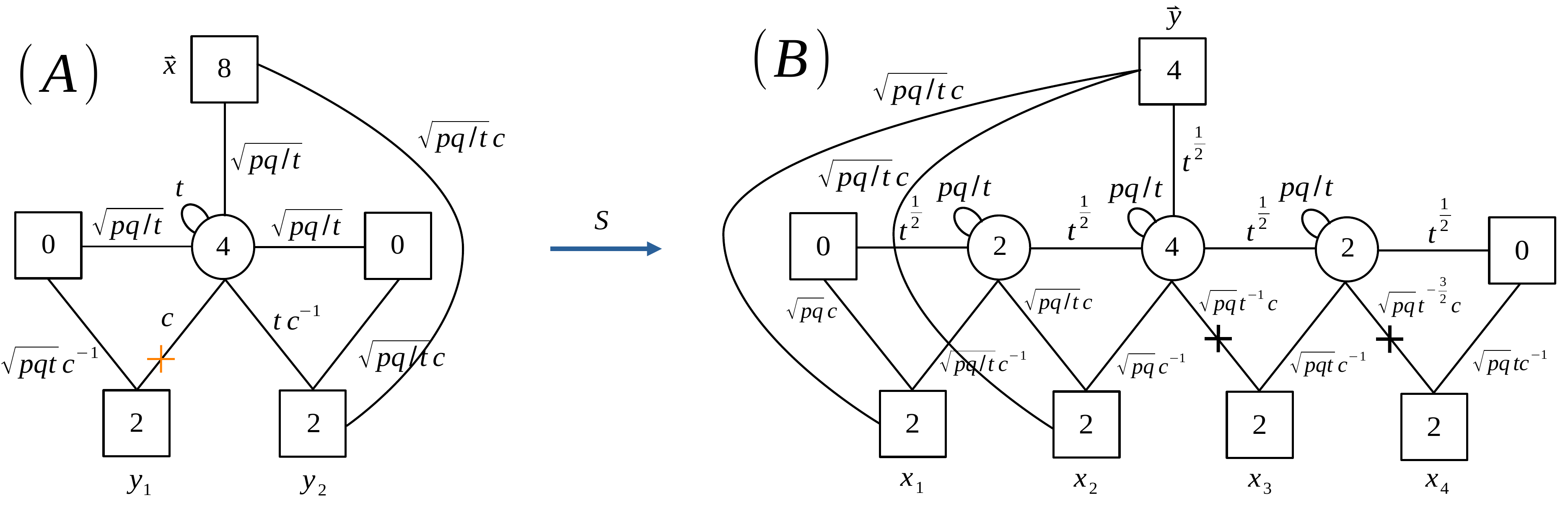}
	\caption{Theory $A$ $\mathsf S$-dualized to Theory $B$.}
	\label{fig:A_dualized_to_B}
\end{figure}

We have already discussed in details the $\mathsf S$-dualization of the SQCD for generic $N_c$ and $N_f$ in Subsection \ref{subsec:Sdualisation_SQCD}, so in Figure \ref{fig:A_dualized_to_B} we just briefly summarize the resulting mirror duality for the particular case $N_c=2$ and $N_f=4$.

Let us now focus on how to get the other dual frames by sequential duality moves.
In Figure  \ref{fig:B_dualized_to_C1} we show how  Theory $C_1$ is obtained by  taking the $(\mathsf T^ T)^{-1}$-dualization of Theory $B$. We first chop Theory $B$ into the QFT blocks.
Notice that, as shown in the figure, the nodes that we are un-gauging have an antisymmetric field with charges corresponding to $pq/t$.   Similarly the QFT blocks  $\mathsf{B}_{10}$ and $\mathsf{B}_{01}$ appearing have $t\to pq/t$ w.r.t.~their definition in Section \ref{sec:QFTblocks}. We then dualize each block using the basic  $(\mathsf T^ T)^{-1}=(\mathsf S\mathsf T\mathsf S)  $-moves (the inverse of the $\mathsf T^ T$-moves in Section \ref{subsec:4d_duality_moves}) and we glue them together.
Since we dualize QFT blocks where $t$ is replaced by $pq/t$,  the $\mathsf S$-walls appearing in Figure \ref{fig:B_dualized_to_C1} are defined with $t\to pq/t$, and we color them in orange to emphasize this replacement.
 Lastly, we recognize the several identity walls corresponding to $\mathsf T^ T (\mathsf T^ T)^{-1}=1$ as in Figure \ref{fig:TSTSTS} and finally reach Theory $C_1$, which is the desired $(\mathsf T^ T)^{-1}$-dual theory.

\begin{figure}[!ht]
    \includegraphics[scale=0.4,center]{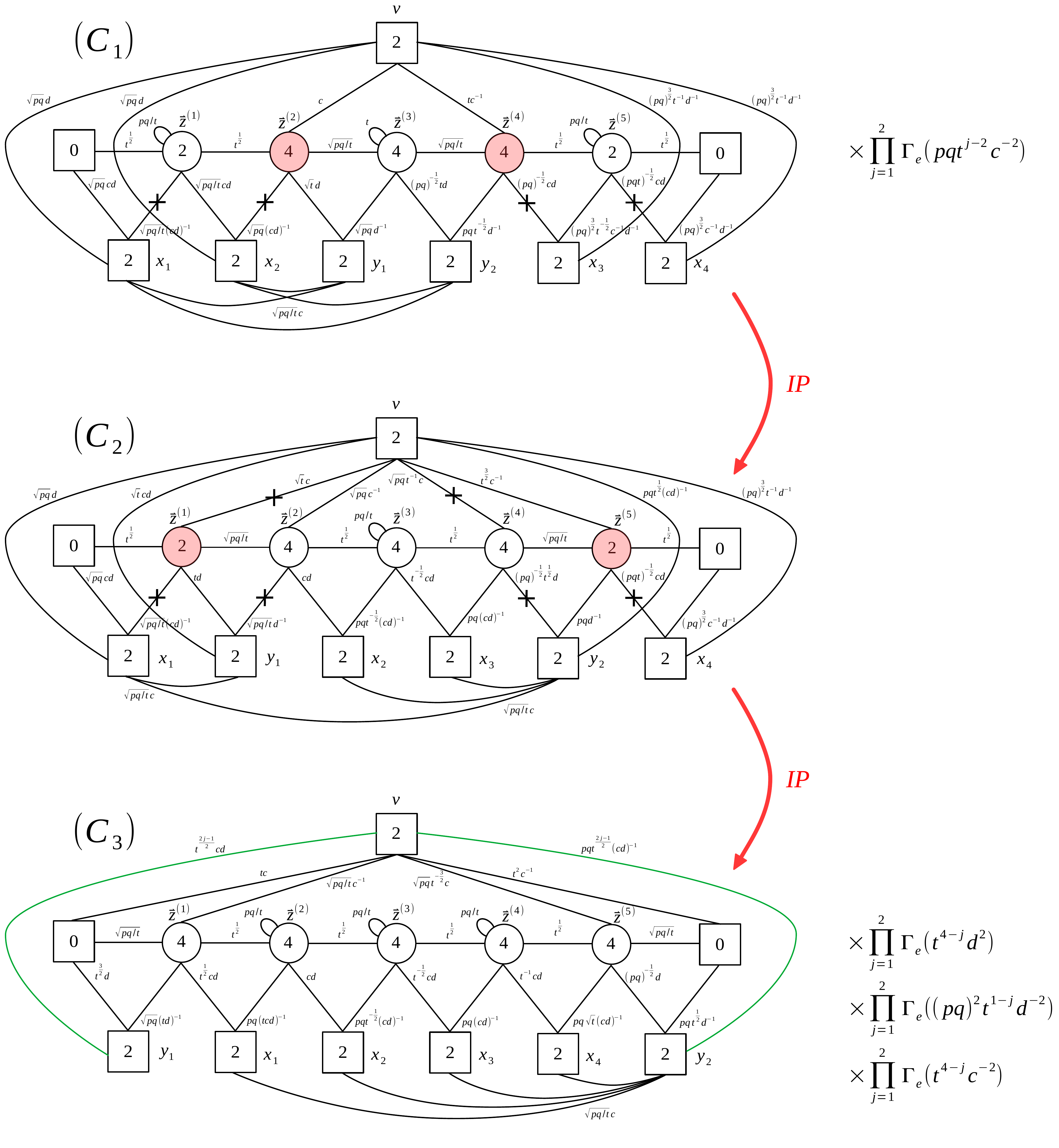}
    \caption{The $(T^T)^{-1}$ dual $C_1$ of Theory $B$ and the IP dual frames $C_2,C_3$}
    \label{C123}
\end{figure}

The $C_2$ and $C_3$ theories can be obtained, as depicted in Figure \ref{C123}, by applying  the IP duality to the gauge nodes colored in red, which do not have antisymmetric fields. 

This operation can be regarded as the $4d$ field theory analogue  of  swapping an NS5 and a (1,1)-brane. 
In this case, as opposed to the swapping of NS5 and D5-branes, it is possible to read off a gauge theory from the brane system both before and after the move, which results in equivalent, i.e.~IR dual, theories. 

\begin{figure}[!ht]
	\includegraphics[scale=0.26,center]{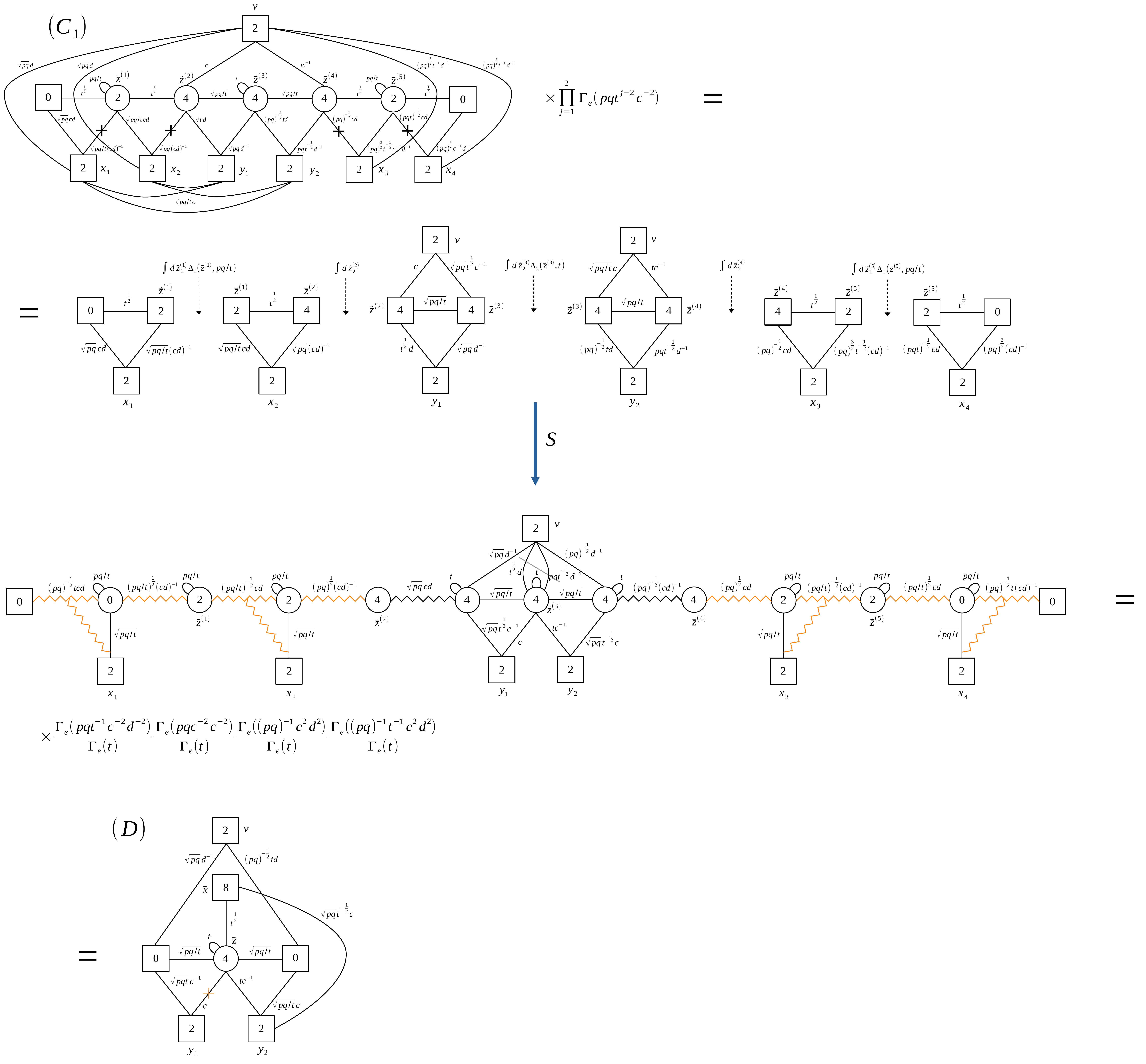}
	\caption{Theory $C_1$ $\mathsf{S}$-dualized to Theory $D$.}
	\label{fig:C1_dualized_to_D}
\end{figure}

We then can $\mathsf S$-dualize each one of these theories to reach  Theory $D$.
Let's start with theory $C_1$, which we chop  into QFT blocks as shown in the upper part of Figure \ref{fig:C1_dualized_to_D}. We dualize each block using the basic $\mathsf S$-moves in Figures \eqref{eq:id1} and \eqref{eq:11S1-1}. Then we glue them together to get Theory $D$ after the Identity-walls are implemented.
One can similarly show that the $\mathsf S$-dualization of $C_2$ and $C_3$ yields again Theory $D$.

One should note that Theory $D$ is actually the same field theory as Theory $A$ but realized in terms of $\mathsf{B}_{11}$  blocks rather than with $\mathsf{B}_{10}$  blocks (if we keep track of the trivial rank-zero nodes and the $SU(2)_v$ node attached to them)
Although Theory $A$ and Theory $D$ already coincide as field theories, we still
implement the $\mathsf T^{-1}$-dualization of Theory $D$. As we saw in Section \ref{sec:QFTblocks}, under $\mathsf T^{-1}$ the $\mathsf{B}_{01}$ block is transparent while the $\mathsf{B}_{11}$ block is turned into a $\mathsf{B}_{10}$ so that we indeed expect to obtain Theory $A$ from Theory $D$.
As shown in Figure \ref{fig:D_dualized_to_A} we chop Theory $D$ into QFT blocks, dualize each block  using the basic $\mathsf T^{-1}$-moves and glue them together.
We now recognize the Identity-walls corresponding to $\mathsf T^{-1} \mathsf T=1$ and after implementing them we obtain Theory $A$, closing the duality web. 

\begin{landscape}
\advance\voffset by -1cm 
\begin{figure}[!ht]
	\includegraphics[scale=0.28,center]{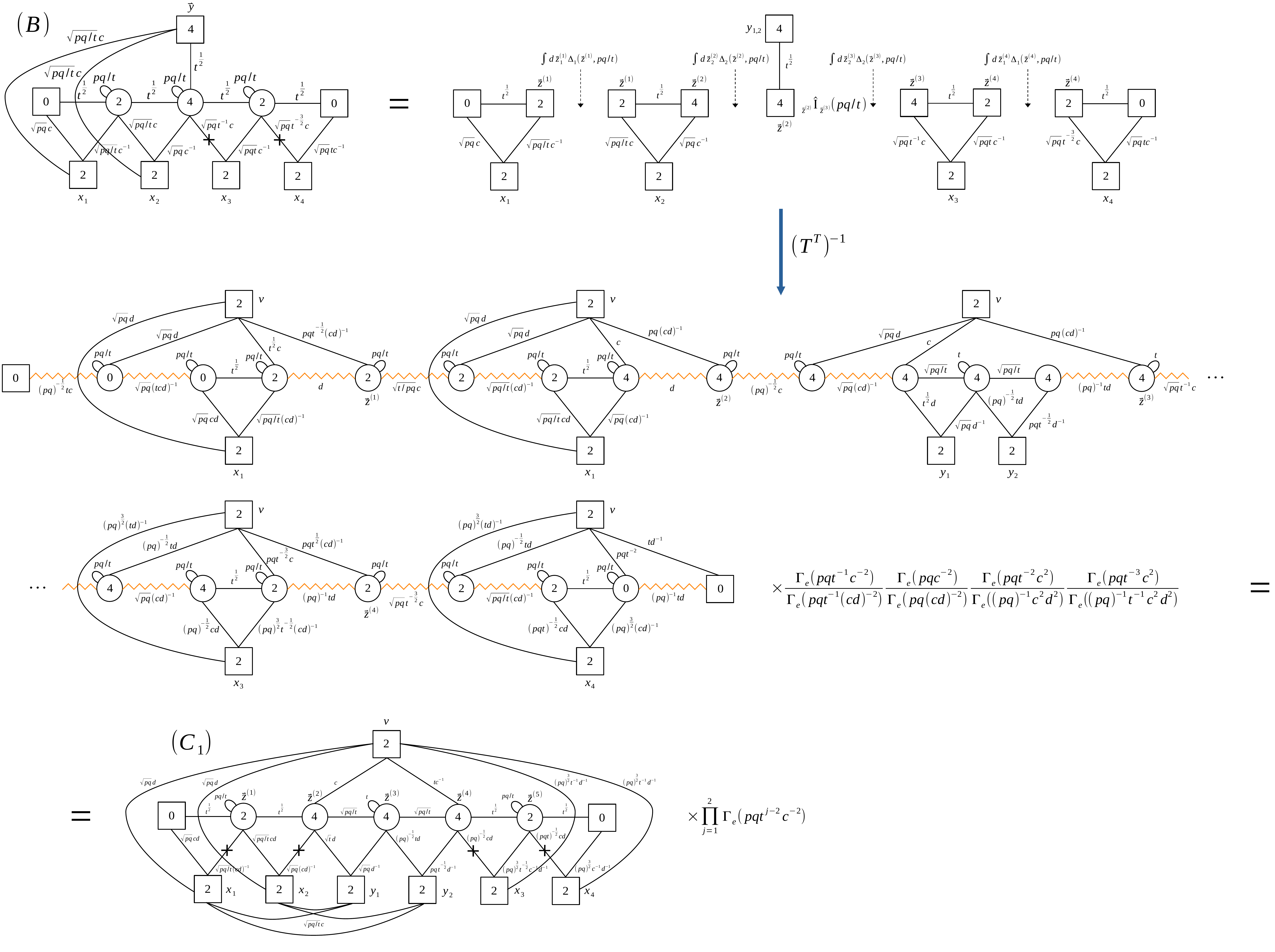}
	\caption{Theory $B$ $(\mathsf T^ T)^{-1}$-dualized to Theory $C_1$.}
	\label{fig:B_dualized_to_C1}
\end{figure}
\end{landscape}

\newpage
\enlargethispage{1cm} 
\advance\voffset by -1cm 

\begin{landscape}
\newpage
\enlargethispage{1cm} 
\advance\voffset by -1cm 
\begin{figure}[!ht]
	\includegraphics[scale=0.35,center]{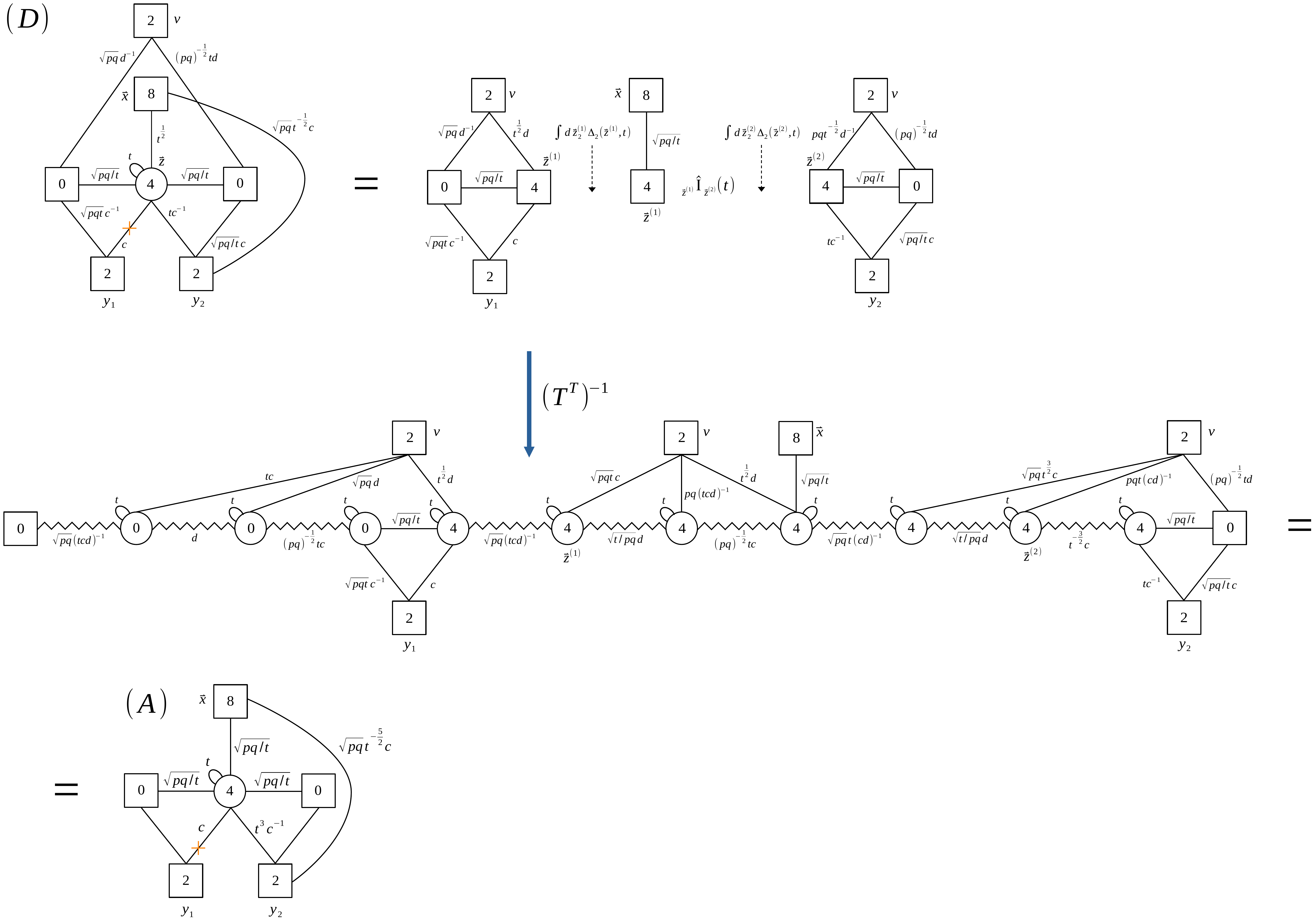}
	\caption{Theory $D$ ${\mathsf T}^{-1}$-dualized to Theory $A$.}
	\label{fig:D_dualized_to_A}
\end{figure}
\end{landscape}

\subsection{$4d$ operators map}

It is worth studying the symmetry and the operator map of the theories to better understand how the duality web works. 
The global symmetry in the IR is given by
\begin{align}
USp(8)_x \times USp(4)_y \times U(1)_t \times U(1)_c\,.
\end{align}
However, this symmetry is not fully manifest in all the frames. The manifest UV symmetry in each dual fame is given as follows:
\begingroup\allowdisplaybreaks
\begin{align}
A&: \quad USp(8)_x \times SU(2)_{y_1} \times SU(2)_{y_2}\times U(1)_t \times U(1)_c\,, \\
B&: \quad \prod_{i = 1}^4 SU(2)_{x_i} \times USp(4)_y\times U(1)_t \times U(1)_c \,, \\
C_i&: \quad \prod_{i = 1}^4 SU(2)_{x_i} \times SU(2)_{y_1} \times SU(2)_{y_2} \times SU(2)_v\times U(1)_d \times U(1)_t \times U(1)_c \,, \\
D&: \quad USp(8)_x \times SU(2)_{y_1} \times SU(2)_{y_2}\times U(1)_t \times U(1)_c \,.
\end{align}
\endgroup

Notice that $C_1$, $C_2$, and $C_3$ have an extra $U(1)_d \times SU(2)_v$ symmetry, which doesn't appear in the other frames. By the duality we then expect this not to be a faithful symmetry in the IR, i.e.~it acts trivially on the spectrum of the theory. In consistency with this claim, we checked that all the anomalies involving these symmetries vanish. Namely, we have checked that
\begin{align}
\mathrm{Tr} \, SU(2)_v^2 \, U(1)_a = 0
\end{align}
for any abelian symmetry $U(1)_a$, and the other anomalies involving $SU(2)_v$ trivially vanish.
Moreover, we have checked that  the linear and cubic $U(1)_R$ anomalies, calculated assuming mixing with all the abelian symmetries, are independent from the mixing with $U(1)_d$, which implies that all the anomalies involving $U(1)_d$ vanish.


We next construct some gauge invariant operators and explain their symmetry group representations and how they are mapped under the duality web. Firstly let us construct operators in Theory $A$, which is the same as Theory $D$. Specifically, we list five types of operators in Table \ref{tab:operators_A}, which combine to form representations of the enhanced $USp(8)_x \times USp(4)_y$ symmetry in the IR.
\begin{table}[!ht]
\centering
\begin{tabular}{|c|cccc|}
\hline
 & $USp(8)_x \times SU(2)_{y_1} \times SU(2)_{y_2}$ & $U(1)_R$ & $U(1)_t$ & $U(1)_c$ \\
\hline 
$\mathrm{Tr} \, P^2$ & $\mathbf{(27,1,1)}$ & 2 & -1 & 0 \\
$\mathrm{Tr} \, P L$ & $\mathbf{(8,2,1)}$ & 1 & -1/2 & 1 \\
$\pi$ & $\mathbf{(8,1,2)}$ & 1 & -1/2 & 1 \\
$\mathrm{Tr} \, L R$ & $\mathbf{(1,2,2)}$ & 0 & 1 & 0 \\
$\mathrm{Tr} \, A$ & $\mathbf{(1,1,1)}$ & 0 & 1 & 0 \\
\hline
\end{tabular}
\caption{ Gauge invariant chiral operators in Theory $A$ = Theory $D$.}
\label{tab:operators_A}
\end{table}

\noindent The operators rearrange into representations of the full enhanced symmetry according to the following branching rules for $USp(4)_y\to SU(2)_{y_1} \times SU(2)_{y_2}$:
\begin{align}
\mathrm{Tr} \, L R \,, \quad \mathrm{Tr} \, A \qquad &\longrightarrow \qquad \mathbf{(1,2,2)}\oplus\mathbf{(1,1,1)} \quad = \quad \mathbf{(1,5)} \,, \\[5pt]
\mathrm{Tr} \, P L \,, \quad \pi \qquad &\longrightarrow \qquad \mathbf{(8,2,1)}\oplus\mathbf{(8,1,2)} \quad = \quad \mathbf{(8,4)} \,, \\[5pt]
\mathrm{Tr} \, P^2 \qquad &\longrightarrow \qquad \mathbf{(27,1,1)} \quad = \quad \mathbf{(27,1)} \,.
\end{align}

For Theory $B$ we consider the  operators that are listed in Table \ref{tab:operators_B}.
\begin{table}[h]
\centering
\begin{tabular}{|c|cccc|}
\hline
 & $\prod_{i = 1}^4 SU(2)_{x_i} \times USp(4)_{y}$ & $U(1)_R$ & $U(1)_t$ & $U(1)_c$ \\
\hline 
$\mathrm{Tr} \, P^2$ & $\mathbf{(1,1,1,1,5)}$ & 0 & 1 & 0 \\
$\pi_1$ & $\mathbf{(2,1,1,1,4)}$ & 1 & -1/2 & 1 \\
$\pi_2$ & $\mathbf{(1,2,1,1,4)}$ & 1 & -1/2 & 1 \\
$\mathrm{Tr} \, P R_2$ & $\mathbf{(1,1,2,1,4)}$ & 1 & -1/2 & 1 \\
$\mathrm{Tr} \, P Q_{2,3} R_3$ & $\mathbf{(1,1,1,2,4)}$ & 1 & -1/2 & 1 \\
$\mathrm{Tr} \, L_1 R_1$ & $\mathbf{(2,2,1,1,1)}$ & 2 & -1 & 0 \\
$\mathrm{Tr} \, L_1 Q_{1,2} R_2$ & $\mathbf{(2,1,2,1,1)}$ & 2 & -1 & 0 \\
$\mathrm{Tr} \, L_1 Q_{1,2} Q_{2,3} R_3$ & $\mathbf{(2,1,1,2,1)}$ & 2 & -1 & 0 \\
$\mathrm{Tr} \, L_2 R_2$ & $\mathbf{(1,2,2,1,1)}$ & 2 & -1 & 0 \\
$\mathrm{Tr} \, L_2 Q_{2,3} R_2$ & $\mathbf{(1,2,1,2,1)}$ & 2 & -1 & 0 \\
$\mathrm{Tr} \, L_3 R_3$ & $\mathbf{(1,1,2,2,1)}$ & 2 & -1 & 0 \\
$\mathrm{Tr} \, A_1$ & $\mathbf{(1,1,1,1,1)}$ & 2 & -1 & 0 \\
$\mathrm{Tr} \, A_2$ & $\mathbf{(1,1,1,1,1)}$ & 2 & -1 & 0 \\
$\mathrm{Tr} \, A_3$ & $\mathbf{(1,1,1,1,1)}$ & 2 & -1 & 0 \\
\hline
\end{tabular}
\caption{\label{tab:operators_B} Gauge invariant chiral operators in Theory $B$.}
\end{table}

\noindent Again they combine into representations of $USp(8)_x \times USp(4)_y$ according to the following branching rules of $USp(8)_x\to \prod_{i = 1}^4 SU(2)_{x_i}$:
\begin{align}
	\mathrm{Tr} \, P^2 \quad \longrightarrow \quad &\mathbf{(1,1,1,1,5)} \quad = \quad \mathbf{(1,5)} \,, \\[5pt]
	\begin{array}{r}
		\pi_1 \,, \, \pi_2 \\
		\mathrm{Tr} \, P R_2 \,, \, \mathrm{Tr} \, P Q_{2,3} R_3 
	\end{array}
	\quad \longrightarrow \quad &
	\begin{array}{l}
		\mathbf{(2,1,1,1,4)}\oplus\mathbf{(1,2,1,1,4)}\\
		\oplus\mathbf{(1,1,2,1,4)}\oplus\mathbf{(1,1,1,2,4)}\\
		= \quad\mathbf{(8,4)} \,,
	\end{array}  \\[5pt]
	\begin{array}{r}
		\mathrm{Tr} \, L_1 R_1 \,, \, \mathrm{Tr} \, L_1 Q_{1,2} R_2 \,, \, \mathrm{Tr} \, L_1 Q_{1,2} Q_{2,3} R_3 \,, \\ \mathrm{Tr} \, L_2 R_2 \,, \, \mathrm{Tr} \, L_2 Q_{2,3} R_2 \,, \, \mathrm{Tr} \, L_3 R_3 \,, \\ \mathrm{Tr} \, A_1 \,, \, \mathrm{Tr} \, A_2 \,, \, \mathrm{Tr} \, A_3
	\end{array}
	\quad \longrightarrow \quad &
	\begin{array}{l}
		\mathbf{(2,2,1,1,1)}\oplus\mathbf{(2,1,2,1,1)} \\
		\oplus\mathbf{(2,1,1,2,1)}\oplus\mathbf{(1,2,2,1,1)} \\
		\oplus\mathbf{(1,2,1,2,1)}\oplus\mathbf{(1,1,2,2,1)} \\
		\oplus\mathbf{(1,1,1,1,1)}\oplus\mathbf{(1,1,1,1,1)} \\
		\oplus\mathbf{(1,1,1,1,1)}
	\end{array} \nonumber \\
&= \quad \mathbf{(27,1)} \,.
\end{align}

We then consider the operators of theory $C_1$ that are listed in Table \ref{tab:operators_C_1}.
\begin{table}[h]
\centering
\begin{tabular}{|c|cccc|}
\hline
 & $\prod_{i = 1}^4 SU(2)_{x_i} \times SU(2)_{y_1} \times SU(2)_{y_2}$ & $U(1)_R$ & $U(1)_t$ & $U(1)_c$ \\
\hline 
$\mathrm{Tr} \, L_1 R_1$ & $\mathbf{(2,2,1,1,1,1)}$ & 2 & -1 & 0 \\
$\mathrm{Tr} \, L_1 Q_{1,2} Q_{2,3} Q_{3,4} R_4$ & $\mathbf{(2,1,2,1,1,1)}$ & 2 & -1 & 0 \\
$\mathrm{Tr} \, L_1 Q_{1,2} Q_{2,3} Q_{3,4} Q_{4,5} R_5$ & $\mathbf{(2,1,1,2,1,1)}$ & 2 & -1 & 0 \\
$\mathrm{Tr} \, L_2 Q_{2,3} Q_{3,4} R_4$ & $\mathbf{(1,2,2,1,1,1)}$ & 2 & -1 & 0 \\
$\mathrm{Tr} \, L_2 Q_{2,3} Q_{3,4} Q_{4,5} R_5$ & $\mathbf{(1,2,1,2,1,1)}$ & 2 & -1 & 0 \\
$\mathrm{Tr} \, L_5 R_5$ & $\mathbf{(1,1,2,2,1,1)}$ & 2 & -1 & 0 \\
$\pi_{1,3}$ & $\mathbf{(2,1,1,1,2,1)}$ & 1 & -1/2 & 1 \\
$\pi_{2,3}$ & $\mathbf{(1,2,1,1,2,1)}$ & 1 & -1/2 & 1 \\
$\pi_{1,4}$ & $\mathbf{(2,1,1,1,1,2)}$ & 1 & -1/2 & 1 \\
$\pi_{2,4}$ & $\mathbf{(1,2,1,1,1,2)}$ & 1 & -1/2 & 1 \\
$\mathrm{Tr} \, L_3 Q_{3,4} R_4$ & $\mathbf{(1,1,2,1,2,1)}$ & 1 & -1/2 & 1 \\
$\mathrm{Tr} \, L_3 Q_{3,4} Q_{4,5} R_5$ & $\mathbf{(1,1,1,2,2,1)}$ & 1 & -1/2 & 1 \\
$\mathrm{Tr} \, L_4 R_4$ & $\mathbf{(1,1,2,1,1,2)}$ & 1 & -1/2 & 1 \\
$\mathrm{Tr} \, L_4 Q_{4,5} R_5$ & $\mathbf{(1,1,1,2,1,2)}$ & 1 & -1/2 & 1 \\
$\mathrm{Tr} \, L_3 R_3$ & $\mathbf{(1,1,1,1,2,2)}$ & 0 & 1 & 0 \\
$\mathrm{Tr} \, A_1$ & $\mathbf{(1,1,1,1,1)}$ & 2 & -1 & 0 \\
$\mathrm{Tr} \, A_3$ & $\mathbf{(1,1,1,1,1)}$ & 0 & 1 & 0 \\
$\mathrm{Tr} \, A_5$ & $\mathbf{(1,1,1,1,1)}$ & 2 & -1 & 0 \\
$\mathrm{Tr} \, Q_{2,3}^2 = \mathrm{Tr} \, Q_{3,4}^2$ & $\mathbf{(1,1,1,1,1)}$ & 2 & -1 & 0 \\
\hline
\end{tabular}
\caption{\label{tab:operators_C_1} Gauge invariant chiral operators in Theory $C_1$ and their charges under the faithful symmetries.}
\end{table}

\noindent  Notice that these operators  transform trivially under $U(1)_d\times SU(2)_v$, in accordance with the expectation that this symmetry is not faithful in the IR.\footnote{There are still gauge invariant chiral operators that we can in principle construct which are charged under $U(1)_d\times SU(2)_v$, but they are expected to vanish in the chiral ring by consistency with the operator map implied by the duality.} Again they can be arranged into representations of the full enhanced symmetry according to the following branching rules of $USp(8)_x\times USp(4)_y\to \prod_{i = 1}^4 SU(2)_{x_i} \times SU(2)_{y_1} \times SU(2)_{y_2}$: 
\begin{align}
\mathrm{Tr} \, L_3 R_3 \,, \quad \mathrm{Tr} \, A_3 \quad \longrightarrow \quad &\mathbf{(1,1,1,1,2,2)}\oplus\mathbf{(1,1,1,1,1,1)} \nonumber \\
&= \quad \mathbf{(1,5)} \,, \\[5pt]
\begin{array}{r}
\pi_{1,3} \,, \, \pi_{2,3} \,, \, \pi_{1,4} \,, \, \pi_{2,4} \,, \, \mathrm{Tr} \, L_3 Q_{3,4} R_4 \,, \\
\mathrm{Tr} \, L_3 Q_{3,4} Q_{4,5} R_5 \,, \, \mathrm{Tr} \, L_4 R_4 \,, \, \mathrm{Tr} \, L_4 Q_{4,5} R_5
\end{array}
\quad \longrightarrow \quad
&\begin{array}{l}
\mathbf{(2,1,1,1,2,1)}\oplus\mathbf{(1,2,1,1,2,1)} \\
\oplus\mathbf{(2,1,1,1,1,2)}\oplus\mathbf{(1,2,1,1,1,2)} \\
\oplus\mathbf{(1,1,2,1,2,1)}\oplus\mathbf{(1,1,1,2,2,1)} \\
\oplus\mathbf{(1,1,2,1,1,2)}\oplus\mathbf{(1,1,1,2,1,2)} \\
\end{array} \nonumber \\
&= \quad \mathbf{(8,4)} \,, 
\end{align}
\begin{align}
\begin{array}{r}
\mathrm{Tr} \, L_1 R_1 \,, \, \mathrm{Tr} \, L_1 Q_{1,2} Q_{2,3} Q_{3,4} R_4 \,, \\
\mathrm{Tr} \, L_1 Q_{1,2} Q_{2,3} Q_{3,4} Q_{4,5} R_5 \,, \, \mathrm{Tr} \, L_2 Q_{2,3} Q_{3,4} R_4 \,, \\
\mathrm{Tr} \, L_2 Q_{2,3} Q_{3,4} Q_{4,5} R_5 \,, \, \mathrm{Tr} \, L_5 R_5 \,, \\
\mathrm{Tr} \, A_1 \,, \, \mathrm{Tr} \, A_5 \,, \, \mathrm{Tr} \, Q_{2,3}^2 = \mathrm{Tr} \, Q_{3,4}^2
\end{array}
\quad \longrightarrow \quad
&\begin{array}{l}
\mathbf{(2,2,1,1,1,1)}\oplus\mathbf{(2,1,2,1,1,1)} \\
\oplus\mathbf{(2,1,1,2,1,1)}\oplus\mathbf{(1,2,2,1,1,1)} \\
\oplus\mathbf{(1,2,1,2,1,1)}\oplus\mathbf{(1,1,2,2,1,1)} \\
\oplus\mathbf{(1,1,1,1,1,1)}\oplus\mathbf{(1,1,1,1,1,1)} \\
\oplus\mathbf{(1,1,1,1,1,1)}
\end{array} \nonumber \\
&= \quad \mathbf{(27,1)}\,,
\end{align}

From the above identification of the $USp(8)_x \times USp(4)_y$ representations, we can easily find the map of operators under the duality web as follows:
\begin{align}
\left\{\mathrm{Tr} \, L R \,, \, \mathrm{Tr} \, A\right\}_{A,D} \leftrightarrow \left\{\mathrm{Tr} \, P^2\right\}_B \leftrightarrow \left\{\mathrm{Tr} \, L_3 R_3 \,, \, \mathrm{Tr} \, A_3\right\}_{C_1} ,
\end{align}
\begin{align}
\left\{\mathrm{Tr} \, P L \,, \, \pi\right\}_{A,D} \leftrightarrow \left\{\substack{\pi_1 \,, \, \pi_2 \,, \, \mathrm{Tr} \, P R_2 \,, \, \mathrm{Tr} \, P Q_{2,3} R_3}\right\}_B \leftrightarrow \left\{\substack{\pi_{1,3} \,, \, \pi_{2,3} \,, \, \pi_{1,4} \,, \, \pi_{2,4} \,, \, \mathrm{Tr} \, L_3 Q_{3,4} R_4 \,, \\
\mathrm{Tr} \, L_3 Q_{3,4} Q_{4,5} R_5 \,, \, \mathrm{Tr} \, L_4 R_4 \,, \, \mathrm{Tr} \, L_4 Q_{4,5} R_5}\right\}_{C_1} ,
\end{align}
\begin{align}
\left\{\mathrm{Tr} \, P^2\right\}_{A,D} &\leftrightarrow \left\{\substack{\mathrm{Tr} \, L_1 R_1 \,, \, \mathrm{Tr} \, L_1 Q_{1,2} R_2 \,, \, \mathrm{Tr} \, L_1 Q_{1,2} Q_{2,3} R_3 \,, \\ \mathrm{Tr} \, L_2 R_2 \,, \, \mathrm{Tr} \, L_2 Q_{2,3} R_2 \,, \, \mathrm{Tr} \, L_3 R_3 \,, \\ \mathrm{Tr} \, A_1 \,, \, \mathrm{Tr} \, A_2 \,, \, \mathrm{Tr} \, A_3}\right\}_B \nonumber\\[5pt]
&\leftrightarrow \left\{\substack{\mathrm{Tr} \, L_1 R_1 \,, \, \mathrm{Tr} \, L_1 Q_{1,2} Q_{2,3} Q_{3,4} R_4 \,, \\
\mathrm{Tr} \, L_1 Q_{1,2} Q_{2,3} Q_{3,4} Q_{4,5} R_5 \,, \, \mathrm{Tr} \, L_2 Q_{2,3} Q_{3,4} R_4 \,, \\
\mathrm{Tr} \, L_2 Q_{2,3} Q_{3,4} Q_{4,5} R_5 \,, \, \mathrm{Tr} \, L_5 R_5 \,, \\
\mathrm{Tr} \, A_1 \,, \, \mathrm{Tr} \, A_5 \,, \, \mathrm{Tr} \, Q_{2,3}^2 = \mathrm{Tr} \, Q_{3,4}^2}\right\}_{C_1} .
\end{align}

The existence of these operators can also be confirmed by computing the supersymmetric index in a power series. For example, we have evaluated the indices of Theory $A = D$ and Theory $B$ as follows:
\begin{align}
\label{eq:indx_ABD}
\mathcal I_A = \mathcal I_B = \mathcal I_D &= 1+(pq)^\frac{5}{21} t^{-1} c^{-2}+\mathbf 5_y (pq)^\frac37 t+\mathbf 8_x \mathbf 4_y (pq)^\frac{19}{42} t^{-\frac12} c+(pq)^\frac{10}{21} t^{-2} c^{-4} \nonumber \\
&\quad +(pq)^\frac{11}{21} t^2 c^{-2}+\mathbf{27}_x (pq)^\frac47 t^{-1}+\dots
\end{align}
and found a perfect match as expected. In the index expansion, $\mathbf m_x$ and $\mathbf n_y$ denote the characters of the representations of dimensions $\mathbf m$ and $\mathbf n$ of $USp(8)_x$ and $USp(4)_y$, respectively, and we have used a trial $R$-charge defined as follows:
\begin{align}
R_\text{trial} = R+\frac67 Q_t+\frac13 Q_c
\end{align}
where $R$ is the $R$-charge shown in the tables, and $Q_t$ and $Q_c$ are the $U(1)_t$ and $U(1)_c$ charges, respectively. The mixing coefficients are chosen for convenience of the computation.

The three sets of the operators we explained correspond to the third, fourth, and seventh terms of the expanded index. In addition, there are other terms only charged under the Abelian symmetries. For example, the second term corresponds to a singlet flipping the dressed meson $\mathrm{Tr} A L^2$ in Theory $A$, which turns out to decouple in the IR by the $a$-maximization. If we call this operator $\beta_1$, the fifth term is $\beta_1^2$, whereas the sixth term corresponds to $\mathrm{Tr} R^2$. Their counterparts in Theory B can easily be identified. The second, fifth, and sixth terms correspond to $\mathrm{Tr} L_1^2$, $\left(\mathrm{Tr} L_1^2\right)^2$, and the singlet flipping $\mathrm{Tr} R_2^2$ in Theory B, respectively.

For Theory $C_i$, on the other hand, the computation of the index is more complicated. We have evaluated the index of $C_1$ up to $(pq)^\frac{10}{21}$ with the simplification $x_i = y_j = 1$, which agrees with the index from the other dual frames \eqref{eq:indx_ABD}. Moreover, we have found that this partially unrefined index of Theory $C_1$ there is no dependence on the fugacities $v$ and $d$ for the $SU(2)_v\times U(1)_d$, which again confirms our expectation that this symmetry acts trivially in the IR.

The operators of  Theory $C_2$ and Theory $C_3$, some of which are listed in Table \ref{tab:operators_C_2}, combine similarly to form representations of the IR $USp(8)_x \times USp(4)_y$ symmetry.

\begin{table}[h]
\centering
\begin{tabular}{|c|cccc|}
\hline
 & $\prod_{i = 1}^4 SU(2)_{x_i} \times SU(2)_{y_1} \times SU(2)_{y_2}$ & $U(1)_R$ & $U(1)_t$ & $U(1)_c$ \\
\hline 
$\mathrm{Tr} \, L_1 Q_{1,2} R_2$ & $\mathbf{(2,2,1,1,1,1)}$ & 2 & -1 & 0 \\
$\mathrm{Tr} \, L_1 Q_{1,2} Q_{2,3} R_3$ & $\mathbf{(2,1,2,1,1,1)}$ & 2 & -1 & 0 \\
$\mathrm{Tr} \, L_1 Q_{1,2} Q_{2,3} Q_{3,4} Q_{4,5} R_5$ & $\mathbf{(2,1,1,2,1,1)}$ & 2 & -1 & 0 \\
$\mathrm{Tr} \, L_3 R_3$ & $\mathbf{(1,2,2,1,1,1)}$ & 2 & -1 & 0 \\
$\mathrm{Tr} \, L_3 Q_{3,4} Q_{4,5} R_5$ & $\mathbf{(1,2,1,2,1,1)}$ & 2 & -1 & 0 \\
$\mathrm{Tr} \, L_4 Q_{4,5} R_5$ & $\mathbf{(1,1,2,2,1,1)}$ & 2 & -1 & 0 \\
$\pi_{1,2}$ & $\mathbf{(2,1,1,1,2,1)}$ & 1 & -1/2 & 1 \\
$\pi_{1,5}$ & $\mathbf{(2,1,1,1,1,2)}$ & 1 & -1/2 & 1 \\
$\pi_{3,5}$ & $\mathbf{(1,2,1,1,1,2)}$ & 1 & -1/2 & 1 \\
$\pi_{4,5}$ & $\mathbf{(1,1,2,1,1,2)}$ & 1 & -1/2 & 1 \\
$\mathrm{Tr} \, L_2 R_2$ & $\mathbf{(1,2,1,1,2,1)}$ & 1 & -1/2 & 1 \\
$\mathrm{Tr} \, L_2 Q_{2,3} R_3$ & $\mathbf{(1,1,2,1,2,1)}$ & 1 & -1/2 & 1 \\
$\mathrm{Tr} \, L_2 Q_{2,3} Q_{3,4} Q_{4,5} R_5$ & $\mathbf{(1,1,1,2,2,1)}$ & 1 & -1/2 & 1 \\
$\mathrm{Tr} \, L_5 R_5$ & $\mathbf{(1,1,1,2,1,2)}$ & 1 & -1/2 & 1 \\
$\mathrm{Tr} \, L_2 Q_{2,3} Q_{3,4} R_4$ & $\mathbf{(1,1,1,1,2,2)}$ & 0 & 1 & 0 \\
$\mathrm{Tr} \, A_3$ & $\mathbf{(1,1,1,1,1)}$ & 2 & -1 & 0 \\
$\mathrm{Tr} \, Q_{1,2}^2$ & $\mathbf{(1,1,1,1,1)}$ & 2 & -1 & 0 \\
$\mathrm{Tr} \, Q_{4,5}^2$ & $\mathbf{(1,1,1,1,1)}$ & 2 & -1 & 0 \\
$\mathrm{Tr} \, Q_{2,3}^2 = \mathrm{Tr} \, Q_{3,4}^2$ & $\mathbf{(1,1,1,1,1)}$ & 0 & 1 & 0 \\
\hline 
$\mathrm{Tr} \, L_2 R_2$ & $\mathbf{(2,2,1,1,1,1)}$ & 2 & -1 & 0 \\
$\mathrm{Tr} \, L_2 Q_{2,3} R_3$ & $\mathbf{(2,1,2,1,1,1)}$ & 2 & -1 & 0 \\
$\mathrm{Tr} \, L_2 Q_{2,3} Q_{3,4} R_4$ & $\mathbf{(2,1,1,2,1,1)}$ & 2 & -1 & 0 \\
$\mathrm{Tr} \, L_3 R_3$ & $\mathbf{(1,2,2,1,1,1)}$ & 2 & -1 & 0 \\
$\mathrm{Tr} \, L_3 Q_{3,4} R_4$ & $\mathbf{(1,2,1,2,1,1)}$ & 2 & -1 & 0 \\
$\mathrm{Tr} \, L_4 R_4$ & $\mathbf{(1,1,2,2,1,1)}$ & 2 & -1 & 0 \\
$\pi_{2,6}$ & $\mathbf{(2,1,1,1,1,2)}$ & 1 & -1/2 & 1 \\
$\pi_{3,6}$ & $\mathbf{(1,2,1,1,1,2)}$ & 1 & -1/2 & 1 \\
$\pi_{4,6}$ & $\mathbf{(1,1,2,1,1,2)}$ & 1 & -1/2 & 1 \\
$\pi_{5,6}$ & $\mathbf{(1,1,1,2,1,2)}$ & 1 & -1/2 & 1 \\
$\mathrm{Tr} \, L_1 R_1$ & $\mathbf{(2,1,1,1,2,1)}$ & 1 & -1/2 & 1 \\
$\mathrm{Tr} \, L_1 Q_{1,2} R_2$ & $\mathbf{(1,2,1,1,2,1)}$ & 1 & -1/2 & 1 \\
$\mathrm{Tr} \, L_1 Q_{1,2} Q_{2,3} R_3$ & $\mathbf{(1,1,2,1,2,1)}$ & 1 & -1/2 & 1 \\
$\mathrm{Tr} \, L_1 Q_{1,2} Q_{2,3} Q_{3,4} R_4$ & $\mathbf{(1,1,1,2,2,1)}$ & 1 & -1/2 & 1 \\
$\mathrm{Tr} \, L_1 Q_{1,2} Q_{2,3} Q_{3,4} Q_{4,5} R_5$ & $\mathbf{(1,1,1,1,2,2)}$ & 0 & 1 & 0 \\
$\mathrm{Tr} \, A_2$ & $\mathbf{(1,1,1,1,1)}$ & 2 & -1 & 0 \\
$\mathrm{Tr} \, A_3$ & $\mathbf{(1,1,1,1,1)}$ & 2 & -1 & 0 \\
$\mathrm{Tr} \, A_4$ & $\mathbf{(1,1,1,1,1)}$ & 2 & -1 & 0 \\
$\mathrm{Tr} \, Q_{i,i+1}^2$ & $\mathbf{(1,1,1,1,1)}$ & 0 & 1 & 0 \\
\hline
\end{tabular}
\caption{\label{tab:operators_C_2} Gauge invariant chiral operators in Theories $C_2$ (the upper box) and $C_3$ (the lower box).}
\end{table}

\newpage
\subsection{$3d$ $SL(2,\mathbb{Z})$ duality web}

\begin{figure}[!ht]
    \includegraphics[scale=0.35,center]{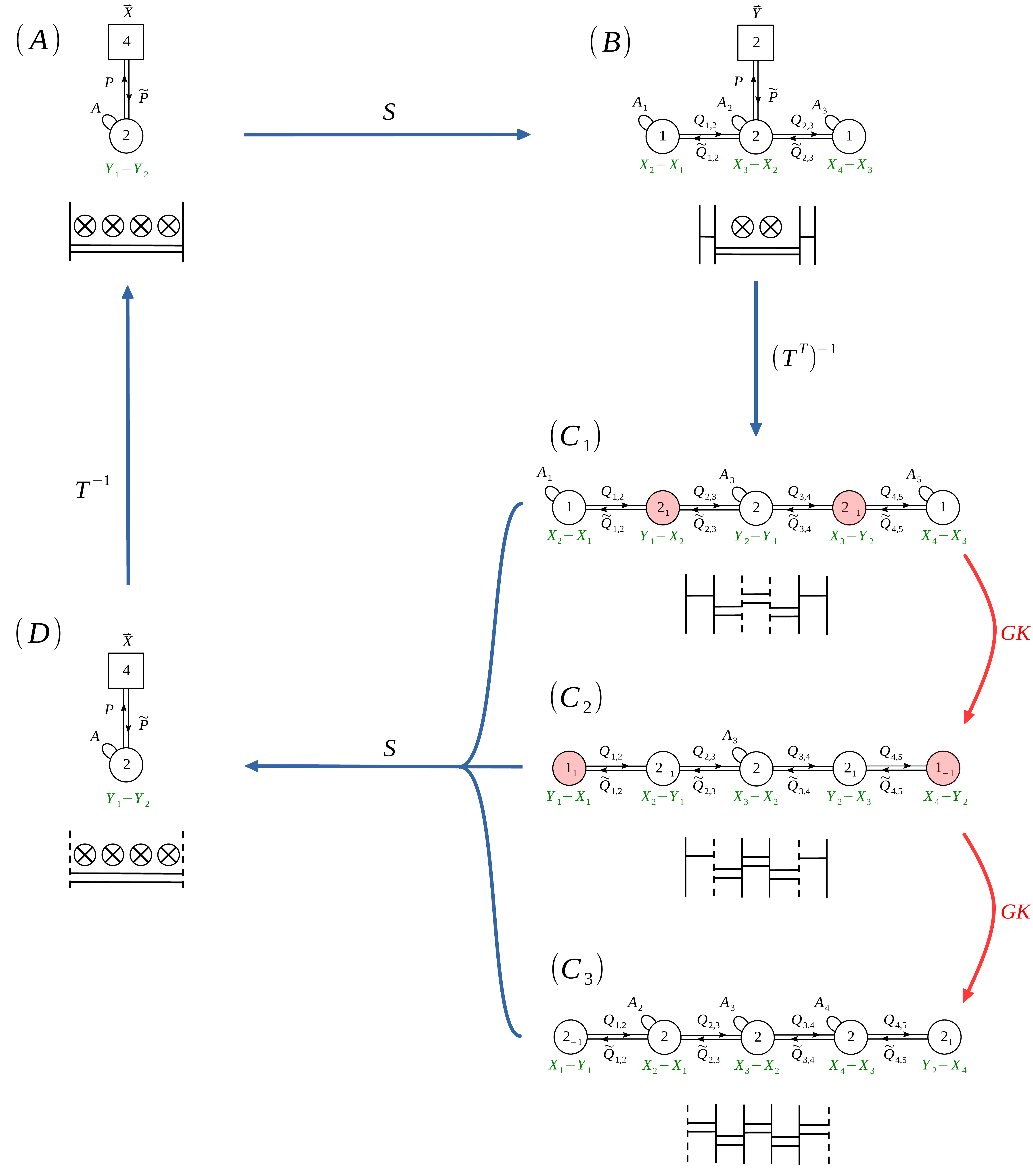}
    \caption{A $3d$ $SL(2,\mathbb{Z})$ web with the corresponding brane setups, where NS5, D5, and $(1,1)$-branes are denoted by a solid vertical line, a crossed circle, and a dashed vertical line, respectively. The FI parameters at each gauge node are denoted in green.}
    \label{3dsl2zweb}
\end{figure}

\begin{figure}[!ht]
    \includegraphics[scale=0.4,center]{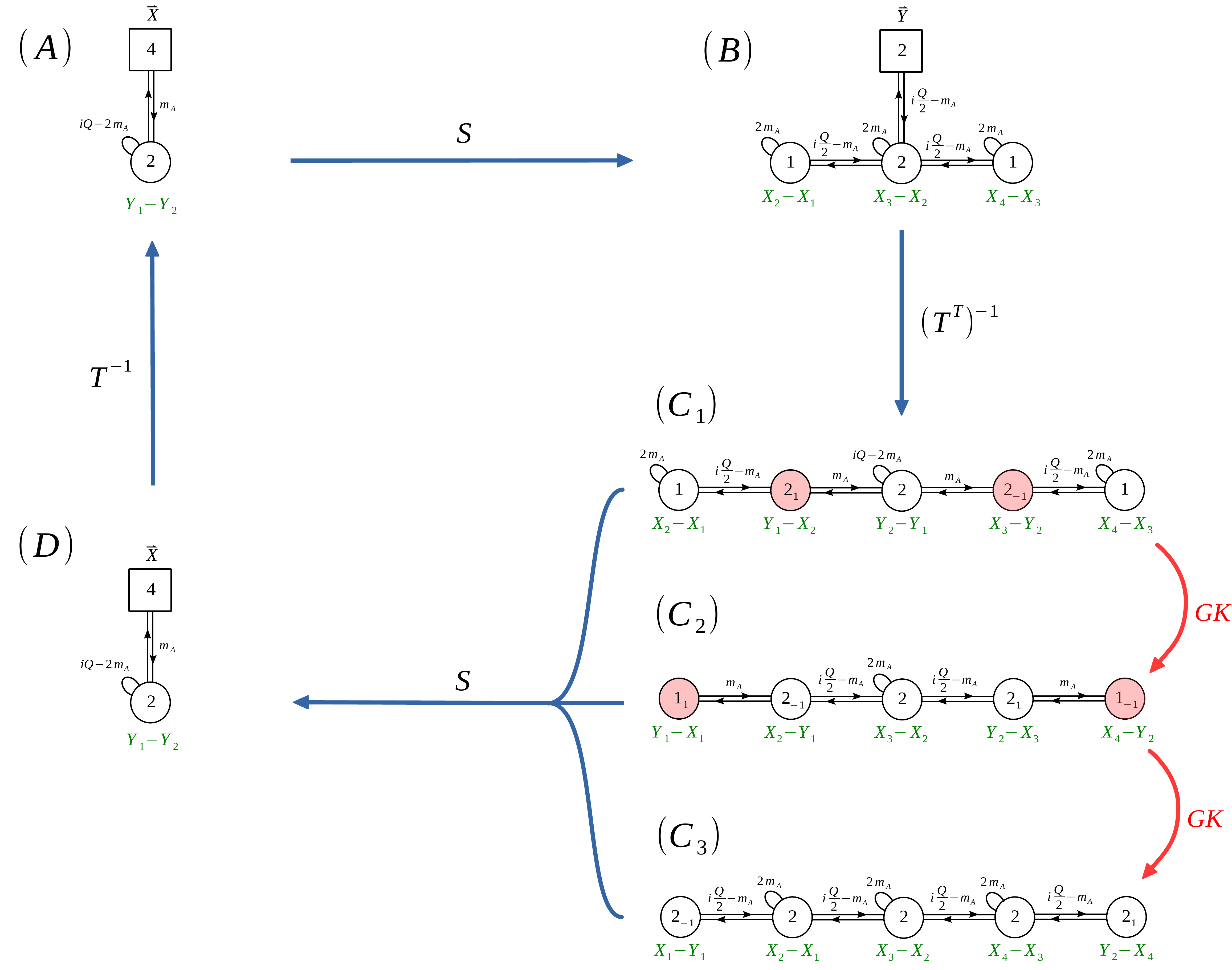}
    \caption{The charge assignment for the $3d$ $SL(2,\mathbb{Z})$ web.}
    \label{3dsl2zweb_charges}
\end{figure}

In $3d$, starting from the $\mathcal{N}=4$ SQCD, we can build an analogous duality web by acting with the 
 $SL(2,\mathbb{Z})$ basic moves.
The $3d$ version of the web is shown in Figure \ref{3dsl2zweb}
together with the brane systems (see Figure \ref{3dsl2zweb_charges} for the charge assignments of the fields).
Notice that we move from frame $C_1$ to $C_2$ and $C_3$ by applying the Giveon--Kutasov duality
which implements the HW moves exchanging NS5 and (1,1)-branes in the brane setup, 
similarly to what we did in $4d$ using the IP duality instead.

The frames $C_{1,2,3}$  correspond to theories with CS couplings. Notice in particular that in the first two frames when we integrate out the massive adjoint at the nodes where the CS coupling is turned on, a quartic superpotential for the bifundamentals is generated. These quartic  superpotentials preserve the $U(1)_A$ symmetry 
which becomes part of the larger R-symmetry group since  all these frames have enhanced $\mathcal{N}=4$ supersymmetry as expected by the duality.

\begin{landscape}
\begin{figure}
\centering
	\includegraphics[scale=.35]{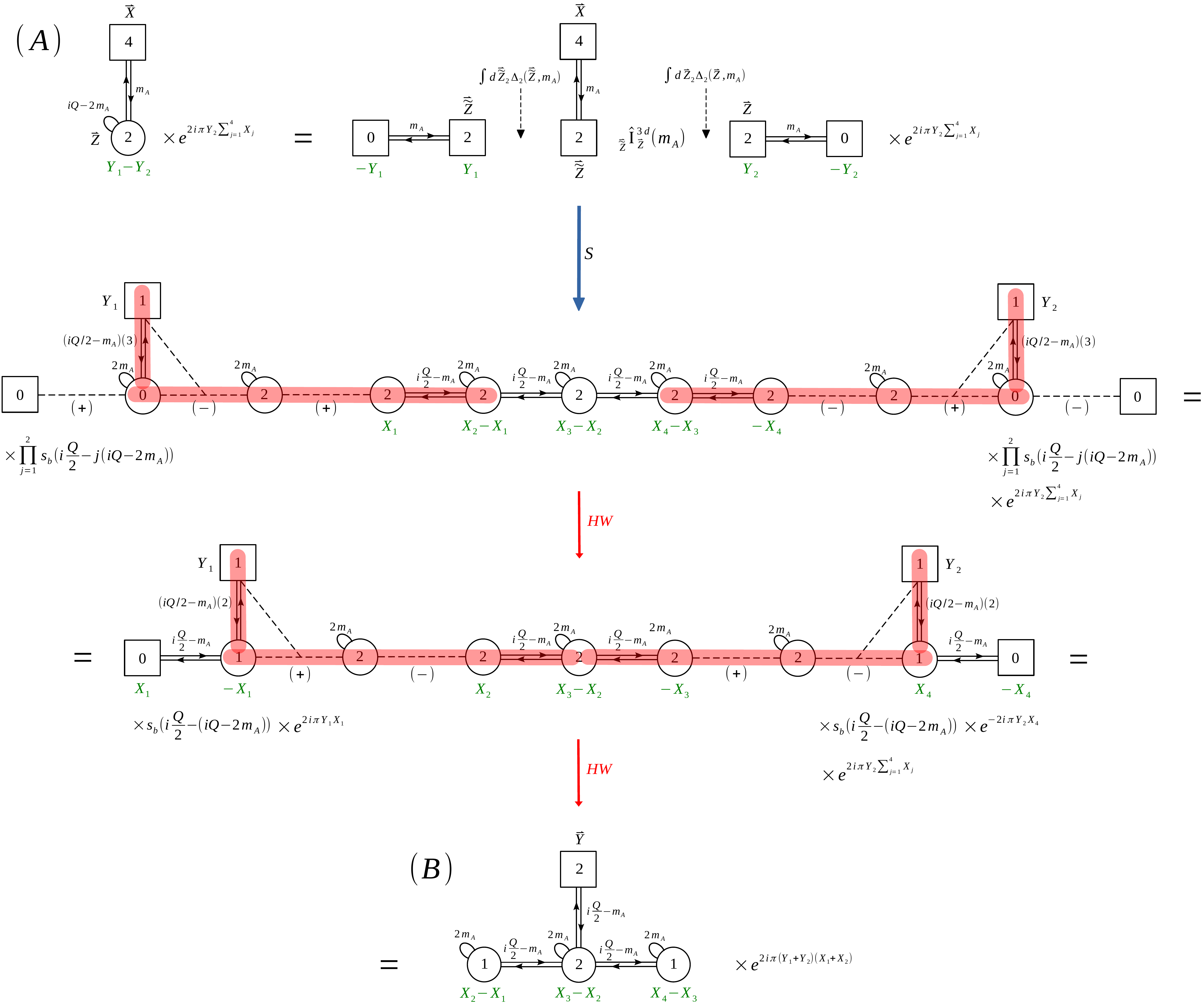}
	\caption{The $\mathcal{S}$-dualization of theory A.}
	\label{fig:A_dualized_to_B_3d}
\end{figure}
\end{landscape}

We can reach the $3d$ web  in Figure \ref{3dsl2zweb} from the $4d$ web in Figure \ref{sl2zweb} 
taking the limit:
\begin{align}
X_i\to X_i+s,\quad Y_j\to Y_j+s, \quad Z_a\to Z_a+s, \quad V \to V+s, \quad s\to+\infty
\end{align}
followed by $\Delta \to -\infty$. 
Alternatively we can apply the $3d$ dualization algorithm. We illustrate below the dualization from of theory $A$ into $B$.
In the first row of Figure \ref{fig:A_dualized_to_B_3d} theory $A$ is chopped into  QFT blocks.
Applying the basic moves in Figure \ref{fig:10_Sdual_3d} and \ref{fig:01_Sdual_3d} we get the dualized theory in the second row of figure \ref{fig:A_dualized_to_B_3d}. Notice that, for the leftmost block, we  apply the duality move
 in Figure \ref{fig:10_Sdual_3d} with a Cartan's reparametrization such that $\mathcal{S} \to \mathcal{S}^{-1}$ and viceversa (if one does not implement such a reparametrization, the first and second blocks would glue with $\mathcal{SS}$ rather than as $\mathcal{S}\mathcal{S}^{-1}$). 
Now, as shown in the third and fourth row of figure \ref{fig:A_dualized_to_B_3d}, we can propagate the VEVs on the two sides of the quiver using the Hanany--Witten move in Figure \ref{fig:HW_move_3d}.
After these two iterations, we land on the theory B shown in the last row of figure \ref{fig:A_dualized_to_B_3d}. 

\subsection{$3d$ operators map}
It is interesting to see how the operator map works in the $3d$ case to better understand the web of dualities. 
The global symmetry group of the three theories in the IR is
\begin{align}
	SU(4)_X \times SU(2)_Y \times U(1)_A\,,
\end{align}
but the UV manifest symmetries in the theories of the web are the following:
\begin{align}
	A=D:&\quad \underbrace{\frac{U(4)_X}{U(1)}}_{SU(4)_X} \times \frac{\prod_{j=1}^2U(1)_{Y_j}}{U(1)} \times U(1)_A\,,\\
	B:&\quad \underbrace{\frac{U(2)_Y}{U(1)}}_{SU(2)_Y} \times \frac{\prod_{i=1}^4 U(1)_{X_i}}{U(1)} \times U(1)_A\,,\nonumber\\
	C_i:&\quad \frac{\prod_{i=1}^4 U(1)_{X_i}}{U(1)}\times \frac{\prod_{j=1}^2U(1)_{Y_j}}{U(1)} \times U(1)_A\,,\nonumber
\end{align}
where we are using the parametrization of the various symmetries that we obtained from the algorithm and that is specified in Figure \ref{3dsl2zweb_charges}, but we should remember that is an off-shell parametrization in $3d$ and that we should decoupled a diagonal $U(1)$ both from the $X$ and from the $Y$ symmetries. In the following we will keep the off-shell parametrization since it makes it easier to identify how the operators rearrange into representations of the enhanced symmetry and to map them across the various frames.

We now analyze each theory discussing its symmetries and the gauge invariant operators with the lowest dimension. Notice that operators with charges $(iQ - 2m_A)$ or $2m_A$ have $R$-charge 1 when we set the canonical parameterization $m_A = \frac{iQ}{4}$, which is the lowest $R$-charge possible in an interacting $3d$ $\mathcal{N}=4$ theory. 

Let us start from Theory $A$, which is equivalent to Theory $D$. The manifest global symmetry group is given by 
\begin{equation}
	\frac{U(4)_X}{U(1)} \times \frac{\prod_{j=1}^2U(1)_{Y_j}}{U(1)} \times U(1)_A\ \,,
\end{equation}
where $\frac{U(4)_X}{U(1)}=SU(4)_X$ is the flavor symmetry, while $\frac{\prod_{j=1}^2U(1)_{Y_j}}{U(1)}$ is the topological symmetry associated to the $U(2)$ gauge node. The lowest dimension gauge invariant operators are listed in Table \ref{tab:3d_operators_A}.
\begin{table}[!ht]
\centering
\begin{tabular}{|c|ccc|}
\hline
 & $SU(4)_X \times \prod_{j=1}^2U(1)_{Y_j}$ & $U(1)_{A}$ & $U(1)_R$ \\
\hline 
$\mathrm{Tr} \, P\tilde{P}$ & $(\mathbf{15},0,0)$ & 2 & 0 \\
$\mathrm{Tr} \, A$ & $(\mathbf{1},0,0)$ & -2 & 2 \\
$M^{(+)}$ & $(\mathbf{1}, 1,-1)$ & -2 & 2 \\
$M^{(-)}$ & $(\mathbf{1}, -1,+1)$ & -2 & 2 \\
\hline
\end{tabular}
\caption{Gauge invariant operators in Theory $A$ = Theory $D$. We denote by $M^{(\pm)}=M^{(\{\pm1,0\})}$
the fundamental monopoles of the $U(2)$ gauge group.}
\label{tab:3d_operators_A}
\end{table}

Notice that the $U(2)$ gauge node is balanced so according to \cite{Gaiotto:2008ak} the topological symmetry is enhanced to $\frac{\prod_{j=1}^2U(1)_{Y_j}}{U(1)} \rightarrow SU(2)_Y$.  In fact some of the  operators  in Table \ref{tab:3d_operators_A} combine to form representations of the IR global symmetry $SU(4)_X\times SU(2)_Y$ and provide the moment maps for it:
\begin{align}
	\mathrm{Tr} \, P\tilde{P} 
	& \quad\longrightarrow\quad (\mathbf{15},0) \quad
	= \quad(\mathbf{15},\mathbf{1}) \,, \\
	\mathrm{Tr} \, A, \,\, M^{(\pm)} 
	& \quad\longrightarrow\quad (\mathbf{1},0,0) \oplus (\mathbf{1},1,-1)\oplus (\mathbf{1},-1,1) \quad
	= \quad(\mathbf{1},\mathbf{3}) \,.
\end{align}
Notice that only the off-diagonal combination $U(1)_{Y_2}-U(1)_{Y_1}$ acts non-trivially and is the one that is enhanced to $SU(2)_Y$, while the diagonal combination $U(1)_{Y_1}+U(1)_{Y_2}$ acts trivially and should be decoupled as we previously mentioned.

In Theory $B$ the manifest symmetry is
\begin{align}
\frac{U(2)_Y}{U(1)} \times \frac{\prod_{i=1}^4 U(1)_{X_i}}{U(1)} \times U(1)_A\,,
\end{align}
where $\frac{U(2)_Y}{U(1)}=SU(2)_Y$ is the flavor symmetry, while $\frac{\prod_{i=1}^4 U(1)_{X_i}}{U(1)}$ after decoupling a diagonal $U(1)$ is a combination of the topological symmetries of the three gauge nodes which we parametrize as sepcified in Figure \ref{3dsl2zweb_charges}. We consider the operators in Table \ref{tab:3d_operators_B}, which have the lowest scaling dimensions and will provide the moment maps for the enhanced symmetry.
\begin{table}[!ht]
\centering
\begin{tabular}{|c|ccc|}
\hline
 & $SU(2)_Y \times \prod_{i=1}^4 U(1)_{X_i}$ & $U(1)_{A}$ & $U(1)_R$ \\
\hline 
$\mathrm{Tr} \, P\tilde{P}$ & $(\mathbf{3},0,0,0,0)$ & -2 & 2 \\
$\mathrm{Tr} \, A_1$ & $(\mathbf{1},0,0,0,0)$ & 2 & 0 \\
$\mathrm{Tr} \, A_2$ & $(\mathbf{1},0,0,0,0)$ & 2 & 0 \\
$\mathrm{Tr} \, A_3$ & $(\mathbf{1},0,0,0,0)$ & 2 & 0 \\
$M^{(\pm,0,0)}$ & $(\mathbf{1},\pm1,\mp1,0,0)$ & 2 & 0 \\
$M^{(0,\pm,0)}$ & $(\mathbf{1},0,\pm 1,\mp1,0)$ & 2 & 0 \\
$M^{(0,0,\pm)}$ & $(\mathbf{1},0,0,\pm 1,\mp1)$ & 2 & 0 \\
$M^{(\pm,\pm,0)}$ & $(\mathbf{1},\pm 1,0,\mp 1,0)$ & 2 & 0 \\
$M^{(0,\pm,\pm)}$ & $(\mathbf{1},0,\pm 1,0,\mp 1)$ & 2 & 0 \\
$M^{(\pm,\pm,\pm)}$ & $(\mathbf{1},\pm 1,0,0,\mp 1)$ & 2 & 0 \\
\hline
\end{tabular}
\caption{Gauge invariant operators in Theory $B$. 
 We used the convention $M^{(\pm,\pm,\pm)}=M^{(\pm1,\{\pm1,0\},\pm1)}$
 for the fundamental monopoles of the $U(1)\times U(2)\times U(1)$ gauge groups.
}
\label{tab:3d_operators_B}
\end{table}

\noindent All the gauge nodes are balanced, so the three topological symmetries get enhanced to $\frac{\prod_{i=1}^4U(1)_{X_i}}{U(1)}\to SU(4)_X$.  
The operators combine to form representation of the IR global symmetry $SU(4)_X\times SU(2)_Y$ as
\begin{align}
\begin{array}{r}
	M^{(\pm,0,0)},\,\, M^{(0,\pm,0)},\,\, M^{(0,0,\pm)}\\
	M^{(\pm,\pm,0)},\,\, M^{(0,\pm,\pm)},\,\, M^{(\pm,\pm,\pm)}\\
	\mathrm{Tr} \, A_1,\,\,\mathrm{Tr} \, A_2,\,\,\mathrm{Tr} \, A_3 
	\end{array}
	& \quad\longrightarrow\quad 
	\begin{array}{l}
	(\mathbf{1},\pm 1,\mp1,0,0) \oplus
	(\mathbf{1},0,\pm 1,\mp1,0) \oplus
	(\mathbf{1},0,0,\pm 1,\mp1) \\
	\oplus (\mathbf{1},\pm 1,0,\mp 1,0)
	\oplus (\mathbf{1},0,\pm 1,0,\mp 1)
	\oplus (\mathbf{1},\pm 1,0,0,,\mp 1)\\
	\oplus(\mathbf{1},0,0,0)\oplus(\mathbf{1},0,0,0)\oplus(\mathbf{1},0,0,0)\oplus\\
	= \quad(\mathbf{15},\mathbf{1})\,,
	\end{array} \\
	\mathrm{Tr} \, P\tilde{P}
	& \quad\longrightarrow\quad 
	\begin{array}{l}
	(\mathbf{3},0,0,0) \quad
	= \quad(\mathbf{1},\mathbf{3}) \,.
	\end{array}
\end{align}

In the three $C_i$ theories the manifest symmetry is
\begin{align}
\frac{\prod_{i=1}^4 U(1)_{X_i}}{U(1)}\times \frac{\prod_{j=1}^2U(1)_{Y_j}}{U(1)} \times U(1)_A\,,
\end{align}
where all the symmetries, except for the axial symmetry, are topological.
The central node is balanced, so we expect the enhancement $\frac{\prod_{j=1}^2U(1)_{Y_j}}{U(1)}\to SU(2)_Y$. On the other hand, if we look at the other gauge nodes, since they have non-zero CS level, their monopoles are not gauge invariant and have to be dressed with bifundamental fields.
The lowest dimension gauge invariant operators of Theory $C_1$ are listed in Table \ref{tab:3d_operators_C1}.
\begin{table}[!ht]
\centering
\begin{tabular}{|c|ccc|}
\hline
 & $\prod_{i=1}^4 U(1)_{X_i}\times\prod_{j=1}^2U(1)_{Y_j}$ & $U(1)_{A}$ & $U(1)_R$ \\
\hline 
$\mathrm{Tr} \, A_1$ & $(0,0,0,0,0,0)$ & 2 & 0 \\
$\mathrm{Tr} \, A_3$ & $(0,0,0,0,0,0)$ & -2 & 2 \\
$\mathrm{Tr} \, A_5$ & $(0,0,0,0,0,0)$ & 2 & 0 \\
$\mathrm{Tr} \, (Q_{2,3}\widetilde{Q}_{2,3})$ & $(0,0,0,0,0,0)$ & 2 & 0 \\
$M^{(\pm,0,0,0,0)}$ & $(\pm 1,\mp1,0,0,0,0)$ & 2 & 0 \\
$M^{(0,0,\pm,0,0)}$ & $(0,0,0,0,\pm 1,\mp1)$ & -2 & 2 \\
$M^{(0,0,0,0,\pm)}$ & $(0,0,\pm1,\mp1,0,0)$ & 2 & 0 \\
$M^{(+,+,+,+,0)} \widetilde{Q}_{2,3}\widetilde{Q}_{3,4}$ & $(1,0,-1,0,0,0)$ & 2 & 0 \\
$M^{(-,-,-,-,0)} Q_{2,3}Q_{3,4}$ & $(-1,0,1,0,0,0)$ & 2 & 0 \\
$M^{(+,+,+,+,+)} \widetilde{Q}_{2,3}\widetilde{Q}_{3,4}$ & $(1,0,0,-1,0,0)$ & 2 & 0 \\
$M^{(-,-,-,-,-)} Q_{2,3}Q_{3,4}$ & $(-1,0,0,1,0,0)$ & 2 & 0 \\
$M^{(0,+,+,+,0)} \widetilde{Q}_{2,3}\widetilde{Q}_{3,4}$ & $(0,1,-1,0,0,0)$ & 2 & 0 \\
$M^{(0-,-,-,0)} Q_{2,3}Q_{3,4}$ & $(0,-1,1,0,0,0)$ & 2 & 0 \\
$M^{(0,+,+,+,+)} \widetilde{Q}_{2,3}\widetilde{Q}_{3,4}$ & $(0,1,0,-1,0,0)$ & 2 & 0 \\
$M^{(0,-,-,-,-)} Q_{2,3}Q_{3,4}$ & $(0,-1,0,1,0,0)$ & 2 & 0 \\
\hline
\end{tabular}
\caption{Gauge invariant operators in Theory $C_1$. We used the convention $M^{(\pm,\pm,\pm,\pm,\pm)}=M^{(\pm1,\{\pm1,0\},\{\pm1,0\},\{\pm1,0\},\pm1)}$
for the fundamental monopoles of the $U(1)\times U(2)^3\times U(1)$ gauge groups.
}
\label{tab:3d_operators_C1}
\end{table}

\noindent These operators rearrange into the representation of the IR global symmetry $SU(4)_X\times SU(2)_Y$ as follows:
\begin{align}
	\begin{array}{r}
	\mathrm{Tr} \, A_1,\,\, \mathrm{Tr} \, A_5 \\
	\mathrm{Tr} \, (Q_{2,3}\widetilde{Q}_{2,3}) \\
	M^{(\pm,0,0,0,0)},\,\, M^{(0,0,0,0,\pm)}\\
	M^{(+,+,+,+,0)} \widetilde{Q}_{2,3}\widetilde{Q}_{3,4},\,\,M^{(-,-,-,-,0)} Q_{2,3}Q_{3,4}\\
	M^{(+,+,+,+,+)} \widetilde{Q}_{2,3}\widetilde{Q}_{3,4},\,\, M^{(-,-,-,-,-)} Q_{2,3}Q_{3,4}\\
	M^{(0,+,+,+,0)} \widetilde{Q}_{2,3}\widetilde{Q}_{3,4},\,\, M^{(0-,-,-,0)} Q_{2,3}Q_{3,4}\\
	M^{(0,+,+,+,+)} \widetilde{Q}_{2,3}\widetilde{Q}_{3,4},\,\, M^{(0-,-,-,-)} Q_{2,3}Q_{3,4}
	\end{array}
	& \,\,\longrightarrow\,\, 
	\begin{array}{l}
		(0,0,0,0,0,0)\oplus(0,0,0,0,0,0)\\
		\oplus (0,0,0,0,0,0)\\
		\oplus(\pm 1,\mp1,0,0,0,0)\oplus(0,0,0,0,\pm 1,\mp1)\\
		\oplus(\pm1,0,\mp1,0,0,0)\oplus(\pm1,0,0,\mp1,0,0)\\
		\oplus(0,\pm1,\mp1,0,0,0)\oplus(0,\pm1,0,\mp1,0,0)\\
		=\quad(\mathbf{15},\mathbf{1}) \,,
	\end{array} \\
	\begin{array}{r}
		\mathrm{Tr} \, A_3,\,\, M^{(0,0,\pm,0,0)}
	\end{array}
	& \,\,\longrightarrow\,\,
	\begin{array}{l}
		(0,0,0,0,0,0)\oplus(0,0,0,0,\pm1,\mp1)\\
		= \quad (\mathbf{1},\mathbf{3}) \,.
	\end{array} 
\end{align}

After we have grouped operators in representations of the IR symmetry group it is easy to see that they are mapped as follows:
\begin{align}
\left\{  \mathrm{Tr} \, P\tilde{P}  \right\}_{A,D} \leftrightarrow \left\{\substack{ M^{(\pm,0,0)},\,\, M^{(0,\pm,0)},\,\, M^{(0,0,\pm)}\\
	M^{(\pm,\pm,0)},\,\, M^{(0,\pm,\pm)},\,\, M^{(\pm,\pm,\pm)}\\
	\mathrm{Tr} \, A_1,\,\,\mathrm{Tr} \, A_2,\,\,\mathrm{Tr} \, A_3  }\right\}_B \leftrightarrow \left\{\substack{ \mathrm{Tr} \, A_1,\,\, \mathrm{Tr} \, A_3 \\
	\mathrm{Tr} \, (Q_{2,3}\widetilde{Q}_{2,3}) \\
	M^{(\pm,0,0,0,0)},\,\, M^{(0,0,0,0,\pm)}\\
	M^{(+,+,+,+,0)} \widetilde{Q}_{2,3}\widetilde{Q}_{3,4},\,\,M^{(-,-,-,-,0)} Q_{2,3}Q_{3,4}\\
	M^{(+,+,+,+,+)} \widetilde{Q}_{2,3}\widetilde{Q}_{3,4},\,\, M^{(-,-,-,-,-)} Q_{2,3}Q_{3,4}\\
	M^{(0,+,+,+,0)} \widetilde{Q}_{2,3}\widetilde{Q}_{3,4},\,\, M^{(0-,-,-,0)} Q_{2,3}Q_{3,4}\\
	M^{(0,+,+,+,+)} \widetilde{Q}_{2,3}\widetilde{Q}_{3,4},\,\, M^{(0-,-,-,-)} Q_{2,3}Q_{3,4} }\right\}_{C_1} .\nonumber\\[3pt]
\end{align}
\begin{align}
\left\{ \mathrm{Tr} \, A, \,\, M^{(\pm)} \right\}_{A,D} \leftrightarrow \left\{ \mathrm{Tr} \, P\tilde{P} \right\}_B \leftrightarrow \left\{ \mathrm{Tr} \, A_3,\,\, M^{(0,0,\pm,0,0)} \right\}_{C_1} .
\end{align}
This operator map provides further evidence that the map of symmetries that results from the algorithm is the correct one.

In a similar way one can construct operators in the theories $C_2$ and $C_3$ and verify both that they rearrange into representations of the enhanced IR symmetry and that they are mapped correctly across the duality.

\acknowledgments
CH is partially supported by the National Natural Science Foundation of China under Grant No.~12247103, the Institute for Basic Science (IBS-R018-D1, IBS-R018-Y2)
and the STFC consolidated grant ST/T000694/1. MS is partially supported by the ERC Consolidator Grant \#864828 “Algebraic Foundations of Supersymmetric Quantum Field Theory (SCFTAlg)” and by the Simons Collaboration for the Nonperturbative Bootstrap under grant \#494786 from the Simons Foundation.

\appendix

\section{The $FE[USp(2N)]$ theory, the $\mathsf{S}$-wall}
\label{sec:FE}

\begin{figure}[!ht]
	\centering
  	\includegraphics[scale=0.55,center]{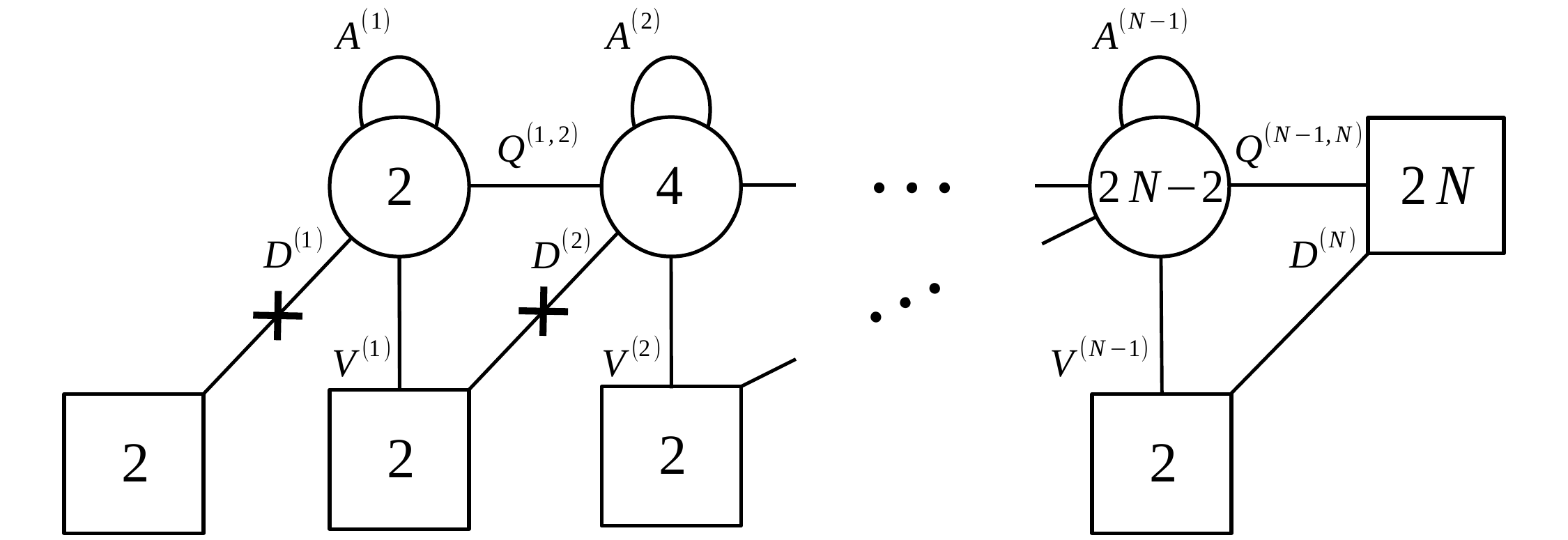} 
   \caption{The $E[USp(2N)]$ theory. Each $2n$ node, square or round, represents an $USp(2n)$ group. The crosses denote the $\gb_n$ singlets.}
  	\label{euspfields}
\end{figure}

We collect here a few definitions following the notation of \cite{Bottini:2021vms}. The $E[USp(2N)]$ theory is described by the quiver in Figure \ref{euspfields}. 
In the  superpotential the  bifundamentals $Q^{(n,n+1)}$ couple to the antisymmetrics $A^{(n)}$, we then have
a cubic coupling between the chirals in each triangle of the quiver and  the flipping terms with the singlets $\beta_n$ coupled to the diagonal mesons
\begin{align}
\mathcal{W}_{E[USp(2N)]}&=\sum_{n=1}^{N-1}\Tr_{n}\left[A^{(n)}\left(\Tr_{n+1}Q^{(n,n+1)}Q^{(n,n+1)}-\Tr_{n-1}Q^{(n-1,n)}Q^{(n-1,n)}\right)\right]\nn\\
&+\sum_{n=1}^{N-1}\Tr_{y_{n+1}}\Tr_{n}\Tr_{n+1}\left(V^{(n)}Q^{(n,n+1)}D^{(n+1)}\right)+\nn\\
&+\sum_{n=1}^{N-1} \gb_n\Tr_{y_n}\Tr_{n}\left(D^{(n)}D^{(n)}\right)\, .
\label{superpoteusp}
\end{align}
Above $\Tr_n$ is over the color indices of the $n$-th $USp(2n)$ gauge node, while $\Tr_{y_n}$ denotes the trace over the the $n$-th $SU(2)$ flavor symmetry and $\Tr_N=\Tr_x$ is the trace over $USp(2N)_x$ flavor indices. All the traces include the $J$ antisymmetric tensor of $USp(2n)$.
\begin{figure}[t]
	\centering
  	\includegraphics[scale=0.55,center]{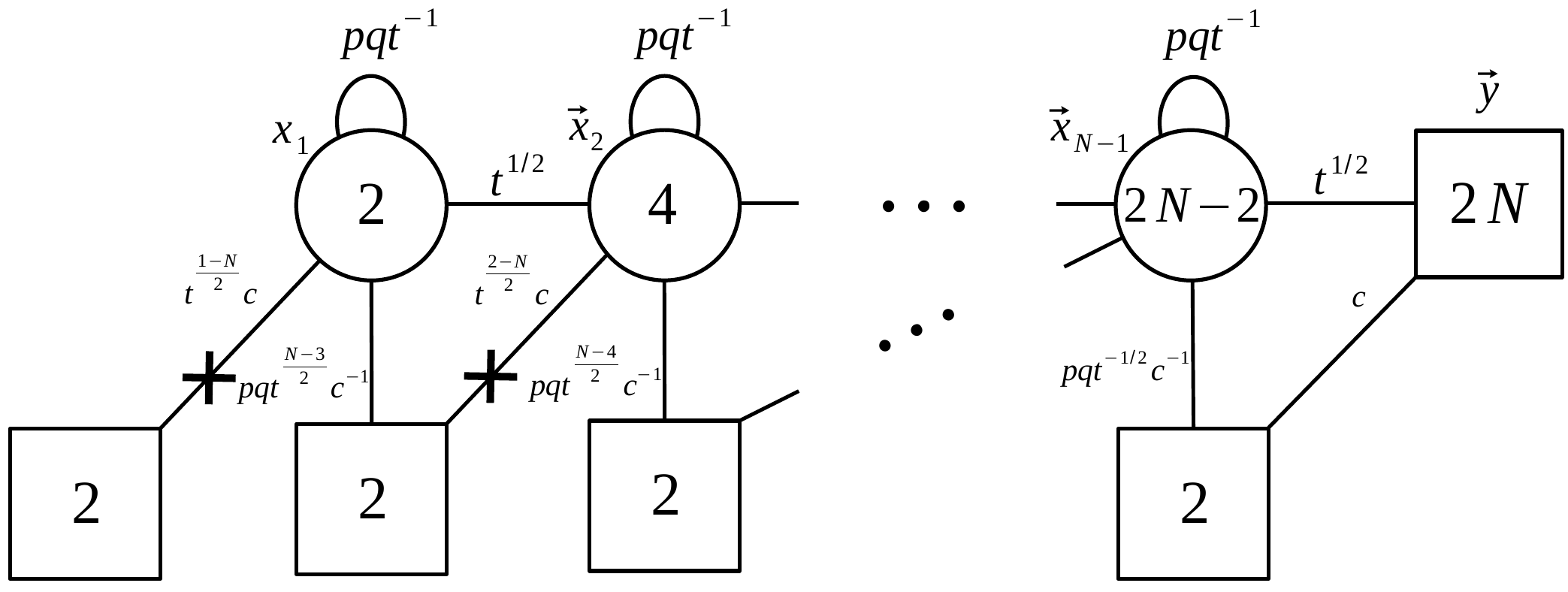} 
   \caption{The trial R-charges and the charges under the other abelian symmetries represented in the form of $(pq)^{R/2} c^{Q_c} t^{Q_t}$, where $R$ is the trial R-charge, $Q_c$ is the charge under $U(1)_c$, and $Q_t$ is the charge under $U(1)_t$.}
  	\label{euspfugacities}
\end{figure}
The  global symmetry in the IR is enhanced as
\be
USp(2N)_x\times\prod_{n=1}^NSU(2)_{y_n}\times U(1)_t\times U(1)_c
 \quad \to \quad
USp(2N)_x\times USp(2N)_y\times U(1)_t\times U(1)_c\, .\nn
\ee
In Figure \ref{euspfugacities} we indicate the charges of all the fields.
Some of the gauge invariant operators of $E[USp(2N)]$ and their transformations properties are given in Table \ref{eusptable}, where we  defined $B_{nm}$ as a  matrix of  operators  charged only under $U(1)_c$ and $U(1)_t$, which include the singlets $\gb_n$.
\begin{table}[t]
\centering
\scalebox{1}{
\begin{tabular}{|c|cccc|c|}\hline
{} & $USp(2N)_x$ & $USp(2N)_y$ & $U(1)_t$ & $U(1)_c$ & $U(1)_{R_0}$ \\ \hline
$\mathsf{H}$ & ${\bf N(2N-1)-1}$ & $\bf 1$ & $1$ & 0 & 0 \\
$\mathsf{C}$ & $\bf1$ & ${\bf N(2N-1)-1}$ & $-1$ & 0 & 2 \\
$\Pi$ & $\bf N$ & $\bf N$ & 0 & $+1$ & 0 \\
$B_{nm}$ & $\bf1$ & $\bf1$ & $m-n$ & $-2$ & $2n$ \\ \hline
\end{tabular}}
\caption{$E[USp(2N)]$ operators and their transformations properties.}
\label{eusptable}
\end{table}

The $FE[USp(2N)]$ theory, which we identified in the main text as the $4d$ $\mathsf{S}$-wall, is defined as $E[USp(2N)]$ with one extra set of singlets $\mathsf{O}_\mathsf{H}$, as well as a singlet $\beta_N$, interacting via the superpotential
\be
\mathcal{W}_{FE[USp(2N)]}=\mathcal{W}_{E[USp(2N)]}+\Tr_x\left(\mathsf{O}_\mathsf{H}  \mathsf{H}\right) + \beta_N \Tr_x \Tr_{y_N} D^{(N)} D^{(N)} \,.
\label{aFE}
\ee

\clearpage
\section{Asymmetric $\mathsf{S}$-walls}
\label{asyswall}
An asymmetric $\mathsf{S}$-wall is defined as an  $\mathsf{S}$-wall  deformed by a superpotential term breaking one of the flavour $USp(2N)$ symmetries down to $USp(2M)\times SU(2)_v$ (with $M<N$). Details about this deformation and its effect can be found in \cite{Bottini:2021vms}.

\begin{figure}[!ht]
\center
\includegraphics[scale=0.6,center]{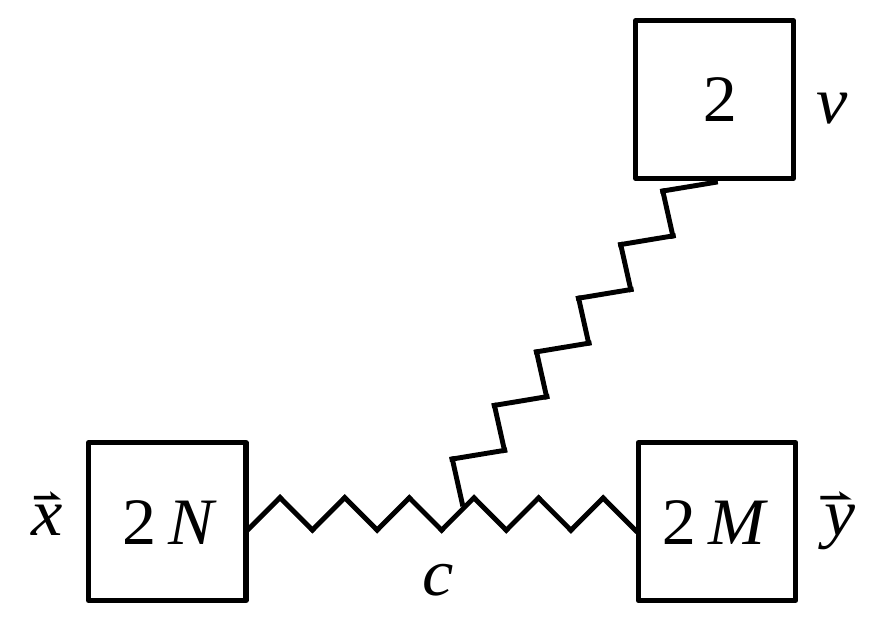}
\caption{Schematic representation of the asymmetric $\mathsf{S}$-wall.
}
\label{fig:Asymm_S_appendix}
\end{figure}

The deformation translates into the following specializations of $(N-M)$ components of the $USp(2 N)_z$ fugacities $\vec{z}$:
\begin{equation}\label{eq:specialization}
	\vec{z} = \{ z_1, \cdots, z_M, t^{\frac{N-M-1}{2}}v, \cdots, t^{-\frac{N-M-1}{2}}v \} \,.
\end{equation}
The $\vec{z}=\{z_1, \cdots, z_M \}$ fugaicities parametrize $USp(2M)_z$, and $v$ is the Cartan of $SU(2)_v$. The result of this procedure is what we depict schematically as the asymmetric $\mathsf{S}$-wall shown in Figure \ref{fig:Asymm_S_appendix}.

\begin{figure}[!ht]
\includegraphics[scale=0.45,center]{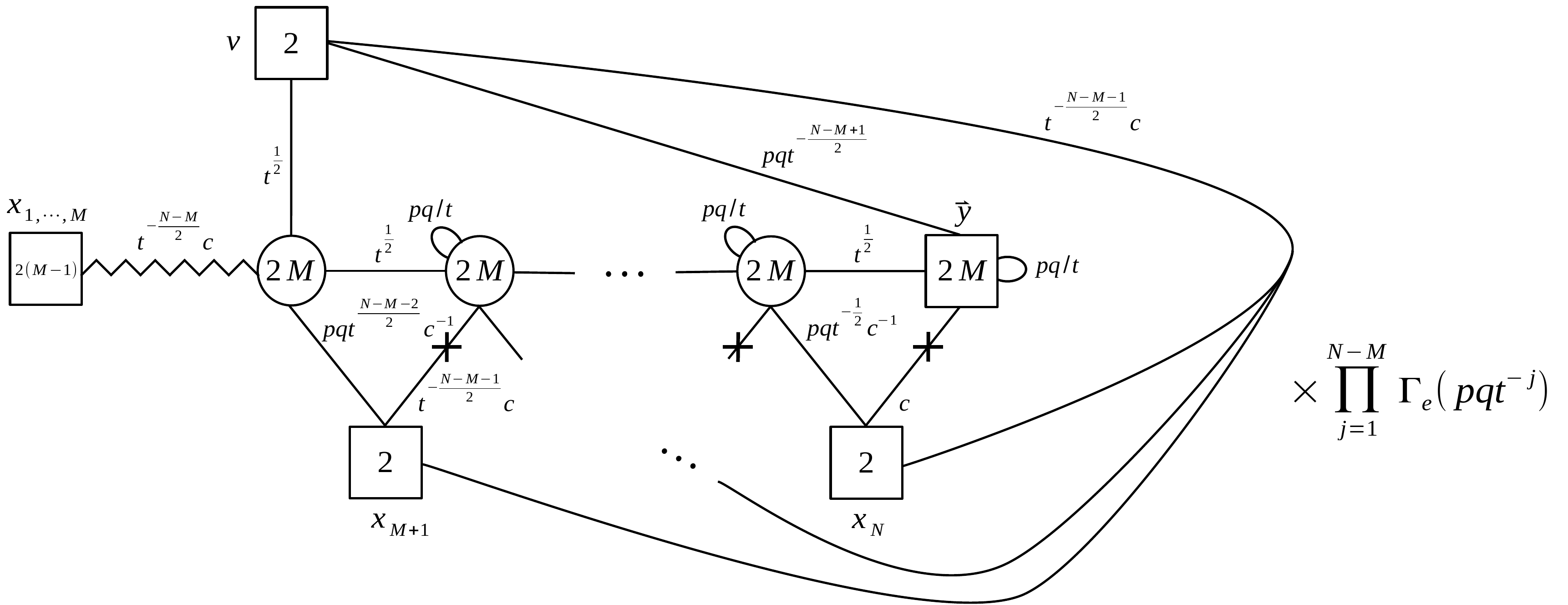}
\caption{An $\mathsf{S}$-wall with broken manifest $USp(2N)_x$.}
\label{fig:Manifest_result}
\end{figure}

\begin{figure}[!ht]
\includegraphics[scale=0.55,center]{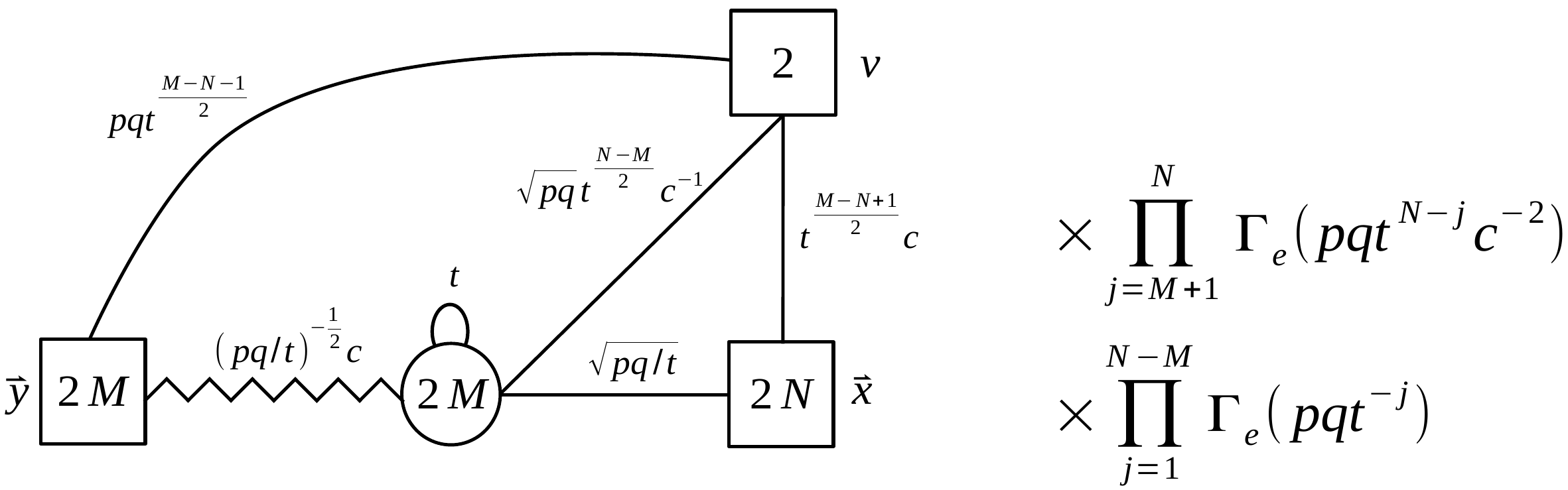}
\caption{An $\mathsf{S}$-wall with broken emergent $USp(2N)_y$.}
\label{fig:Emergent_result_FEclosed}
\end{figure}

It also useful to give a Lagrangian description of the result. In order to do that we have to choose whether the broken $USp(2N)$ is the manifest or emergent one. This choice gives rise to two different descriptions that are related by mirror symmetry. As it is shown below, in the case of breaking the manifest $USp(2N)_x$ symmetry the result is given in Figure \ref{fig:Manifest_result}. If instead we break the emergent $USp(2N)_y$ symmetry the result is given in Figure \ref{fig:Emergent_result_FEclosed}.

In fact, up to singlets, the theory in Figure \ref{fig:Manifest_result} coincides with the $FE_\rho^\sigma[USp(2N)]$ theory of \cite{Hwang:2020wpd} with $\rho=[1^N]$ and $\sigma=[N-M,1^M]$. This theory is related to the one in Figure \ref{fig:Emergent_result_FEclosed} by mirror symmetry which swaps $\rho$ and $\sigma$.

\begin{figure}[!ht]
\includegraphics[scale=0.35,center]{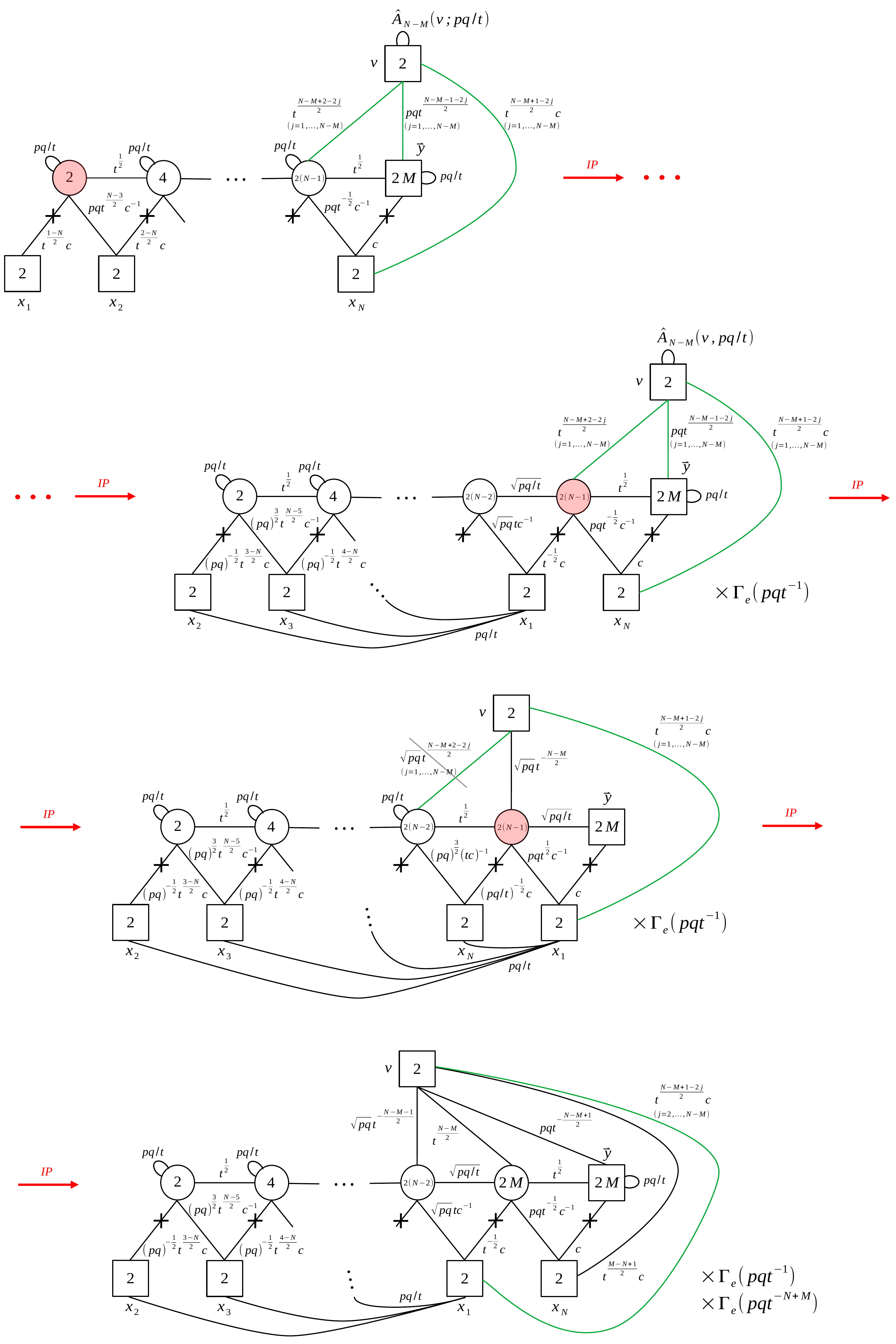}
\caption{The first part of the derivation of the Lagrangian form of an asymmetric $\mathsf{S}$-wall with the specialized manifest symmetry. The grey dash in the third quiver indicates a set of massive fields.}
\label{fig:Manifest_proof_1}
\end{figure}

\paragraph{Specialization of the manifest symmetry}\mbox{}\\
We first consider the case where we break the manifest $USp(2N)_x$ symmetry with the specialization \eqref{eq:specialization} to obtain the first quiver depicted in Figure \ref{fig:Manifest_proof_1}. The factor $\hat{A}_{N-M}(v;pq/t)$, defined in eq.~\eqref{eq:spec_antisymm} contains $(N-M-1)$ vanishing $\Gamma_e(pq)$. This fact, together with the pattern of the $t$-charges of the green chirals attached to the $USp(2N-2)$ gauge node, indicates
that some of the green mesons acquire a VEV which Higgses the  $USp(2N-2)$ node down to $USp(2M)$. The first Higgsing then triggers the Higgsings of the nodes on its left and the VEV continues to propagate towards the left of the quiver. 

Instead of studying this sequential Higgsing, we proceed using sequential IP dualizations. We begin by IP dualizing the first $USp(2)$ gauge node and continue applying the IP duality moving
towards the right. These dualizations do not change the ranks of the gauge nodes but at each step the antisymmetric of the next node becomes massive such that we can then apply IP there, and so on.  The result after the dualization of the $(N-2)$-th gauge node is depicted as the second quiver in Figure \ref{fig:Manifest_proof_1}.

Now we IP dualzse the $(N-1)$-th node. Here the duality has the effect of trading VEVs for mass deformations. More precisely, the duality leaves the rank unchanged and has two main effects on the fields:
\begin{itemize}
	\item After the dualization, the $USp(2N-2)\times SU(2)_v$ chirals read
	\begin{equation*}
		\prod_{j=1}^{N-M} \prod_{i=1}^{N-1} \Gamma_e \left(\sqrt{pq} t^{j-\frac{N-M+2}{2}} v^\pm z_i^{(N-1)\pm}\right) \,,
	\end{equation*}
	where $\vec{z}^{\,\,(N-1)}$ are the fugacities for the gauge group $USp(2N-2)$. From the above expression 
	we see that all of them become massive and only the  $j=1$ chiral survives. 
	The flipping fields produced by the IP duality cancel exactly with the $\hat{A}_{N-M}(v;pq/t)$ term.
	
	\item A set of $(N-M)$ legs connecting the $USp(2N-4)$ gauge node to $SU(2)_v$ is produced
	\begin{equation*}
		\prod_{j=1}^{N-M} \prod_{i=1}^{N-2} \Gamma_e \left( \sqrt{pq} t^{\frac{N-M+1}{2} - j} z_i^{(N-2)\pm} v^\pm\right)=1 \,,
	\end{equation*}
	where $\vec{z}^{\,\,(N-2)}$ are the fugacities for the gauge group $USp(2N-4)$. All of them are massive.
\end{itemize}

After integrating out the massive fields and keeping track of all the singlets, we get the third quiver in Figure \ref{fig:Manifest_proof_1}. Then we apply again the IP duality on the same node. Notice that, since we integrated out massive fields before the second dualization, we don't go back to the original theory and obtain the last theory in Figure \ref{fig:Manifest_proof_1}.

\begin{figure}[!ht]
\includegraphics[scale=0.35,center]{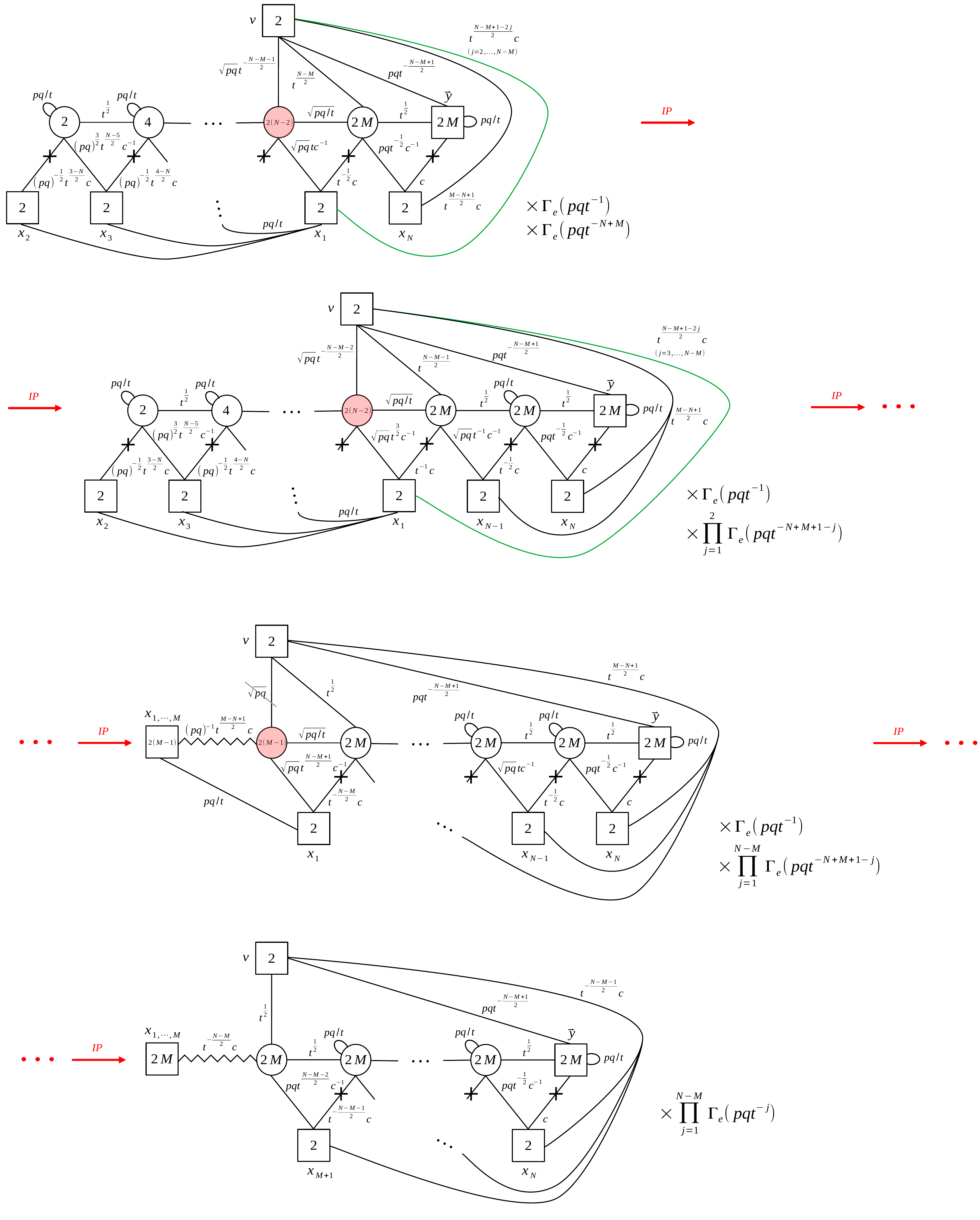}
\caption{The second part of the derivation of the Lagrangian form of an asymmetric $\mathsf{S}$-wall with the specialized manifest symmetry. The grey dash in the third quiver indicates a massive field. }
\label{fig:Manifest_proof_2}
\end{figure}

As highlighted in the first quiver in Figure \ref{fig:Manifest_proof_2}, now the $USp(2N-4)$ node has no antisymmetric so we can apply the IP duality and reach the second quiver in Figure \ref{fig:Manifest_proof_2}. The $USp(2N-4)$ node becomes an $USp(2M)$ node and the $SU(2)_v$ flavor {\it moves} to its left. We can keep applying IP dualities towards the left of the quiver with the effect of  creating a sequence of $USp(2M)$ nodes and moving the $SU(2)_v$ flavor to the left. After dualizing the $M$-th node we reach the third quiver in Figure \ref{fig:Manifest_proof_2}. Notice that the $SU(2)_v$ leg attached to the red $(M-1)$-th gauge node has charge $\sqrt{pq}$ and hence it becomes massive. Now if we keep applying IP dualities along the tail the $SU(2)_v$ flavor does not move anymore while the ranks of the gauge groups remain unchanged. 
After the IP dualization of the last $USp(2)$ node, we reach the last quiver in Figure \ref{fig:Manifest_proof_2}, which coincides with the result anticipated in Figure \ref{fig:Manifest_result}.

%

\clearpage

\begin{figure}[!ht]
\includegraphics[scale=0.35,center]{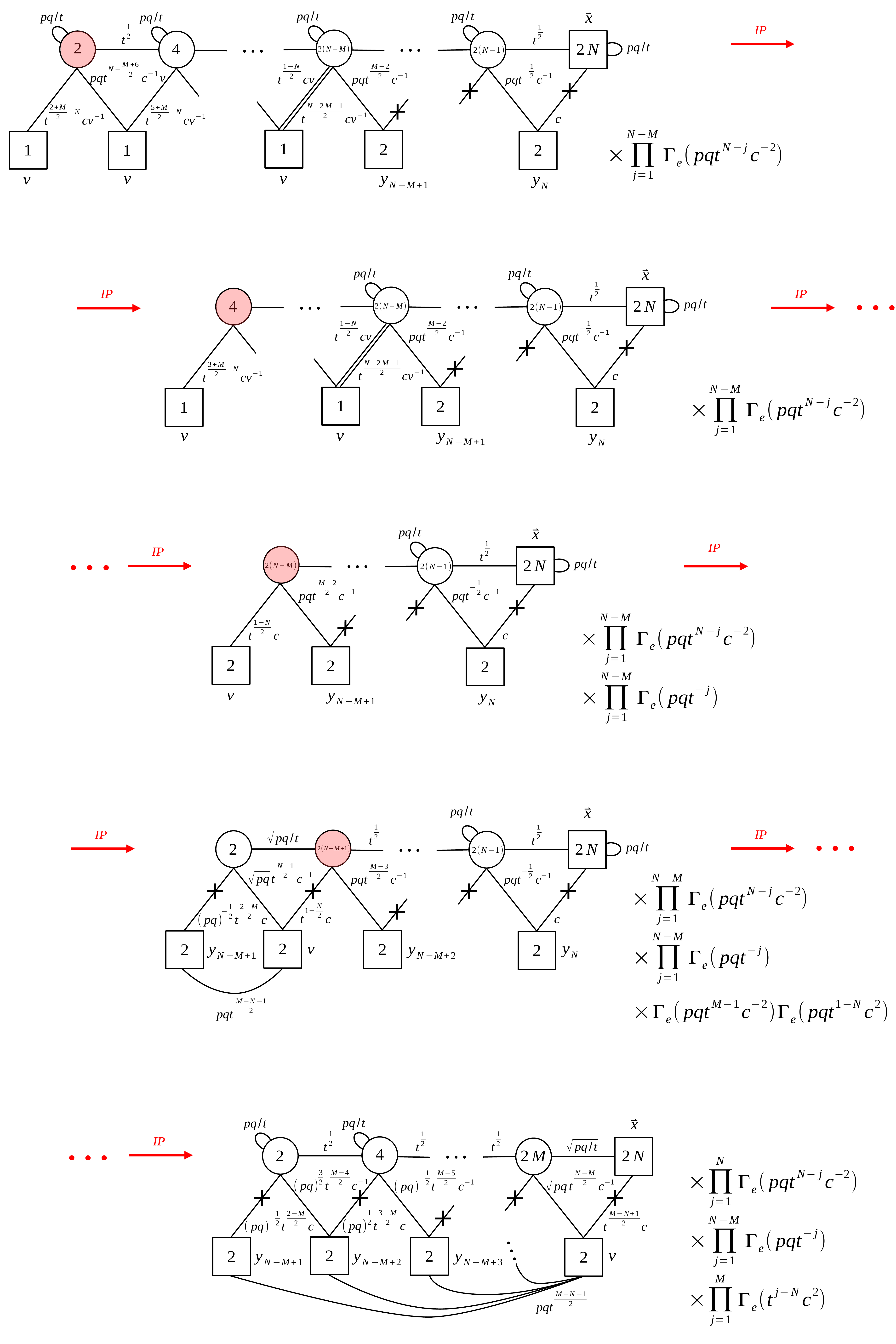}
\caption{Derivation of the Lagrangian form of an asymmetric $\mathsf{S}$-wall with the specialized emergent symmetry.}
\label{fig:Emergent_proof}
\end{figure}

\paragraph{Specialization of the emergent symmetry}\mbox{}\\
Now we consider the case in which the broken symmetry is the emergent one. We choose to specialize in geometric progression the first $(N-M)$ fugacities of the $SU(2)_{y_j}$ symmetries from the left to get the first quiver in Figure \ref{fig:Emergent_proof}. 
The specialization has the effect of breaking the $SU(2)_{y_j}$, for $j=1,...,M-N$, down to $U(1)_v$. Indeed focusing on the contribution  of chiral fields of the saw charged under $k$-th gauge node for $k=1,...,N-M-1$, we see that the specialization yields
\begin{align*}
	& \prod_{i=1}^{k} \Gamma_e\left( t^{\frac{k-N}{2}} c (t^{\frac{N-M+1}{2} - k} v)^\pm z_i^{(k)\pm}\right) \Gamma_e\left( pq t^{\frac{N-k-2}{2}} c^{-1} (t^{\frac{N-M+1}{2} - (k+1)} v)^\pm z_i^{(k)\pm}\right) = \\
	& \quad = \prod_{i=1}^{k} \Gamma_e\left( t^{\frac{3k-2N+M-1}{2}} c v^{-1} z_i^{(k)\pm}\right) \Gamma_e\left(pq t^{\frac{2N-3k-M-3}{2}} c^{-1} v z_i^{(k)\pm}\right) \,.
\end{align*}
For $k=N-M$, on the other hand, we get a slight different result, since no chiral of the saw connected to this gauge node becomes massive.

We then apply the IP duality on the first $USp(2)$ which confines, since it sees 6 chirals, to obtain the second quiver in Figure \ref{fig:Emergent_proof}. Now the $USp(4)$ has no antisymmetric and sees 8 chirals, so it also confines. We continue sequentially confining the nodes until we reach the third quiver in Figure \ref{fig:Emergent_proof}. At this point the legs can be rearranged in such a way that the $SU(2)_v$ symmetry is restored. If we now apply the IP duality on the $USp(2N-2M)$ node, the effect is that it becomes $USp(2)$ instead of confining. After this dualization we get the fourth quiver in Figure \ref{fig:Emergent_proof}. Notice that the legs in the saw have been rearranged by the IP duality such that now $SU(2)_v$ is exchanged with $SU(2)_{y_{N-M-1}}$ and also a bifudnamenal between the two is produced. If we keep applying the IP duality along the tail, we get increasing ranks starting from $USp(2)$. After we have dualized the $(N-1)$-th node we find the last theory in Figure \ref{fig:Emergent_proof}, which corresponds to the anticipated form of an $\mathsf{S}$-wall with the specialized emergent symmetry of Figure \ref{fig:Emergent_result_FEclosed}.

\begin{figure}[!ht]
\includegraphics[scale=0.55,center]{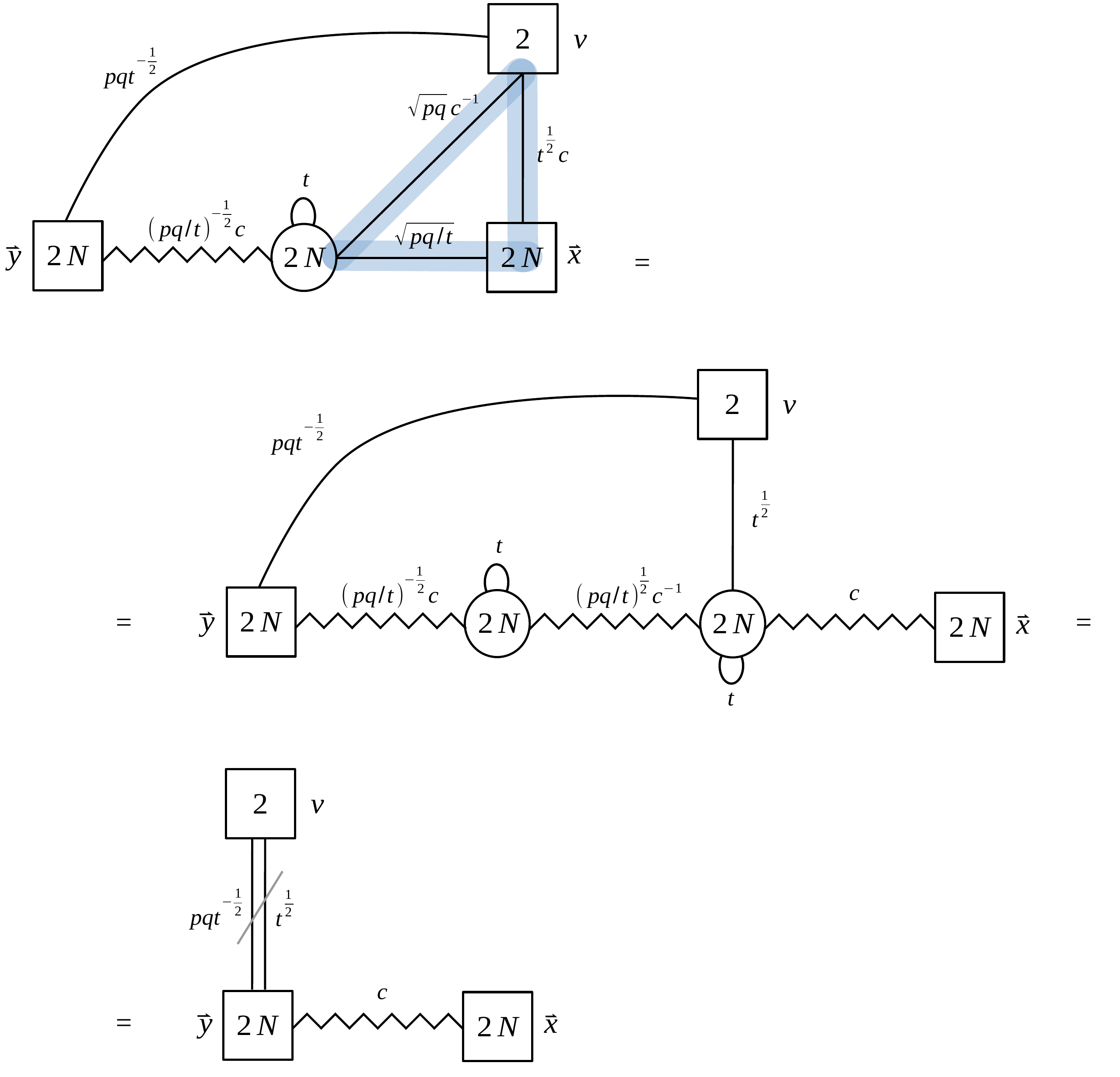}
\caption{Starting from the asymmetric $\mathsf{S}$-wall in Figure \ref{fig:Emergent_result_FEclosed} for $N=M$, we apply the $\mathsf{B}_{10}=\mathsf{S}\mathsf{B}_{01}\mathsf{S}^{-1}$ move to the highlighted part of the quiver. After implementing the identifications imposed by the Identity-wall, two flavors become massive, and we obtain the symmetric $\mathsf{S}$-wall. }
\label{fig:Emergent_limit_N=M}
\end{figure}

As a final comment we check that we recover the  symmetric $\mathsf{S}$-wall in the case $M=N$, in which 
there is no breaking. In the case of the breaking of the manifest $USp(2N)_y$ symmetry, as it can be seen from Figure \ref{fig:Manifest_result}, this is trivial. On the other hand starting from Figure \ref{fig:Emergent_result_FEclosed}
we need to perform a couple of steps to recover the symmetric $\mathsf{S}$-wall as shown in Figure \ref{fig:Emergent_limit_N=M}.

\clearpage

\section{Derivations of basic duality moves}
\label{proof_dualities}

In this appendix we show how to derive some of the new duality moves presented in Section \ref{subsec:4d_duality_moves}.

\subsection*{\boldmath $\mathsf{B}_{11}=\mathsf{S}\mathsf{B}_{1-1}\mathsf{S}^{-1}$ }

\begin{figure}[h!]
\includegraphics[scale=0.4,center]{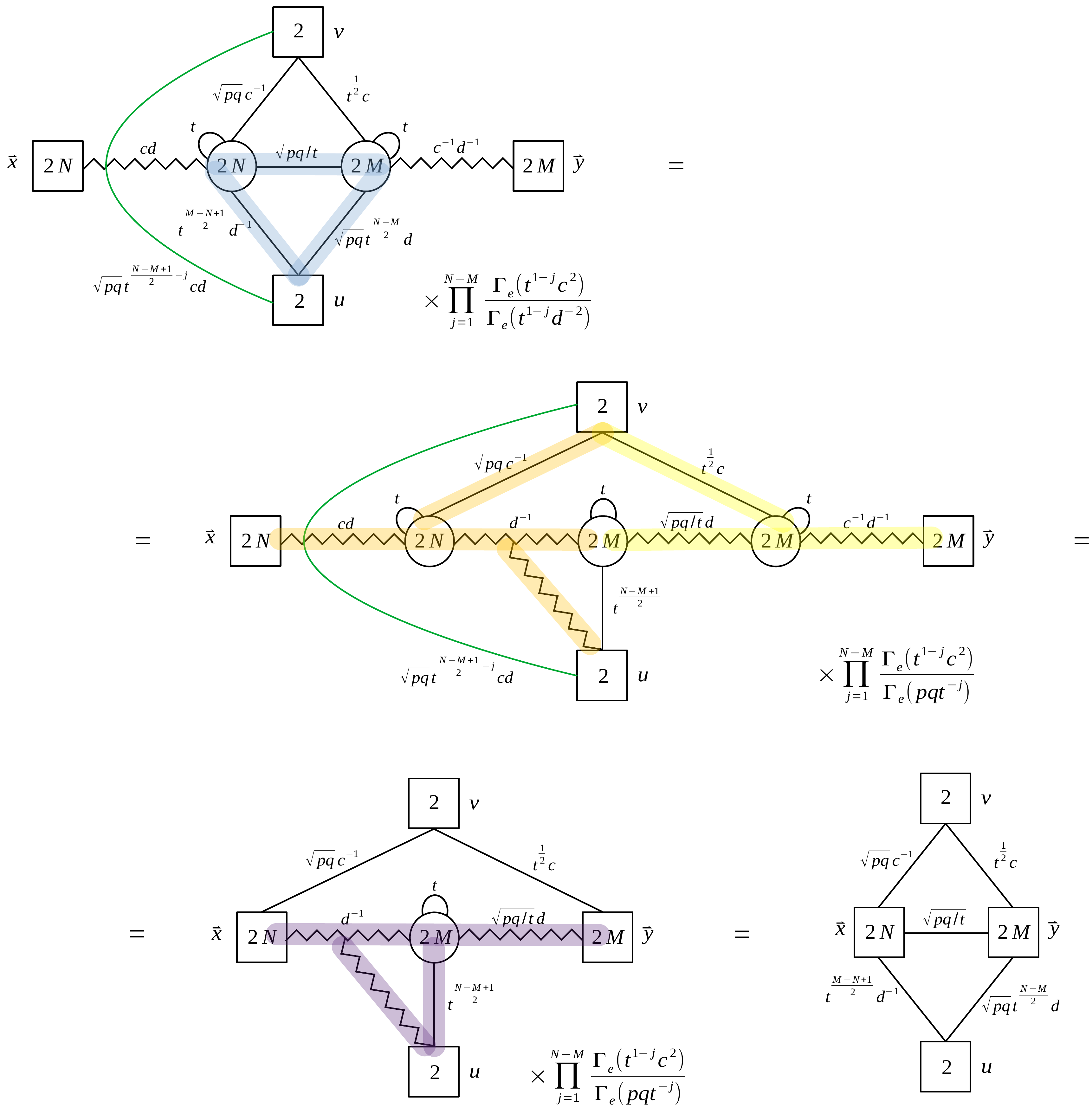}
\caption{Derivation of the duality $\mathsf{B}_{11}=\mathsf{S}\mathsf{B}_{1-1}\mathsf{S}^{-1}$. }
\label{fig:11_Sdual_proof}
\end{figure}

To derive this duality we proceed as in  Figure \ref{fig:11_Sdual_proof}.
In the first step we apply the duality $\mathsf{B}_{10}=\mathsf{S}\mathsf{B}_{01}\mathsf{S}^{-1}$ of Figure \ref{10S01} to the blue part of the quiver. In the second step we apply the braid move of Figure \ref{braid} to the yellow and orange parts. In the third step we apply again $\mathsf{S}\mathsf{B}_{01}\mathsf{S}^{-1}=\mathsf{B}_{10}$ to the purple part.

\subsection*{\boldmath $\mathsf{B}_{10}=\mathsf{T}^T\mathsf{B}_{10}(\mathsf{T}^T)^{-1}$}
\begin{figure}[!ht]
\includegraphics[scale=0.5,center]{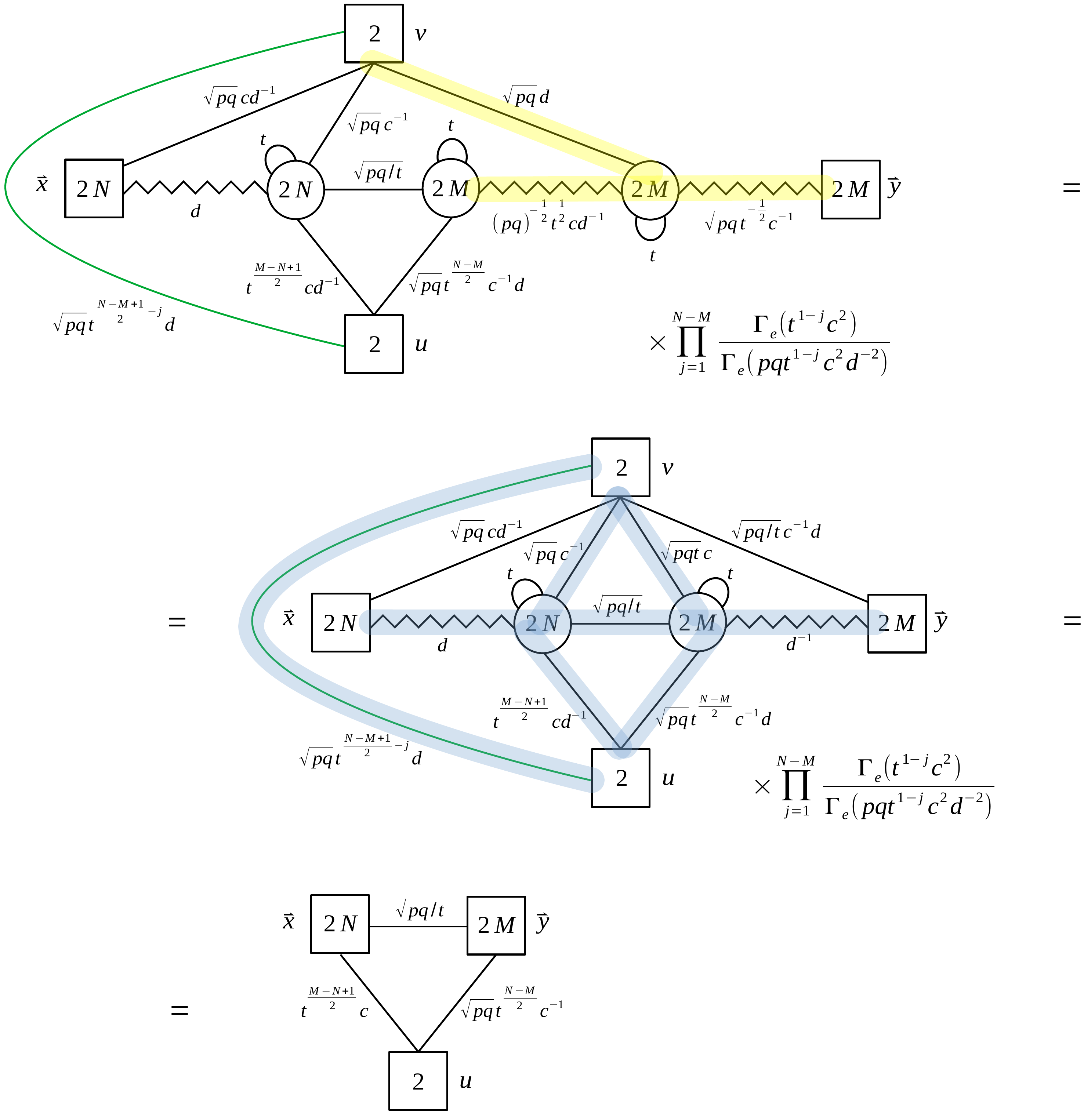}
\caption{Derivation of the duality $\mathsf{B}_{10}=\mathsf{T}^T\mathsf{B}_{10}(\mathsf{T}^T)^{-1}$.  }
\label{fig:10_Ttrdual_proof}
\end{figure}

To derive this duality we proceed as in Figure \ref{fig:10_Ttrdual_proof}.
In the first step we apply the braid move of Figure \ref{braid} to the part of the quiver highlighted in yellow. In the second step we apply the duality move $\mathsf{B}_{11}=\mathsf{S}\mathsf{B}_{1-1}\mathsf{S}^{-1}$ of Figure \ref{11s1-1} to the blue part.

\bigskip

\subsection*{\boldmath $\mathsf{B}_{01}=\mathsf{T}^T\mathsf{B}_{1-1}(\mathsf{T}^T)^{-1}$}

\begin{figure}[!ht]
\includegraphics[scale=0.55,center]{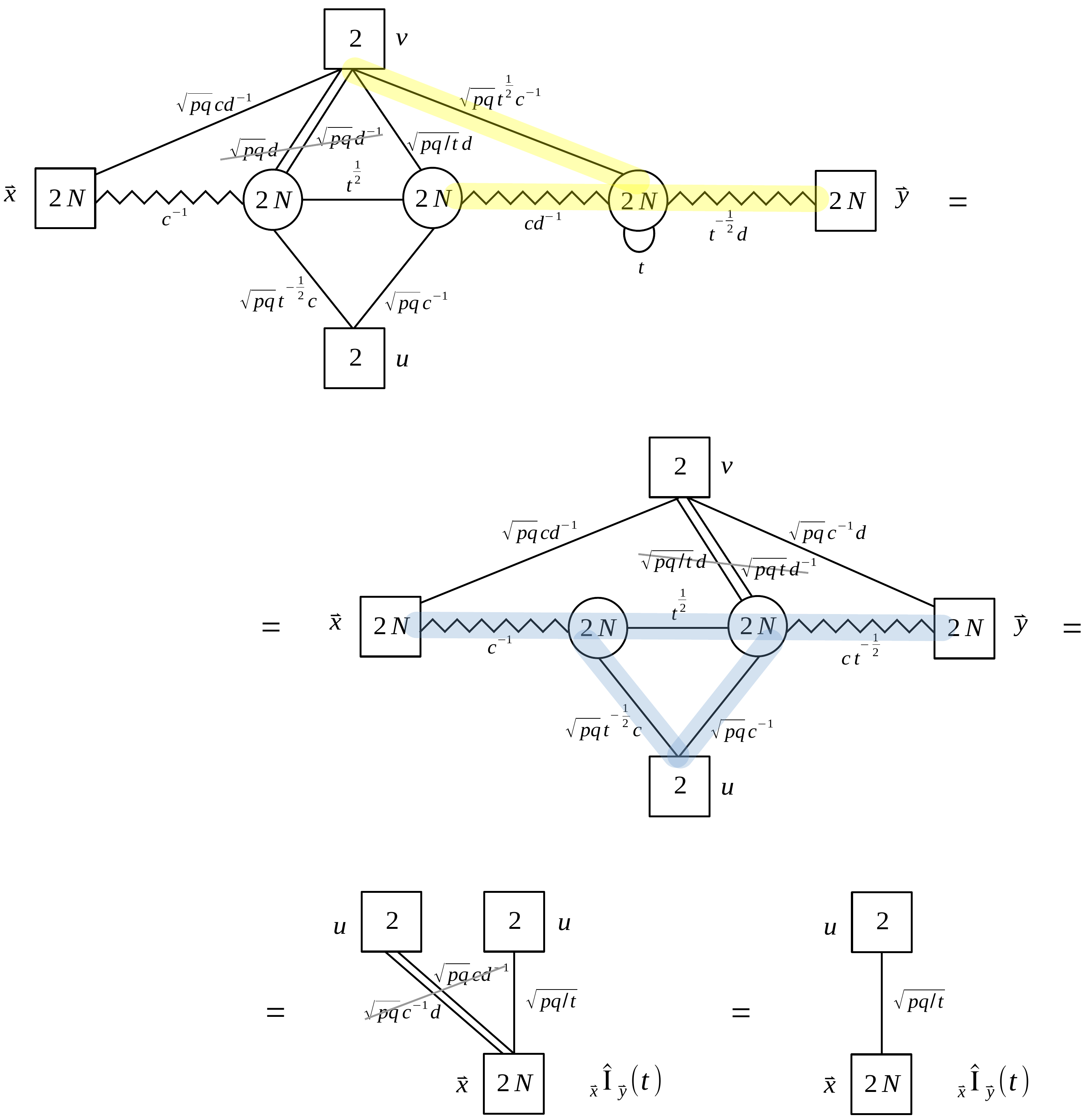}
\caption{Derivation of the duality $\mathsf{B}_{01}=\mathsf{T}^T\mathsf{B}_{1-1}(\mathsf{T}^T)^{-1}$. The pair of lines with a bar on top denote fields that give mass to each other.}
\label{fig:01_Ttrdual_proof}
\end{figure}

To derive this duality we proceed as in Figure \ref{fig:01_Ttrdual_proof}. In the first step we apply the braid move of Figure \ref{braid} to the part of the quiver highlighted in yellow. In the second step we apply the duality move $\mathsf{B}_{01}=\mathsf{S}\mathsf{B}_{10}\mathsf{S}^{-1}$ of Figure \ref{01S10} to the blue part.

\subsection*{\boldmath $\mathsf{B}_{11}=\mathsf{T}^T\mathsf{B}_{01}(\mathsf{T}^T)^{-1}$}
\begin{figure}[!ht]
\includegraphics[scale=0.38,center]{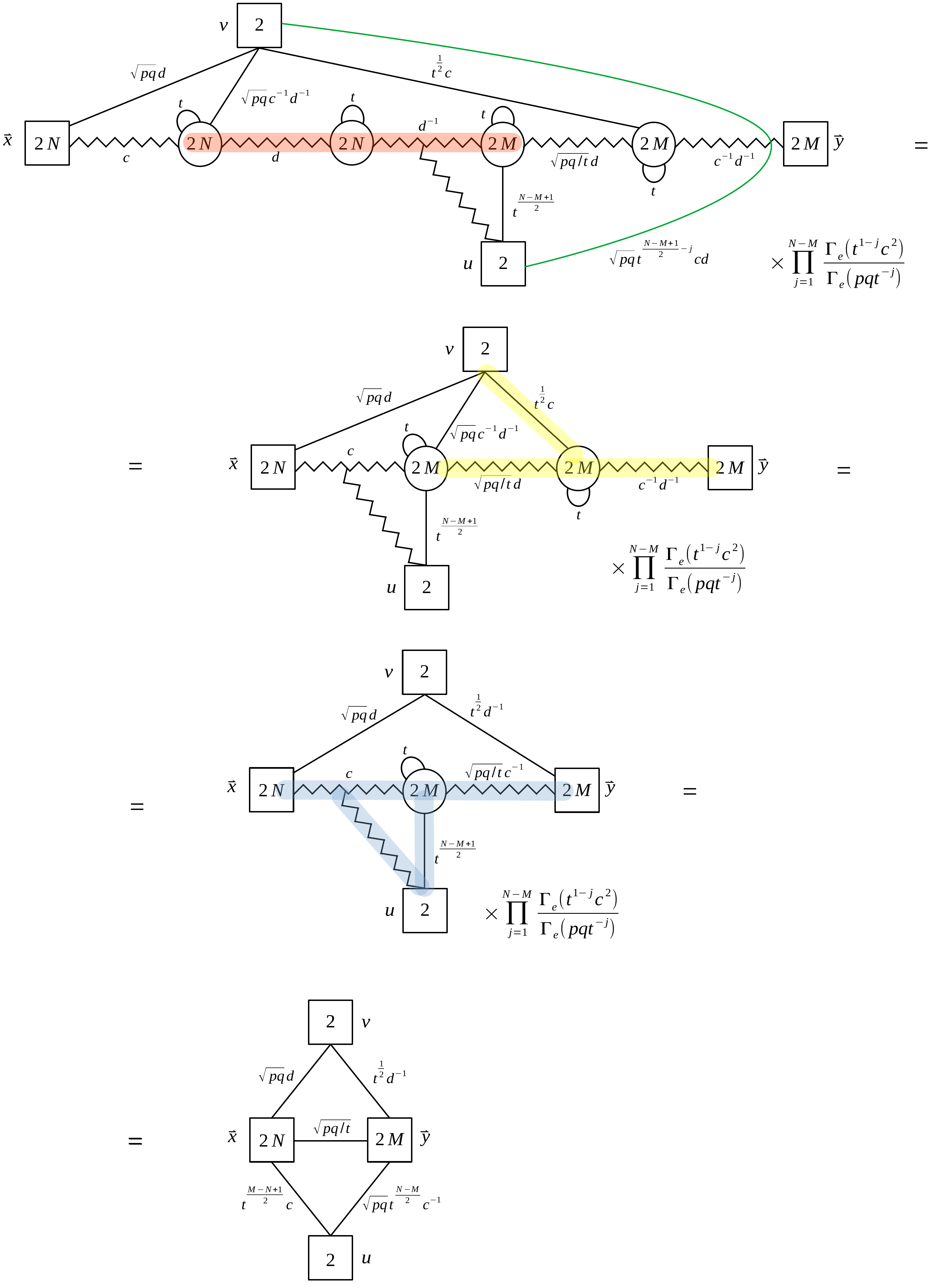}
\caption{Derivation of the duality $\mathsf{B}_{11}=\mathsf{T}^T\mathsf{B}_{01}(\mathsf{T}^T)^{-1}$.}
\label{fig:11_Ttrdual_proof}
\end{figure}

To derive this duality we proceed as in Figure \ref{fig:11_Ttrdual_proof}.
 In the first step we collapse the two $\mathsf{S}$-walls highlighted in red. In the second step we use the braid move of Figure \ref{braid} to the yellow part. In the third step we apply the duality move $\mathsf{B}_{10}=\mathsf{S}\mathsf{B}_{01}\mathsf{S}^{-1}$ of Figure \ref{10S01} to the blue part.

\clearpage
\section{Proof of the Hanany--Witten duality move }
\label{HW_proof_IP}
\begin{figure}[!ht]
	\includegraphics[scale=.35,center]{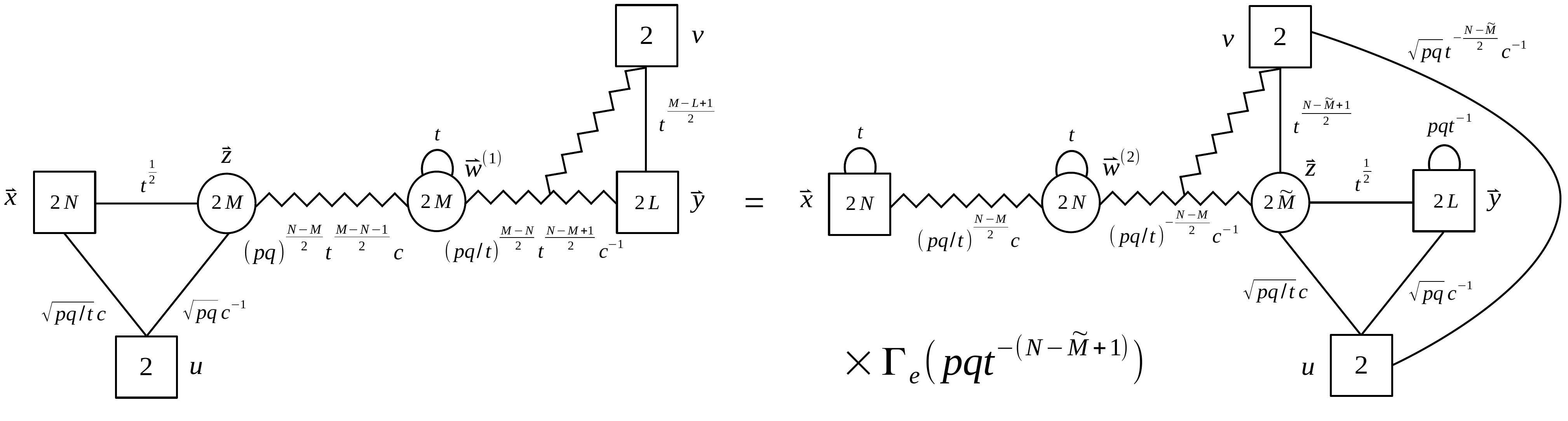}
	\caption{The Hanany--Witten duality move. }
	\label{HWapp}
\end{figure}

In this section we derive the Hanany--Witten (HW) move  shown in Figure \ref{HWapp}.

\begin{figure}[!ht]
\includegraphics[scale=0.45,center]{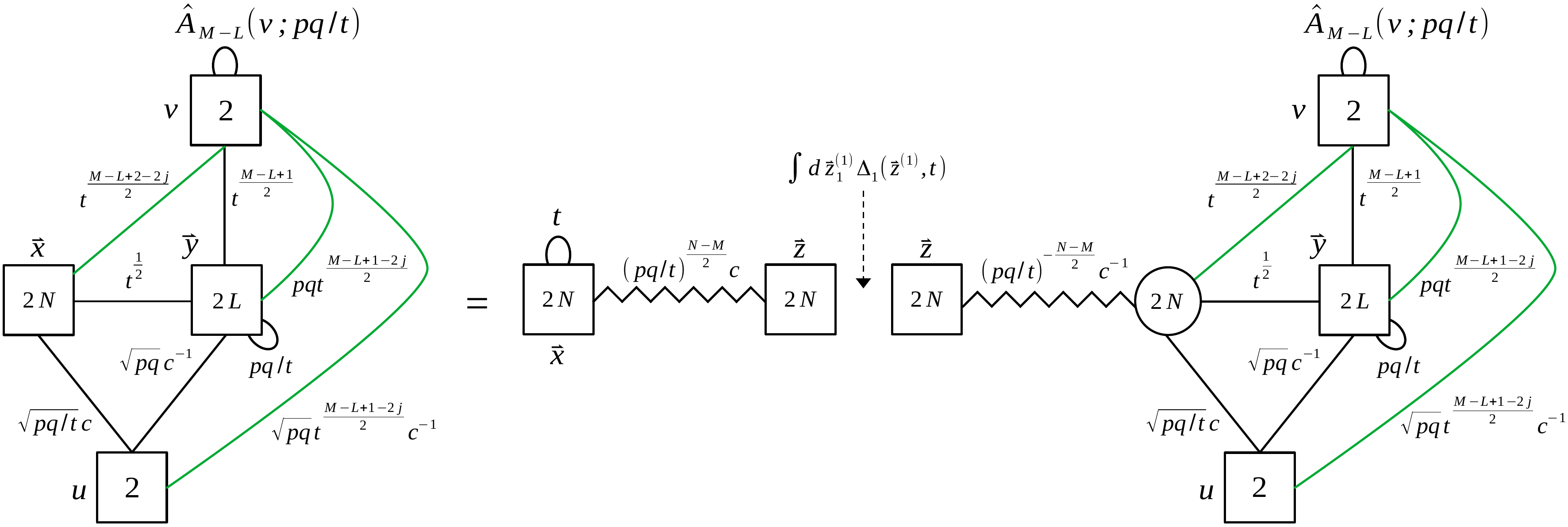}
\caption{On the left is depicted the starting point with the asymmetric Identity-wall implemented. On the right is shown the theory with the addition of an Identity-wall written as two $\mathsf{S}$-walls. The green lines denote gauge singlet fields labelled by $j=1,\dots,M-L$.}
\label{fig:HW_step_0}
\end{figure}

On l.h.s.~of this duality, the two $\mathsf{S}$-walls fuse to form an asymmetric Identity-wall. When we implement the identifications imposed by this asymmetric Identity-wall we obtain the WZ model on the l.h.s.~of Figure \ref{fig:HW_step_0}. On the r.h.s.~of Figure \ref{fig:HW_step_0} we also add an Identity-wall obtained by the fusions of two $\mathsf{S}$-walls.

\begin{figure}[!ht]
\includegraphics[scale=0.4,center]{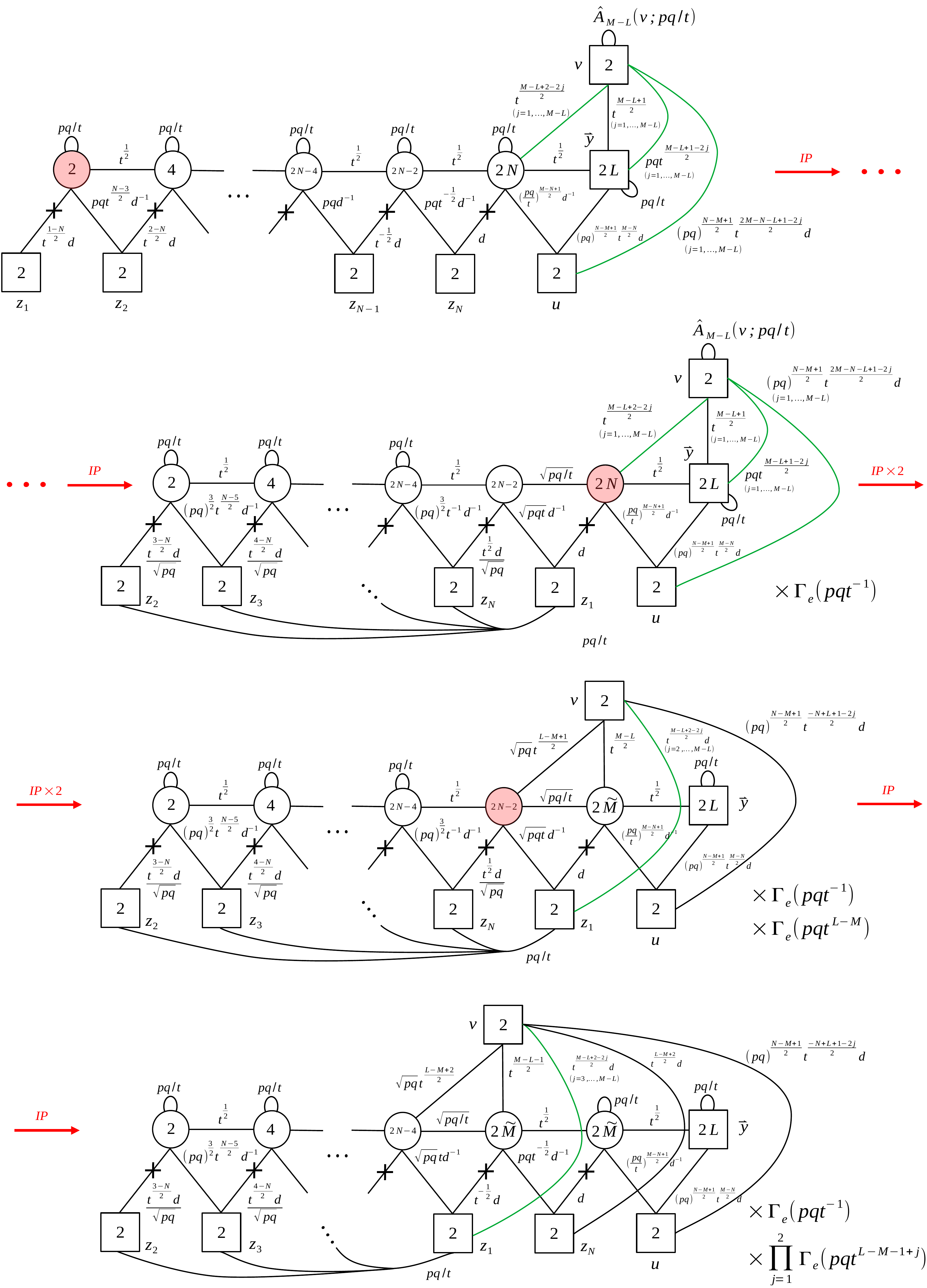}
\caption{The first part of the proof of the Hanany--Witten move via Intriligator-Pouliot dualities.}
\label{fig:HW_proof_1}
\end{figure}

We now  focus on the part of the theory on the r.h.s.~of Figure \ref{fig:HW_step_0} that is composed by  the second $\mathsf{S}$-wall glued to the rightmost part of the quiver as shown  on  the top of   Figure \ref{fig:HW_proof_1}, where the $\mathsf{S}$-wall is now presented with its manifest $USp(2N)$ symmetry gauged. We also define $d=(pq/t)^{\frac{M-N}{2}} c^{-1}$ to simplify the figure. Now we apply the IP duality on the first $USp(2)$ node: the rank does not change but the antisymmetric of the following $USp(4)$ becomes massive; furthermore $SU(2)_{z_1}$ is swapped with $SU(2)_{z_2}$. We then continue  applying the IP duality along the tail until we dualize the $(N-1)$-th node to obtain the second quiver in Figure \ref{fig:HW_proof_1}. We notice that here the mesons built with the green chirals connected to  the $USp(2N)$ node acquire a VEV. As in the previous section, we can trade the study  of the Higgsing induced by this VEV for two  IP dualizations of the same node which take us to the third quiver in Figure \ref{fig:HW_proof_1}. Notice that the resulting rank of the node is $\widetilde{M} = N + L - M + 1$, as expected in the Hanany--Witten move. Now we apply the IP duality to the $(N-1)$-th node highlighted in red to reach the last theory in Figure \ref{fig:HW_proof_1}. Notice that the rank becomes $\widetilde{M}$ and also the $SU(2)_v$ flavor is moved to the left.

\begin{figure}[!ht]
\includegraphics[scale=0.35,center]{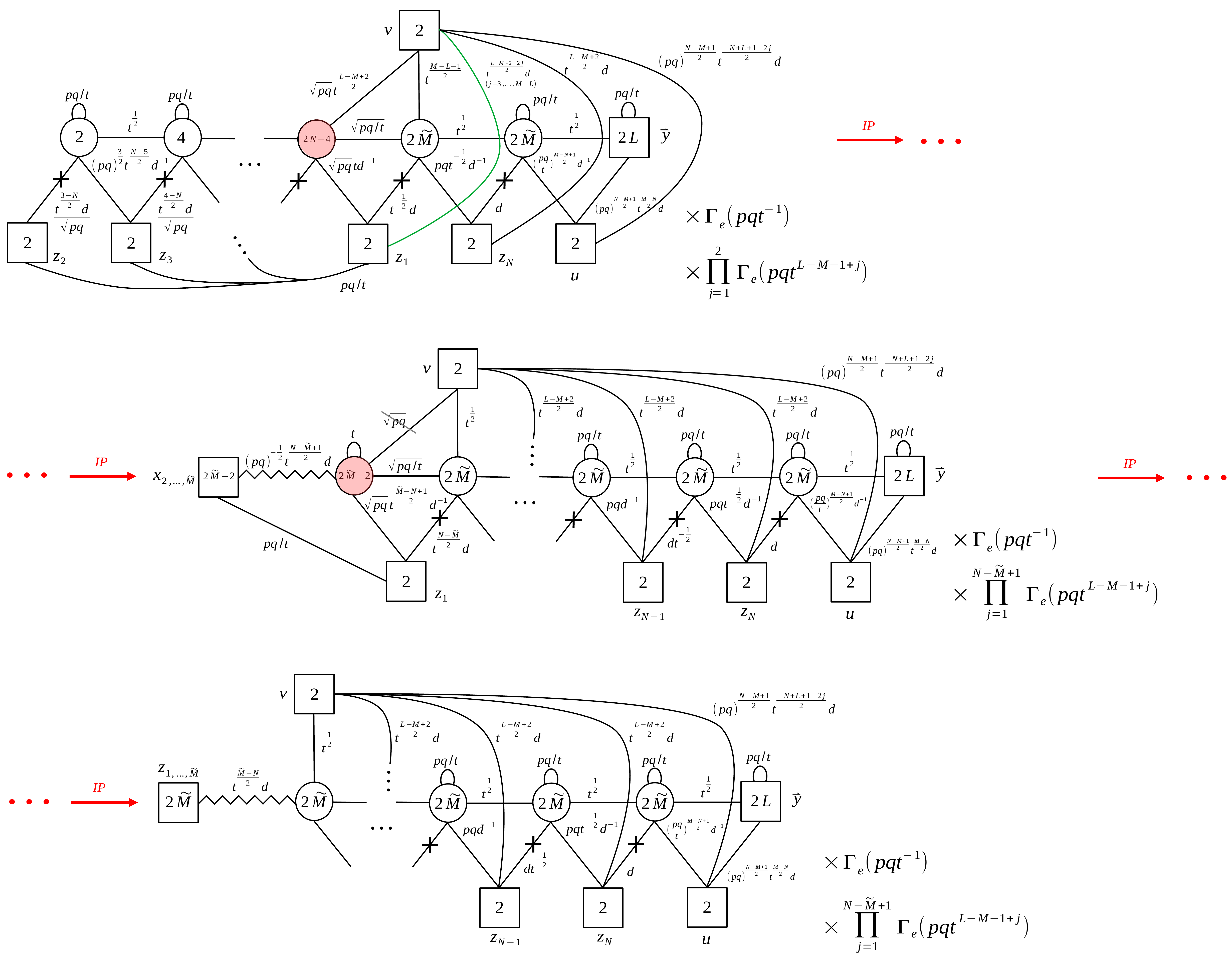}
\caption{The second part of the proof of the Hanany--Witten move via Intriligator-Pouliot dualities.}
\label{fig:HW_proof_2}
\end{figure}

We keep applying IP dualities on the nodes moving towards the left until we reach the $\widetilde{M}$-th, resulting in the second theory in Figure \ref{fig:HW_proof_2}. Notice that the $SU(2)_v$ flavor connected to the $USp(2\widetilde{M}-2)$ gauge node has a charge of $\sqrt{pq}$, so it becomes massive. Now we continue applying the IP duality moving to the left inside the $\mathsf{S}$-wall.  The ranks remain unchanged while  the $SU(2)_v$ flavor is  no longer moving to the left. When we finally apply the IP duality on the $USp(2)$ gauge node  of the $\mathsf{S}$-wall we get the last quiver in Figure \ref{fig:HW_proof_2}. We can  recognize this as an asymmetric $\mathsf{S}$-wall (with extra singlets) with specialized manifest $USp(2N)$ symmetry, which is depicted in Figure \ref{fig:Manifest_result}. Now if we glue to this the $\mathsf{S}$-wall which we added in Figure \ref{fig:HW_step_0} (and also revert the substitution $d=(pq/t)^{\frac{M-N}{2}} c^{-1}$) we obtain
exactly the r.h.s.~of the HW move in Figure \ref{HWapp}, as desired.

\clearpage
\section{Limits of the $FE[USp(2N)]$ theory for $c = 1$ and $c = t^{\frac{1}{2}}$}
\begin{figure}[!ht]
\includegraphics[scale=0.5,center]{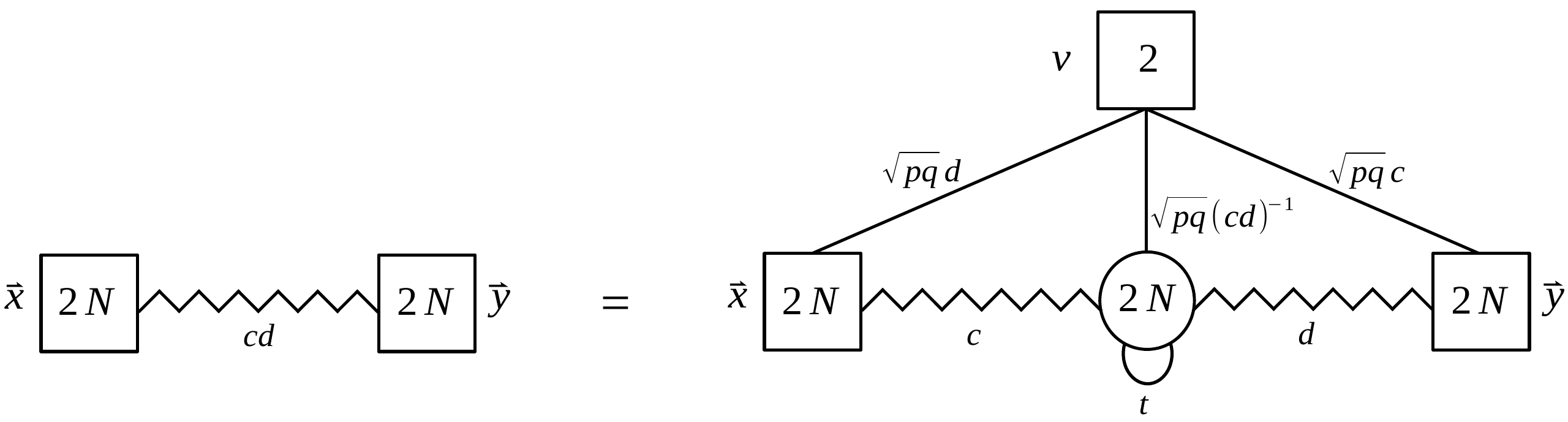}
\caption{A flipped version of the braid duality move.}
\label{fig:Braid_flipped}
\end{figure}

In this section we will study two interesting deformations of the $FE[USp(2N)]$ theory. 
First, we consider the effect of introducing a  superpotential term linear in the $\beta_N$ singlet flipping 
 the $ \Tr_x \Tr_{y_N} D^{(N)} D^{(N)}$ meson. This deformation has the effect of 
giving a VEV to the meson $ \Tr_x \Tr_{y_N} D^{(N)} D^{(N)}$. Since  $\beta_N$  appears linearly in the superpotential, it must have $R$-charge 2 and charge 0 under every other $U(1)$ symmetry, which implies that $U(1)_c$ is broken and the $c$ fugacity is specialized to 1. In order to analyze the effect of this deformation we consider the variant of the braid duality  in Figure \ref{fig:Braid_flipped} (related by flips to the braid duality in Figure \ref{braid}).

\begin{figure}[!ht]
\includegraphics[scale=0.5,center]{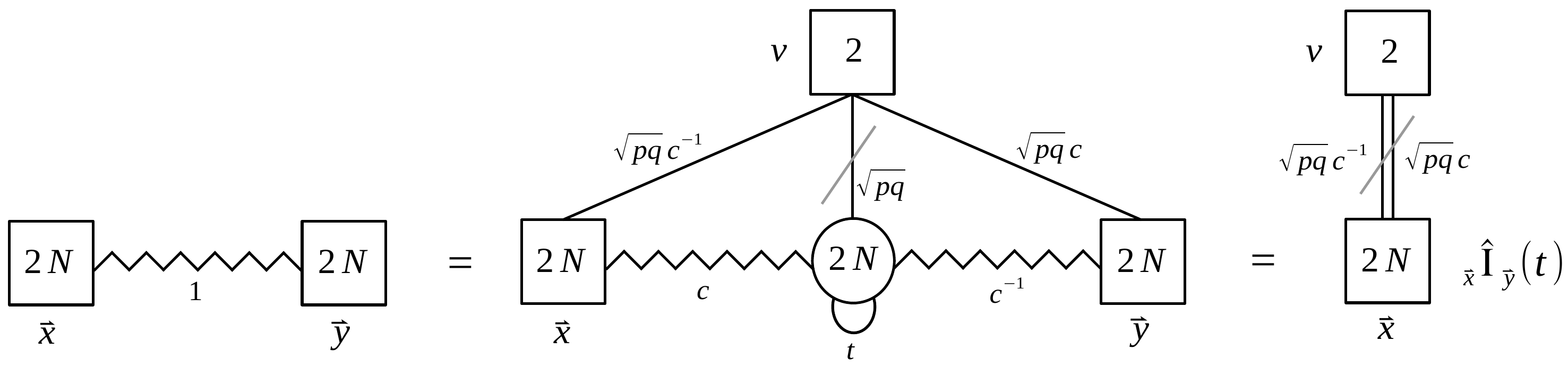}
\caption{The result of the $c = 1$ degeneration limit of the $FE[USp(2N)]$ theory.}
\label{fig:Limit_FE_c=1}
\end{figure}

In this version of the braid, the limit we are interested in corresponds to $d=c^{-1}$.
Indeed, as shown in Figure \ref{fig:Limit_FE_c=1}, on the l.h.s.~we have exactly the $\mathsf{S}$-walls deformed by the linear term $\beta_N$. On the r.h.s., instead, the vertical $SU(2)_v$ flavor coupled to the middle gauge node becomes massive and we are left with two $\mathsf{S}$-walls giving rise to an Identity-wall. After implementing the identification of the $USp(2N)_x$ and the $USp(2N)_y$ symmetries due to the Identity-wall, we have that also  the two other flavors become massive. In the end  we are just left with an Identity-wall. At the level of the index this identity can be written as:
\begin{align}\label{eq:FE_c=1}
	\mathcal{I}_{FE}^{(N,N)} (\vec{x}, \vec{y}, t, c=1) = \frac{\prod_{j=1}^{N} 2\pi i y_i}{\Delta_N(\vec{y},t)} \prod_{\sigma \in S_N} \prod_{j=1}^{N} \delta (x_j - y_{\sigma(j)}^\pm)\,.
\end{align}

\begin{figure}[!ht]
\includegraphics[scale=0.55,center]{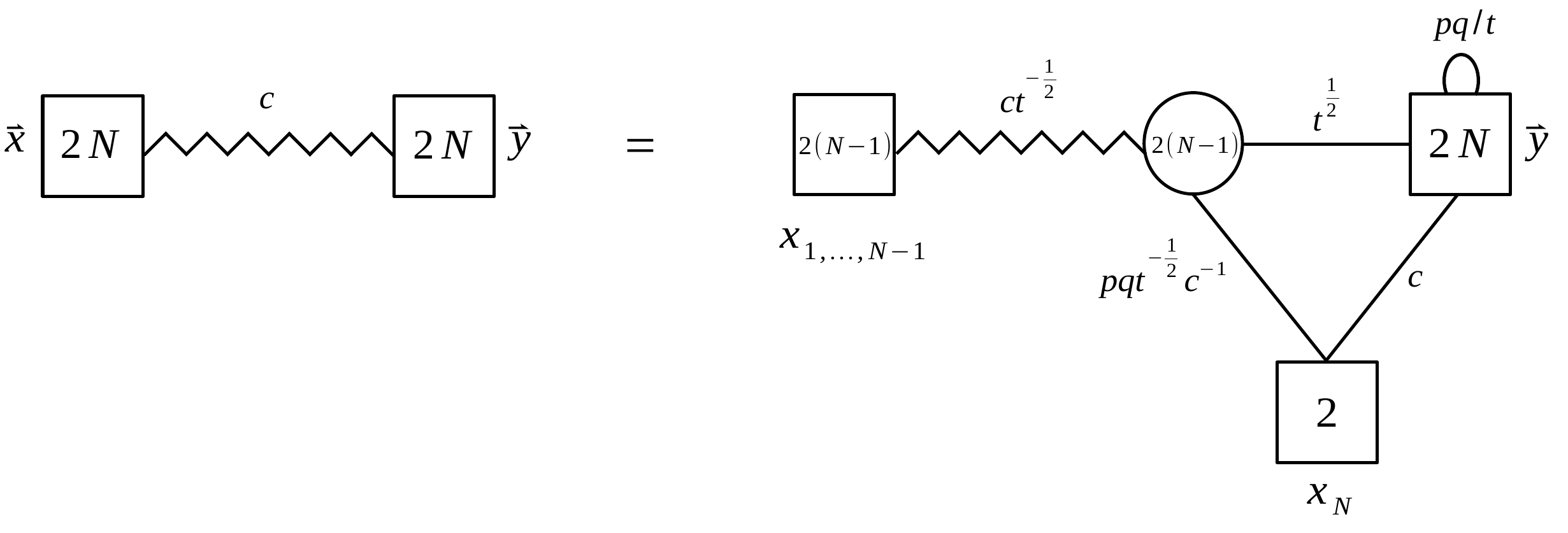}
\caption{A quiver representation of the recursive definition of the $FE[USp(2N)]$ theory.}
\label{fig:Limit_FE_open}
\end{figure}

\begin{figure}[!ht]
\includegraphics[scale=0.55,center]{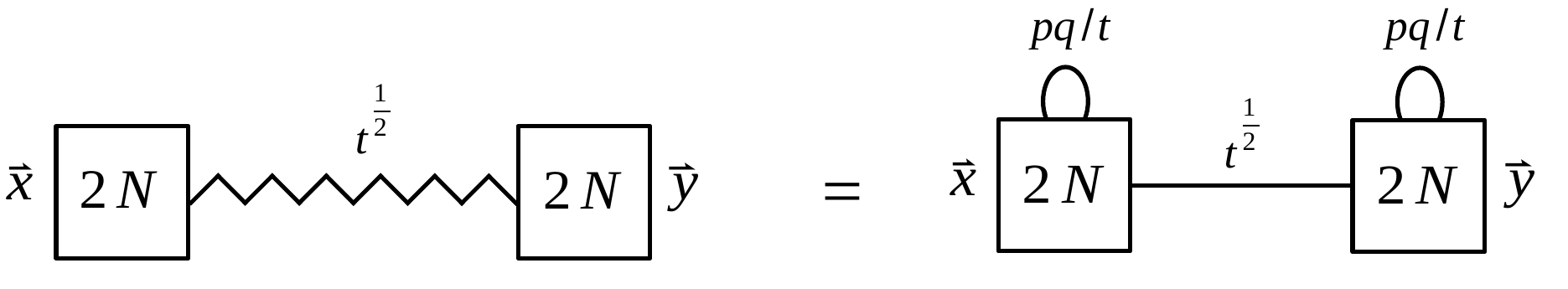}
\caption{The result of the $c = t^{\frac{1}{2}}$ degeneration limit of the $FE[USp(2N)]$ theory.}
\label{fig:Limit_FE_c=t12}
\end{figure}

Another interesting deformation of the $FE[USp(2N)]$ theory  corresponds to  turning on a superpotential term linear in  the singlet $\beta_{N-1}$ which now breaks a combination of $U(1)_t\times U(1)_c$, which at the level of the fugacities fixes $c = t^{\frac{1}{2}}$. This limit has already been studied in \cite{2014arXiv1408.0305R,Pasquetti:2019hxf}, but here we present a new derivation of it.
The effect of this deformation can indeed be easily understood considering the recursive definition of the $FE[USp(2N)]$ quiver, which can be represented as an $FE[USp(2N-2)]$ quiver coupled to an extra block with the $USp(2N-2)$ symmetry gauged as in Figure \ref{fig:Limit_FE_open}.
\\
When we take the  limit $c =  t^{\frac{1}{2}}$  in Figure \ref{fig:Limit_FE_open}, the $FE[USp(2N-2)]$ theory is subject to the deformation of Figure \ref{fig:Limit_FE_c=1} and reduces to an Identity-wall, so that we obtain the 
result in Figure \ref{fig:Limit_FE_c=t12}. At the level of the index this identity can be written as:
\begin{align}\label{eq:FE_c=t12}
	\mathcal{I}_{FE}^{(N,N)} (\vec{x}, \vec{y}, t, c=1t^{\frac{1}{2}}) =A_N(\vec{x};pq/t)A_N(\vec{y};pq/t) \prod_{i,j=1}^N\Gpq{t^{\frac{1}{2}}x_i^{\pm1}y_j^{\pm1}}\,.
\end{align}

\clearpage
\section{Derivation of a relation between WZ models}
\label{proof_N_to_1_chir}

We now demonstrate the relation between WZ models in Figure \ref{fig:N_to_1_chir} that we used in the main text to study the Higgsing that can occur in the dualization algorithm after applying the duality moves.

\begin{figure}[!ht]
\includegraphics[scale=0.52,center]{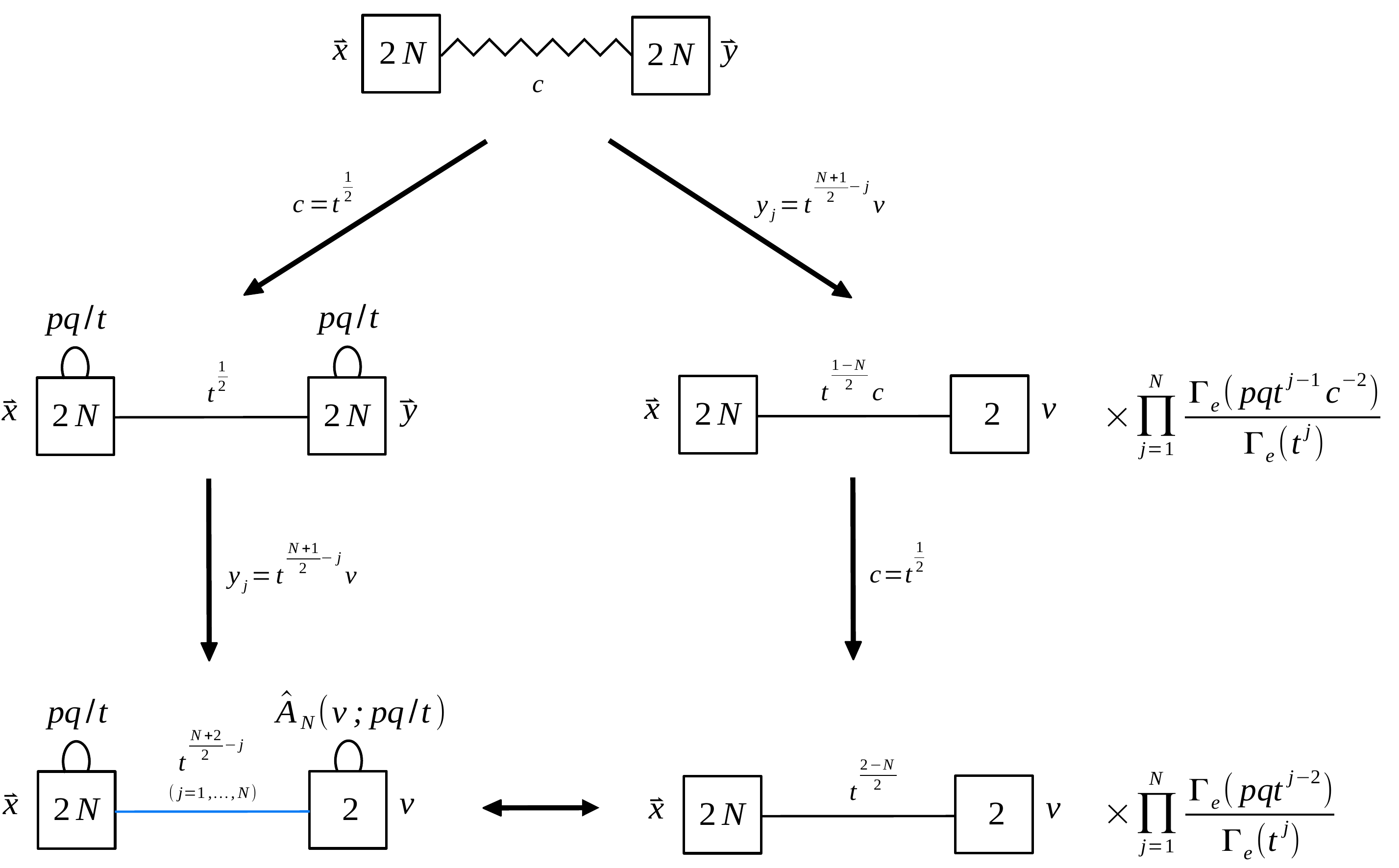}
\caption{A relation between deformed WZ models obtained by starting from the $\mathsf{S}$-wall and implementing  two deformations in  different orders.}
\label{fig:Nto1_step2}
\end{figure}

As shown in Figure \ref{fig:Nto1_step2}, starting from  
the $FE[USp(2N)]$ theory we can first turn on the $c=t^{\frac{1}{2}}$   deformation of Figure \ref{fig:Limit_FE_c=t12} to obtain the left theory in the second line. We then turn on the deformation breaking  $USp(2N)_y\to SU(2)_v$ by specializing
  \begin{align}
	y_j = t^{\frac{N+1}{2}-j} v \;\;\;\; \text{for} \;\;\;\; j=1, \dots, N;
\end{align}
to obtain the the left theory in the third line. This is a theory of $N$ chirals in the $USp(2N)_x\times SU(2)_v$ bifundamental representation, a $USp(2N)_x$ antisymmetric of chirals and extra chirals whose index contribution is encoded in the $\hat{A}(v,pq/t)$ we defined in \eqref{eq:spec_antisymm}. 

Starting again  from  the $FE[USp(2N)]$ theory, we can also implement the same deformations but in the reversed order. Namely, we first break $USp(2N)_y$ to $SU(2)_v$
moving to the right theory in the second line. 
Considering either of the lagrangian forms of the $M=0$ asymmetric $\mathsf{S}$-wall of Figures \ref{fig:Manifest_result} or \ref{fig:Emergent_result_FEclosed}, it is easy to see that we are left with a $USp(2N)_x \times SU(2)_v$ bifundamental chiral plus extra chirals. Finally we take the specialization $c=\sqrt{t}$ to reach the right theory in the third line.

\begin{figure}[!ht]
\includegraphics[scale=0.59,center]{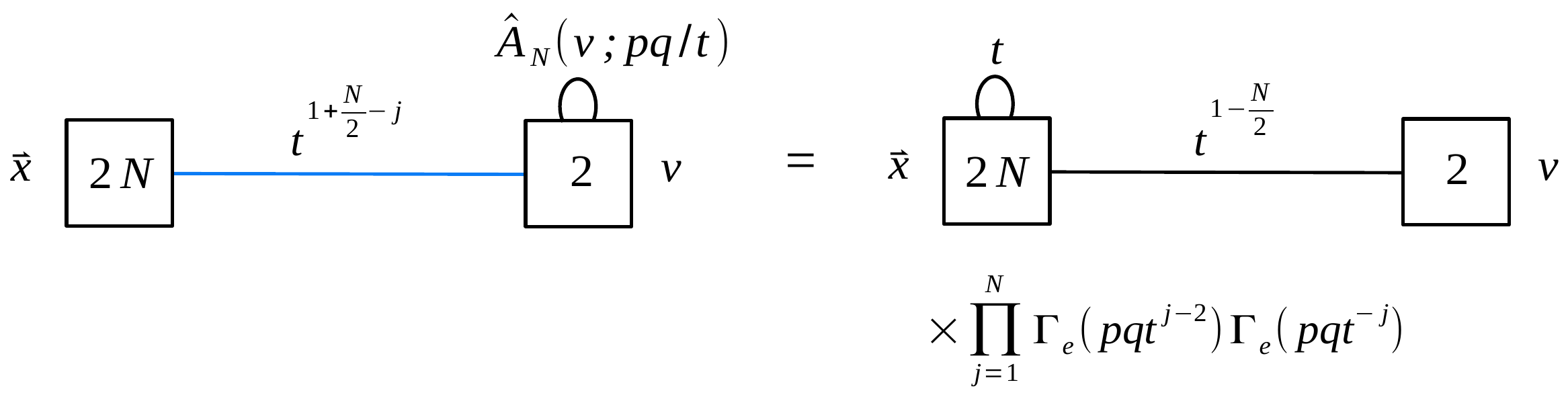}
\caption{A relation between a theory with a set of  $USp(2N)_x\times SU(2)$ bifundamentals with $j=1,\dots,N$
and a theory with a single $USp(2N)_x\times SU(2)$ bifundamental field and a $USp(2N)_x$ antisymmetric field (plus extra singlets). }
\label{fig:Nto1_result}
\end{figure}

We then have the relation shown in Figure \ref{fig:Nto1_result}, where we flipped the antisymmetric of $USp(2N)_x$. The relation between the theory on the l.h.s.~and the one on the r.h.s.~can be neatly understood by looking at their indices
which behave as singular distributions having equivalent actions on test functions. On the l.h.s.~the charges of the   bifundamentals are
\begin{equation}
	t^{\frac{N}{2}},\,t^{\frac{N}{2}-1},\,t^{\frac{N}{2}-2},\,\dots,\,t^{2-\frac{N}{2}},\,t^{1-\frac{N}{2}} \,,
\end{equation}
and so, as we saw in Section \ref{sec:Higgsing via IP}, if we try to gauge the   $USp(2N)_x$ group the mesons constructed with these bifundamentals take  a VEV Higgsing the  $USp(2N)_x$ group down to $USp(2)_w$. Indeed 
these legs produce $N-1$ colliding poles. The residue can be taken for example at the following poles:\footnote{There are also other poles that we should consider, but they are related to these by the Weyl group of $USp(2N)_x$ so they give identical contributions that cancels with part of the $1/(2^N N!)$ Weyl symmetry factor of the partially Higgsed $USp(2N)_x$ gauge group.}
\begin{equation}
	x_j = t^{\frac{N}{2}-j} v \quad \text{for} \quad j=1, \dots, N-1\,,
	\label{eq:spec_Nto1_residue}
\end{equation}
where $x_N=w$ parametrizes the remaining  $USp(2)_w$, while $v$ parametrizes the other $USp(2)_v$.

On the r.h.s., instead, the meson built out of the bifundamental dressed $N-2$ times with the antisymmetric has zero $R$-charge and thus takes a VEV. Hence, also on this side the $USp(2N)_x$ is Higgsed down to $USp(2)_v$, but this time it is not due to $N-1$ mesons taking a VEV but rather a single meson dressed with $N-2$ powers of the antisymmetric. Indeed looking at the contribution to the index of bifundamental and the antisymmetric
\be
\Gamma_e(t)^N \prod_{i<j}^N \Gamma_e(t  x_i^\pm x_j^\pm) \prod_{i=1}^N\Gamma_e(t^{1-\frac{N}{2}} x_i^\pm v^\pm  )
\ee
we see that there are colliding poles from
\be
\prod_{i<j}^{N-1} \Gamma_e(t  x_i^\pm x_j^\pm) \Gamma_e(t^{1-\frac{N}{2}} x_1 v^{-1}  )  \Gamma_e(t^{1-\frac{N}{2}} x_{N-1}^{-1} v  )
\ee
and the residue can be taken for example at the following poles:
\be
x_1&=&t^{\frac{N}{2}-1}v\,,\nn\\
x_2&=&t^{-1}x_1=t^{\frac{N}{2}-2}v\,,\nn\\
&\vdots&\nn\\
x_{N-2}&=&t^{-1}x_{N-3}=t^{2-\frac{N}{2}}v\,,\nn\\
x_{N-1}&=&t^{-1}x_{N-2}=t^{1-\frac{N}{2}}v\,,
\ee
which is  exactly  the same specialization as \eqref{eq:spec_Nto1_residue}.
The effect of taking this residue is  the Higgsing of  $USp(2N)_x$ node  down to $USp(2)_w$. 

In conclusion, if we try to gauge their $USp(2N)$ symmetry with a test theory, we see that these two theories lead to IR dual models.


\bibliographystyle{ytphys}
\bibliography{refs}

\end{document}